\documentclass[a4paper]{article}
\usepackage{graphicx}
\usepackage{amssymb}
\usepackage{amsmath}
\usepackage{amsthm}
\usepackage{bm}
\usepackage{textcomp}
\usepackage{stackengine}
\usepackage{float}
\usepackage[top=30mm,bottom=30mm,left=30mm,right=30mm]{geometry} 
\usepackage{fixme}

\newcommand{\slfrac}[2]{\left.#1\middle/#2\right.}
\newcommand{\mcnt}[1]{\mathsf{#1}}

\newcommand{\alignedM}[1]{
\ifthenelse{\equal{#1}{}} { {\mathrel{\stackon[0pt]{$M$}{$\leftrightsquigarrow$}}}}%
{\mathrel{\stackon[0pt]{$#1M$}{$#1\leftrightsquigarrow$}}}
}
\newtheoremstyle{remarkstyle}{1pt}{1pt}{}{}{\itshape}{.}{.5em}{}
 
\renewenvironment{abstract}
{	\begin{center}
		\bfseries \large\abstractname\vspace{-.5em}\vspace{0pt}
	\end{center}
	\list{}{%
		\setlength{\leftmargin}{0mm}%
		\setlength{\rightmargin}{\leftmargin}%
	}%
	\item\relax}

\theoremstyle{remarkstyle}
\newtheorem{remark}{Remark}
\newtheorem{proposition}{Proposition}
\newcommand{\pdfinlimg}[5]{
\makebox[#1cm][l]{\immediate\pdfliteral{
  q
  #3 0 0 #4 0 0 cm
  #1 0 0 #2 0 0 cm
  0.885 0 0 0.885 0 0 cm
  BI
  /W #3
  /H #4
  /CS /RGB
  /BPC 8
  /F [ /AHx /Fl ]
  ID
  #5>
  EI
  Q
}\vbox to #2cm{}}
}

\title{\bf\Large Harmonic-aligned Frame Mask Based on Non-stationary Gabor Transform with Application to Content-dependent Speaker Comparison%
\footnote{\footnotesize This work was supported in part by the  Austrian Science Fund(FWF) START-project (Y 551-N13). }}
\date{}

\author{ Feng Huang, Peter Balazs \\
Acoustic Research Institute, Austrian Academy of Sciences \\
{ \tt  fhuang@kfs.oeaw.ac.at, peter.balazs@oeaw.ac.at}}

\begin{document}

\maketitle

\begin{abstract}
We propose harmonic-aligned frame mask for speech signals using non-stationary Gabor transform (NSGT).
A frame mask operates on the transfer coefficients of a signal and consequently converts the signal into a counterpart signal.
It depicts the difference between the two signals.
In preceding studies, frame masks based on regular Gabor transform were applied to single-note instrumental sound analysis.
This study extends the frame mask approach to speech signals.
For voiced speech, the fundamental frequency is usually changing consecutively over time.
We employ NSGT with pitch-dependent and therefore time-varying frequency resolution to attain harmonic alignment in the transform domain and hence yield harmonic-aligned frame masks for speech signals.
We propose to apply the harmonic-aligned frame mask to content-dependent speaker comparison.
Frame masks, computed from voiced signals of a same vowel but from different speakers, were utilized as similarity measures to compare and distinguish the speaker identities (SID).
Results obtained with deep neural networks demonstrate that the proposed frame mask is valid in representing speaker characteristics and shows a potential for SID applications in limited data scenarios.

\end{abstract}

\section{Introduction}
Time-frequency (TF) analysis is the foundation of audio and speech signal processing.
The short-time Fourier transform (STFT) is a widely used tool, which can be effectively implemented by FFT \cite{soendxxl10}. 
STFT features straightforward interpretation of a signal. 
It provides uniform time and frequency resolution with linearly-spaced TF bins.
The corresponding theory was generalized in the framework of Gabor analysis and Gabor frames \cite{Gabor1947, ma09, Grochenig2001}.

Signal synthesis is an important application area of time-frequency transforms. 
Signal modification, denoising, separation and so on can be achieved by manipulating the analysis coefficients to synthesize a desired one.
The theory of Gabor multiplier \cite{feinow1} or, in general terms, frame multiplier \cite{balsto09new,stobalrep11} provides a basis for the stability and invertibility of such operations.
A \textit{frame multiplier} is an operator that converts a signal into another by pointwise multiplication in the transform domain for resynthesis. 
The sequence of multiplication coefficients is called a \textit{frame mask} (or symbol).
Such operators allow easy implementation of time-varying filters \cite{839987}.
They have been used in perceptual sparsity \cite{xxllabmask1}, denoising \cite{majxxl10} and signal synthesis \cite{Philippe2007}.
Algorithms to estimate frame mask between audio signals were investigated in \cite{Philippe2007, Olivero2013}, where it was demonstrated that the frame mask between two instrumental sounds (of a same note) was an effective measure to characterize timber variations between the instruments.
Such masks were used for timber morphing and instrument categorization.
In this paradigm, the two signals were of the same fundamental frequency and their harmonics were naturally aligned, which vouched for the prominence of the obtained mask for TF analysis/synthesis with uniform resolution.

This study extends the frame mask method to speech signals. 
One intrinsic property of (voiced) speech signal is that the fundamental frequency ($f_0$ or pitch) varies consecutively over time. 
Therefore, the harmonic structures are not well aligned when comparing two signals. 
We propose to employ the non-stationary Gabor transform (NSGT) \cite{BALAZS2011} to tackle this issue. 
NSGT provides flexible time-frequency resolution by incorporating dynamic time/frequency hop-size and dynamic analysis windows \cite{BALAZS2011, Holighaus2013, Ottosen2017}.
We develop an NSGT whose frequency resolution changes over time.
We set the frequency hop-size in ratio to $f_0$ to achieve harmonic alignment (or partial alignment cf. Section \ref{sec:pansgt}) in the transform domain. 
On this basis, we propose the harmonic-aligned frame mask.
To demonstrate feasibility in speech, we shall evaluate the proposal in the context of vowel-dependent speaker comparison.
Frame marks between voiced signals of the same vowel but pronounced by different speakers are proposed as similarity measures for speaker characteristics to distinguish speaker identities in a limited data scenario (cf. Section 5 for details).

This paper is organized as follows.
In Section 2, we briefly review frame and Gabor theory.
In Section 3, we elaborate frame mask and the previous application in instrumental sound analysis.
In Section 4, we develop the non-stationary Gabor transform with pitch-dependent frequency resolution and propose the harmonic-aligned frame mask.
Section 5 presents the evaluation in vowel-dependent speaker identification.
And finally, Section 6 concludes this study.

\section{Preliminaries and Notation} 

\subsection{Frame Theory}
\label{sec:frame_theory}

Denote by $\{g_{\lambda}:\lambda \in \Lambda\}$ a sequence of signal atoms in the Hilbert space $\mathcal{H}$, where $\Lambda$ is a set of index. 
This atom sequence is a frame \cite{ma09} if and only if there exist constants $\mcnt{A}$ and $\mcnt{B}$, $0 < \mcnt{A} \leq \mcnt{B} < \infty$, such that
\begin{equation}
\mcnt{A}\lVert f\rVert_2^2 \leq \sum_{\lambda} \lvert c_\lambda 
\rvert^2 \leq \mcnt{B} \lVert f\rVert_2^2, ~ \forall f \in \mathcal{H}.
\label{EQ:FRAME_BOUND}
\end{equation}
where $c_\lambda =  \langle f, g_\lambda \rangle$ are the \emph{analysis coefficients}. $\mcnt{A}$ and $\mcnt{B}$ is called the lower and upper frame bounds, respectively.
The \textit{frame operator} $\mathbf{S}$ is defined by $\mathbf{S}f=\sum_{\lambda} \langle f,g_{\lambda} \rangle g_{\lambda}$.

Given $\{h_{\lambda} = \mathbf{S}^{-1} g_\lambda :\lambda \in \Lambda\}$ the canonical dual frame of $\{g_{\lambda}:\lambda \in \Lambda\}$, $f$ can be perfectly reconstructed from the analysis coefficients by
\begin{equation}
f=\sum_{\lambda}  \langle f, g_\lambda \rangle h_\lambda.
\end{equation}
The dual frame always exists \cite{Casaz1}, and for redundant cases there are infinitely many other duals allowing reconstruction.

\subsection{Discrete Gabor Transform}
\label{sec:dgt}
We take the Hilbert space $\mathcal{H}$ to be $\mathbb{C}^L$. 
Given non-zero prototype window $g=(g[0],g[1],\cdots,g[L-1])^T\in \mathbb{C}^L$, the translation operator $\mathbf{T}_x$ and modulation operator  $\mathbf{M}_y$ are, respectively, defined as
\[ \mathbf{T}_x g[l] = g[l-x] \text{~~and~~} \mathbf{M}_y g[l] = g[l] e^{\frac{2\pi i y l}{L}}, \]
where $x,y\in \mathbb{Z}_L$ and the translation is performed modulo $L$. 
For selected constants $\mcnt{a}, \mcnt{b} \in \mathbb{Z}_L$, with some $N,M\in \mathbb{N}$ such that $N \mcnt{a} = M \mcnt{b} = L$, we take $\Lambda$ to be a regular discrete lattice, i.e., $\lambda = (m,n)$, and obtain the \textit{Gabor system} \cite{Gabor1947} $\{g_{m,n}\}_{m\in\mathbb{Z}_M,n\in\mathbb{Z}_N}$ as
\begin{equation}
g_{m,n}[l] = \mathbf{T}_{n\mcnt{a}} \mathbf{M}_{m\mcnt{b}} g[l] = g[l-n\mcnt{a}]e^{\frac{2\pi i m\mcnt{b} (l-n\mcnt{a})}{L}}.
\end{equation}
If $\{g_{m,n}\}_{m,n}$ satisfies (\ref{EQ:FRAME_BOUND}) for  $\forall f\in \mathbb{C}^L$, it is called a Gabor frame \cite{Christensen2003}. 
The discrete Gabor transform (DGT) of $f \in \mathbb{C}^L$ is a matrix $C=\{c_{m,n}\}\in\mathbb{C}^{M\times N}$ with $c_{m,n}=\langle f,  g_{m,n}\rangle$.
The associated frame operator $\mathbf{S}:\mathbb{C}^L \to \mathbb{C}^L$ reads
\begin{equation}
\mathbf{S}f=\sum_{m=0}^{M-1}\sum_{n=0}^{N-1} \langle f,  g_{m,n}\rangle g_{m,n}.
\label{EQ:GaborOperatior}
\end{equation}
The canonical dual frame $\{\widetilde{g}_{m,n}\}_{m,n}$ of the Gabor frame $\{g_{m,n}\}_{m,n}$ is given by $\widetilde{g}_{m,n}=\mathbf{T}_{n\mcnt{a}} \mathbf{M}_{m\mcnt{b}}\mathbf{S}^{-1} g$ \cite{feistro1}, with which $f$ can be perfectly reconstructed by
\[f=\sum_{m=0}^{M-1}\sum_{n=0}^{N-1}c_{m,n} \widetilde{g}_{m,n} .  \]
Note that the DGT coefficients are essentially sampling points of the STFT of $f$ with window $g$ at the time-frequency points $(n\mcnt{a}, m\mcnt{b})$, with $\mcnt{a}$ and $\mcnt{b}$ being the sampling step (i.e., hop-size) in time and frequency \cite{feistro1}.
In non-stationary settings, the hop-sizes are allowed to be variant (cf. Section \ref{sec:pansgt}).

\section{Frame mask for instrumental sound analysis}
\subsection{Frame Mask}
Consider a pair of frames $\{g_{\lambda}:\lambda \in \Lambda\}$ and $\{h_{\lambda}:\lambda \in \Lambda\}$.
A \emph{frame multiplier} \cite{Balazs2007}, denoted by $\mathbb{M}_{\boldsymbol{\sigma};g,h}$, is an operator that acts on a signal by pointwise multiplication in the transform domain.
The symbol  $\boldsymbol{\sigma}=\{\sigma_{\lambda},\lambda \in \Lambda\}$ is a sequence that denotes the multiplication coefficients.
For signal $f$
\begin{equation}
\mathbb{M}_{\boldsymbol{\sigma};g,h} f = \sum_{\lambda} \sigma_\lambda \langle f, g_\lambda \rangle h_\lambda.
\label{EQ:FRAME_MULT}
\end{equation}
Here $\boldsymbol{\sigma}$ is called a \emph{frame mask}.
In the considered signal analysis/transform domain, $\boldsymbol{\sigma}$ can be viewed as a \emph{transfer function}.

When Gabor frames $\{g_{m,n}\}_{m,n}$ and $\{h_{m,n}\}_{m,n}$ are considered, we set $\lambda = (m,n)$.
In this case the frame multiplier in (\ref{EQ:FRAME_MULT}) is known as Gabor multiplier.
The corresponding frame mask $\boldsymbol{\sigma}=\left\{\sigma_{m,n}\right\} \in \mathbb{C}^{M\times N}$ is also known as \textit{Gabor mask}.

\subsection{For Instrument Timbre Analysis and Conversion}
The application of frame masks in musical signals was investigated in \cite{Philippe2007, Olivero2013}. 
Based on DGT, the proposed signal model converts one sound into another by
\begin{equation}
f^{\mathrm{B}} = \mathbb{M}_{\boldsymbol{\sigma}^{\overrightarrow{\mathrm{A}\mathrm{B}}};g,\widetilde{g}} f^{\mathrm{A}} =  \sum_{m=0}^{M-1}\sum_{n=0}^{N-1} \sigma^{\overrightarrow{\mathrm{A}\mathrm{B}}}_{m,n} \langle f^{\mathrm{A}} , g_{m,n} \rangle \widetilde{g}_{m,n},
\label{EQ:FMASK_SNOTE}
\end{equation}
where $f^{\mathrm{A}}, f^{\mathrm{B}} \in \mathbb{R}^L$ are two audio signals and $\boldsymbol{\sigma}^{\overrightarrow{\mathrm{A}\mathrm{B}}}$ is the unknown mask to be estimated. 
An obvious solution is to set $\sigma^{\overrightarrow{\mathrm{A}\mathrm{B}}}_{m,n} = \slfrac{c_{m,n}^{\mathrm{B}}}{c_{m,n}^{\mathrm{A}}}$, where $c_{m,n}^{\mathrm{A}}$ and $c_{m,n}^{\mathrm{B}}$ are the DGT coefficients of $f^{\mathrm{A}}$ and $f^{\mathrm{B}}$, respectively.
However, this solution is non-stable and unbounded as the DGT coefficients in the denominator can be $0$ or very small.
To guarantee existence of a stable solution, it was proposed to estimate the mask via 
\begin{equation}
\min_{\boldsymbol{\sigma}^{\overrightarrow{\mathrm{A}\mathrm{B}}}}{ \lVert f^{\mathrm{B}} - \mathbb{M}_{\boldsymbol{\sigma}^{\overrightarrow{\mathrm{A}\mathrm{B}}};g,\widetilde{g}} f^{\mathrm{A}} \rVert^2 + \mu d(\boldsymbol{\sigma}^{\overrightarrow{\mathrm{A}\mathrm{B}}})},
\label{EQ:MASK_EST_REGU}
\end{equation}
with a (convex) regularization term $d(\boldsymbol{\sigma}^{\overrightarrow{\mathrm{A}\mathrm{B}}})$, whose influence is controlled by the parameter $\mu$ \cite{Olivero2013}.
As the existence of a stable solution is assured, such approach in general can be applied to arbitrary pair of signals.
However, it might be difficult to interpret the estimated masks (e.g., the mask between two pure-tone signals with different fundamental frequencies).

Given that $f^{\mathrm{A}}$ and  $f^{\mathrm{B}}$ are of the same note produced by different instruments, the frame mask between the two signals was found to be effective to characterize the timbre difference between the two instruments \cite{Philippe2007, Olivero2013}.
Such masks were utilized as similarity measures for instrument classification and for timber morphing and conversion.
Rationality of these applications roots from two aspects:
\begin{list}{*}{
\setlength{\labelwidth}{8pt} \setlength{\leftmargin}{12pt}
\setlength{\topsep}{2pt}
\setlength{\parsep}{1pt}
\setlength{\itemsep}{0pt}
}
\item[ 1)] Instrumental signals of a same note possess the same fundamental frequency. Harmonic structures of the signals are naturally aligned.
\item[ 2)] DGT performs TF analysis over a regular TF lattice, and consequently preserves the property of harmonic alignment in the transform domain.
\end{list}

\section{Frame mask for speech signals using Non-stationary Gabor transform}
\label{sec:pansgt}
Similar to audio sounds of instrument notes, (voiced) speech signals are also harmonic signals.
Analog to the above-mentioned applications, this study explores the application of frame mask in speech signals.
In particular, we consider to use voiced speech as source and target signals and to estimate the frame mask between them.
We are specially interested in the case that the source and the target are of the same content, e.g., the same vowel.
For such a case, a valid frame mask could measure specific variations among the signals, such as speaker variations. 

Nevertheless, attempting to use (\ref{EQ:MASK_EST_REGU}) for speech signals, we immediately face a fundamental problem.
For speech signals, the fundamental frequency usually varies over time consecutively.
Therefore, harmonic structures of the source and target voice are mostly not aligned.
To address this problem, we propose to employ non-stationary Gabor transform, which allows flexible time-frequency resolution \cite{BALAZS2011}.
Within the framework of non-stationary Gabor analysis, we intend to achieve dynamic alignment of the signals' harmonic structures.
In the following, we shall develop NSGT with pitch-dependent frequency resolution to achieve harmonic alignment in the transform domain,
and shall propose the harmonic-aligned frame mask for speech signals on that basis.

\subsection{Non-stationary Gabor Transform with Pitch-dependent Frequency Resolution}

We consider analyzing a voiced signal $f \in \mathbb{R}^L$ with a window $g$ that is symmetric around zero.
As the stationary case in Section \ref{sec:dgt}, we use a constant time hop-size  $\mcnt{a}$, resulting in $N = \frac{L}{\mcnt{a}}\in\mathbb{N}$ sampling points in time for the TF analysis. 
However, we set the frequency hop-size according to the fundamental frequency of the signal (see Remark \ref{rem:2}.1 for discussion on pitch estimation issue).
Following the quasi-stationary assumption for speech signals, we assume that the fundamental frequency is approximately fixed within the interval of the analysis window.
At time $n$, let $f_0(n\mcnt{a})$ denote the fundamental frequency in Hz, we set the corresponding frequency hop-size as
\begin{equation}
b_n^{f_0} = \left\lfloor\slfrac{\frac{\mcnt{p} f_0(n\mcnt{a})}{\mcnt{q}}}{\frac{f_s}{L}}\right\rceil,
\label{EQ:BF0}
\end{equation}
where $\mcnt{p},\mcnt{q}\in \mathbb{N}$ are a pair of parameters to be set.
$\lfloor \rceil$ denotes rounding to the closest positive integer, and $f_s$ is the signal's sampling rate in Hz.
With (\ref{EQ:BF0}), $\mcnt{q}$ frequency sampling points are deployed per $\mcnt{p} f_0(n\mcnt{a})$ Hz.
The total number of frequency sampling points at $n$ is hence $M_n^{f_0} = \slfrac{L}{b_n^{f_0}} \in \mathbb{N}$.
Consequently, we obtain the pitch-depenent \textit{non-stationary Gabor system} (NSGS) $\{g_{m,n}\}_{m\in\mathbb{Z}_{M_n^{f_0}},n\in\mathbb{Z}_N}$ as
\begin{equation}
g_{m,n}[l] = \mathbf{T}_{n\mcnt{a}} \mathbf{M}_{m b_n^{f_0}} g[l] = g[l-n\mcnt{a}]e^{\frac{2\pi i m b_n^{f_0} (l-n\mcnt{a})}{L}} .
\label{EQ:PD_NSGT}
\end{equation}
It is called a \textit{non-stationary Gabor frame} (NSGF) if it fulfills (\ref{EQ:FRAME_BOUND}) for $\mathbb{C}^L$. 
The sequence $\{c_{m,n}\}_{m,n} =  \{\langle f,  g_{m,n}\rangle \}_{m,n}$ are the \textit{ non-stationary Gabor transform} coefficients.
In general, due to the dynamic frequency hop-size, these coefficients do not form a matrix.

Eq. (\ref{EQ:BF0}) features a time-varying and pitch-dependent frequency resolution.
More importantly, it allows harmonic alignment in the NSGT coefficients with respect to the frequency index $m$.
For example, with $\mcnt{p}=1$, for any $n$,  $c_{\mcnt{q},n}, c_{\mcnt{2q},n},  c_{\mcnt{3q},n}, \cdots$ naturally correspond to the harmonic frequencies of the signal. The parameter $\mcnt{p}$ allows performing partial alignment wrt. integer multiples of the $\mcnt{p}$-th harmonic frequency.
\begin{remark}
To satisfy $M_n^{f_0} b_n^{f_0} = L, \forall n\in \mathbb{Z}_N$, zero-padding for $f$ may be needed for an appropriate $L$. 
If an extremely large $L$ is required, it is always practicable to divide the signal into segments of shorter duration using overlap-and-add windows, and obtain NSGT coefficients for each segment separately. 
A practical example for such procedure can be found in \cite{Holighaus2013}.
\end{remark}

Now we consider the canonical dual $\{\widetilde{g}_{m,n}\}_{m,n}$.
Denote by $\mathrm{supp}(g)\subseteq [\mcnt{c}, \mcnt{d}]$ the support of the window $g$, i.e., the interval where the window is nonzero.
We choose $ M_n^{f_0} \geq \mcnt{d} - \mcnt{c}, \forall n\in \mathbb{Z}_N$, which is referred to as the \textit{painless} case \cite{BALAZS2011}.
In other words, we require the frequency sampling points to be dense enough.
In this painless case, we have the following \cite{BALAZS2011}.
\begin{proposition}
If $\{g_{m,n}\}_{m,n}$ is a painless-case NSGF, then the frame operation $\mathbf{S}$ (cf. (\ref{EQ:GaborOperatior})) is an $L\times L$ diagonal matrix with diagonal element
\begin{equation}
s_{l,l} = \sum_{n=0}^{N-1} M_n^{f_0} \bigl| g[l-n\mcnt{a}] \bigr|^2 > 0, \forall l\in\mathbb{Z}_L.
\end{equation}
And the canonical dual frame $\{\widetilde{g}_{m,n}\}_{m,n}$ is given by
\begin{equation}
\widetilde{g}_{m,n}[l] = \frac{g_{m,n}[l]}{s_{l,l}}.
\end{equation}
\end{proposition}

\subsection{Harmonic-aligned Frame Mask} 
In this section, we present a general form of frame mask based on the above pitch-dependent NGST.
For two voiced signals $f^{\mathrm{A}}, f^{\mathrm{B}} \in \mathbb{R}^L$, denote their fundamental frequency by $f_{\mathrm{A};0}$ and $f_{\mathrm{B};0}$, respectively.
Using  (\ref{EQ:PD_NSGT}) with the same window $g$ and the same time hop-size $\mcnt{a}$ for both signals, we construct two Gabor systems $\{g^{\mathrm{A}}_{m,n}\}_{m\in\mathbb{Z}_{M_n^{f_{\mathrm{A};0}}},n\in\mathbb{Z}_N}$ and $\{g^{\mathrm{B}}_{m,n}\}_{m\in\mathbb{Z}_{M_n^{f_{\mathrm{B};0}}},n\in\mathbb{Z}_N}$.
Denote $\alignedM{} = \max\left(\max_n (M_n^{f_{\mathrm{A};0}}), \max_n( M_n^{f_{\mathrm{B};0}})\right)$.
To simplify the presentation of the concept without losing the frame property (\ref{EQ:FRAME_BOUND}), we can consider extend the two systems as $\{g^{\mathrm{A}}_{m,n}\}_{m\in\mathbb{Z}_{\alignedM{\scriptstyle}},n\in\mathbb{Z}_N}$ and $\{g^{\mathrm{B}}_{m,n}\}_{m\in\mathbb{Z}_{\alignedM{\scriptstyle}},n\in\mathbb{Z}_N}$ e.g., with periodic extension to the modulation operator wrt. the index $m$.  
Under such circumstance, we can denote the NGST coefficients in matrix forms as $C^{\mathrm{A}}= \{c^{\mathrm{A}}_{m,n} \}_{m,n}\in \mathbb{C}^{\alignedM{\scriptstyle} \times N}$ and $C^{\mathrm{B}}= \{c^{\mathrm{B}}_{m,n}\}_{m,n} \in \mathbb{C}^{\alignedM{\scriptstyle} \times N}$.
The \textit{harmonic-aligned frame mask} (HAFM) $\boldsymbol{\sigma}^{\overrightarrow{\mathrm{A}\mathrm{B}}}\in \mathbb{C}^{\alignedM{\scriptstyle} \times N}$  between the two voiced signals therefore acts as
\begin{equation}
f^{\mathrm{B}} = \mathbb{M}_{\boldsymbol{\sigma}^{\overrightarrow{\mathrm{A}\mathrm{B}}};g^{\mathrm{A}},\widetilde{g}^{\mathrm{B}}} f^{\mathrm{A}} =  \sum_{m=0}^{\alignedM{\scriptstyle}-1}\sum_{n=0}^{N-1} \sigma^{\overrightarrow{\mathrm{A}\mathrm{B}}}_{m,n} \langle f^{\mathrm{A}} , g^{\mathrm{A}}_{m,n} \rangle \widetilde{g}^{\mathrm{B}}_{m,n} .
\label{EQ:HA_FMASK}
\end{equation}

To estimate the frame mask, existing methods \cite{Philippe2007,Olivero2013} for the problem in (\ref{EQ:FMASK_SNOTE}) can be directly applied.
For both Gabor systems $\{g^{\mathrm{A}}\}$ and $\{g^{\mathrm{B}}\}$, the parameters $\mcnt{p}$ and $\mcnt{q}$ in (\ref{EQ:BF0}) need to be appropriately set.
We set $\mcnt{q}$ for both systems to the same value.
However, depending on specifics of the source and target signal (as well as the application purpose), the parameter $\mcnt{p}$ may be set to different values for both systems.
\textit{Example} 1: If $f_{\mathrm{A};0}$ and $f_{\mathrm{B};0}$ are close (enough), we consider $\mcnt{p}=1$ for both Gabor systems. 
This leads to a one-to-one alignment of all harmonics. 
\textit{Example} 2: If $f_{\mathrm{A};0}$ and $f_{\mathrm{B};0}$ are significantly different in value, we may consider an anchor frequency $F$
and set $\mcnt{p^{A}}=\left\lfloor\slfrac{F}{f_{\mathrm{A;0}}}\right\rceil, \mcnt{p^{B}}=\left\lfloor\slfrac{F}{f_{\mathrm{B;0}}}\right\rceil$.
This results in partial alignment of the harmonics, i.e., only the harmonics around $F$ and its multiples are aligned.

\begin{remark} \label{rem:2}
1) The proposed approach practically depends on a reliable method to estimate the fundamental frequencies.
A thorough discussion of such topic is beyond the scope of this paper.
In the evaluation, we applied the methods in \cite{Huang2013,Huang2017}.
2) It may be a false impression that pitch independence is achieved in the frame masks by the harmonic alignment. 
On the contrary, the resulted frame mask is essentially dependent on the fundamental frequencies. It equivalently describes the variations between two spectra which are warped in a pitch-dependent and linear way. It contains information related to the spectral envelopes and also highly depends on the fundamental frequencies. 
It is our interests to utilize the proposed mask as feature measures for classification tasks. 
\end{remark}

\section{Evaluation in Content-dependent Speaker Comparison}
\label{SEC:APP}
We now evaluate harmonic-aligned frame masks for speaker identity comparison in a content-dependent context.
In particular, the source and target signals are of the some vowel but pronounced by different speakers.
In this setting, we estimate the frame masks between an input speaker and a fixed reference speaker.
For different speakers, we compare them to the same reference speaker, and use the estimated masks as speaker feature to measure and distinguish the speaker identities.
It can be considered as a task of close-set speaker identification with content-dependent and limited-data constraints (see the experimental settings in \ref{sec:evaluationSetup}).

To estimate the harmonic-aligned frame mask, we adopt the approach (\ref{EQ:MASK_EST_REGU}) and use transform domain proxy \cite{Philippe2007}.
For our case, the first item in (\ref{EQ:MASK_EST_REGU}) can be written as ${ \left\lVert \sum_{m}\sum_{n} ( c^{\mathrm{B}}_{m,n} - \sigma^{\overrightarrow{\mathrm{A}\mathrm{B}}}_{m,n}  c^{\mathrm{A}}_{m,n})\widetilde{g}^{\mathrm{B}}_{m,n} \right\rVert^2 }$.
With diagonal approximation on the covariance matrix of NSGF $\{\widetilde{g}^{\mathrm{B}}\}$, i.e., $(\widetilde{g}^{\mathrm{B}}_{m,n})^H \cdot \widetilde{g}^{\mathrm{B}}_{m',n'}  = 0$ if $(m,n)\neq (m',n')$, we estimate the mask via

 \begin{equation}
 \min_{\boldsymbol{\sigma}^{\overrightarrow{\mathrm{A}\mathrm{B}}}} { \left\lVert %
C^{\mathrm{B}} - \boldsymbol{\sigma}^{\overrightarrow{\mathrm{A}\mathrm{B}}} \odot C^{\mathrm{A}}
\right\rVert^2 } +  \mu d(\boldsymbol{\sigma}^{\overrightarrow{\mathrm{A}\mathrm{B}}}) ,
\label{EQ:HAMASK_EST}
\end{equation}
where $\odot$ denotes entrywise product. In this evaluation,  we use the following regularization term
\begin{equation}
d(\boldsymbol{\sigma}^{\overrightarrow{\mathrm{A}\mathrm{B}}}) = \left\lVert \boldsymbol{\sigma}^{\overrightarrow{\mathrm{A}\mathrm{B}}} - \boldsymbol{\sigma}^{\mathrm{Ref}} \right\rVert_2^2 .
\label{EQ:REGTERM}
\end{equation}
With (\ref{EQ:REGTERM}), the objective function in (\ref{EQ:HAMASK_EST}) is a quadratic form of $\boldsymbol{\sigma}^{\overrightarrow{\mathrm{A}\mathrm{B}}}$, which leads to the following explicit solution
\begin{equation}
\boldsymbol{\sigma}^{\overrightarrow{\mathrm{A}\mathrm{B}}} = \frac{\overline{C^{\mathrm{A}}} \odot  C^{\mathrm{B}} + \mu \boldsymbol{\sigma}^{\mathrm{Ref}}  } { \lvert C^{\mathrm{A}} \rvert^2 + \mu } .
\label{EQ:HAMASK_SOLU}
\end{equation}
Here $\overline{\raisebox{.5em}{~~~}}$ denotes complex conjugate.

\begin{figure}[!t]
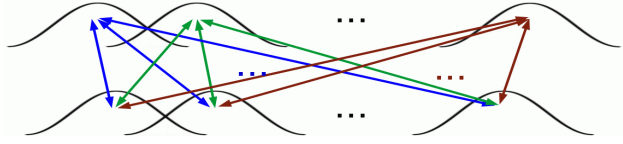

\vspace{3em}
\hspace{10em}\pdfinlimg{0.23}{0.2}{1137}{287}{
789CEDFD7F705465B62E8EFFF19999F3F10E7E9CF11C75C67BA2DF7B0A81336618F5467ED4505489
225043A9853828E58FC18AC60F0341F107770272AD839AC4E83DF7E40CC122478E224322C21D4085
1920B938584084D233E38560C4296B1C28116B8AB1D09090FAD4F7D9FD74AF5EFDEEDD9DEE4E77EF
FEB1524FA53A9DDDBBDFFDEE77BFEF7ADEB5D6B386FEBFA12183C16030180C0683C16030180C0683
C16030180C0683C16030188A1BFD5F7D157A1B0C0683C16030180C0683C16030180C0683C1603094
16FABFFA4AE3F4177F21FA4F9FF17EE3CDD367F8AF21E58C30AF448540EEBB0C0F0E8C8411A2107A
830D0683C16030180C86FC41ACE2137FFEF458E4A76BCFEE37DF7C73C3860D2FAA9FA6A626E7377E
36447E70F0C18307F9599CC40CE9F280E64DCEF000366FDEAC47C89A9656E2F9B645F2667BC72B38
0CC303C77384E02442B842BF4083C16030187288E8BA79FA8CAC98FC39E8FB917FD9B26830943468
24F33187B94B96545B5B3B7BF6ECEAEA2997457EBEFD77175F70C105FCCD9F4B467D47BF29FFC2C1
575D31161FBCF3CE3B7192A79F59DEB9F6255AD11EC38AF8B342BF64439A705604DC4750270C8FC7
963D86E13165CA14DC6BDCF10B127F3018FC034686CDE55755D5D4D4DC74D34D181E2B56ACC07813
1AAE87878D1383C1603014339C750A7F72ADC4A286A50D0B1C0C21AC953513A7C2281A376EDCD8D1
9E81446029E40BBC897FE1057EE330FCE053F5F5F5309F366FDE8C65D7591C0D0643B141EC643CFB
E450307449A0C4FA95871D13C2ECC80F0E5BB4A0AE5EFDE09D3B233F30923115E0604C14DACC865D
8D77F0AFDABA47F045717A65F34311439606B26CDC3B3F81227DF6864784227118603CE060FFF020
05E3C2A187197EF00E06188EC10A427A65C3C36030180CC50F2C55DA8EC25228CB1C8C28673B5156
4CFD43B6C58FE87D482E94F897902CACC5F82EBAB142BF7083A1C2416F321D527836F184E251D50F
3E8D64797E31456CDEBC991E8453C73F65824CE0396103E3809EDEC3B0C03BD7BE840FE2E3DC9C01
29D3340DB30ADEC77F79667C6AC85C1245036152DC61C39DC278705C902450B88338A0BDE315DC71
DC77DC47462CF84F286E500E3C86088A9F0BEB8543D33C0756DD23384676E7866C84180C0683A168
80454DEC28A1519A40E11D066360B1636604370CF18315D31F07C8107A860945CDA7C82EB7B38789
37F12F6E3F7AEBA3AD8C0643C141535966006DC7C260C66B3CA7DA4DE0B78DB3004ED277A01BB304
8D73C70BC62F85F18CA986DF68F1C38587244991F5E066617880083B3339A7712C0DE050F4228DFC
7BF98DDCDCC3D873BC6058923826316249EAF598B4716230180C867C43AF35624771CDA249A35D51
B0AC68476125DD75B08B8C294D8D262DF1A4B39585B539C13F582EB9B789EF326E6530140062B882
B6C01ED64F259914EC58BC2FA466E45F17F87A486DE9D4D6D6D235A67D1F30D741BBF4CC60F343BE
A1676FF01AAC021C1EDA65893B45672526F61106E3A5FE2CFE4B071647A9E31A93516A310F0683C1
60C837DC24A9D367B842795B8E3535DA2175F95555A436B4A3445F2257CDD0F9599ACA39EE305850
8CEB086CBFC1601821C42D8567100F9DE37400B5E95CFB529A9B1BB9A539BA557AE3457B256C42C8
2B24041463A0B6EE91EAEA291C187AAB8DA177F9B81129A8F7506478449D6589610F58C540BDC9EF
72E24535180C068321054E1DFF14EBD1F2554F310C5EDB51DCEBE356708A309B5C6D45EAC40D722B
6DD751160C769D58506644190C3981F8A6F524F0EDBFBB180F20A370932900E4EF19F42BB961A682
6DBC62C50A7A251CD7391A6F3205F9BA17A7CF305BCA3F27638DC0F0D09B5DFE3B5818D0AD290B87
702B463B70470E43C88687C1603018720E09EFD17135D48B10E6421D63FF67F3BD30892606C34B9C
7CE7643BA2B65C1A0C69421CC47E9E82C78D0EA0A85B2A680618F6E4B96DAA7EADCD7B6D393326D9
88550E010E4217A1B3E146AE0DFEE244D685D8EDF2D5B8FB58D74491D24FACFCECCF6030180C862C
409BC49F56CCF01E2C3AD485084C282E0093D22F24B78B4BA43F9F42823ACC823218D20758129E1A
21537A1280B5CC04466AEB458F2F82E74B07137266F0132B99168C588DA49F9933254E9FCBAFAAD2
8E4BEE6805EAF5850E4DACFCD180683F8915FE65C3C36030180C598329BD0C86D70B8DA627C5B9D0
C002ECE93DCC25523291A906C6C024B63CF4761A0CC50F6E53E8585F3E4A4C987AF3CD3765122892
F0DA6401872456981684587166D0A18026A99D513FB3A38E1D3BC6921912E687BE6539669229FFC0
28AA1ED6C40A4B8388FFEBF58E1E2BB2C2A26ABCC16030188A1CDA33A5359A6EBAE92606D115A7AF
47FBADC4BFF6D8B2C7443E5788956C4D87DE6683A108C127484B3D701EA0670AEF904C85DECE2CAE
8B31CCB804ADAA4DC5755C6C4FEF61478CB40827BA50FA4DBF1032450956F41EBB91159C31C1B268
60C9F59E44B73AB5001829BAA6A55532734BEBBA0C0683C150789C3AEE69DE8286804CE94DE92953
A648C254E88D4C1F42AC162DA8BBFCAA2AACF894779688945C55C93118CA06FAA91132C50707D632
E5A64B9A71085B9438465E23A80135764C9A2035A853B462C50AAC0B7A78A03F25672AF446660DA9
A2A579374B75E01D67FC1B0C0683C1A021E2E44C97D0761473A6A2B5E64BAAAA8B8E47922C75B718
686D6D7489B4981F4305436BE5ED3AD88547BEBA3A418886AE07BD05518A0F8B6E33FD11A00012E8
45270B85E9D0094EA057295E6FCE2125C0C846313CB84F25CB4469EDB9A580F06E8C7C6155923FC8
02C143362A0C0683C110019703AC92B0A3B876E86807AE1D641C43A5BF76E81C6A1DD141F9F737DF
7CB338A3190D86C2809BF38E621B3DD414A2296932A5A1DB4F878B44AFC16C1641F8F44B6B5502A8
8EAE7D37928CC66542244ACA292E8EC22CD49225E9A6902C1E8A32F0C7190C06832157E0DA875552
3428B84A8AF678A947F8E896F3F7893F47231B75CC8F94D4E1F596EE951A0C99429E0E320B311D2F
505559692D97F4A391ACE53A2D488BC353491B1345347534260E5FBA3D9075A779F7FDF419FA2E1D
BA2D79A9E5DD3918000C7260D962526FB0AADADA5A3C35B21757AE976F30180C86149070387A6DB0
4A4AFA39BD368CF029CB9542A7A86B95AA9A895371E1E2952BBF0B3718FCC02067FC1B1E070671E1
71B8FCAA2A2616A5D86428F5A7C3A960452FB6163865A097245D16C9B517F8AB9DDA5E3A6A1AEFB0
DE5FA98F8474FA99B1F12095B870118AC7F322E5ADB3A8C86630180C86328068FA3932E31446AE84
FA8612072899148E1EA0457418CA1BB4966115833A49C42F9F026AFA9562D85BA60DF6CB8472B345
5CD8B2D95269855F4933598F8C93247B43A45F870D7B2BB9C193BACDF2BC60DD64579074E387D5D9
CA2342DE6030180C69E2C49F3FED3BD00DD620512EF44F6981F4D01B993F387BAA422D6503167DC2
0D5813FE329431F098B3002EE701EEAB309C491769AA349047804E9247E8A2C6CCAED2C5F8CA7572
E024892B25BBA45089A8CD33F3D42F4351AEBDE15C1D652BF08C603C5C10FBC1F010B5C3CA7C700C
0683A1A2C09078AC92B5758FE86A8614BE6BEF78A5384BF7E61B74573DFDCC725176E212198FE8A8
BC3E3194316830C32A66E694149E13D65036321423E9224C86349B9DD421BAABC28D76CBD357CB45
9135704AF4674E5594AB2E5987E31DF24DA9C9851FACAA78A64AB4649BC1603018D2842EDCA983FD
B0222C5FF51443382AD7823A7D86B215CC179008A8EAEAA818A0655719CA06A2DBC698250E75CC09
CCA0346B50C0B03746474B6A155ED36C2ED7282F294DA82BA48F1D7D1933A72A7A9948F44EB2EC88
D35152BB4AF6E22AB6BB0C0683A1FC80299DB5662814AC771D29585481BB8E81C012C9888E9A8953
45AC835BF751F1438BE830942CE8A466ED39CE03E2930D5460A870687F0DE68499F36ED5515E9822
F026DDFA656339EB5850D959D2B1A09A6E97C1F5E604E8166E50C8F0A0B40B9E322DF16430180C86
928684C4D3390533408C2829C654AE7A4D59F758A0BC1558150B593AD53F0D86E287D8FC0CFA65A8
12A782EAEA294C1BB4219DA2DF3085A28B58B64FE6044AAC9741D7C96552A8446AF83AD534426F67
318361B43A0810ABEDA2057578E2C8434B7D90180C064325835C89F2C83A4B88650AB14A62AAB7A4
89C07E93801F9153BEC057F8D860281568213B5D2E013FACC556DE8A343901332E57AC58A1A3253D
477FDD2365A00428427652C3978A7678C72F58648B851F2C938DA596521E2C463076B4B717C725C3
F6E20C0683A144C19DD5282F8838A7C40CD089008664A005C51818ADFA25EAC1160468280930D80F
D469CA9429E26461B01F333E426F6131C3B1811DA937CE09E8D8352DAD253A274818037D703A8C81
02B06513D698A7DED37F3A0AFCB281C92040A70E5AE88D37180C06C3B090E8B5DADADAEAEA29A2E8
45B92ACEEDFE3A86A137BB48E048AC47632663111D0CF86110A0759AA198417FEBE6CD9BF1D4C3AE
632817153E19CDE53F3EF4361733845C48B611FB53EABD965C050A704092445679161620D9A3A1B7
B0B4C011429923ACBCEC4F3C77AC6B8627D1B45F0C0683A184408D05EE3AD239C50804D94AB584D9
3421B24E3A2446B6A61912538AB5500D95007A1F60DDD5D6D6EAA18B1FEEAB68E3DFC670600726FB
D7A9E3DE1C8B19409C119C136AEB1EC1FB25A104C8E181D63A8275582C58EDDD9428B206BAAEA7F7
30882A8604E327F19B59691483B2FE34180C8622875E2541A0F44289B53E3009C8E6F6343B962969
3A08B0BA7A0A8300B5EA97C1500CA09F9A216A74A6D045C5BA421512EC97BFE24D7C8139E1E96796
CB9C40AD0FFA778ADC1941F7255604B4B6A6A626EEBE9C38155744CDA2D01B59EAE83BD08D913073
DEAD229C3876B417E1C0ECC5D4E194D6FF0683C1102E4E1DF78ADD33798A511C5EC8414DCDA20575
251791525410E1446DA39258DD79E79D0CF819B275D0501CE83F7D66D7C12E8E55B196AFBA622CCC
39CC0FF86F190FD442DAA892858439F6F2ABAA1801C839416721E5BC01233C1565E13BD7BE44AD12
D976C35560CC601131CDA25CDD2686DD4AB964FA88A74C99C225A3C879B7A19861CFA6C19073C42B
0F4632D0DB3B5EA1F4B7287A719574827C72FFED235EE273729E02F43683006B6B6BB5C870757534
FB98EEAAD0DB69A84C90F8D3A35A5F5F0FCBAD9C44698A2AB5DF49B794BAC9C25E6B264EC59C807F
1595D9CCE141F725860769A0D62AB198B49C83C1B78F2D7B4C6FC461C9102945EB704311227058DA
58355402A418A576A060B9C42A8937F364447DFDF11F8F6E7AB5EF4037CE7FAAF7BDCF07BE1CCAEA
892B2D950C340F570A0B8A266B3CFBB8A68645938BCA8232541A30FCA27EEA9A1A511BC040A59F3A
F4E6150CFE59257F41805A914007D149E1BF7C242565741EBDF386F680EB69AD122C19E45342C943
BF7DE504F62733EFB8CB2131A258327434BEF5BC2127C8727248B9B3ED7FD362560D6509FA4DB8EB
C83458FC964C58F19BE47C9C9FECD9F768D5A8E7ABBEFB4F5517CAEFA66BAF786EC2F8B6699336CE
99D579EFDC6D0DF55D0D8FBEDDF2ECFEF6B6FD9BB77EB46B073E75EAF8A7127A24AD22232B09309C
83AAD42C4122111D64558E58AEC15000D04F2D95A7C48B0A8399D26D1512F7DB77A2FFD09173DD7B
06B7BC3ED8DA767E65C3F9C5B50377CC3E8F37F3FA54724EA0D4830409701E068BF126BD50A5C8F1
BD6883E761AF7BE4F2ABAAB85288FB324F610C060187076344A5E61763719D921CA512B361280364
37C66C641ACA0CB23A33A860D182BA9A895319AD4D234AEBB5E66B7BF6F49965D75CEA91A9D11745
5175A107F993EF8CBEE885D17F2BBFF561E05F205F5BEAEB4AD4B3435945984C5AF28BE9FFC3661F
1B0C398416A3907940927AB41845690D48E721C28BDEBEC183EF7A74695DFB6073D3E0D22706EEBB
6B70C68C8109D58355177EF5CDFF6B301984D4E4B66D0ED0D54F3FB35CA7B055574FC15DA0D46A58
1D3814D979A35482765F522A41DC9766CCE7A9FFA53FA524873CA4AC058CBB50218A31869CA3189E
D6626883C13042707E067B92FA174C7D4D56792AE75873DB8D514A254C2AE2AB72DF4C6458F8AF47
AFAA2E7CECAA6F6F6BA8EFFFEC139EAD243888B63AE8AEEA5CFB92235821156A4AB1EEA7A1E44091
4FF1534B8A1FC528BC640DDF382CF2A74C1A8996CF993B386ECC30746958E00CF96EEA909A13302D
537055042BE88C08C519C4CA53B4E475B617560D8C198BF42BDC8D880D0FC6884A7D138C100A2D9A
6085210BE464276424EEAA92B0DC0C86149028021AF398933133EB853BDF33331F9FEEC686286FF2
FBA78452A9DF8F568DA2BB0A2F40C786DE7B27F49E1C39E82864851A6EFFD25C71D2286CC231E41C
5A8C42FBA92FBFAAAA6CEA8A62AE5B5C3B30123E05DC77577E29950311AC10FD07CCD285DCEC12D0
7D49A577A1785A8C22F4FB5B8140B7FB052BF0FCD6D63D4295485B2C0CE90003E982A09F703F6EA3
D7502A90B12AF5EE75FC405EC528FCF87CE0CB7D6DCD3AB42F1993C26FBAA578C063577DFBED9667
792D226AF1EBD5AB4AF749D41A6BF41170E311EFC06E2903B3D6509C3875FC5398EEB0CD6A6A6A74
F4696D6D6D918B51AC6B8F729CC0A7BEB76F10D0EFEC7E6760DC98EC29957C5DC180EBA233428A82
51C486559F0A53459769B6F8465145401B98DE857929AF5F9D431CFBFA637FDF0E45B8769A372227
7733E7D7C5F291181E4EF978D6020EBDDB0DC58F11929A701999C1500C60908FAC929C8DFD621479
8277F2F7DED9DFDED679EFDCA66BAFA0D72921434A33293FBDAABA109FDA3867D6A685F3055D0D8F
FE6ADA94B6699342EFDBECFB24725F98CCA2052BC873B9449684F5622815F49F3EC39237BAB4107E
9CA209453BEAEE987D7EFDC68114074CA81E04B6EE18D40A362B1BCE6747A9C0C872228093517FE2
1690D1686784105E7F75B0DCDE2CAE14B5758F7046924909667CA90498EDFAEA3F26FDB2B6F78B5E
7FE7E0CFBBBA9FD4F714BDBDF378D78BBD9B81B6C31DFCDD7C643DD07868DDE33DAD1A0FED6D14CC
7B63197EE36C33B72F01266F7BF0C0E7EF17E0EA186AC2785D191E8CD7C5732DE2EA45FB081B4247
5E39D1B0B393512A43A903AB3056C3450BEA24509F4614B7B6FC7C2AFD20DB1407785251BDEFF574
B49346C5A993A44D555D486E959042E56355FCAF77A48A125C76CDA5FC0D56157AF78E1CE23DE4AD
F9F6DF5DCCE474F11EDAFA68C809688C891384FB2A2C215A125BDCE3C60C5EF9BD4198C4C9047B41
A6C8867E5875AEB969505C12BD7D9E184516AC0A5F071E87531D7C37EF1E2B7D517446889FE89251
DF21ABA2CE9BFFF8917C9DBC60CD774FB628E6BEF4568A88DFBC08DD9781FACCCB3F68FF56D38F26
EC5EE83F9E4CEA5B2F4D5DB6F705FDBEF79197A67E6363CD37FEEDBA6FAEFEAFDEEB7FBB8EC09FF8
CD77F85ADE973FF9E2A28E9B0B233FCBABC69DC2832C61997C9067CEBB754D4BAB095618528031BD
81A466845E2AFDF114A7B2C03F4329428B0531C847FB4174C5C034CF96CE80D7346AF5F5572778A0
62C17B6E5C5F32E754A24E8516AFC08B35B7DDE89D7FF445F81D7A57E7E46651590B4BE4D8D1510B
8AC26BF41D586A956184A058BA082030D094BA34ED1DAFC0422B09356CD29C950DE7E3D7E57B221C
EAB4B8764084D0D7B50F5E7CC9B9ACE30049AF709264F42A878FA7DC2FB29BCB623FB367CFF697D6
CDC9F74A2617DD975C2CC0E3B8B153CCC38397DF73F68349FF7E1F39CE8BBD9B037B06AC07FCE8B2
75338E7DFDB16694073E7F1F6F0A4B0229D364CAE15381C46AE6F62505BE642C0AB85FA051B44879
CBE86ED6E2EA068383427AA972FEED064358E0ACCB8A33986925C807461466DDF4F9D4D07065DD40
A3F6B535C7BD518E6A9F4A86626E145D547C331A049824E48F148CC747E9D8E88BBA1B1BF62CAEE5
EBF2A054044D9ADABA47E44E31397DC58A1556FFC5901D741E250692B3AD4DB1F49288E61A8A948E
1276E3E44CE9EBC5BFFC6C68F2C4812DAF7B1FC17334676EFA34EAF7C3D2AB642DC9093027AC6969
85D9ACE7044A83E6D0AB285AFADA7D79C9A8EFD0EBD1D37B38F45B1FD8E62135BC1B0FAD8BB29B8D
35A3564F672295DFBA03752215BAABFB49FFD966752D05871247151993E390D2EFE8C33AFA76EA53
150678723BD7BE545B5BEB28C752E3A8F0ED31940446E827C25292CEC793FD695E2A43C9A13F5678
9D12B8986F452159827CD231D15344657FFDF11F7B3ADA372D9CFFDC84F1129EE7CF81D27952E453
2D375CB7ADA17E7F7B9B4791E483499C53C2A718E9F7D8D4711FEDDA7174D3ABFAC8D07B3B87778D
4AB95822B56E00599524A7DBCC63C808CC8EC110D2FB2A980A30CC30A8641E28FE7175F0DD61B4F8
C499BE6451708CDFB8315E51AACF07BEDCBA6330A56C4552261578CC0FABCEA13DEB370E80F4F9DB
33429D614A838A9239355AB92D96939D16114B07DD96E14119588AE4246B7F910C98DE2F7A276F7B
507B91B4C3C86924888F44F739A94F3CB2F9C87AB22AE14D4EBC9FF39A87E123FA461752981103C0
CB7DABADD5896F58F1B91177EAB88537185C8CD04F843921D9C7D34915312F95A1E4A0837C2896CE
412B12B8D94DB0A4515BEAEB48A3448E0FE428EE6CF27BA9465F84E3F1A923ED6D2C26E5A5427CF6
49D3B557B0D494B89F0245FF24D8AFF3DEB96800E07D7BEC5F3849E8BD9D5B506E0B0BA21061066A
2E5A504756658BA3217D505512169764E55CE02BEA5D2A58D79E406A408BE41AFD579DBA22D5FD0F
78D18029642B2EBEE45C6BDBF919330632AD6C35A1DA3BF996D70773E8568EFAAF139D11A03C0CDE
1E216523DDD6E1A0A4DBC9C4D28B6AFE611A14498DF01DF0A6648D64A6D53736D6E06010B1C0AB63
10A03EA113ECE7FCBEA8E3E6EFFFFAD6C044AAC2F41537E2B86722C303AFA97114FA3D32141BC255
41374A65282D6060EF3AD8D5D4D4347BF66CA97984B598413EE9F0297D00280CD6DC3D8B6BDBA64D
22F7892B9FC7842674115E2FF62FF2DF961BAEDBB4703E2898D4E41D8A840882586D9C334BE75825
C4FE695F953AA6BBB181AD02B1125D0B1E167A87E7E30EC25292E474E6556189ACAD7BC48AC21886
850EB160ED33473B0EEFF4F41E2E2ADB381D3437B9E4C57FC90210A261E90F18D3F26706A6FF38E0
48FC4B4E05F2053637676EC679583FAC3A077A05EA97035F528C1A8B4E23A60596B1C3FBE28CC868
7868B17492295929306CE80109FDA62743EF17BDF3DE58366AF57487E9D06194ECC2EFEA7E529C50
38FEAD93BFF31F46F19309BB17E248F02FEDB4D224CB090E040B7BE0B74FA560733944E01E021E76
B02A2DE3299E685B320C44B20C8E2C06ED088525F5C74B6E2532540EB0B6EA201F9659C76F0AF0A6
E9E30009FA68D78E6D0DF5A0457145BE20A6E338989E9B301E340AA4E964CF3EFD45673FFC033811
48594055DF14F2E9310F17CEC6F3EC6B6B96B2BF724C593E8F3066FA0E74D3CF28BE2A8AAB47FD8C
967D6C480E86216DDEBCF9A69B6E92A2DE1744D4DB1830568A4FCD7D77B99C2575DDA8ABAF4F8BF8
8C1BE31128872E8191059A1F07DF1D7C7EA5A7E59E11B702D098117AAF8401696704A541750067FA
C383F1843A1C94EECBF4578AC2C02F31E1C5EF6DACF1FC4D2A188F2FC06B5228EFFD60EB7C7DF0F7
7F7D6B8A0490C643EB4094B40CA00EF6D3DFAB4502C9AD02C95A7E7BE9F4999EDEC30CE0E4DDC423
8F1B8A3FF1668AE208C573A30D0683A178C05572D1823A1D153FB5FA9AFAFA7ABCEF8FF34FF8F3F4
99A33B5FEB6A7874CD6D377A894B11B213652E31C4894CE4B750ADB5F7DCB2BFBDED54EF7B6EE2C0
7BEFC49994FA605CF74FEB0126BEC03138FFDA9FDF2D2704B14AE077B1CF961FB9D03139580D67CF
9E4D1940D94386555C9C39E3862281E4513AE27EBAA877C999529327BA1952E041528847207F8AA0
7A1AF83D598FBC33AC643AECF64347CE815E053AB9860D0E5CB2C8F35E653177E1F6613EC7ACAE13
6764A725F567A5674E1DF7F445A5D633468857643C5256A318E876A017862FE89CD21E22477F4F64
220221E17C9203257EA5C04C2830238F55450205E37A8011C575EF4FFE566713D90ABC18B57AFAAC
AEA57EBF555EBB175C18CF386EA5A8D0503916B7DB2FAE1EFA8D36180C86A2822C073AC8074B249D
53975F55857573D7C12EBFE1C14F81A7BCDDF2ECAFA64D89B3A4241C47A3E9DA2B40A3F6B535079A
677D07BAC1A45A6EB8EED1AA51D400D4A17DE46B4CA1C26F7CA3AEFCAB3D625ED0A02C739F7D92E0
E45292803AB0B0CCC0DB0A6389319CBCA7D474E2126982150607B0D229C8AFB363F01328BE5D5AB8
F27B01F404DC2469577CF55526E27E51545DE8E561655463080783822D7F66208BD257A457F8789A
F70533B9E38C60C92ACE09E25D4AE68F60381FC5D2759A6DCDC4A96B5A5AFDB245051E2DBADB9D0D
3ABCDE79BCEBA28E9B1D5AA45D45A03F296E1CFEA54B4DF1E3635EBE5D3E12C8AAF0DF99DB97504B
507BB8BC7CAB989B2C9E78152B6EE5E461E10CE05685295C454B40AA1C320210246BC58A15A5522B
C16030180A0967A5A3982A6651293E851FACB94F3FB3DCD97584C505CAB3BD79E59ADB6E74F4CC9D
0C29A76814A8103EB23F31A84F9F19EF6F6BA85F7DFDD5D1732A468653C5CF16791FFCE8D7AB571D
DDF95A54593D31AA10FF3DD5FB9E9C192B11CE1C9C6F35FAA2D24AB1CF026455B5758F88AF0A4B24
6E2E96C8AC93D30DE50AC6FD9283CB68F18BA597DCB0418393B1924347CE3947CAEB4041F514983C
D153ED633C61765D0493B57BCFE0D22706FC3EB574BE1D1FC4C753B869F8027302E6764C02975F55
A59D11FE4AAFCEA9A8CA4EF7A57C902B855355BCF02324FED531E79D7090635F7FECA441C543FE62
AE22C091467770E0F3F7B5A74978D9F20FDA876D5BF391F5A3564F0F94016C3BDC81EF753C59AE42
60846AE10CE0562FFFE937C9B895C328B3EB3DBEE6469C2451725385519D25F7EC1B0C0643C18035
B4BDE315CAA84A900F9326B82B85291434AAABE1D18D736631784F084E423294D68B88FD091AD5DD
D8908C4601A0455BEAEBE88D8A93A3D839712A7C45349230C2A4F6B5350B5DC2A9E26181B16F077B
72BE0B5FC156E9142AA1549A7C9525FA63E2EADC5896801F8AAB679A9C6E2857609030459D8384D1
5C1744743E991D137A0B4702F0A664017B9492482AF2F64CBAD46642F5A0131D9D1DA425B823BBDF
19C88A5EFD1E17B5B2E13C3E9E6CE24553C51941FF359D11FECABC51BF55440616FF656D32592928
965ECCEECB9DC7BB98D32475780395F7FCBAE80E5EECDD1C18318893A7F0DD88D7EFAD93BF1BDF39
5F0403799E09BB170EC51C6A3B4F7F086EC592C1F1DA554D3F72B8182306E7BDB10CEDC99FDF8AA5
136015882B13EC1BF71DAC0AF75AE7CA15ED7D37180C86C2807338564998D36B5A5AB12C6AFF14B7
2BF76FDE0AFEB2F69E5B0282F7B417C9A70B411A051696E2DB3FDAB5834C2A81E6C4D2A084B831B4
8FEEAD40A30E2D94EA54605E3D1DED43CE865B446E3DA0D9B1D7654FA908DC688AABE346EB708EDA
BA47989C6EE239958CFE20B1740C95FAFAFACD9B379741B44FEAC4A8EE3D49539F60B506460C3AF8
61D5B99CF0A964A0F76AC9A28CE955D5855F815E353779CE385AE03A250A6C68D1823AC7E5446784
1E1B43CA7D5942B59ED1F205071A5DB53DE50FD214E9A28E9B1D4F8D73B6C77B5A9DA2BDF21AFF72
BED7FF273A1F37115448B2A880E623EBFD87817C915BE970417FE3996F95C398403797F9FF1B622E
00E60411ACA899381503067382AD140683C1A041B1F499F36E95F08FA9D5D7FC6C6A3595FA9C92BB
128FC71ABB0E3101EBD9BEEA61D0A814A9496070A069E293F233359D15C573EE8FD5A21A0A12CCA4
FB4C0E3EFBE11FFC5FCA00C504FF57A24E600AEA574EE072893BCE25520B29D36C36A5DC8A05B5E0
300CB4C2365EE01D7A1F92994F256456A5164507214A961183DFEB370ECF6274F460AEBA25992E01
B8DB96D7076B979E4D5393D0A157CFAFF47C763C1B6E3DE684DABA4774955EA9F41ACDB58C14F3D5
B59EE9D5CA422A30AF706EDCF181BF5EF5AF3FF1673F252B11B56CEF0BA9CF1C4D89522AE8FA6C3D
673F48B39D0C0264201F1A19A80C39A4B8D5A5BF9C2CF9567C217E2B510BC4EFC9DB1EC4997332F6
B49F14BC1BB75E8607DD551457C7BF4A680630180C863C816EFDC7963D36E3C67F1C7BF9A5B3FFD3
A8BB2FFFDB04EA14A42CE105E0A937DBA64DEA6A78F4E8CED752084FD1177674D3AB9B16CE2793D2
E7273BA32B4A6AEF6E9C334B33A961818FA01981FF7ABBE5599C1CCD0EBC28FAB62A8452E9FBCEE8
1D2DD5A5779B6D95AC28C06A621EA5686BE387E27EE059E1AA0DE4104B160D438BC0B9FC9F92EB1D
96BC80AA14463DC0691EE9D57D77795AEE19D1AB8B2F393767AE77D5DD7B0653547AA5FB12AFFDB5
9E593ABC38C7061BB3F3F4871775DCEC54837282E8C467E48FFA73AE687CE77C4D6D488BE4540FFC
F6A9F49BB7F378D7F77F7D2B4E98E6B588DF2A9DF2C1E0562FF66E3EF6F5C7B9EA4C0C00CC062258
C18DB82277501A0C064301408E0323EA899FDD0946F38BB1356E7A91DF7FA4DC552D375C372C8D22
BEFEF88F60529DF7CE8D1231471D429D9FCA7E3812C7A75E97030333025381F05FBC0FD224AC2D58
8AB0EAC28F76ED08FDA6E4F576FB5F30395DC7F05C7E55D5A205756B5A5A4D5CBD72208290B08E74
1E250C660C8FA22A2D3472A423A6E74F209217BBDF19FEE3D4A6703EEE74788AFF8E1C23A157E89F
19B3765E56354B722DE9875ABEEA29CA16E9E181B9C2A9F5EC77ED857297FDDF0B9EEBA9A6C7F2A7
3427D24174635EBE3D75B3F12FA7B654602A56FA178E86651A18E0C604068520EAD7E45659307DE7
CE321B17AB834EC56582407BC72B5E684A194D140683C1900E60231D696FDBB470BE47A314B390DC
A540CF14001AB5ADA1DEA35169788E40647A3ADABD703B25001858E117EFB3B6AFC7A472AD640EC6
C7903F9718FAD822BE3DF45B5360D0967EF1C51767CEBB5502BD58D591C9E9E557ABCBE000638035
68A64C99028359D4B045603FF416E61613AA8767168B6B8305D56931FA2B0507061016C9BEFDA123
E7D66F1CC8825E45243BDEFEE6DFDC030EC54050FCD662E92059181EE9645F1683A58D3680532CFF
A0DD49447208D1B0AA7DC7BEFE986E29E7B35A36905EA7F429CC48DC9AA06F0FFCF6294F895D113D
DDB0684060D38F1813A8FD56D97D2F483408941657E79281F150CCE2240683C19005028573BFFEF8
8F20385BEAEBA2E2E42A5F89348A82E70E87024076F0A9FDED6D3883FF5BFCDBAD380C076F9C33CB
4DC212CAA6584CD3B557E0E41FEDDA91224D63249D80DF38BF34C349D172F81DA862E8F72E94D1C2
F243B5B5B5A2947BF95555B0A8B5B8BA2D9465094673319847A2B970F741AF3024CAF2A6A7492852
94E8ED3BD15F756182127B6BDB793F61C131A9EBFC16B87BF175BD7D83EBDABDD0C48B2F09943D4C
856FFDCDBF905E79D204353522961EEE450DDBB7CEF43573FB12C733A5FF04E9E8FDA237F525EC3C
DEE5F752F955047158FA8DCCFA1AF519C0AD1EDADB38E6E5DBA3AD4AAC7EE5902CEDB74A9F0BEBCE
A4B8BAD4E3A09E0983846D23CE60309416524B120DC54CE5A39B5EDDD650DF366D92F026F10D6957
11DF89BAA8224C67D3C2F97E6DBD14C4E754EF7BFBDA9AA9642EE7E7D789D03ACBF2E20568DD9EC5
B5645279ED195C3EF3A7842DB24E5654DD5D2A0247AE9A228195036D75E0465348194B24037B2803
C87A6445B2DF6EC82D442C1D375A223FF103720DBB481BCC456227E7E49287F3CB445F5350DD0F76
85165467EE1578D6F41FBBC21760555B770C16671F925ECD993BE8D0C374BC5793277ACA81384371
5E5AB2263DDED39AC07DFED768ED631AB57A7A3A394D60229A9B383217C2B67EB0D53B553AE422B7
BDC798C0657B5F100D762D15E84406D2A1D678681DFD56196D6C627EA0948DAE5B575D3DE5B1658F
D17719FA1830180C86110253D9D19DAF914669EF4CBC6E540C9AE9F0C8B5F7DC021AE564240586FD
C7D397844929774F82B74B05D781A939A5A9F2077E05BE0BDFE8E91646F40901061FA22551B942D5
3C5CFB50519A0785E92BB22A2C883A397DCA9429B5758F3039BD027BA68CC10A65B8DD37DD749316
4B079FDAB06143E8CDCB130EBE9B017758BFD165555A008D6EA9E757261C1098A825E74926DC5760
38DF0B23FCD091732086197AAF3CFA894EB8FF81015CA0DF842E9EE982BC46D7E725A718F3F2ED8E
428556324F8687F636FAFD53642E0CAE9373B61DEE28CC1D4C86DE2F7A1FF8ED53BC4C1D9D189875
056E85CB4F53CB421A80390433862C19DFFEBB8BB164E8820BB670180C866243EAE02B8F46C5BC51
8EFB2941E0CEF14F450805D8D6DDD78E6E6A6AEAFFEC13990303BF9D60A8009814BE6EF5F557FB45
1EE4B4F22FF0171CCCAA4FFE5325BBD89CF498BC3EFBE11FC098A2495541CA1BE81970C3D06F74B8
C0003878F02016449D7D8CD778074BA7B82D6C892C69D029D9B56737EB1069D1EC152B56B07C67B2
0F86DEF811625D7BBA7CAAEAC2AFEEBB2B55D81E78C4924571CE259D038AE13F5B7353C2A9D2EFC9
D4B371161F4F7D185E8075665553D8A357E8902DAF0F862E0EE9DFFDEB39FB0115C8A5BA2E880616
A097FFF49BA8C2C3FF1A0D42747CE0AFC33678F2B6071D8F8F76FA44BD42916FE15714F8DA09FDBD
B89C784C2078DF4B531DE75A82E661A4671EEF69051D1B4A2444814608E60A6A1CB1E8F325A3BE83
DF63475F469548135737180C25012C5B7D07BAB7AF7A98FA0FF11C25553DCA150C574EABB1975F0A
C088AAADAD4DB314913737BEF70EC8D17313C64B6E94DF1B1515428FD5F90DAC9F9B8E9D908FA918
ED711A293D8377F0DFD06F6B88901593C935B367CF9648302DAE6E4B64A903760E6E25C5D219E489
9F9B6EBA89A973A1372FB77086EBCA86618A52811081FEEC7EC715424FBDD99ECEB7907F85656327
6BB03CF5FA12366CD8505BF748CDC4A9175C32FB9B7F738F2752A14222D3C184EA4174C2D61D717A
15CABCD11FABEBC44A528CCA2383206500767DF51F93FEFD3EBC33AB6B693AED1491BD005F4F24C4
8EBE2A10139CD0AFC79EFA46E41B68CFB2BD2FF07A3574A563E15C208F121318D86C69F3C18307C1
AA588FC31157B72A870683A138D11F2968FB76CBB36DD326912E496E942BBC20117DB1883B7C64E5
6D37DF7DED68D02896EF94822358715278C1FA3FFB045FCAB2BCF8169C53D3379D8BC417F8967D6D
CD694EA479923E48B64E796490BA1C91CE39D9B30F9D09EAC777F03A1F8D2939F447640029012782
151452EE5CFB52F949C0550EFA4F9FC19DA55A97244F712AC0ED16E9B632466AB13E7FA45FD6082C
282C6EAFC2F47316DF82E1F1E69B6FD6D7D773C38DC3034305EFFCCB9AF7264DDA9A29B722BD5AFA
C4C0C177070B36BA9C2F0289204D10ED888EBE9DFA008CFC87F636A619A7A75D3CFA05DF0707016D
29FE4C2210A5E51FB46B7A287CD3218CE83470ABC77B5A53D0435CEF893F7F0A262E35CB58039AAA
A11854C5DF210683A112002B880941B0FC13E4F2121D525A5CE29F62D57837CE99B5BD79E5FECD5B
C19B162DA8E3A634F7A5612A73533A6990CFE9331FEDDA0126253E29F77B13FD53689E5F153084EE
4ABE6A9FFDF00FEC2B96C75A7BCF2D5AA5F0487B1B3A2AF4DB5D3C0075C2B01121384600B2F02BC5
D5CBDEFC2E3390292F5FF5D44D37DDE43065D842BACE7519DFD93B667B194374A060EA738A4CC1F2
CF493FF3C52B6F7CE9E717F8762D59503C5DDD1F2939040318C3438A4FD17D8937A9FC893981077C
F36FEEF1040033A75793277A9DDFBDA740F40ADF02F6E42410813D65DDF37D27FAC93846AD9E0EA2
01B236AB6BA966256F9DFC5DE8B73223F49CFD007489C58BFD9956E489C222C7BC7C3B2ED9E156D2
933DBD873BD7BE545B5BEBD4E3A06085FF7883C160280030E78046ED6B6BEEBC776ED3B55724E4FE
F8D94DE2FB0CB793A27BDC946602A9946864900F66397FBA74FF679F1CDDF4EAA685F3BDEF55E78F
7BA312BF9D8A16A133A97480FED42A7F14A330F821AE439A58ACEF2916B823AE6E2866683905C673
E2D997D24230787037C1A78A595B20B7FDE05C172C6487EFE4F67B41D9FC7A7A13AADD6CA3D0BB65
28B285C2879D310CA24E8331231931DE46CAE933386CC68DFFC892D000E8D5B871BFBCF27B99712B
F676739327569FA23746380E71E66F6CACA1F61D8B46813B8C24F6122704A1D0B1701400141A823F
43BFADD90117056E05D2A4830035C9A2A78F3C0B5DAAFD5612388ADF6FBEF9E6CC79B74A0933CF09
3E71EAD3CF2CE7FE6D994D2C0683A13881A986EA792029426744EB5BE88CC8F40970F0AFA64D79BB
E559A1519CDFB829DDB9F625AF78C4E597CAFCC68A338ECC29385102934A849333856340DC7A3ADA
9395E52DCE69131DABA5080393BC0C1AA26340A55C47B0823ADBC579AF2B1CCE4D219FC22DABA9A9
A1254CC563AA8E683F7559DECDD4799A9A0B5C7CC9B91CF60027E14347CEFDB0CA95D1C33B5420CF
53E47346DD428501D1D217B9128685FB4569181C585B5BCB812461E40C0E5CFAC4C0D5D767C6ADC0
3AA7FFF83CE9D548BA4277267FFF60EB7CED6719B57A7ACFD90F723B9C7ABFE8C569857D2C38D098
6CD41527C45A90068328812EB1EB9C5040BF0CBB9F5B6138C1BAC070D2AEF0A9D5D760C060D89800
A0C160C81348A3F6B7B7456994643F293D3A09E793EA517C7FE39C59DD8D0D4777BEE697E9E3ACC5
690D7C8A523C34A29C8C51CC7E60469DF7CEF5AB5B443D5322B15E75E1B26B2E453BC1BC86ADB551
847326D89F741DBAB16DDAA4D09B5442A0DA36C8B856CAA5B87A2524E094342846413EA593A7488A
E5B0F2BE89294400EE989D90F4D477A23FE75F8A73FA0B0183CAA52E045C983EC1648EA7182B853C
DDF811C536ED7D887F3032FF634E58B4A04EFCD7205654CFF682484FF4635AD8F2FAE0E2DA01FF85
A706BA057784F42A597FA67969AC42A5B51772EB4212ED0BAD7AC17254E501D0CFE623EB276F7B10
BC896968D4DF882B5AA86AC23866D9DE17769EFE50CC0F2E19C2D0C9AAD6B4B49A6085C160084416
7608691489CC7313C65397CF913AD7BC46E815A8CDEAEBAFEE6A7814342AA97B28B2D88128751EDC
BA7CD55398D044DA142F66CEBB15EB204C2CCC69E07120655291CAAD211523778F568D02D7435353
7C694960FFE6ADF1F0C5D117A11B436F524980239C35619F7E6639DD1C34CB998603BBCB1F416A28
06D04F4D310ADE2F4E05308369D81449F859BE3B21F501BA682F20957947FEA57A431E5D1D284BDE
BD2740B022DF4F93EB6FAA7B84CE292A60E3058CE19EDEC3A90B7CE3BF7D07BA31278C1B378E1180
00CE430959AC7132BAC0B040AFEE7F6020D3E0401C3F63C6C0BAF6C14347CE0D7B2D8EFBE3E53FFD
465C2AF422D17F940F85BDEFFFFA569D4E95D77B97F39190DA87CB03401BC9AD749A95131C28F48A
7EAB9DC7BB30BA5E7CF1C5EA9B27714717BF2FBFAACA042B0C06C3C8C1C8BAB8DA83AE8BA4F3A162
82DEF27EDBB449F8D447BB76A488B2D31323D62FEE1FD2FA65BC1F5E83616DD8B0615F5B735C743D
A83C938EEEC3F79EECD9E77C57E06BFF9FA1C369CFA685F3F5D539D7654807926D21CE0E661F53B0
22B501662830C8A7C44F2D71BF15AE6CEC9FA6C0A172AB50916262CCA81070BE215AFA12A0851F8C
163CCE1413086C8F4349761DEC62C1050E30EEDDE14FED8CD0E781218DEBCD825E555DF8D51DB3CF
835EF5F60DA6A69FA40097FE72B276A9800E480A55CEFB79E6F625606DC23224D3AAD8D6C48C10D8
785C1AB95582E0A1F255E9C8C08B3A6E9EF7C6B285ED0FEB3186E1E117AC30180C8640E889481294
A8DDAD03EA1CEDBE78B25284DDE078D099FD9BB766EA18E2BE10E331E84DB8FCAAAA2953A6ACBCED
E6B8CC452C8C50BE5DD32B32293038B99C14BB58C5BC6438CBAE66A9B8C6D09B575AD0821520E67A
89A460C5D3CF2C679C7CE84D350CC592A73CAFE2C4A9B475E9419068AE0A0CD774AE572A47380A15
D37F7C3E1F5F27EF0416026E6D3B5F987A55E27486414B2A24E26C58296A6A6AB07CE0BFE9D4EA95
53F59F3E83390176B2B841A980446A96A22E157A1E1469CE5C2F852D53EF153E457AE53F397AF207
5BE753FD5B62D2028B2B8DFCCEF275E3A1753ABCB0E444FF52DCE564407FB61DEE20978CC604C6E4
011D07160E00C6FCB709DF9F39FA1B632F92C0E3152B563091B3D2262283C1903E30451CDDF91A58
C9EAEBAF161AA5F94B3C1F2A51AC8F340A14ECD4F17477FB9DDD3FF2A999F36EC5E288596BECE597
CEFE4FA3827502A97AA7DEC4B76F6F5E79B2675F59CE6F6088D2F9B870F473E84D2A5D6098C15C7F
6CD96392C9CE2512EF608914C18AB21C48C50F6F0FE4F419DC08EA348A9F7AEC682F9A8B8A22A137
B2C01D32EC315A940FE67ABE9BE4841A3ADEB10244FD31CD16745B6A93810A2D5A509766CD773FA4
9A959C90048DF97A0C474F7D5D2047A4577E81C4D46075E62DAFC793E01EDADB182DA814B3F05FFE
D36FFC0DC8613F33C85078041856BE8750BE4748FA87815BBDD8BB79F2B607A5B735B1D220F9BAE0
C1EAEF4EFE7B8C37CC4EB5B5B59D6B5F32C10A83C1A0E1098F476854DBB44981FC2559885DD3B557
6C5A38BFA7A37D2422E434A2761DEC82C9347BF6ECE9FFF91232A984E2BFBECC2C8617B6DC70DDB6
86FA328B82F3CFCFDB573DACD3C470B3426F6449033D4C5625D9C7645558222B39A8AC18C07D15C7
8D88DBC4481B335D02E1C4E3E550A12219401FFCEC00D420DF3788F558595343B4F4C97DE8BECCE2
9CB285425605A6268A049C13F08D814180C99C5FA057AD6DE7EF987D3E537A356ECCE094957B1C85
3A91E0CB1FD0E6684AD1C61AFC9ED5B534C505962B3E1FF812DCEAE1CEA7C55BE797B3D0AAECE056
DF9F399A71C8160468305432C8626099838F8095D0E9130F9F4B991BF5DC84F19DF7CE058D1AA188
B74E31867DFBC4CFEE7C62C27509A2EB3EB085F80DEAF7EBD5AB844995FDCC9F107839FAA261150B
0DC3E2F4177F8179F6F433CB2981A243CBB8445A2717FE8E309A8B77E4B2D8CF4D37DDD4DEF18AE5
83A7C0CA8604D1BF9C28540442CFB44E0E17E11402F67F2AFBAF3E7D863535F0848A062C7742A2C5
BB47FC2D4CDF5BD3D22A3B2D14BBC85A910056FA9E635F3CBFD2E3BCE9D0AB5117FF4682FDF8BB66
FD630518F6B8F08B3A6E16D74C3989FE658AE3037F458777F4ED9CB97D09CB813972163A269018F3
DF26CCFDE7FB761DECD2E5E3538CC6224FE236180C69E2EB8FFFF8D1AE1DDB573DBCF69E5B12B852
A0C2432240A3F6B535A7507B48E77DE7BF94405F73DB8D5E30618C2FE035FD533A438A617E605268
43E5D4636247E17AA5D816FA04DD157AC3CA06E213917D69C93EC612C9841D5BF2F20D76322B4F09
C365F015E5D74C392435B6BC9E60992F7F6660280FA69AFF8407DF0D88739342C0396C8004FB6178
605488FB526B9564F1A806E68B61B0E19C1878E227C50BFA491904984E6070E099D15DCD4D8349E8
D5EFF1FB8246AF9492548902BDFAE6F78FB34B972CF28203030B8EA4F3EDC31EC0EC2DC91ECAC7F8
292D306153732B51B1884606E2CF88530FB789EF4CFAF7FB707CFE54440C0643E8C073DD77A0BBBB
B18142792E7B4A0CF3D32A1330E337CE99B5BFBD2D6B0A936C4AC14C0566C7742D27258A7AEC0ED1
43CBC1A4C4ACAAB4990AD7AE65157147426F523981993B0CF811593909253A75DCB28FF3D6F3CA4F
DDB9F62518B142A6E41630D8CF6E416AF4F625D8E777CCCE8D42C5B037AE3F5208D8AF7D376E4C5C
78C1393E8BAF63A21393A744291D34077C0AEFE776DF43342B580B18DFA2C30BC1AA864DD74AE762
FB4EF4A3DF48AFD863175F72EE1B77FE4FED10F1C2CFFE618FDF933579E2C0CA86F3DD7B06B5F7CA
FFA599EE76DED5FDA476C4F47ED19BD721543CF0F3537F171DFBFA6372AB4B7F3959DC881EE195FA
5689AAECF3DE58F6626F7018AA4D650643C9018FEDC99E7DA0515E6E54A0132AF99F9DF7CEC5073D
9D8758FC86DF4F9DDD2E99E8077AC2178ACA455D54BE84A9B5F7DC02FAA037A84B42A92F5777505E
83D86A4A95BEFA8721CD4ED62167925AA5EB4A5BC8599E40511AF4FC4D37DD24E581D0FFF8D37A3E
7D60246B5EF3C3AAA4559072F25DCE3B818580AB2EFC8AB56EB398B4F5473006C06218EC27AEE4EA
EA298E1C5FCE2F8A7302830085E6535F9D2333576BD0E7035F821FCD7E76030510C42CF718D67081
8213AA3D5590DDEF0C642AB7E86F7CDBE10E2D210E0691BF2154BA383EF0D797FFF41BD04F2913EC
28046A7A85D76061E0568177A7EC6D1883A1A4411AF576CBB36E015CED874AACC02BA9496B6EBB11
1FD4346AE48DD1AF8FEE7C0DFCA8E9DA2BA428306894A3BB4E8DF465D75C8AC688DE4546D23D6509
ACEC3A00121CB912AE3A14D0B687F126914530F261412D5FF5540E8D37836C05531060D1823A1DEC
873E17418061B76E0C82E93F4E48A72ACC7095ED35FCF6170206ABF267756544AC18EC47797389CB
C5130AA6D3D37B78C8B7D0E4EAA2E4855F2C85C55EA34180E97DFBB00D3BF6F5C7713B3C124556BD
F2B18C742DD0D5E8FFE7570E815E65B7F1F8D6C9DF691FD9E33DAD05183FC5834C07128EA1DF0AC4
4A642B243850C829191698328E04172B4CA10183C1302C923EF2EFBDB3AFADB9F3DEB981C2E66E8C
5FEC056954D41B95EB9528FAE7679F1CDDF42A1A4646E031A988438A4C2AA1AC55E4BF4F4CB80EB4
6E24CA816509B0517D2BB7AF7A38F426952B183CCF809FA9D5D7E8FCF774027E0C19411BABD2D578
AD95FD2AC72B3D12B073963E91C068BAF7E45D4ADDDF8C3B669FF75BFB2C049CA6992AAFE99CF2C2
716B6A3836383C74B09FFF5359F45BEA7F49965F6DDD23F876EDC246DBFCC43FCDD3EA7760668F79
F9F6A81D1EC9CDB96CDD8CDE2F7AF13E08296EEBD5D767C0AD4424A4B9C98BBD4CDF86C7915AE90E
14A0C0E3A7D8906C80E9281D2E19CDBB374CFA65ADF630321B2EEE708CA55CE1CEA2632DDFCA6028
2A9CEA7D0F346AE39C59D1EAB73A072A49929424258146F51DE8CEB402EF5012991A7FEC313811DA
E6E56DF9C50313DB4677D5DD97FFED929FCCD06B53A68B6FB277CA03514E1AEB3ADCBBD09B549690
F173E2CF9FC28262289A0501E6A38719EC27A931E280400F07DAA8867400E6A28D6A58D4856F036E
5CEDD2B37E0B7F654306B95D780049B767CF9EAD83FDA64C994229CE14B57747D2F214E7949A8998
137410202B0BE35F23914F99D5B5D411E8F657DA4503B6BCEE2954644AAFAA2EFC0AF4EAF995430C
C24C7DED30F8855281E5157EFC942E7A7A0F6FDEBCF986076B2E78B05A0A070742A8D6E46D0F268B
0934180CF9C6C99E7DA02A6BEFB90506B6963497DF7403F9AB4AB172D347BB766441A3D20798D4FE
F6363029B2BC686392B3BCE9FFF992B1975F3AE3C67F6409F27C041C9607A2796731DF62E8EDA904
0406FCC082A2EA1787AB692664DDB7E04DE8588A81D03F555D3D857E4053F6CBBA574F1DFF54DBD2
F73F30105663C0E6FCB63DA8569A1742E14727D8AFB6B69642FA7258E17B5897C4D2222A78C79B13
32D18590D7CD47D6D3854156050C5B66B7EF443FE815EEAF3F7F6D587A75C7ECF3B83BC9E89517C6
16B3F947AD9E4E6BDF66B974C60681190CF64CF5CD93BE31F6A2EFCEFB4747809D375A0202C581F5
83ADF38D5B190C05C0A9DEF77A3ADA372D9C2F1ADA6E149F7203451D5511C31B763868D4D19DAFE5
7B3EC43AFE76CBB360528F568DC2F73AF27D5AA71DCD5B76CDA5775FFEB7605230A5F01B8BE69A96
569CC1B6FD93A1EF40B7A49EA11BD7FEFC6EBE6FCB5CBE21418094F516771518D6A20575DABA33A4
0F58CB4F3FB37CF6ECD9622DB38CAFAE2B64633B3BA0DF2EBEE49CD8CFB0B743EC4CC765F64DA543
18B87BC63731DD711F03143BCE59264E95C4A5D0C706E60434529703161736E532F4ED08BC47FA5F
6F9DFC1D980BAD6BFE9EBCEDC18CEE38E8D52B6F7C9905BDC250C1ED58D73E78E8C83969CF82038D
5A6FE1C0E7EF87DBDB25070C0F002301ACEABB93FF1EC3C3FBFD60F5058DF3FDF5AD1CE0CDF19D59
722BFD91D09F1183212CF8D708FC4959BC3D8B6BC515255A13F1A4A4A01A526DD3269146E563A7D7
D9613BFBE11FBA1B1BA4463000D29750A557B510FF7A62C275744B710DC20BAC4A58252DC82735B6
AF7A58FBFB3030426F524581AA5F8C41D2ACEAA69B6EC2BA89016C5E1507C9F24A981A03362AA15C
0CF9C33C90731DEC8A85231051E07D6F872F04B22AB430F0B3788E300C409D6A6A6AB4730ACF5DB2
58D0918F968CCEA00F466B2942A87706980688AB902259A94F8831FF83ADF329C44D33FBB2753346
9217D6DBE7A5ADDD77D760A6F4EA8755E7E6CC1D6C6D3BDFBC7B8394C4427B9A8FAC1F79BF5520FA
3FFB849E568C0AC912FDFECCD163FEDB044F833D2615A8E300F53B0098353AFFF8C05FADCF0D0641
FA69AAA78E7F4A1AE54579A5A81B15794D770F89D57313C66F5A387FFFE6AD59A83A649AB2D41F51
14046BC397C6F525226D2399D2293F681E08D7DA7B6EF9EF8BEB66DCF88FD5D55338B75C32EA3B58
25D7B4B4EA0D3D43609FE385A781AFC23B6D5E0D05A403B4F79C809FA8BD17D963B7BBA3A173D3A4
48ABC8FAE18762D4A244611839962C4A603129D267F20AB9A181858061F073238287E1359D5337DD
74932C13CC9C6A6A6A22DD764E9B93013312F24217B6945DD0D955E2721D76B365DE1BCBE2A27091
90307F0A55D61772E8C83950241025EDB84C13A3A6BE73C1FC5F5CD038FFA1BD8D43411BBFFEEFB5
475877057B837B71E0DD32E38D1D7D59F5CD93E6FFE2FE9FBC567FD5BFFE44CB039259472B082B37
D6CCED4B1A0FAD4BA1C1EEF7ECDBBD3054080204F176BEF6E2B3B5F164994495098FA738C42A7218
691428583A342A3BA9070760525D0D8F46E99E42420B63ED0493EA6E6CF02E2D5280BE66E2545925
6985C23AD57A4D86643705F757B3E935B7DD187AAB2A16204DB4A066CEBB55EFA2C30E0459E8DAB3
DB8257E37DA54CDF53C73FA55D216C343A0FD43DC27920F4D69613D6B527D8C63951A84873A24EE6
45826DFFC32AD7AABFF27B837D27FA697CEEDFBC95D949F25851459FB20F9936231F4876693ABB2A
2152B1A606031EC33B411A2E71D70576B2B827B44B6858F222EFE8A72C7526726FDF2006C68C1903
FE8ACC41F87DFCF5F78F2FAE1DD8F2FA205347B31E1B95064D739CBDB84B467D07AFB99BD476B803
A415DC2A5AD6399652A7FD563A261063A672EA2F1B0CE90034EAA35D3BB6D4D7B54D9B1455998838
77E21E1FBFE679E44DD0287CAAA7A3FD54EF7B056AEA575FF51DE8C697D227E557998856988AB513
57042675F6C33F300365F9AAA7306F0899BAFCAA2AC64A25134CB3C9D9DF1BFBDBDBF490C09FD669
E1DE1410046657890588DFB0A64010A4FAA7DD94A84571FA0C45B0B5CE006326D157B045B5F721AF
9D563977E4E0BB09E671880A151A786A020B01FFCB9AF7606D32A456E2A3A64C99B262C50A799452
8C8D22B9AD14035CD3D28A3981F5EC6036E34230CEA35236BE4B900A50FCFDADA61FCDEA5A9AAB8B
4A7D067AAFEE987D3E99F7CAEF5514DF2286D3D61DC1F5CE8AE45E141BC876D163ED1DAF609C8BD2
11330431375290E7C0E7EFDFD5FDE4451D374B996057D74245091AB732542012FCB0116FD4B686FA
D5D75FEDAFC09BF04EA2430A5CA6F3DEB930A4B34E38CAE253F808491F630B7591597F239BAEBD62
E39C59FBDA9AC1A486623BF9A293463BEAAA2BC6C2E6FCEF8BEBFEF7CE5ECBE8CF086B6EBB5186C7
A355A3D8C923BCBF86EC201B8F404FEF610E728C6D9A82632FBF94B5AB2A3CBB4A7A49642824D28F
DC135DA463A26C00E71CDA249E505DA0C0BF61EFE3E7035FFA0B01037FFF0F8B844C818F30C116C3
A3F8B7269CE631BA15C35B9BCDF811D90AB9A2635F7F3CE6E5DBC578069FFAFEAF6F1D491462D61F
04016F6EF2BC57990607FEB0EADCE2DA81F51B07705B8BFC3685351E9C7758170053A2ECC561D893
772F5FF5147837A744AC2C8FF7B4EAE1A1B501A57030BD5AE056CBF6BE80E134C291603014336487
F664CFBEEEC6062F172616D11755ED4BA454CE9F3886342A3B6F54D6F9B6E47D7B16D7460B5D09D7
4B6CA158F86BEFB9058DE4DA3714139DDEBC79736D6DADC8A361D2606546D89FE88D11B6B6D2803B
E224A6598F150F984D8F2592A39D6621463B4804652B2A33B4D5B3D3229B2A33E7DD2A7C13BF5917
55F86605F64CC110AE42850327C563CEDC2013FD6FEE21F5D8BF796BE9AA159131A1FDF4626BD90A
EA01B677BCE211ABD367666E5F1277404482BB7ABFE8F567C48CA4ABD3FF971C70E8C839D2AB4C13
AF40DB973E31D0BD27D87B556948DDD5A78E7FBAEB601779F725A3BE23C383AB068687F0230C0970
2B90A664BE2A6158F80D0AD678689D7CD660280FD0BFB3AFAD394AA3123D3B3A46CEFF5FD0287CD0
ABC05BD8050533FCD14DAFE2DBA56470C26F9F663B98544F47BBAE6FC57504C612260A9897975F55
25BB8E982856AC5841190AFFCA5EA24B6781EECB575F81E1CA5DC0E0013D0FBD5506C29F49515353
C325928942BA2870858C73CC24BA92973629993E59BAD67209013DBCB8B628142A867CE9F3DC739B
3469ABDF2C9F316B274CCDD4617EE97C573180BB0AACBCC62040C6BDB3F8C24FD73DA18B13012FF6
6E2E9EC6031D7D3BBFD170EB37EEFC9FDFFCFEF12CE8D5CA86F3A0574575454505CE93E4DD634747
2749BDF9AC1D9AC081CFDF6FDEBD01DC2A411E30B164B0BC633181865207BD516FB73CBB71CE2CBF
AAB9E38D12E909D2AB35B7DDF8EBD5AB40A372D698B4E73170A2FD9BB7924945252622D63B5504E3
EF28C719E5BB135262232625237C74109484914BCEBE4DB05960D3C2F9BC29DEB0A9BA3087E3C490
2BC8D6B4534C877A6592629FAB32D6C509874C49A41F8CC94023C19057380A15EB370E14ACE7035D
2DDA77E325E95F32FB9B7F738FDF1A07134C7DDAC2B43627A725B0FCAD6969D509C5DF9DFCF7A356
4FD7F6F0437B1B2994AD17D6FC5D723A800D2F56FA058DF3273DBB79E9130393270E24CBB14A067C
041FDCFD4EE1865FA9A03F2617C93993F977A2C32F73A678FDB81D8DFB227E2B1DFEC7D74E89AB31
2FDF8E83777DF51FF28DA15FB5C19002A051FBDA9A4134965D73A943408498043AAA7E356D4A7763
4301BC5181533498544F473BA85CBC914A01C39FD5B5F6E7777BC5827D0621F3F4771DEC6A6A6AD2
9BD29E9A4D24E972C3860D4EFA79E8B7ACB4801E7B6EC278A94805566B7D58B49025F2E96796DF74
D34D924C81C701C4EAB1658F5168BDCCB61798792D617EDA33851F6D1894CD2597041C858ADAA567
C36A09FD358C0607D1D6FAF9E3C6FDD26F84CF99EB068F95EEC861CB610C333C18D3C237C65EF4AD
A61F49152AFCBEF497933B0F6E95DD0609E408FDAA2F5B37436A274DD8BD90EDE93BD10F7EB4B2E1
FC84EACC5C57175F726EFA8FCF37370D1ABD72A08B1EF2E9C092211E2B169AF167E6825B351E5A27
3181FE342B7183B2D219B85546259BED1E19F2073DBA4EF5BEB7BFBD6DED3DB7C4EBED6A3212317D
F92F276D0A1466FBAA87937919F2B14BE6BE234CCA49DD72543262095FAC17ECCCF3727290A9AE3D
BB49A6F48EB428DFF6F41EB6B0EA11029C9D3785841DE43DF4261952A03F56618729EAB0A0B42C03
F719241430F4D6667D81025E2925E5359962DA8890A912BDD89206666C6DD05E7D7D08817FC2B517
2DA873B8361E0D500C108DAD3B02CC6FD8DE85B4E8F2F15D4EB8237B032B26B5B22FEAB8396A0037
FDE8FB3347E361E95CFB92B69CFD3EBEC2DF3EC75CF71F803186DB077A85D19511BDAABAF0AB1933
0640AF0E1D395781A6BB935A3814DB8E635D366D505D10ABD091ACA03C6EC1B1AF3F5EFE413B53F3
9C1257FEFA56B8A70FED6D04B77236DB53B4D060C82138B440A37EBD7A95A41A89D740B312AD7F0E
4AC2302DF0174D4C527F4BAE1AEC9CEDEC877F882676C534CF03D3B824BA6F4B7D5D0AF7197819E3
37287BABBDD59E88748C4C392E2D7B42B30338789CAA575D8881147A930CC3223A694484D6590354
7BAC58948D7443745DF4078B1F0CE2DAB06103A5D113FD0EE3CC3355247014CB7378E6A4AB43EC7D
0E7E56D0C0909098372C1360DF181EFF7B67AF84C6252B04ECE8C895EE58924579D9DE1760E84AE1
21BC009FE2B4407582152B56E0B142EF85DE66E0AEEE27BDD646E2CAD0DAD4B9397D27FAB7BC3EB8
64D1805F277F587A45EF55B274BFD2BDEF9982C40AE613F8756DDD23FAC1F1B464233B72A4DEBA67
A47FC0AD5EECDD3CE9DFEF8B578EF6255E896CE098976F27B7D2DF5E395D6DC837FC9C9DDEA84D0B
E7934649F5A8781A548C9E4473A362AE9F961BAE038DFA68D78E003F510103A4BFFEF88F6FB73CFB
AB6953E2095C8EC6A00A53649B71C9FECD3179D0B8232D256674E90D268C30AE29F5EE8721230811
E6CDB22E2D09E8DBC4A74684C4B9448A33977B8FA5A2D8C0AC4998CA30FC682A6BF5091612A200C5
D0708F7F495C6FA9E3BEBB126CD7DEBEFC3AAA24F095A10BBA686F74C0D73D42BD3BE1DAF21B6DF3
D79FC53B7B8E7D117A378EA443F49F9EE0036BB9C60C5D58BFAC0EAC2927656D3063888C7CB213E6
1BCB3F68D736F9CB7FFA4D9A1FC4FD5DBF71E0FE07D2AC299C70C767CC18686D3B5F99DEABA1D85D
665E2AE652CCA8A29C2CCFD1CC79B7EA9976C867BE7E3EF065F391F53FD83A5FEBAEEBE040119964
4C20B815785CE8176E283F9CFDF00FFB376F058D7A6EC2F8686E944E8FD250CE1D10ABD5D75FDDD5
F028685446F5EF723569E807AAEF4037C8114C71AFFD7E3F54E29F386CFBAA8745D83C90037A9EE5
885B6A4D4BAB5FC58B021422E4157A526D39E1D4F14FF58DDB386756E84D320C8BC0C14F6225E215
B0A0A8000692851553F61E8BB3108F5EDFC5541661435C8EB8A57440A36DAD840E98A6DA5E5DD79E
3DA54A7137FB6395AF39BC319EFD614B2408A9E35DF12F7F7A4ED5855FED7E6720D957FB9B573C70
A850CFD90FC6BC7CBB94F485AD0B8B97558AA85FC17203BADFE8CEC64387BE450F17FE1A771EEF62
2E15EDF0657B5FC8E2DA4196B7BC3E0876FFC3AACC8A5E815ECD99EB0DDA7C6F05142DBCAD862F7A
75894F3FF5A65B33D993858FB71DEE98BCED4172ABA88AC5C61AA6F2C93BA4F91775DCBCE04063FA
C4D960D090A7FEEB8FFF7874D3ABA451BAEE92D463920C2901DD52F4ECF474B417434DCF53BDEFA1
31B88468CBFD4A83E24D8BB4FCED9667C52715B84A72BF11B33D1E676A34E17116293F1188F6072F
25EB6743A6D8D7D6ACEFA0A75A6F9D599AE08DC333456EC234ABB1A32F93824D52CA0AEB239E2958
50E1E61F91463159925E69EEA5C89ACE068BA98CE3CD2D556C70D294962C4A2AA69729FA632AAF5C
1D3012C0AC35D17676DB7408688A2CA1C042C06055B89034C74FB10D33690F55D3A3B92D91178CB9
920810CE0C94B0D049678CA8C79B4CC314A77601AE14B783817FC45DDD4F8EB02BFA4EF48322815E
655A5318F4EA8ED9E71D7A556CF73A4FE88F892A8BFF57C2C82F8895C6C6F060CC03D87720BD3AF6
F5C7CD47D6835BC5BD5431A79553E20AB86CDD0CDC6B10EA0AE961C3C8018BE5E8CED7BA1A1E05B9
70BD518E4F27F1F7EAEBAF06F9020503110BFD2AE893DA525FE7914127C32BA89E54DBB449605267
3FFC837C5C9F4ADCCD98B4294FB468419DDFE9CC786FFF4269C8393AEF9DAB6F9F78F90D250DEE57
D01C7576F5855B89F984073CF596C5481B73FA0CCD39EE9F80D09143B1610E8DA275877FE9B5DB66
80A205EE8EA3F990C519B056CAF0C0B42FC303561C06C3D8D19769FB1FEFC80231EC6E1BE18F5972
E21593B9D89C0F661120E7041F0EFBC14C8F94170FED6DD455A846AD9EDED1B7D3DF545ACEE456CB
573DE5C484687A857FE12EE05EE048DC173A88B90F935BB605EB5AABC665D4BDA9417A05A29429BD
FA61D5398C902DAF0FE20CE9B7A40CA629191E4CD1E514ED0C0F0C18BCC91504265C7BC72B385886
071E2E10F9CE835B67752DC520748A5B39D58435B70AFDDA0D85C7F053F7679F7CB46B0783E25C02
95F84272A3E8A80261018DA237AA181ECCFE48296130A9A66BAF88A777290E25DE285EE39ADB6EDC
DFDEA6991435B8703978CA7A7A0FE38963D886EC484BD2874CE37C48B9855E0C9D50F64027EB510A
2E1F7A930C3907539358D08D9E20BDCF5F53533373DEAD8F2D7B6C4D4BABB69DF029D626C8D47CEA
8FD50682D1BB68411D1E6A7C2996607CD1B8C80F1E7CBD466312187BF9A5A45168463C3A31E2932A
C218458303AD1500DB755822203AAEB5B5B5B3233F182135911F0E0F8E10BD3AE07D3AA424F873D8
F8D5408F957C04037BE913AEAF0A58FE4C80F8769E528D467E2A3903D89363BE3EF0DBA7F4E3E377
DB89F1CCA02F2A41E9DC34F63C7E703BD0FFBC417890712370640E9334276F7B502B1BE4F67997B3
F5F60DB6B69D07BDCA885B7D33A26172FF0303AFBCF125E855E08D2BCB094AFBAD609271F970026E
6584607860029F5A7D4DCDC4A924E3181E1851605BCB3F68C7FDD58393AF35BD624C20B9D55B277F
97D16D359410D2DD86FAEC93A33B5F038D6AB9E1BAC010384986725C3CF446818968C5863CCD27E9
1E7FFA0C7D5222979150524A5D9D5410A6378D1A325496C00308BB08161A1E433C5678B8F830326C
C3BFDDE1EC48877EDF2B011C1818B79A2977353C1A7AC30CF983845145CDA789532FBFAA4AD32B6A
B0E35F6437E0447830713C9E68615BE9ECF9C066D60124B08D1DDB8C4F3D9338707ED9E4D427B745
B3F8C17B945AA1C27F1F199B8AFB8E0180F1C611282304CB841866FF7D719D333C52BB2C03FF95E2
F8E6A600131AF6B37C2A1D32E51FB18126778A53A5EF650B3CB8F78B5E5DD5974A6B4E4592141FE7
FEA7D02BEE7C32A5115301176E3A0AE5C789C91FE1285A70A0512B1BE860C59120C5190E1D39477A
95694D61F15EA55351BD6C6630CDBEF5C6B86C86C846199F658930DF75B08B097AD4099CB97D49BC
A0554CBCC2EFBAC2609EF7C63270AB6463D8507EC014F4D1AE1DDD8D0D1BE7CCA21347E716B94C44
3977965D7369E7BD73F7B5359FFDF00F81656DF3D1DA6123CCF1A2A7A35D985440449F6282380657
0D26C8B8443E2FE053B0BB18C547234A52A2E407D33226675A538C2848274FCA90FA9E8E0460F420
FE522D5A24440C6509FDBCD391447A45DB09762C1E4FAE8CFC91585CEE7E485A533A4694EC70828B
0929C3379297A56921971CCAE95AD287434CD66F4C9A4EA54720870707618AE191DA4F9A9A4325E3
32FA8468ADDF729E3377D0FF29FF4986ED9961A9538A33A77FF0F8CEF9629D72B71F2B7246A183EE
394F9F1192D577A01B863103F5719B5E8CFC382197231CF6CD47D68BDE3B4C6ED8DE19B5390B48B3
61B41F7CD7734D4EFF71C6DEABC91307962C1A6070A03E734604B94421C343E275DF8CFC70846078
60D804A6E3A1C35FFED36FC0982EFDE5649D43C7BBAFA50219130816D6D1B7F3F8C05F43BF6443CE
D1FFD9271827A0515E1DDB98C2793C494A05F54575F0987914E320A0517EFDF0C2353E6881C0CC79
74D3ABA0785A28C3F149C5C5DB237270383E78155399E60CE7B833F243FF2FB7BB25E6966B65EA46
1AF20DF436F3E378AF314AADFF2B10DCA3E6C38BA5704D4B2B1E553CB0F5F5F5F42F9370D19DE445
E40EA711A1CFAC216F867EC96586D0BBB47B4F82A9B9F48961142A1C5E23A4A93F8D38CF8CB855E0
97FADFD9BAC32F65F0FB193306F42295A66B2C7D3759FA772DF5253FB4B7516AAD32A4CA9F42352C
3B4BBFCD9C2ED2210EE963E7E90FB597EDF19ED65C9D39D3DE06BD6A6EF224D633F25EE1E009D5DE
B0C783D01F146919FA139AF3CEF4B37BBEC32A6FFE82A181CF1D06AAF8AD1CDD751D198803845BD1
6F1548E4CBAC93CB15181B277BF675353C0A1AF58BB1350CDEA3430ABC495E27B08F982B87DE28AF
A06D483EE264E704A3E9E968073FD2C95CBAA494F8DD7875B810AF8E7092AB48785822E616F72E74
2246198FF652BC34B61923535764C65D0EBD61458852BCBF23BFE4A8E114D9ACE6E3CC3F436F9BA1
D8C0A1A22D4C58A4A1B7CA69E1B06FEE39F64560124D916B6B7B55A8627BFBB43F47CE470A3FE3C1
4E26A5E225C0780EBD6331A477BF33B0B2E13C067316DE2BD02BB033895BCB82D29607D2B94CF412
86B157F1991AEC11E97547C2425EE01872ABD02FCD302CE2E3FFB34F40A318D4B7EC9A4BE987D215
780364CF639E1D302F7C101F17CE9E511BF2FAA09DFDF00F60526DD326C51D522A1CD1B93A50C22D
F5756052E9B733871E79430180812AC30083FCE8A657436F92A16028FE07ADF85B6810E872AB785D
54491069BABDFA4EF46B9D0D6155878E9C0BFD1202D173F683A87327520308D6E64F5EABF7F77C49
3C47635EBE5D8CE78B3A6E2E9E4BE0FE52F79E41D02BBFF6FEB0DE2B7C041FA4F72AF44E2E124857
386315EFEF3CFD21B8D565EB660881D2BAEBE2B422C39AD5B5F4C5DECD4535D51804B89BA77ADFDB
D7D60C4E0436C1803DBA69A2D42989D81D6C51D228CF1BF5D927FA8429BEAB909786EB7ABBE55951
20749960A22C21AE7DD3C2F9DEB59853B5AC811BFAAB6953640C78517F6AF41A0C0643FA7084D44A
4B6548B32ABFD97CF125E7601287DE483FC677CE179B93BEAA635F7F1C7AABB28316FDC385149BBD
21EDA1F76AE91303575F9F99EB8ADEDBE7577AB185C57675C506B0A4B74EFEEE81DF3E456EE5C0AF
163873FB12E356C580FE889A2B6854E7BD73E9A68906F225AB1BA5DE018DDAD650FFD1AE1DC306C3
E4CFA1931A673FFC8330A904A5C12045F7D5D75FDDD5F0E8C8D5096CAE287E44E3903FFB84D19EA2
DC187AC30C06438902B6A2B61E5F79A354CD1B4C8F81B15E5B77242D59558026F9DFBCABFB496D5E
8E5A3DFDE53FFD26F4DECB1A8FF7B48A9D8CDF692A69870B10708C0AD0AB09D5D9D3ABD0AFA29841
1DE9595D4B59B92C814CA9126CDC58C023F093D7EA9B8FAC376E5548607602D738D2DEB669E17CF0
8804374D2CAB28AA389128D400E313F40434AA6BCFEE62A8C09B0C0C59D4B2030E8D8ACB4D44CAF2
8249F51DE80EBDD98602637F7B5B9C6B575DB8AFAD39F426190C8612C596D7132CC6E5CF14573A55
468021976621E0B0D076B843271FC1BCCC87A44321D1D1B75367D088E85FA96CD262CCBCF286573F
3A537A5575A147E19B9B06855E5980901FF45B2D3810576211B540C76925E9781842814E5BEBCF4C
3B243013F0D4F14F8F6E7A1534CAE31AA41589316F227B9EA07D37FAA2961BAE4BD31B156E9F8049
6D5FF570BC30962FC94B6A639149BDDDF2AC247C192A106BEFB9458F93A1F7DE09BD490683A144D1
77A2BF98152AB24060216011332CFCD229DFD873F603BD690F0372F2B6074369520E71E0F3F7C51E
C6EF87F636065E7B4900CF02E8D5FD0F0CF853F352E3CAEF0DDE31FB3C987B6F5FC6C181A5D54559
0397B9EBABFF78BCA7F5A28E9BE99F0A8C0C94FA5678349271ABCAE9B45CE1EB8FFFD8D3D14E6F94
54DE89CA9BC724D0C1AD24B788EE1B1C40DFCDD19DAF49F995E2EC79B03C703D303ECFDD16BB22A1
54F13F636A846B6EBB717BF3CAFECF3E29CECB31140C18009A743F3761BCF3DFD05B6830184A0B5A
771AC661E8ED19395ADB026A15DD77D730858CF30D2A3944B7EB37D6B00A55282DC92DB498F60FB6
CE0FBD3D3901C8D1FA8D0318333FAC3A9729BD9A337710235017BD320CA971BEF3F487CBF6BE70D5
BFFEC42156F10443FED9F4233C23A9B9552523B5EC030B2D6DA9AFA3BFC6A14BF16C295FAA14AC4A
7C6AFFE6AD92A41F560ED4B05F8737C1A4D6FEFC6EB439693D2975816088FBDA9A8B3958D150601C
DDF91A9F05E6528195F3FD525F940D064358708AA552A1A274A714B67C5D7B80B97BC7ECF3A15C1A
BE5152A8249DA424D28ED201EC5E615578117A7B4682C0B1017A85E1047AA5E53133F55EA5387FC5
028FC0437B1BC7BC7CFBA8D5D375F96047809D6C0BDCAAF1D03ABD0BE128105ADF8204C59D35FE98
37059D46441AB569E1FC9E8E76D2A862E050C9BE9D6C119709DED774ED1501695F4172EE7269C573
45866200867D340A343260462E4B6230182A1C4EA49C5FCFA1E4C0B5B27B8FBF10B0577BA8F0AA86
CD47D68B738AE57B1EEF692D9B051D6C91CE055C2350720E858C6E04E9D59CB98319D51426BD5A5C
3BB07EE34032EF55D98C8734AF48FF17DC0A4F04B9B98405FAE52CC8ADC677CE07B73A3EF0D7D02F
B018E079A376BE061AE579A35495A804D294584C8AAF41BB48A34EF5BE97E62D0BE7021599C295A2
CD6EE928B9A8448FDBC639B3707532DB272B0120FF2DC26B37E40FBCDD542FC1E0C1F3E2C9A7976F
9D7783C15018804369DBEFF995E137295738F86E80E93B6ECC60C182B2302DF79CFD4087C6F94BE2
96F4D48DC6C3BED50E859DC7BB426F556170E8C839D22B3F734F8D1F569DBBFF81812DAF07D42C28
E9C1302C86BDBA039FBFFFD0DEC6A8DF73638D96B370442D0072ABDE2F7A43BFAE02F79EA40EADB9
EDC6E7ABBECB3D763FA7D03944B01869348268EC6B6B068DD25425EBFB5598AB0693F23404FC9585
237F462F33F6AFCE7BE7E2F86268B9A1C871B2675F7CABA1EAC2B53FBF3BF426190C8652072C436D
EFC1440CBD4939446FDFA05F6DE0CAEF0D164607FBF3812FA518AEA450959958F45B277FA7152A60
E586DEA4824122D0309C9A9B3C75974CE9150627BD57A555122EDF00B75AB6F7059026874CF98303
C56F5572EED18C8031D677A07B7BF34AA1516452716F944FD44E4AD682461D696F1B7AEF9D12221A
781CF6B7B7E162499AE47ADD38C6C83BCBAEB9D463529B5E4DD3CB5042FD60C81FF034E9E705E327
F426190C8692061797F253A8D0C0EAEC97C82E4C21E0796F2C8B4B52448C40D88AA177484E206C02
0CD17325C402B4EEEA7E32F4B6858BDDEF0CB4B69D775214D30146E9CA86F3189622B056E1C0E802
517ABCA715A4893B12C2AD1865AAA916FD56CB3F682F1BBF15695477630338914BA0C459A3C894E8
333C37613C3E024A828F97167DC0C847B379BDA9CAF22A258D92BB464391A06DDA248EB147AB46E1
D9B1516430187202C7FC13374AD94C32B8A2C042C05B5ECF3DAB924E6B3CB48E76A0B8A896ED7D21
F4AEC8ED35F2753CF965630DCCDAA1321A3923C4A123E79A9B06F17C65947B05BE7FF5F59EF2FFD6
1D19ABB2972B4094F040494CA0D6AFD0AF49B53008F1AC056E5F1479BA042B2BBDDDF22C6885264D
D102528E1F4A518CA66BAFA036782956593AD5FBDEBEB6E65F4D9B229713258F52434ABDC6958249
39975972976C08175F7FFC47CA9B7068E171B321643018720247A1A200EE9BC20313666021E0D6B6
F339FFA2A158389CD6857652A8CA0993B73D288158A3564F0FBD3D45058E07907A3C56CB9F19983C
3180DA0FEBBDC2138A8F070607569A25006E45BF5560652B11B5E09F635EBE3D19B7CAF71D1FF6EE
E87FF51DE8669C5BD4CCF329814735CFD53BDC5A070DF9F5EA55F4D494DE4878EF1D70C0B66993FC
97E64F0A5B7DFDD5DB573D6C3E29434E70A4BD4DEF4E80D187DE2483C1501E58BF31C1CC6B6EF228
5559AE5C290A01E7F07A712A2F852AA6EF07945F0A95C6437B1BE36E82A61FF59CFD20F4261533C0
8C76BF338051E78F471DD67B3563C60083039DE154964F6BEAABA3DF8A1AECDCBED0397D3AF10AFF
C20308221658B92094AEC3979EEA7DAFA7A35D6854BCB2923FDA2D31430A1CA4BBB1A168BD51C3B6
0A2DEF6A78F4B909E3BD0B4FAC8A258284E299C2C59249857E5D867242E7BD73F563A5752F0D0683
212338AB5E6F5F82E5E654C52D03E84236E91402CEEE2BE45B666E5F226148B4E8767DF51FA17742
FEF062EF66AD74DDD1B733F426950AC08CB2A35755177E3579E24073D3E0A123E742BF8A70010A0F
6EF593D7EAC19B1C090B27D92A454C60FEE889D6D63BFBE11F40A360CEC5BD51118B2E41AF4F45BE
49FCDBAFA64DD9D650AFB5EC124A351525B77200AB15E4C893AD4E5499F044D17DD952421B436FB6
A1FC80E7453F6BABAFBF3AF426190C86B201661827D028F426E5E31AE585E39523E6CC7577FEB3C3
F20FDAE38A6491A48FE623EB9D3694197A7A0FEBF2AC8FF7B486DEA45244DF89FEAD3B066B979ECD
C27B457A55181DCBA20588129EB51F6C9DEF6554C51CC44E28A060CCCBB763A01660AF0354E2E8A6
57372D9CCF22382E934A569337422BBA1A1EF568D4E933E94F4DE9A8A3E7157AFF0ABF3FDAB563CF
E2DA68F561474CC39723864B7EBBE5D9D4A5B2CA751635140C1893DA4585A72CF426190C86728293
E251C6516A84538D8B98FEE3F323145BDB79BCEBA28E9BF5C6782528E061B448E495CE1A33E34790
69579CFEE22F20FEF73F30E02F0130ACF7EA8ED9E75BDBCE1F3A728E4F7125DC05E71A8F7DFD31B8
D5E46D0F4AED60D77B15D9EB00ED1AB57A3AB8D5437B1B036302B306A611D0A8CE7BE77A54C2AF56
A7C3F97C342AEA8DFAEC13C7FDA4A94A71427BEA4FF6ECDB525F97E0834B74C6914C31476CCD6D37
7A4CEAF8A78177D339B9C13042E011D39A276058A137C9603094136A979ED58619A389CA7B09EBED
F314E31D8B14162CACD92C2E9C72E230CFB4F1863FCB9E9C12209292CC82AB0EBD3DE584BE13FD5B
5E1FCC825E6178835EAD6B1FC4502FEF673919C8ADC0F175606A607D2BE0B27533E6BDB10CDC2ABB
67F6EB8FFF28DE285D6657270AF9752780961BAE03FBD8BF79AB708A1245FFE9335D7B76EF595CDB
74ED1554CF100EA515FC845B8149ED6F6F03790CBDE5868A82D6FA33F97483C1905B604A81DDA58D
B19CEBE0152760ACFACDD48B2F3927F929694EB63C8CC277824B7F3999420D9520F62BE9630C77AC
1022596078858A4EF4D37BE5DF0D18965ECD993B88C7BC328B5E9DFEE22FE2B76210A05F805D243A
C1ADEEEA7EB2A7F7F0B0A7F568D4CED7B635D48316A5D294D0AC2A529A1646DDDA9FDFDDD3D18E33
84DE39230498143AA1F3DEB9B8AE8482BC89FD20518EC6A40C21E264CF3E5DF26CD3C2F9A137C960
3094190EBE9B60802DAE1D08BD4985016CD4405DEBDDEF447B204D12F4784F2B381455FE68A1BDD8
BBB9129805FB87A27F5286B56C2A1A171BF468ECEDF3281288D2C5979CCB885E8D1B13A55738839C
B65CC9BE83635F7FDCD1B773D2BFDF87A7554A5CE1B5BFBE155E5CD471F3BC3796ED3CDE95B02B12
6110A0516DD326D1F92235765D0F5424B04DEAF33E37613CEC37B089B31FFE21F47E1839C089C8A4
E44AE97B8A7BE8129D71B8765048E9C94A981B0D4588B75B9ED594FF487B5BE84D32180C650647A1
024657E84D2AE4B5FB0B01C34C5DBF3129AB72DE79EBE4EF989A21C16F959042A501FE28B628E964
8598E8C50058A77D27FA5BDBCEDF313BB39AC27CD2EF7F6060CBEBC145AFCA1252290CA3F427AFD5
073AADB4CB1543FAAA7FFDC9DC7FBE6FEDCFEF068D12FAE044F709B1D2257A9BAEBD02A4635F5BF3
A9DEF74A2B312A19BEFEF88F472225B4A23D10A4AD2182F09E33EE9E5BB450A1C1102EE85096BD0E
F3961A0C867CC0D119AB9C4590570AC3D26F703A01904E9FF0CF635F7F7C51C7CD3AE46F7CE7FC4A
DB833DF0F9FBDCF0A72DBAE04063E84D2A7B247B420FBEEB95969B3123E3E040A157206829CE5F66
C065825B4DDEF6A008CBC8301EF3DF26CCF8E97F796CEA38169FD5027DDE3B4E689F72CD088D0AFD
EA86BDF6748E01C0A458454B3814FB248157C67A83FE3830A9F4BFC560C837BC91FCD927E254650C
AA0D4E83C1900F389CA2A2EADD705E0D2A04FC7B29049C6C6319C698E8A5F30548D65065A450E90B
D4BA6AE893D09B6400AFC753BCFC9901D0AB4C8303274FF4AA656DDD5156DEABD4F18DF45BCDFDE7
FBEEFCE9A4BBAF1D4D671389832E3EAB11251431778CDF1B552A08DC2C024E1DFF1417854B0B102D
D4BA13914E0093DA525FD777A03B70E75F777E297691A17421E3EDE8A657F54E08C676E86D33180C
6509A706EEBAF60A7254099A9B02CC4BA710B0E6560FED6D64B41B53A858E836997D52DE18F3F2ED
92E67FD9BA19A1B7A7C2E18C3AF0057AAFA6FFF87CA6F46A42F5E0CA86F3DD7BA2DEABC0F3972EA8
F8DDDDD8B071CE2C378ACFEF844A4C92FAD9D4EA1B1EAC59D8FE70E3A175A15F484EBA822F4EFCF9
D3B75B9EF5A21C63974FEA84ABD6A9520C805C7DFDD5DB573D8C3E4C3624CA66A8184A143202BD67
5CC5E816BF1FD9603094281C858AFB1FA814850A4266DD57DEF8D26F55DE31FBBCBFCA0F53A874A5
1B30ACA14A3521E6BDB12C5AEE27E2AEA2ABCE50302435684F9F710E38FDC55F76BF3340EF5546DC
AAEAC2AFE8BDC2C725B455CE5F42F0B4130F74EF6B6B8689154F7D8A9126D007ED7C71D2A3EEBE76
F48C9FFE978B1AAA9D942B5664EBE8DB993AE83759D1A5D03B64285295787BF34AE68B25F04A5D99
37C62BC1A4BA1A1E05932A9EAB301852A0FFB34FE2F2E95517B6DC705DE84D32180CE58100C985D3
67B4F934FDC715A1A31E08875D4A2894563C0665888A86C55C333FD89A2A85AAECED8DC77B5A7595
9F9DC7BB426F926158805E75EFF19C50575F9F59E215A708D02B3C2CA5C2AA4019F6B7B7C5BD515A
5A21D103E584F9ADB9EDC6179FAD9DFBCFF7511BD02968E5E85A5CB66E463ADCAA4840271D98D473
13C62768F7F96B13C7AA68BDDDF2AC6652064349E0A35D3BB839F0D855DFC6EF6D0DF5A5AB126330
188A1FBA4853D58595A2AB1C88648580A93B0D634957A1FA56D38F6047816455728F8143457B23E2
A56A3EB23EF426193202E915585260598114B8F89273A057CD4D831EBD2A322BE5EC877FD8D7D61C
D5FAF6978EF245F4C99FA05130BA60863999412FFFE9370FFCF6293CEFA262A15F08B7E29C3061F7
42702B7F726558D0D94C7D07BAF72CAE5D7DFDD5013D208EB918996A9B36A92404370C063F38ECB7
D4D7C5B5FEAA2EC4A31D7AC30C06431963CEDC044BA9A2142AFC007B4A5608F8F19E56D84EA3564F
E7EF6FFCAFD1FE14AA4A8327FAA704A8EFEA7EB2C23BA414A18303B7EEF0E89523049A0E66CC1800
BDC263125652212CFF9E8E76D0A8689C8F8EDCF3692CC463FC22C4214AA3864B0BC28B9DC7BB30C8
13D43E637AECE2BC960AC293B73DD87C647D3A7EAB7CF7129814AE5122A024494A3AC7FB574C97E3
57D3A61893329401F05845EB71C706B9C9A71B0C86BCC251A890C24C15081A36302CFDF926A3A6BE
A3E37C80077EFB54E80D2E0670EB9E9D33BED3AAD29709FA4EF46F797D70C9A26CBC57A45707DF1D
1C964D644425FC0753E57B4B7D1D19814BA0821C3112C9068AE1D54E5226568AC6F8753F3AFA7652
A6C6A9172CDE2B3D57805BBDD8BB993A8A3911054D53081D17B86771EDB26B2E0D74CFC593A422EF
AFB9EDC623ED6D8ECD693B2486D2C5C99E7DA2DEF9C2E8BF5D7BCF2DF22F1BD8068321E7C0C4B2FB
9D0493696543E5A653C9340B93E98ED98A697EFF38C5FD3CE754AC0A95FE4825CFCFE80A312041AF
2AB92BCA03FE3B487A75FF03033FACCA4C36F0CAEF79CA99AD6DE7253830D957F8DB90ECC9228DDA
B470BE17BD26BCC91FCE9748199EAFFA6EDBB449205FFB376F4DBD539DE9007EEBE4EF96ED7D0123
5F73281D13A869D784DD0B9B8FACCFB98A4B42DF7EF6C9D14DAF46831E637213F1883E9FDB0E4C0A
FD297D62CFAFA16CB0AFAD590F758CF3D09B643018CA1BA00F8E2043E84D0A11DAA2585C1B259B17
34CED7325F4CA10ABDA94582BBBA9FD4BBF4BD5FF486DE2443AEE02F2AD4DBE739B241AFFCF1B1C3
D2AB3B669F07BDD2AAEC6902340A34615B433D689488210B3B88CA4AF87C52789FF567614A9DF8F3
A781D795CEB5A7899DA73F7C686FE398976F775C54D1820B91644389099CB97D49DBE10E671AC97A
7F866579C9A49E98705DBC48B1AE50AC9854D3B557E0481C9F0EB70D7D101A0CE9438FD8B66993F4
74A195A60C0683214FD09A0C175F5289B954C92C87950DE72F98FF0BADC0004C59B927F406170F9A
8FAC8FA7E7BF34D544FF2A07A057EBDA07E7CC1DCCB4E855D5855FDD7797F759309D001952328BCF
3E39BAF335D0A878A5A444253AA10C4C3C977F7915787F7E3768542816D4E7035F1EF8FCFD406EE5
970A1CB57A3AB8159E2087F1A50FFAECD6DC7623F3A4D02751ED3EE58D923F854995532967832110
9840F4B4806724F426190C864A80A35041813B034CBB97FFF41B2757E21B77FECF6FAA42C016F8C7
425DE49BE8A2E51F587045F9C33FE0B3A657E3C6782185F82CD841FFE933A0512F3E5B0BE35F5426
342900A2C911E27F89FCB7E586EB362D9CBFBFBD2D505421CDC733E74FB1702B3FB1D285EDF8F88C
EF9C8F6727D06FE56FA730A95F8CAD11879DAEAB45E909F61B7A129DE3258E0D972C96FA804A9EE5
0C258723ED6D7A1F665F5B73E84D32180C9580E5CF24A4536D79DD289587DE2F7A61F3307F2A9A75
FEFF2EFBE6F78F8BC4594E32CD4B1D300235EBBCABFBC9D09B640811781040AF5ADBCEA74FAFFEE1
FFEE9E75C9E29F554D91FA50FEE83E9D24C5FFD2E7021A75B2675F913F7DDA6FE508B03B7E2BBCC0
618D87D60506D09EFDF00F6FB73C0B26A53B449724D6C965CF4D18BFA5BE2E8592A1C15096C080FF
7CE0CB780D85C804A2375BEC8930180CF9C3D61D09164E652A54E86D58BEF8FEAF6F95603F6F27B9
E947C2A724EF2CEB889D720273F3D94B3FD86AA27F86380EBEEBD1AB3B669F77E855CDB7DE99F4DD
FF114DF3F10BF4F940871468D4C639B34A57E2BBE7EC078FF7B48EEF9C1F2F6F1773EF3AEF8C5A3D
1DF30FB8153E82490697ECC941C724D0FD9453DEA1267CDF81EED02FD660080B58C19DED05A35106
83A130E83BD1AFAD9D19332A5AA182108564B1739A776FF0A7E44B21E0618BDA9431580159FAAA12
2ED990028103E0F4177FD9BF79EBD3B73DFD8BB135BA2C5480CA840A5D23997AF8CA6BF1C10D0DEF
E024E531BA40944097A2F956B1F0BF787471A46AF0F79EAD99F1D3FF129740974A523182A9BD5420
5CDB573D0CA6D97FFA4CE8576730848B8F76ED68BAF60A3C2F7852F0806CA9AFE3FBE5317B180C86
228756A8A8BAB0E2A61D67A67DB177B3B3870CFB672849216074D7C1770342252B67F65E70A0311E
1BB9FABF1EF8FCFDD09B642806C0BC3FD9B36F7BF3CA8D7366C1C2911A31C1AE285D9F77F445775F
3E7BF67FFAB77FF8BFBB9DC76DF2C481950DE7BBF70C96B4C6824C0E785830B7D021CEC7874C4A5C
518C847412CA844CB54D9B84EEED3BD0EDAF9C15FA351A0C61011C4A2284F1020C2BF426190C86CA
8153DC360BA1E3F2809709F2456F5CEEB8E947DFD8583379DB834392C17DFA8CBFF82958D5D61D95
CBAAC840A5D33AFA7686DE244358C00372F6C33FEC6F6F238D8AFA56FCBCC9A7820E12F1B3AA2993
BEFB3F148DFA7DEA24AC09D583A457C964034347FACA186FBEF9E67F5F5C97507E37B1D3E2FF425F
4D1DF7C4CFEEDC75D055D72C92AB3618C245344A36F61099EBD660301412B04CB4AD1248102A019F
0F7C49852E06E4E0C5451D373B325C38C661A029643D2AC1C8D979FA4371517DEBA5A98FF7B486DE
2443FEE02F56059CEA7D8F34EAB909E31D7F930EE7633E94F6B6ACB9EDC6EEC606FA594E7FF117F0
23CC45E04A55177E95912AFBF41F9FC70777BF3350B44F9C5FCD06BF3FDAB5635B433DE50DA30EA9
58EF3D5A354A4748E2F5DDD78E9EF1D3FF725143B544097EFFD7B72E38D0E8F8858BB6070C860200
7391ECD8E039EABC776EE84D32180C1505D0016DA23CBF32FA7EE5ACCEBCD2595D4B854F792EAA7F
BB0E7C21B013EE7F20805549BF39A72D6F78D5A25532C8CCED4B426F92212748ADAA7DF6C33FF474
B4AFBDE7166104E24CC16F4A4FF04D2F212862E1F04F2F5C6DD5C34777BE26F1697E8D6ED02BF0A3
A54F0C5C7D7D0692EC920DDADC34E80FC72D9287914C6ACFE2DAA8507CA220BC14DB12F1C3C7A68E
BBE1C1E874C47D1E9D75253A818FF7B4825BE92E0DFD4A0D86C2E3ED9667E3B1B25517628ED2FFB5
E7C26030E41BBD7DAE4D127A930A8FB6C31DCC6810738529544341F330DE71C4E789DAA56743BF90
C24317DFB96CDD8CD0DB63C80730E64FF6ECDBBF796BE7BD73575F7FB5264D09E2128A1A88970A34
AAABE151AF3AD2E933A98B1CF9FF057AB5E5F5C12CE815BD57A057878E9C0BDD94F28A176F7AD5D3
7696EE8A758E244CC5E32123FE3B5883A77ADFEB39FB41F391F533B72F896E5C445895E65381DC2A
F4D1623084826871F0D87364AABC0683A1F0D02AC7577E2FBEC11BBA295218C00811C9F461BD2D12
B7D3DA76DE6FC84921E00A01FA61C2EE8512F887DF1532662A042C2CBB69E17C2FA8CFA14B2A6141
927D247AADE586EBB635D48347804DE46448E0247D27FA41AF6A979EF50BC5A406E6B4193306F0C0
068AC9E40FC2A4A2AEBA405D8ED86B1C83233DE2F9D927FE531DFBFAE3B6C31D93B73DE8C803EA59
4BDEBC6CDD8C87F636BE75F277A18F1F83A16038FDC55FF4C3057AE51C606B93C1602800EE989DC0
0ECA55A1227046FD7CE04BAFB8522CD88F295469AA66393193E2E62B6945B24CF1784FAB6CA1E3B7
1972A50ED2A83D8B6BA51C92CE874AA1DA07DA05F285CFE20CF96E24E6A8F51B0716D70E68C1D274
70F125E7E6CCF50A66B102423E70E2CF9F82497965791369A656E79054A928938A70CF744E0E6ED5
7C647DCDFAC75888DC890374DEC154F6C06F9F9247D24C4A431903338F76F8FE7AF5AAD09B643018
2A104B162584B1ED7EC78BFD2BCBF5D7B928FC7957F793E26161C2424FEFE1F4CF73E8C839A79229
70F5F5A5ADF39C1144F48FBFF167E84DAA6464F7D862B87EB46B4757C3A3ABAFBFDAEF84D2906249
640A4DD75E011A75A4BD2D57DEA82C0072047A75DF5D833FAC729FC461E9D51DB3CF835EE5641309
4492321D921D9F827EA2DFB6D4D7793EA96C3B8DDC6A7CE77C1DB1ACB9952E258CD798E89CED8ED4
4198064369011391DEF639D9B32FF426190C860A84E36D59FE4CA5A453351E5AE798229242952660
938055F9B7CAA510F050108F0BFDC27308864D8A51F7D0DEC6F2BBC6B204481045E7E88D9262B2E2
43898BCE2556E6258D3ABAE9D50278A39236DEA7A147E0A10345BAFF8101FF46476AE08105290335
4BF6D826C3A9DEF7BA1B1B248923EE9352DE3D4F68029D19E9BA3D8B6BD1EDC93CE6597405B85547
DFCE99DB975CB66E864C650EBD12750B1CE3E75619F5B6C15084C010D5E5A8565F7F75E84D32180C
950947A162CEDCCC8C8A52042E0D5CC0494C98F7C6B2A1B40D1BDD3927FEFC69FA8580CB0F9A528D
EF9C9FA2A30C61219A03F8D9274777BED6D5F02828008DFFA811A2294050741FE3D38EB4B7814164
A42F51F86B14905EDD31FB7C46AAEC42AFB6BC9E1002EDF8744E1DFFF4ED966761B9C5E526843AC5
3A4D52CF9E9B307EFBAA8703995416AEA264C763E27AF94FBF99BCED41C6043ADE2B8128C9905BE9
E9AEBC377F0C658C933DFBF4E6CF9EC5B5A137C96030542CB4D5F1C3AA73A1B727DF38FDC55FBC14
2AD9CEDD5833E6E5DB4762DBE0B5BF10F0376375BE020BFA940DD8930C3432D1BFB0906C687DFDF1
1F41A3F6B53547B37B9CC428658788E0B904A76D9C330B1F048D4AFD15C506A79D600DA45773E666
967825F40A4F31A9C7B163C7B635D47B3E292D6F18798DDEA31CFA3FC56A1C8349E160F65EFEBACE
7F6672ABB84E604C3D46CBC888303B1E581CE9702B83A1C8E10C7B3C68F17D8CAA0B31E385DE4283
C150B198FEE304858A324E05E2540C2B42EABCD0EAC889F8B023F4415FD5FA8D651E4839AB6BA9CE
4773EA231B0A06CDDC4FF6ECEB6E6C008D8A964052EC29C13395A8E0877FADBDE716D2A8FED36732
FDDE92C0C177079B9B3C1999F483036BBEF5CEAC4B163F7CE5B5F19ABCAAC7E4358B4979B5B79A57
0A0F1DF2EDA814ACBBC8ADE6BDB14C6B59481CA01325486ED5D1B793DC4ABBCF4AEBFE1A2A100949
A0A32FB2FD0183C110226A979EF52B54942B1EEF69752C8AB6C31DFA809198109142C0BF774C3298
70A15F75FED07C64BDB6CD60C585DEA40A041810F37A3C8504CDA174265424432A1E9C16A9CF4B6F
143E081626342A87C93EF9BDEA113CAAF82CE8D5F32B8740AF0299D4A88B7F4326E5E49AE9DA5292
71D672C3756FB73C7BECD8B1F41B3672AA922C6ED0EFA763BE9554337732AD1C5D0BEDB7323E6528
7244A3FE6273DAA6856EF0B9C160301412EBDA136C89D6B6F3A137291F8079B0F3F4879A4CC1BA60
0A550E1158B26AC9A2B265A9E0503A2B0D0CCBCCB0C280DEA87D6DCDE044716F5450D92381508335
B7DD081AD577A09B34AAC26F192E7FF73B03F45ED5FCEDAFC1A4DCB2BC4E97C6D2D040B8EEFEC1F3
8B6B07BAF714CBB649EA5B09AE84194F2AF13965ADF4838C6370A4F8AD0C86E2C4AF57AF92FD223C
B33D1DED153E9B190C867071ECD8316DFF977AC9DA1419DC4CFC918DD949BFACCD87C1E050D4F2E8
D56438FDC55FB42536AB6B69E84D4A8D881ECBEFFD08EFE36FCBC787350670C0A9DEF740A33AEF9D
CB283E5D81577BA01214FC148D4AA63857C94087A05B362D9CEF315347A923E2CB4BE8D5AA0BEFBE
7CF6A4EFFE8F447FB4F77AFA8FCF2F7DC2A357C5DFC33B8F77DDD5FDE455FFFA9364E583E30C6B63
0D6302436FB3C1A0C1A7EC57D3A6E824D01095480D06836128C235B442C584EAF231FE7546C0E46D
0F6AFB01F46AE7E90FF3F48D5B770CFAA5C6020B0117BFF5951A4254D9B7E8E4D09B941A1152F3B6
7099D8EF8C3891F3F1B7B3A254197C3B69D4C639B3BCAC01ED407162FC74C2D4E88B7E31B606F646
57C3A37D07BACBDED208D4D30B7C3F7EC0679F1CDDF4EADA9FDF1D65528A4C45232423DE28BAF61E
AD1AF5B3AA29357FFB6B7F646F20F0B03FBF7288DEAB74C2F342011EDEB74EFE0EDC0A8FB088043A
D1805ADDE227AFD5835BA5992F590C1768286F605AC3C32BBB466DD32685DE2483C1607004EB4A3A
D82370297FBCA7556C836F35FD082F7258973630957BF73B037E5675F5F5F1925565831F6C9D2FDB
DAA3564F2FF2C193DA4F94EF8F9F3AFEA96652814E2EE6C81C3B766C7F7BDB9EC5B58C31131A15AF
1895E89CA23C025EB4DC70DD96FA3ACF1BF5D92772C2D0BBBDC04871C9A7BFF8CBD19DAF75DE3B37
5E905739F5A28EBF98F4C4B26B2E5D7BCF2D605EE84CF0A3258B0602B53D53E38ED9E741AF0E1D29
5E3155B0A49EDEC3516ED5F4A36F6CACC10C89173A469AE0633E73FB92B6C31D814F7A050E364358
E8E96897471813E0DB2DCF86DE2483C160B8EFAE041BA0446B2A25AB01DAD1B7D349C17EE0B74FF9
0FCBE1B71330A27E58E50A8BE11D875595BA11F2D0DE46BDA72D9BD8C5795D3E3753D6946A241F77
1A10FDF8D90FFFB07FF3D64D0BE7B3E6116994AE76A4F3A1B4970A346A5B437DD79EDD42A3FC28CE
DB916FC4B73B3EFB041475E39C5951C7932F42525E3C5A35AAE9DA2BC0B9C8A4FCBD075EBCE5F5C1
FB1F18B8FAFA8C85D967CCF092B6C29D6087CDB702B78AEA04C6B42C447A5D2761E1F7E46D0FBED8
BBF9F8C05F873DADC1907330FE59589556DA34180C86B050960A155CE27BBFE865588BA4068CEF9C
9FEFD55FCEDF77A2DF5F08F8CAEF0DFAF3D94BD120619B4B4BF4CFE766CA3AF02F8119A5D957C918
59349147EB9C3B067F22050029689B3669CFE25A98FD27FEFCA9FE8ACA84B39D227FA2737A3ADA59
962B2E20EFEFD858352E7429D3CDD274B6E20107BDBAEFAE41FFE6496A545DF8D59CB99E1CA878AF
8A41B75C2E1CBF777DF51F0FED6DD425FC2402D02F153873FB1270AB643181591435361802A11F13
3CB0E29D7F6EC2781B630683A11870F0DD84E57E716D9928D4C130C05ACF5016C90B00C9720ECB49
C998641FEF3F7DC61F2C04834A0A0197FA42F0D6C9DFC5F7AE37D63CDED31A7A9352205C2F55E441
F313BAB725E4ECD1AA51FED03E6101A051205F6FBEF9269842A90F9BBC82D967E82EFAA448A612FC
7D4A830286D996FA3A4FFF70645D9A35BDBAF27B83A057EBDA0B14159CE9656A6EE5401E7CA96FF5
83ADF3C1ADFC73ACC1905B44E5D3634FF1B686FAD09B6430180C4365A450E1580B4CA192F0954B7F
39B9A36F675E6B6E260B3E84C9E437A5604485DE632307068FDEB206870DBD492990072F55403294
F32912E7A39B5E85E9EEA763009854DC3315D397E09FABAFBF1A34AAA7A3FDD4F14FB31BBA9543BE
C8A4D0635A085187066915C4E7268CDFBEEAE18F76ED48B3A3325299C03859BF71E0FE0732A829AC
BD57AD6DE741D00A7FE3FCD3977E87DCEAA28E9BE339564948D6B75E9A3ABE737EE3A17556FBDB90
27EC595CAB9FEEA33B5F0BBD490683C14038F169A1B76784802520259324F87FC181C6743E98936F
F7BF192904EC5A50CB9F296D8720AFF4B275336847E1374CA9D05B9502415EAA91A8A0A752FCF334
E576BED6D5F068DBB4495CF77F5635C5A74DE1FD9960F347FC26A051FBDBDB2C3B6058B03ED7DB2D
CFB6DC709D2B87E80F9E8C302996362E580B3166D6B57B221599D2AB2BBFE7F9BC0AE6BD4AA7ABF9
E2C0E7EF0BB77274D7B57220FF1CF3F2EDC6AD0C3987277FAAB64D2A67E3C86030143FCA43A16228
B6EE6305C752AEF751276F7B305C313A342CB01070EDD2B3A177DA083173FB12BD475DE4A27F7945
FFE9331FEDDA010B9FF93B6E892847FC3CA6558517A051F8C8BEB666A3516902B408E40814C95FD4
D84FA960806D6F5ED977A03BDC36935ECD981120073A2CBD5A5C3BB0E5F5417F218610416E859976
D4EAE94EB1607F05611CD67C647DA973ABAF3FFEA3EBF78CFCCEF2B311A4F9D5A78E7F9AF5579719
F0EC6B577EE7BD73436F92C16030089A9B125670ACDDA137296B80BC8CEF9CAFA3532EFDE5E4E303
7F0D7D230BE6D0FA8D01BEAA3B669F2F2A4B29532C38D0A8CDA79EB31FC88D08BD6DE920A398AEC0
B83ED8EAB0D849A35209CA45D4B9A3C640C41BB571CEACFDED6D85749A9434E893EA6A7834AAE6E1
58A74E8747B2CFC052A9E0912C2837147C3EF0E5C177BD00BF2CE8D50FABCE915EF59DE80FFD8E10
BBBEFA8FE51FB4FF60EBFCC068C0B83A508C5B351E5A27B34469C1E335890F351FE7B43EDBFBDE48
3891512A01E6CC787664D585F853FF37F4A7DB603054381C858A258B4A2F204D82FF1FDADBE8C4F6
872E43A727F9AD3B02CC245856252455E1B4F3C5DECDDA82CA61CDAFA245008DF20B4A28A709567F
3966ED3DB7C0CE1FB91E42E5001DF5D1AE1D5BEAEB9E9B305E34BE68CD8ADABCCE47C34D79BBE5D9
52D14244F30E1D39D7DCE44D0299265E8D1BE3CDD5A057E068C9EA1D170C68C3B1AF3F7EBCA7757C
E77C96FFD34E2B27030BFF9DBCED411C5C5A5A16099E26F522E3CF66E1A51A19232B27446300623D
50F6A5CC0D064369A1FFF419BD584FFF71A9EAA8833DB19EAFA85214A1061D08EC95DF730DA409D5
F1A81EDC8ED01B993E6014E9509F657B5F885E45719BB26942CBF68207810D8113C56994CF33E20F
3C83E5FFAB69537EBD7A5560E05979F4D2083B36F880CF3E418F8149797913B1BEF5A853AC80948E
A26CBAF60ADC97FDED6D81C5B94A68BF62F73B03A457997AAF30812C7D62A07BCFA070C9147D9E6F
08B74AA662A14B09E3301C7CE0F3F743EFFC61E1F752A54F8B12DC4CEA0C197CF5083E5E36C003FE
F095D78AAFBF6DDAA4D09B6430180C0EB4420556F3D2CA88912A54DF7A69AAC499B018658AE3436C
6D6F5F80D8F2C597B885804B02B81C4DA97EF25A29E9D90E2BA377AAF73D18EA30D7196C26A2DCD1
3428BF7D1533FE41A3DE6E79F664CFBE142167A562E7E7A9DB03DFE93F7DE6E8CED7D6FEFCEE65D7
5C1AE7ADB1BEA52925C40AC7ACB9EDC69E8EF68C76AA4BA2DB4F7FF117A1575978AF56369CC7C739
8D17F27A9DD14E6EC5CC5681A362212F705891732BD74B9509AF39FBE11F2C976AE4C0C3FE4FAA7E
1FE6D8D09B6430180C0E1C858A92B3EDB17C3BC1FC97AD9B215125459549C136C0649A50ED9A4360
553084426F5EA6802D24D611BABDB4F8B8FFD6904675DE3B37C1AA57B63DDEC4BF1CC50970AEB669
93B6AF7A188C40C65849774501BA5ABF3EBAE95550D75F8CAD89D3D5C4F8C968745F8458E1EEE078
F1E7861EF696EF8EEADEE3B1247F91BB61818FE08307DF0D59DA0244A9F1D03A2F265005046ACDC0
684060C47B45BFD55B277F177ACF3BC899972A8B5C2A0BFC8B60D3C2F97AE7EA64CF3E9B630D0643
B1C151A828ADAA4998541FDADB28FB9F5CA6779EFE90FF2D2AE34A3706CD0EB491B6EE182CFCF6F2
4830AB6B69945245E22D4B51D70B4B734F473B0CF5D5D75F1D10D4E7CB9312171569D447BB76388E
92ECEE5DA9DCF15C019D4627A0BF9FE36953319F20E5E53D2655C1C5B93033EC7E6760E91303575F
9F19B7BAF27B5E4437E679F15EE503C36E5E6172683EB2FE075BE77B15D85511764EDD42B8988D35
E6E5DB171C682C1EBFD54872A972E9A5AAD4C0BFFED36728F2C93C4ABC0EBD490683C1E047F79E84
F5174B76E84D1A165CB2611EEC3AD8E568F6361E5A57B49B573AD80CBF4BB710B0984CCB3F68D701
3CC56302A5068C1CD8E7A051512D6E87406921EE444308C76F6BA8F7BC51D966BD9593919F05C8A4
D6DC76A38453C6E3FA7C36278ED9525F3792DE2E579CFEE22F5B5EF7BC577E7FF7B0C181A457878E
9C0B6B28F67ED1DB76B803DC8AA1DADA57C5CD192D6A016EB56CEF0BB24B161602BC5469F39A5CE6
5255AA97AAEF40B778AB31696C5A182D83586C51280683A1C2E1285450832EF456256DAD6ADBB1AF
3F76CAA03085AA98DBEF5CCB924501BE2A584AA1B72D1D80BA76F4ED141715003E1B7AAB9275358C
79D028ACC54985B8FDDC2A82D5D75F4DC33E45FABF2175E783C3EE6B6BA66017F799136E412CA88F
0AC92D375C87DB74F0E0C1527990C345DF897ED02BCC249906075EF93DAF8E03E85558E508C1AD9A
8FACC7A4CDFA564E8E95A36B71D9BA190B0E3416614CA0A100C00CCCA9831AAA988D436F92C16030
0442CBD05D7CC9B9D0DB930C4EEC1C16599DF2CC5C9EA2755125BB96C042C08E9A7DD1DA9614FD93
DDE687F636E6AFA3B2380C34AAA7A37D5B433DACF4E0703E7F685FC4B607EDDAB3B836998E9C21CD
DB71AAF73D963FD662F25E265AD08DA007F0A35D3B1CE590A21DFCC5064C7DC707FEBA7EE3C0E2DA
01BFB2686A60DABFEFAEC191D3ABEC6E16630227EC5EE8F0297F1D6172AB077EFB14B855A0C28C8D
96B244DBB4499CA2317B607EB669D96030142D9C08B4E22925990277753FA9AB49E2C5CEE35D4325
B5A4A2A92480EBDA038C9C3B669F2FFE6B41FB47AD9E2E6911E33BE7E7AFAFD23DF2B34F58CC28BA
0AC7E892E79CF265976B47094C7AD0A84C45E40CCE0D62595E3029AFFFB51F2A31254D12D3C8A440
BEB2BBDD866437E2C49F3F7DE58D2FEF7F60406BBAA6EFBD02350B942A1AC916C7B038FDC55FE8B7
D26956787DE92F27C71D58B1C8C0312FDF0E6E75E0F3F74B651BCD901D3C810E356974DE3B37F426
190C0643322C7F26216264EB8E624FE7C1B2EBD4915CFE41BBFCB7145758F4B9BF1ECDF41F9F0F57
AD2B1D78F9E6B16D64D0ABA11CD9C3199D04D6E3D19DAF9146D11B22493A6E6E94CF1B85057A7F7B
DBD07BEFF0548183C72CFC74EE1498547763C3EAEBAF5E76CDA501BC35F1CF961BAEDBBEEA617C24
2703C090029843408ED6B50FCE991B50176F587A854FE1B3DC67CB535D00FF67D1E68EBE9D33B72F
F1C7013A7AECF45B3DB4D76202CB16FBDA9A399F53B5067F86DE2483C1604806D8F3A5A250D17FFA
4CCFD90F18F227593CC9AA50152D02CD8FDDEF0454F99C3C71C06FCC1415EEEA7E5207E7144CF48F
DEA86D0DF5BF9A36259A9E2382DBBE703E49DE111A8575D9F18C18B243DF816E90A37868A583C484
351CF676CBB3F403A618D2453BDACB00984F5ADBCE6741AFC68DF102929379AF7202FF7DFF7CE0CB
177B37CFEA5A2AD1C532DB501B504F3E5817301D315CC15036600EA64CE016F56730188A19A7BFF8
8BA350117A935240CA477225C59F5A78BCC88DB1D45BBBB076FC513A78A7988B85351E5A27B19778
91D7BD6206957537366C9C334B2FB27E71E3C7AEFAB6C3ADD6DE734BFA34AAC8475128D0A17D00A3
2B59C04BCB4D44D53F122513DBA64D029392CE0FEC5EEBF30203D3262616D22BFF664E6AFCB0EADC
FD0F0C6CDD91AF28F1C07263F45B696140ADBE2E2239C2AD70BCE374B6315672E83F7D46EF8F6126
09BD490683C1901A7A492D4E850AAE860CB317037ED4EAE9BBBEFA8FD0DB967EFB87FD5720ABBAF2
7B5E4D99E2B407C0A1E289E41B6B1EEF69CD79BF9146D11B8555155C098C295E6FD7892E8B59F2B0
EDC1BC40A3524497A5FE5E22CD9B58AE7062BDD0997B16D7FE626C8D0EC589475A263229D0D8FDED
6D3A376DD80EACC01E2E12608621BDCA885B0113AA071D7A957369114798E8E53FFDE627AFD53356
81EEAA785EADD26027CF9AF7C632702B1B57258AA39B5E15550ABCC042107A930C0683213566CC48
48A72ACE141E49A1E2BA899514EF042EDF25BD80C266706E4731E7B861A8C8FE306E0A0C98919F53
68543CE443F126C72DE2A4EAC08C7FBBE5D9BE03DDB94DEEA864F49F3EC3325E090EC140CDF958FE
784F47BB69CE972E0EBEEB79AFA6FF38408C74587AB5649147AF0AB0821CFBFA63702B8909D40A81
CE9F245C989A5EECDD1C7ADF1A32C2A685F3F5DC82893DF426190C06436AAC6C385F54D6BBDF3BF0
D6C9DF79311E11EB9DABE4CCED4B42EFB7FC2170BB78FDC6628CC9D4190DE33BE767CD47B05CEE6F
6F83411E0F1EF3299CEBD0BE7F8A55896D9B36697BF34A5B6D738BFECF3ED9BF796BF476043A04C5
3F156352605E7ED79EA174815B097AD5DCE449E5641A1C3879E2009695EE3D83F91E0F9F0F7C89D5
E1AEEE27C56F95A2BE15262BB0307F4CA0A138A1B771301185DE1E83C1601816B0D5F56A883534C4
C60426298F79F976A908C968F9B25F13973E11E0AB7A7E65D25E0A0B93B73DA8931AD2F98824BE9D
FDF00F6FB73CBBF69E5BA2BEA74002152419B7E6B61BBB1B1B1C6F54F1F449E9024CAAA7A31DDD0B
030677246AD224F2592D4DBFF6E7771FDDF99A95912A7BF49DE807BD024BCAC27B75F5F583A4577E
EF550E39B8702BAC145ACE22EEAE8AEDC8710599B97DC9CB7FFA4DEA7CABB25F658A1998DE195ACC
0DB43D8B6B436F92C160300C0B2C977A05BC63F6F970DBE3AC6BCC4A96BD47AC86073E7F3FF44E2B
001C7D7BA276E9D9C05E0A0B0FED6DD4DBC2C3DE9A933DFB86F746054594B54D9BB4ADA1DEA35111
D1A722B9FC22879387A2FFA5FF3CF1E74FC1A43CDD0F11FD48D49F77B68BB7D4D7914905269D19CA
129A38831F2D7D22E3A25755177E3563C600A6B5DDEF0CE42F3890F9569897B05268B975A94BAEDF
C13193B73DD8D1B7B3606AA5863481D95E4F414737BD1A7A930C0683211DE8B80E2C946135C36F9B
2DFFA05DF4D2B91A565448FC96D7032C139825C563C4361F59AF93C47177FC66F6A9DEF760B1C30E
4F88228BE41DBBECC947A3F0A98F76EDD0DAB9C573EDA502AD87A97BEFEB8FFF087A2B4C4AA7AD3D
5A350A348A37887C8A4C6A84796A865244A0041F5F8019811F815E4DA8CE865E35370D1E3A722E4F
230AA7156EE504043A2F2876046E85E9CB3C534582D5D75F2DCB81279F7EFA4CE84D32180C8674E0
847384A850A1F7427B7A0F0B9FA2D13EAB6B69E87D556074EF09304870BF8E0FFC7528A21E10E26D
C2EFDE2F7A254F012F44F4CFABC01B9135C0CA9810D4E74FCC498CF1A337CA7382D81A9AA3DBE400
4CEAED9667A3959195D11258550AB78FCE41FF098D5B1904180C5835B6EE181C29BD1AF1539F4C83
7DC18138B79258652E2B5A2D90B50EDB0E77388BA051AD42E2D4F14F313B452319AA2EC43A127A93
0C0683214D380A1530E3436F129630AE805AFD40AF6B9563D11D7C37A034E7B831F9AA0B9311C0EC
98E6061305F6C96D0D93F62CAE7D6EC2F884E0B198E32321C04FCCF898370A140C2B69D62DA99CF1
905DE79CEA7D6F5F5B73CB0DD7793742A2FBD41D893B0D236579C1A4F011E1CE569CD7903E303581
5EDD77D760A6C18198E848AF30E9A5F08E65077CFCC0E7EFC7FD56B1FA5697FE72B24441481E16FE
0B6ED57C64BD131368A3BD00D8DFDEA63776F067E84D32180C8634E10498890C4258C0B245DD0301
ECF6DE2F7A855255C8BA269799AC64556F5FDE35B552B7EDD8B163A051337EFA5F58F815863A7D1F
C1417DCA9207EDDAB470BE096EE71BA04520475E148D2E2095C876E56681496D5FF5B014F3AA90A7
CC902B040E18CC5D585FEE7F60E0EFFFE1FF6444AFAAFF9FA13B669F6F6D3B8F592E7D2791FFC8C0
56F5F41E5EB6F785312FDF9E2C2650BF0EF45B19F2075DCFBDE9DA2B46B2D56630180C0506D62CBD
96DD7757C8E9548FF7B43A42B858D1FC875514403D264F74052BAA2EFC6AF73B851657EFFFEC93A3
3B5F83A10E0B3C208ACF2F34117B8DC591340A767EE8FD59DE002D122695B49E54EC66B54D9BF476
CBB3673FFC43F0EDAEBC67CD902760A159D7EED1AB1F569DCB885EE1F83973BDCFE20CC94E9EC540
05FF3AF0F9FB0B0E34825B25AB6FA59721702BAC4DFDA7CFD843913FA07B9FAFFAAEDEE709BD4906
83C1903EB04068850AAC5F2136E6AD93BF7336097FF1D63F5B283B6F53A088F12B6FE4BD7340A33E
DAB503563ACC6F86B873C913C787F76790821FACFACE7BE7EE6B6B361A95F77BF4D557B847940191
8A519A46493D2FDE35D82A60525F7FFCC7D05B6EA840905EDD31FBBC3FAA79587A75DF5D09F42A27
0467D757FF21F956C92A5BC9EBF19DF31B0FAD339DC07C60FFE6AD3A9EA1ABE1D1D09B6430180C19
C1F1801498C2C89A88EF8D47B34756312785CA105808B8B52D63E9FB145203D10C9AD36760A26F6F
5EF9AB695360A5330747B8925F74426BC48146ED6F6F3B76EC58E8DD55F6C0CDEA3BD04D26150DBC
74AA7AC532D778CBD6DC76634F47BB312943F1E0D09173D9D1AB71633C9FD7FA8D03C326960E4BBB
246150B8955BD92AA88E3096A7E51FB4F77ED11B7A1F9634F4DDD9B470BEDE9AB332EE0683A1E4B0
645102A50ACC0E2E00B042518B892BD7A8D5D36527D0622D048E9C08B1F489914600D2A8C012F676
CBB330BCB94F989080E32F23C503222F7E36B57AC64FFFCBF36D8B929D3CF47E2B4524EB37BC7F74
D3AB303FB44F8A9476D935973AB709C7B44D9B0492EB6752765F0CC5008EC3CF07BE147A75F12519
070762157BE58D2F4536708463BBE7EC078D87D67149D21B7D2242CB37B14801F45B818E85DE9325
0DDCB278AD8DC8C4651394C160283960151BA1D76324E0B4B96CEF0BBA04157EFBAB5055F8042B97
8F1BE4372A16D70E64D145A451BF5EBD2A9E141C5425CA55D88E383ED6DE730B3E38E6BF4DB8F497
93696CFC60EB7CBB5F3947BCB8C0679F509D3EAEA9A8EE8813EF4777218ED785BD0C8662C0B03303
E8D5C177079B9B3C0DC04CE9D5B831DE64B87547F60541B4CAE581CFDF7FBCA715A4C99F60E5D40E
26B76ADEBD21D06F6593E1B0C04A24731A66B32DF575D6750683A1E480C5ABF00A15BAFC6847DF4E
2911C2E5E9A1BD8D433697FABA4B5EAFDFE8AA550077CC3E9FBAC724A8EF64CFBE7D6DCD6B6EBB91
B9514E95A284F8B198B94E331E1FD9DEBC52C763487237158943EFA8F2C3893F7FDAD3D10ECECB9B
851B118DC3F405F8115126A5AABCA51E0F06433103A394F46AFA8F33F65E4DA81E5CD9707ECBEB69
C55DA43E06DCAAF1D0BAEFFFFA56FAA7024301B984E100CC8ACBF6BEF0D6C9DF85DE7B2501F67C77
6383842863423BBAF3B5D01B6630180C9902139A5EAAC68DC93BA5D28BD7B1AF3FD6B1EB5892C677
CECFD3F7963A74BF75EF19F41B1893270E04EECDE283A051DA32D7B16171135D97288A416854A095
7E57F793DAB4601962FF6186F4EF2C71AAF73DDC2FE6B2B9DA7D414A209B16CED74CCA60283F7C3E
F0E5EE77BCDA5598E8B4AA529AF46AE91303F87846CF88FFE0DE2F7A977FD02E1AEC5A2AD0F993DC
6AC18146D0B1D4E73400CF4D18AF3DEF7A21D31BB00683C150E4D00A1558AA9CD90CC03BC4893F7F
EA07DFC76F0999489106A2FFC41239E9DFEFE3027451C7CDDF6AFA914EA132A406CC037F4EF70FAB
CE492D0FD6785D7BCF2DF11875271F4A0AEFAA7F2DBBE65256293ABAF335110D0E54B4C08BE6DD1B
F486EDAE835DC78E1DC360401B80BE13FD78FDF5C77FE4E0E1D952886318C8A4A2E96CBA2CAFCF1B
C574836D0DF59AED1A0C1502CC27DD7B3C2794BFC0C4B0C047F0417CDC9F7B15F828054E59E056CC
B70A96078C445CB0A6305ED06FE570AB0A8437FF9F3EA3CD09AE1798C4F43284350B6F0E6B4E180C
0643B101F3D5FD0F24AC4A5D7B761F7CD7FBBD79F3E6175F7C714D4BEBD3CF2C5F11F9796CD963F5
893F7807E07F71D88B919F37DF7C131F3F16F909ACE5C1771EDADBE8445048B08469FDA5037016A7
CECB3FFCDFDD53FEA173E39C59B2E927A1145A59C22F7BDE366D12EC738F4625CFBEC15D2389EEE9
3D8CFB8BBBBCB0FD61311BF0FBB68649F37F713FC6C3A20575325464782C5FF5545353537BC72B1C
1E070F1EE4D96CC53CFBE11F407EBD6A5F91FB95A0A9E8ABF3B5FAFAAB71A73EDAB523F4661B0CC5
004C205B7778F4EAEAEB33E356C08C191EBDCAD47BA5016AD07C64FDE46D0F7EABE9477A394BD00C
8C4C923800146CC181465DBCBE8C81A59FA40953BD98135802FCE6C4929FCC88C64844A6B895B7DD
8CFFE2481CBF61C306AC173803C917776E43BF3483C160F0831EA8E5CF2450AA4993B6CE9C77EBD4
EA6BC68D1B77D965975D10F9F9F6DF5D7CC9A8EF5C96E4C739A0BA7A0A7EEEBCF34ECC969C18312B
F61DE83E75FC53594A3AFA764AFE145722AF9662A217C3696AE8DD95DB9ECFC967C1AAA68C3F5CF3
B7BFBEF3FF37F7E12BAFF597D995D7529C08BF1FAD1AF5C2E8BF850D1FA551B1DDDAC02FE29A883B
C80511F7147716F717C3E3F2ABAA7466C1450DD518061800632FBF14FF92E18137398AF8273E888F
CF9E3D1BCC8B345C16CD1403A0FC70AAF7BDB75B9E05997532DADC7A529177C0A4BA1B1B4EF6EC0B
BDD90643D142BC5713AA33E35655177AB5FF9A9B82356FD399948E0FFC95DC4A3BAD24F74A7EF39D
CBD6CDC07A97DA6F55723321CD09AE172050582C409A6A6B6B6FBAE9A69A9A9AABAE18CBD5C16F4E
78C1E7A2315B7521DEC17F79003E85CFE20C624EE0CCDCB005C30AFD920D068381EE064C4AB493E7
CE795E2F2EDFFA9B7FC18C07AB185359F5CD9360FA6256ACAD7B84EE06D8C04D911F1AD8DC6EC25C
87436869E35334B62F88FD7056044DC391F814E6C3CE835B47AD9EAEA3D027FDFB7D95B07197CEAD
49F6A6FED7D71FFFB1A7A37DD3C2F909AE0D558748EBC2C5A3C862DE287CB6FFB34F522CD91C1EB8
53B85FB86B18035C137943B9D8E12E63A58BEECD46A8F1058DF3313CD6B4B436C57EB82149BF1586
87779E895349D585AD9383E354388CF40A5F2DE18B6503B989A045DB573D1C7523FAF51563AF794F
79BFAC5CB2C1903EF8ACF59DE85FBF7160E91303E3C60C66947B75F125E766CC1820BD0A3C73EAEF
3DF6F5C7E05633B72F0994B020AB126D7670ABBBBA9F7CEBE4EFD29194294EA0C19A46C112C0248F
C542CFF0F8C10A02F300F33C0ED0D10B582FC43FC53210DCB5C362417382BB735C77009A133840E8
15168B8ADA8B33180CC50032294C7D1B366CA09D0CE31673140854E29AF27B1AB7ED1DAF30400B53
96932DA5CFC9BD29EDE267B82027464E89F453D014074DBBEA5F7FA255D341AFC0A72CD12635D0C3
4777BED6D5F0E8EAEBAF16AB5BEC706EF4F105BD1B7445E145D3B5576CA9AFDBDFDEE644D9394C4D
E234406AB05AF1DE691F13460BD732EE1632844F32B5691EC09CF08F1046CECBF0C0F0F396D1658F
89B70B0343165FFC89F7313E711883038B7924A4E9542593EA6E6CA04FCA113C77DEC1EDA32A88BE
F662EE0483A1C8417A75FF03034EA4F4B0B8F27B9E986A6BDBF94347CE65FAA5980C3B0F6E9DBCED
412C70DC77729C565A83FDA28E9B1FF8ED53E9EB04863521E81949B6DDB022D09C101A25DE2510A8
E5AB9E62089FC4EFF57ED12BA11147DADB74B6EFB6867ABD5E70D717CBC1A20575DCD973766BF915
DCAA65A241D6FAF90683C1302C923129721C4C5098A6F426DEC5979CCB49C4B2F67460CAC5D4575D
3DC59B721FAC66C14459627EF1D63F33E86BC8E47D12F1F5C77FEC3BD08D5546BC51518D3EBF6A41
6298DFC3575E3BEB92C593BEFB3FD22959C5F50B778A4C4A0F0FEE2E62CCE00ECAB2A5D72C8AFE45
635A9A7EE4571749413A3042383C48AFC8E06451166E45BF5529722B56FE029F8D0BF7456E629446
F98A7F8170BDDDF2ACBF2CAFC160C815845E8D1B937170E09CB983F86C6F5F802E6E8AD909B3E28B
BD9B6BD63F269A154E70A0CEBD12BFD5B0A70D05E293EA5CFB129775895E90A06ECCDB4F3FB35CF6
C4F47A7174E76B8E7E2CA6479DE47BB2675F60E025B8524FEF613AC248AF604EE8BD382C1FE456D1
8DB824D1EC0683C1903E1C37044DE5352DAD9E3B3E62B2EA29887632A629AC147AEDD03B723999D2
4F1DF7E815BE68D22F6B25EC81EBC877E7FD232661B1D82BD69E8CAF329F7DF2D1AE1DA0519E4743
C58049E9A8B8411EF9CDA809BCB9FAFAABF191D9FFE9DFBEF97FFD3E82E8DD0CAC35A6F3A4D0F358
A46A264ED52B14162CAE8CB86BA4518123A1ED7087AED5925D35160E547AC738506599FEF6DF5D4C
6E25842EF43BE5BF65CE9B014CCA1FDDA798D4DA7B6ED9DFDE16280C62012D06C30891625707E408
1409336416DE2BD22B10B4F4BF17DCAAA36FE7CCED4B2EFDE5645DDA5EB2AEB838F2CDABFEF527E2
B72A8679002445C73048F009B7BFF4222EB3B43F2802EBDA2FC6D66062DCB3B8F6E8A657B1DC63D9
F2E2FD223321DE4FB1A72A6F722FAEBDE395C7963DA6BD63680FFD566809962D2C6D46AC0C06C3C8
411B15B30AE61CBD8F841F98CA9885302B72F39FD3546BDB79BD5EAC6BCF5975AAFEC462F4584A12
72759B7E24E939B49C97AF7A6AD7C12E676BABCCE05F2FB0B2804675353CBAE6B61BA381E58ECA44
CCBB112D23154BE6C51AD479EFDCB75B9EA570C1E7035F6295F7270EDC31FB3C8B46C98205928B01
80DEBEF3A793D0F35C8FC85FB03862D890490D9BA98D7BAAE358D22CD39C62DDE4AA8D86396B255E
E09D78C38A6FADC4E5C048D8B4707ED442D0859263F7312E125275E1C639B37A3ADAFDE35C920A2D
0ED660C815867D82408EB00E8228655AF46ADC186FDB0A8B66B20C50FF57E31927B7D2FB514EEE95
BCA6DF6AE7E90FC34A37661803E812F7BBB4DC1043C1E91B4A67C9DE525F17CFFC65F294DA74C2E4
9951C3187EE335ACEE11BD638C1666DA3083C160F0835B498CB5836DAC639B692A6FD8B0C189DD1A
8A1439D26BC4924503396B4F6C3581493FBE73BE1698C56A827632E24B88D5E5575581033EFDCCF2
686874D959923A77093C685B433D689498D90916B8884B444C71D17F038D5A7BCF2DFBDA9A4FF5BE
17E8DAD8BA23B8103005EDE9B89CFF8BFB67CEBB9564CA1F3891FE1A8423B555002361247DA24FCB
50550CE39A8953C75E7E29D771B43381F11501B1421B8EEE7C0D77245910A693398523A52C6FF90D
6F83A15410B865C117BD7D1E4502BDF24FA4C3D22B7CEA9537BE4C36853AB428CEAD2291F00EABC2
D41ACDC38ABD8F235FFED36F7892C0BD97DC82F3B02653A211317BF66C89CA1E76BD9016763736E8
1932219334E2AFFF68D78E745C72CE014EE0BADE40A69005A301F3DA570683A19C205B49B4424524
07244566BF640929989CF5BA30FDC7E773DB369CFFE1CEA775E838B0F37817752D6839C35466E019
5D122056326397CD3408F3BBEF40F7DB2DCF0A8D4A48AB09147F8B1DD679EF5CAC474E207AC2C9D5
9F07DF1DF41702C672FF2F6BDE031F4157EB0078C66CE02E9C3A1E750E6624650F4B80FA8DB8A763
5EBE3DB30E497E0943CA9B869180F14071DD28019C38956D4E6741CF07306E8FB4B7914971DF35A0
9E54EC3E2EBBE652DC3E8F49A5C701CB66C01B0C250D3C89984B5BDBCEDF31FB7C46DC4ABC575B77
0CEA092AD9A38D25F2C5DECDF3DE58A6399456B188F3ACA61F5DB66EC684DD0B771DECCA93DF4AF2
AF9D3D4F4971159292E94CB5BD7965C254190B5CD7FB510CBD605860462777BC69BACD30819E7E66
39B5B6421F540683A198C199A473ED4BDEECA7C894CC245D7B763B7A7DFE0D3A6D81575DF8554673
F5B0E161CD47D63BE539967FD02E8771026798223D6B6239332E3A4A068BC02591CEB5BB0744BC51
A0511BE7CCA2C279807AB6936213B3C3C1BC1C1A953E7AFB0683F2AF7F7F59D52C919E40F7FEF7C5
75E8792D9434ECDD74DE642916F13CE6A39399F34571159041A1DE4C85966CAF02DC6ECAD7FF6ADA
94046762925429D8069B16CE4F9F49190C8662C6A123E79E5F393479624070756A4CA81E5C5C3BB0
E5F580A2570EE8B7FAC96BF5E2B4F28B04CACE247ECFDCBE04C7E7D0EB4D5D2990262AEC4900C9D4
EA6B461E4AB7AFAD59B327EAD3C6CB7C244EA49869F9A9F49D563472D0FEF68E5740ACB040F88915
CC21AB69653018FC60981F4C4A7A79A48203FDF242A6D23995A35011286A34945E610E87AF1DF8FC
7DD979E38A8055C04FEE289D0AF6C46B616540218630A7391386BB819F667A0B6914968FCE7BE78A
422C23F774FD561DDA27B9516DD326814631FE219D3E4F7177FA4EF40796BC04AB420FA34BF76FDE
9A826B0F7B7EFEB96CEF0B52B519D0F52B7378B3FA2362EC18034CB3A2978D0A81928FAC79776EC7
C9A9E39FEE6F6F0329967D547D1F136872E4BF5BEAEBF41DCC797B0C06438800F139F8EEE0F26706
66CC18C8885B31067BE91303DD7B060397663D15BFFCA7DF3CB4B751C42BA2AC2A36D3CA2E1617D6
595D4B855B059E7058A03D98E8E89992AC019D97D4D37B7884F398A6549C36994E455645315B2E88
6B7F7E77468D77AF3A5213841E2B4DAC2414508A05DBCC6C3018E856808549D11B2DE5C7AAA9696E
25713EC102D1DC9430EDAFDF38A00F1836124CEFC33B96E465EB6688C9CDD830BAC0E4AB9D1362A2
03B1C24C88D98F533A49A2DE222BB00252EAF83479537BA3FE499583C74AE14A67C72A10E1C5C357
5E4B6F148CF0D4B72CCD4BE6618CDC000109DC507DACE1580A0D908CD8DC8BBD9BF5DE29FECCB4C1
194186BD932AC83D043A34D3E9C3B486D07BEFC00600C9D53E29AEFB9A52D1F94826959D4BD16030
940A9C058EC181CFAF1C02BDCAC27BB5649147AF86522EB26F9DFCDD5DDD4F5ED47173B4B47A2C84
5E5895265CF45BC9C29A4EC009F733DB3B5E91146C4602705E05010199D231E1594F71FBDBDB8452
C5EBD12BB73EE7D8961BAE4B6651A47F6B8662E6843F655B1287C38A1B37180C45028A0CC05A8601
495534C93FA22EBA787352CC7E7EC70466753DD5AF6C88A653F59DE8DFBA6310840B070CEBC298F4
EFF7F59CFD40BFC3BA45D1097F630D5684039FBF9F8E1F1F73F8860D1B986D4ACB99652F6AEB1EE1
4C98D760AAF4C9CBA9DEF7E88D7A6EC2787A2858025EFC5051D973EDD188BC86A1BEADA1FEE8CED7
72450FE576A367441A1DB41463E35B7FF32FFED5BCB5ED7C4657EA1F3F5CACB5E81F7E2FDBFB020F
603596C77B5A737E5F84778358458B9DC58895C4F9A71EFFA9817BCAB2BC0982548929D5E299C2EA
BF7DD5C3F848FE46A3C160287E704AC472B9F48981C91333F65E55DF3C846537709D9577C8ADA2BB
94E456AAD015457445867DF2B60731036B4AE55F6BB837BB79F366E6AB8A9ADFD8CB2F05B762FE91
6C52F9D959A6B36B4F47BB5342424FADD1928B49AA5365774786629911581770815A9129B5B6ADC1
60286FF0A9973861510D25D780319966B67EE0EC01EAA4A7F72BBF373863C6804EB04AA1ACCE1376
1EDC8AE9FDD25F4E669E14D078689D9EED81E623EB9D36A4583E18D4BDA6A5953AF08C03949990CA
15C92E67E4FD9CA279A78E7F8AA561D3C2F9ABAFBF3A2E09CB15C1975FA3DF84F91DA551414A7D39
693345C8D13F336EFC47AE8FF87DF9555553FEA1D3BF88DFFF407021E04C7378F5DEE9451D376329
67E19551ABA77FFFD7B7E6F64AE57A451550F36E8C13EE40F259C8686C601D0739F2A8B1DE418D38
13E33439F6266E6557C3A3D4AE1F4A6F2BD86030540E30F9801F2D593410187D9D0255177E85C5F7
F9959ECA50E06284D966E7F1AE595D4BC1AD64EED5C95612104892F593D7EA1D6E45700A65163666
4E3127646F9653680E03988F6E7A3541984244FF14CFDAD7D69C8F359DD5ACC01C459D8975ACE886
937CF3D0878DC160280C60E832495F873CB15E0F23E2926DCBA770543126BCB9C99BF953CFF338CC
7F4E7D66CA1450FF6DDE1BCBDE3AF93B5D171E2FF0E65010994AD63CBEE04CB87CD5530E85E455A7
A8429B43E0FC5F7FFC472C075BEAEBE88D228DD2DAE6D1D03E9F7A76DBB449F8146854B280B42C72
A682CF13734EF97D975C32963F13708BC9AA74630E1D398735DDB9DD7E60E41CF8FCFDB6C31D9E23
72638D2EDFCCD439FE895191EF5B83ABC605CE9C77AB5332805BACFE1D06D74278EF1D9023DE56F7
0E464442A48A0A5EE06E52BE3E457B327ADF60309404327D84F5F19885B6EEF016D92CBC57A05758
A06542769A41BF5574FA8D45D7C71996F8B022FFFAC1D6F9E056C7BEFEB83F26A60AEA84C5420BC0
723FCA499B1A498A9606D64157D236B656327E63CD6D37A6239098352884C5AB1673A2BA7A4A76BB
700683A11421D5A6EEBCF34E3CFE74D6306F54D4D1033E957272C0C4152B19FFFB7426F6D48A7030
B0C54F21348A857D01BC600A959B399546F09EF8E6986A4AB2C0D2189E3FA2EE115C3E56871CCE84
C216715A166FA5372AA188862EABE1139D68B9E13A7CAAA7A3DD7F5FFC2197FE2ECD34184FEADA4B
C543FAA7BC9522122729FDB3E5F5E07469B9B97D27FA392ABC3A5649DA899B78D5BFFE4482F94541
5D4B51C9E2CEE2BFF983847670A114DEADB3AA65AF55F7D847BB76EC595CDB74ED15D168135F5529
AAFB924F8149BDDDF2AC4E67B695D76030A403C7D1833976FDC681FB1F180852644D05868E08BD72
A6A0039FBFBFE040E365EB66E82558178294F7316363399EF4CB5A96269460069181F293A97814C4
D71F4B2A7416732066DDC0C956CA2FCA6ED5C827D8147B95CC9BC0EA20BB70634747EBB614830A96
C160C807FA633A030CE562691EAD08BAA6A555EFABA43F0FF04847E52F752842E0F9E59DC643EBC4
37C1ED323D93C3CCDEF5D57FA4D3BC643B639CC03113FAFB011323A6C75D07BB986536C20E67E5D6
6D0DF5B0A2E3B57723DE8AC0BD352159CF4D184FC56CA79AC648A22686BDB312FFC6C43A9D56CC05
4267150DC56251FC39D4205054AB98FEE378F995FFBDB33790F3F64752E474A914798D1BFDADA61F
C97E298684847AE61BFD11ED5C6AFFEA8079DA099E507CE4E662C5249372B3DB12EBF3CA96E991F6
361DA5694BADC1601821740CC62B6F7C991DBDC2F2BDAEDDD3E67526A59DA73F7C686F23489353D0
CAD1608F265E35FDE8BB93FFBEFAE649A0571B366C484194B871FAD3754FFCE2AD7F06B1F25F4B3A
88522A91A7D0E954551762F5748ED74A5623E967FFFB001605EA018A394177954ED6B609DF60281B
D04AA4734A5CF3975F5585D7D4F41B617985BE13FD692A144DA80E88FAD3F8C1D6F9DA3915275611
490A10AE9C74084B03732614C578FA23EAEBEBF76FDE2A043323760912449F0569942E1A9550BFD5
C9A88D78A33AEF9D8BEF2DA43A81BE3491D067180323FDD033E89F14959D0FBE3B187350C68135DD
094AC13B4E6AB3BCEEFDA2D72934E6F751F2C5CEE35D39BCDE618F6498289E0E1DCD82174B7E3203
FC08BC18EB78825B4A575B8EDDDCB5F7DCB23F9149190C06435E81E518F4EABEBB02EB090E43AFEE
987DBEB5EDBC2E7DD21F295F026E45BF95A3B8EEC7F8CEF98FF7B4525D4A6FC139F2BCDE1936D6E0
E0D48131C926ED933DFB5C17552CCC63D3C2F905EE7009FE61550E9DAC2DEEAAD04785C1601839C4
FB80475BAAF7B20A4FFA3214C37E057E07E6D7048673A738CFF181BF3AD6B58E37C0248C59BDA36F
A7F3D5D9EDFFF4C70ABB53C087D320CB57E14FD6CED0D955C962EA60301FDDF4EAB6867A58DA9A43
B9642A31D20F33FF7313C6D3E40EA45105DBD4D205199959CCAE60270CBB1C60F9F62FDC175FE2F2
ACE6A680007EBEC662ED7754E9081302E42BBB9199E2DE0DF3F1584ED99D3F9D34F6F24B451E4442
37657734815E8DBE88B19A8E93D16030180A8C4347CEAD6BF7889256884A339E64CE5C4FCD15333C
4F856973E7E90F1FF8ED534E34A0E7AB8AC9033278DB53BA88D0A5C643EBFE65CD7B9476D7C0222E
5BA6F894AE973194DEFC8C45D3CDA58A2CB5585543D9C2EA57E2B1A2B28EF594E5ABA26280E6AE32
184A167449E331A70E8FAE52475F0C439872F5759F0F7C99CE9ED8D22792522A60F907ED81D6B50E
330028AF3D94E8CDCF9A589D3A1EED2289F2FAC6D88B58F5D58B065741803A83A6EF40F7F6E6951E
8D4A2CD7EE88F5694B1BAF31E16F9C332BB52E81BF85391F18F2823591172DA8F3C740826E072670
F95F80080F9B2B8D053A99FA07B3E7441B3F41C837265581953A53293CA7A9F88D6EFF68D78E34BB
973EC723ED6DA0BD51314627565385F66100345D7B854790376FB515D36030840EFF44D4DB37487A
E5DFF24A0DACECF7DDE57D168BC59A96562F11FBE649DF9DF78F9CAEFDF1F9CEEB0B1AE757373CDF
D37B585A32AB6BA9FE203073FB9214EBB87FF27729556C4EC6BA1C6227F7C7348E9CEC2A8A013AC1
F30683A184D09FA860C6071CD632B8154B7E07EA1B8C04EB370EEFA84AA1A00E8CEF9CEF8FD9D612
70DC10C3EF1F6C9DAFC3B033EA16E71D4CD752F5558A1331EC2DBEBF84BEFAEC1318E46FB73CDB72
C375AC2498903BA3FD1789A21334B661CFE7AA46460E8707953A74601BE8B6D477CE28520E2B75EA
5BEF4802EADBF1784FAB4478EA603F1903635EBE7D4417FBD927B805B8173A262430A18F4CAAA7A3
3DC0E7E8C8DA47A8346FAEB73B11DB1AF52FB2C573D30D0683013312E8556BDB794CDA99D61406BC
EA847F730F6B6AC0BAF8E9BA2744CBC2515FA7BB4AB6C52EEAB879D9DE177ACE7EC07A5809076CAC
C19B073E7F3FCD4BC02CEDDFE0DADEBC32C42E95179253E08801721F5B6B13190C86E2073D2FE008
7884F1204BE9043CD78B16D44919A69C7F297E6B5182406CDD919452613AD5D5DBB5244534AE2092
51C5E917E42B57C5A4A29B63B1282FAC118C0004C68EF6C2A157DE76B318D8F19AAD8906B6F64691
466D9C330B337CDF81EEC016866B6953451FCCBA66E2541DFB2D7AF25908801C7C77F8B53850509D
61F65ABCD7D5EF8DC9E667715B872272BB4F4CB84EEE4BA0F608D7C1FD119F94D0A504B76322BDC2
9B775FFEB7632FBF746AF535BA28765837D4603018B203AB9F60EDCE945EFDFD3FFC9F193306FE65
CD7B7D27FAA913482D0BD6C2D07C4A224C64533420C63BF22916A31C36541BD378426E724402A878
6660E6143CFDCC72D1ACF0576C09BD910683615868250AF02909F6A36E1BF8549EBE942F76BF338C
A34AE7BD3A58B6F7051DEBA5B9955397EAAEEE27733279FA230DF07AC3860D4FFCECCEBBAF1D0D9B
998E27291DE5E443F923C130AB6F5FF530BD51C9041992B52445045A6EEF1475C26BEB1EC16C4F59
3F2F1674E25450035C7B3A53BD3F1F0A4B6A3AB1FA2932E91EEF690D94A7E00B09F5CCF462B735D4
3B91998E1814D6BE7D6DCD6DD326499DE578065CA24F0A786EC278DC620C0F09A31589F5CEB52F25
E3F8C5B3D01B0C064332F4478A096641AF70F0846AAF5AD62B6F7CB9E7D817A0451EB78A89B53ABB
64B2C447E9556C1B0D47926DCDDCBEE4F8C05F75AB029AEAA3543A96BE48A65CB1C424A7800ABA8F
2D7B8C0AF345D24E83C110082A51E8505EFCA6AB058F766136D2EFBB2BD5DC9BA2010C0690A86C5D
9348B6B6B43C85E87233586B84CDC6EC07D3BAF3DEB97E0F145F905B694B9B9E29D8D8DD8D0D12D4
97ABB8AF3CA550617880377178804F490DDFA79F599EE9D6990E75C07A9AC6CAFBFB4037A5684061
15F687E2F3B7D62449B7DFDE7B07B70644187C4AE8126ED996FABAA148283E633803023823BF7581
30102E5033DCE2A1C82306FEB8A6A59585ABD88DBA4C76B2FA92B67A1A0C866243327F106B0ADFFF
C0C0DFFFC3FFC9343210CBC1CA86F3A057D50DCF7FA3E156ADB5AEB3AE44363021D23BB2E863A17F
EBE4EF923592EFE8D5D9AF9A5E0C906C65270810AFC1AAF2142F643018468EFE98ECB357A53451D9
8F456C031FDE5C45CD69C0E04C36D35EF9BD54517FBAC06B40C9D77FBB6E56D752BD7945C036DE38
675676BAA9F82C69543C37CA1165750A48C55EC356EF6A78F4A35D3B4A481F1B733B95FD303CC0B2
45379E4A14CC9CCD6E000C9B45A5316E4CD231D07C64BD5E6435BDD2CB6B3AA3112BAC17ECA7AB27
C732DDF0E7EAEBAF7673A3128FA4570B840BB42B70F3B3A7F730174A0A2402E8524943B3204083C1
50BAC0F485757CF3E6CDF4B06096FBE6DFDCE3A55045B6C532C537FE61CF05333AC8AD02EB5B393B
69B2FA3FDED39AA29E9456582DE6CD2B666A63E5D54180D5D55366CF9ECDD8067357190C450506FB
C1305EB4A08E9BE7786C2F19F51D3CC2DC0C91C30AD39EA54F0487FFA588FB9292AF121890101EF0
D25456A4722E01840866304C6570A2343B0A16724F473B2661D2A8807C1965694BA5211C004CFFCF
97FC6C6A35934C33526F087D7830D80F2DBFE9A69B24D88FE508599631EB93373765BCC2B6B69D0F
6864C4E7F8FD5FDF2A3A15D190FBC8A665FA0AEAB816B05D575F573365F5225E494A05FBB54D9B44
49C661BB85ACCAAFA5892E3D75DC5895C1602819E8A8038975193B3A5AF39D8EF8A79F598E7F6DDD
E129F7A617999088EF1F07B1D22B7B60B5144DB2BCA4E9243A549CBD5B6EB8AEF8675AEE78AF6969
9D39EF565D1854BA74A8F898A0C15099A0AE42E7DA97B4B21FCBCC89B25F819BF4F9C097819935F7
DD95D44371D9BA19224391B091B5B166D2BFDFC7BA811AB0782919110DEB4AA99E8AD98C34EAB909
E313AAB23A017E0ACC9C62E9A8277EE6AD2C52B80AAF6136D3F157FCD320E9360588643297D80374
DAC85D2A07DFF5E243D25F583136924518361F591F8F12898D048C8D1457A75F7CFDF11F19CBE72F
01E9B22A470B7DF445BF9A36458BDBA7D32771C75F4D0DD6470E0F0A453208B0F88787C160300844
5781653518D54CE3DF5FC512AFBBF76446AF40A946AD9EAE3994685251BFC22F529410ED1F9B5161
63789BA2A32F2A2A1DDD14E062C15D38F4A7F42D631BBAF6EC3E75DCCA011B0CA181D38894F1C504
28CA7E7821657C033F5500AC6B0F984EA5DEABD3244C9812F1A503FF30F73ABA046CBF38A7748955
BCA98FC404051AD579EF5CD0A2A83A4160D68C047AC50CECD5D75FBDA5BE0E9F95A03E6ED9614DD1
FE1D56C215565584B33AC309A48CAFC47233D88F65324638FC3430145736A45B5012AB7032017F30
68273E647CE77CFF91FE06ECDFBC75D93597729D4DCAA47C946AE39C59B8D7D955B7671B2407190B
A51422C1F0E0F623EF42110E0F83C16010F49F3E23650AA9042B8B9D56FF1E0A4A106620C4FA8D03
4B16A5A457DF3FCE655DC7F3FB118F548929FDE2F582038D43898A526DD32675373684DE6F19813D
0C834D621BC8AA68B031F4C5160B832114B0FC011E4F4C809239C53AAD786CE5D90CE509C597FACB
BF62CA0D3CF8A1BD8D8EC21BF7A69841A3DB2FCE297F220CD81368D4D14DAF6E6BA8C7649B50254A
7BA67C76350E038DC2C7F159BD6A445F4434B7F11B6633EC64742F2889262694EEC9531F8EE4835E
085C24841B7C8A6D66613292C13C4DDD9F0F7C89BB9CCEA665DF89FEC096E3A63BF2147E05757741
3F7D06375D9443B4B24430AB8A44F7814971C755B419B3CE26430374DE81B88919766B7AB90683A1
68215B43B0EA6BEB1E91358E650AC1A792D5A84D3661626E0F2C52E96554893645AC4CB0933FE517
AF1037D698976FD711E06BEFB925451B8A13B2CF4956A52B424A6C43E88D34182A0D30E1E844A615
C707937C0A8F2A13E4937EB6505350F71E774675CA12D194C56FEAACEA091624CB29988E666F6F5E
194D808AF0234A0DB8415C0E638AE549915BC58F1F7DD17313C6C30E773C1429888678EDB5F487A3
48902799BE2C8EA756098B176378483C3C8B28E94C9F3C8D87DDEF0CA496AD98337750C680D398C9
DB1ED4D9CAAC5492EC7A4FF6EC93603F27372AC03F15FB73D89ABF995E2F8607FDC5BA0C1C860706
0C868D373C6202950683C1503CA03941573B958BB8C6512838BB00E6DEBEA069FFFF5D162D4D15F3
3DF90BA64822AD8E09143140980ACD47D6479BFDD927A5C5A734447A57168BCBAFAAAA99383545D5
AAD2BD5883A198D1AF2A4FC16613F7B1C41AE923C36DEA8C19095B5581138544FD5161F5A28E9B77
1EEF72AEF754EF7B9EE349D7600DE450811A7DEA1D3032D8D2A0515AC62D453FFBBB91220F096589
264E85D9CCB8887CF3943487075A82E18156D17DC99F9C6B95F82F3650FE3199560900D2EDFF141D
5551B1C7C84EA656507762FF8EB4B7FD626C8D1E007195FB403215D3C3F76AFEC6023B7378B3F0F4
2D5FF5948E11A5D7980B655843C2603018341C310A499E12FF14672DEE37663443F2E0AD3B822915
71C1FC5FC8EB64B8E6D99FCDDCBEE4AEEE27E7BDB10CBF67752DE50BE0E53FFD2674DB66E49D8FDF
ED1DAF505F5DAC3859A64BFA020D865201B3471984C60910C61BEBC7E14D7096D45146057E4E1D41
F5C063306D8A58FA4F5EABA74CBA0E597CBBE5D98418BF440B99E951D13ABC5AC62DF62FFCB9F69E
5BB4869B131299669FD09FC224532C40E87366988A22015855888215D16B89EC374AE5299D031B58
792AEB50B78C0E43D7B5B605A4594DA876EB94C99F1815126F7FE0F3F703E8DB679F74DE3B5752EA
E2859845AA315922554CC2F1E8CED786BD964C3D8F783C4F1DF7F82CF572E5C7A95A65CBA5C16008
1D0C66F072072242C17497E087CB596AE5A264FFE2FB4B16652058940C5832FC750C257CA5D403AA
FB55E11BBD58C0B4A06085ECC2D97A6130E41CB4EE18CD2551B8B4E729855D9CDBE032B55E7D7D00
A562D4DF3736D65CFACBC96D873B9CFFF61DE86656945F914F02F924D62B6A45476CE6A66BAF101A
E5F461D69DAFFF24ABA5D79EFE8871E3C6C16C1689C59CCC8199EE0DF6C7C46FC1AFABABA748B01F
DAB9A6A5B5A7F770FEEEF2B0C4842FB6BC3EE8382ED7B50F069E01344AA2E8FD27FF68D78EA878A3
764EC5023B13823C93A453E1181DFBE744998E10742263A1D4A1F2786619946B09C8068321444435
734E9F61EE00F77FB85E805B619A920CD0AC67AA343500AB2EFC6ADC186F5198337710A642739357
6203CBC4C177070F1D3917B8E1566693278D3A7A09B958E04660F9D65E4283C19073E0E1C244073E
2595A758300E7C0A8FA4DEB129AA39070DBBF89273923BE33492D55D7FF25A3D4C68FD3E33A7FC4C
2A69685F2C95069F3AD9B32FAF5724EB11663CADA4475F21E6460A56E476FE1F76B750EA09A2496C
0FC59AF0677BC72B4565C663AD14D1F52BBF37988CCECC7B639928A86B2F1598F2B26B2ED58E4819
15097F260739174E42ED911CDE267D532858A1B9AD982B7E151483C160281818CCC06D1F31273841
8D7CDB078B51F5FF33F4C3AA7393270E802E2DAE1D58FA844797D66F1CE8DEE3D1A5BE13FD36F509
B00C6145F08A71C48416C970290368C5380C861C824FD3A9E39FD242239F920D25264F15A77F4AD0
DAE6C914D42E3DEBFFD7C39D4F3FB4B7D179332A38E06752FE9499C454A9C7AEFA36A3B99CDECBF9
ED20A28215758F0885A13382821505AB05DC1F11B0A5FB92C383EDD1F5B3421F037E6055C5220B4A
F5FCCAE003C0B2BFB9FABFCEDCBEC47FBDF9E8D81C9E93A7D241355AB082829CB6FD6830180A0C6E
0632D99602B69214CCDC01094E1E3902F7CA8C1D24BB2F92202F795558C1E3D5454DDDC860C81158
5A08E6B1A89B528C821B4AA1376F58606A9D50ED39F4E51DCEAB789FD1687115EBD367BA1B1B92FA
A4945B2A50289BD15C059EB4611EE3D6C05416A1F29A8953C16E9C421EF9F86A31DD299A21EE4B4F
3423227E2B62144505A737D66F1CF0BB8A3824161C68BCABFBC9149F1D611B0A30542803C8D01A71
1DD6D6D68A6085D918068321DF907986C1E14CB615310A267B16E7E65BE5808215D45DE43A0EE0D6
60059148245B2F0C86AC4131046E288983FE9251DF71EACC12C5FCACC16CDEFD8E5B94CAD9C5FA68
D70E3AA7E21EA820D5BE60B9BFD8EFE7268CD7E7CC6B9F08A3A16212082F6E0D9D11D479638669FE
DAC0AD2DBA2F75FC068C7656F22D9E04DE4C152D88635F7FFC62EFE6D467C8557E5C9EC047784D4B
2BF56464FB516C98B24C1030180CC5062C55D456C202C15817AE56F9D0564A21036B48D15DA78EC7
052B24621CF74B6AB584DE5483A114216A032B56AC60C28E53124E6F2895C4AC95CC732D52753D1D
EDFBDBDBBA1B1BB635D4036BEFB965E39C596DD32679B5A852702B45B228FB968E40FAC8EF8EF38E
DC29714630C42B4F5B7FFDB1E28C0C15D0156661BA834FF51DE80EFD8E0FDB75A913C4C49599FE5D
285AF04ECD9C77AB16ACC09F64BEC6AA0C06439E20D30B6CF5450BEA646F878B14B5AD0A90F36B48
1F6455522287355070A73AD7BE24BE2AEB6A83217DD0F7E125624C9CAA6BCB4A2246A93C50593B17
1CDB1BF8FAE33F822E812C1CDDF91AF8D7BEB666F2AF4D0BE777DE3B77CD6D37927FE15F197D51AE
2E9366B3847851BD5C9C11B9DD5F0203C5F010DF876C67F1EBF05DC5E39FCA55F706FE59E42E2AFD
758CCFA43F319E5A35716AB1F9130D06439941C4BA753003C5BA6DF2294230B661F9AAA7B41619EE
57316B3B1B0CC509D2074FF72052AA55B6B5E9FC8555565A13E0B03BF0FA80119ACAA18741925549
F13E98CD94018C3A23729161CAE1B1EB60179D62323C186AE89F6F8B9F7A07B6B060712385EC1FC9
7AE3F020E9C6F0C03B0C952FFE9B6530184A08DCCCE95CFB12D547C53F3565CA1458EC14A71D2A85
65A202C1ACB764A12F165A693024837E3A4E1DFF948A073A9A0BAFF18E5FDDD49E26079F0F7C1962
9F8034C13086D94CB223C1EAD4ED010FCADA57251715D79153EE4BF229161329DD21914ECB4BF7EA
087A9F9DA77BECE8CB58DB11CF7E795CA6C1600811429474E577D1B682654E1DB9219B6A8A15B4F4
B0A033F445C7363885E30D064320F80451F2452C2EFCD0CD211B4A8622875410D3197060408B16D4
E1FD8C7605DDE8C74871469CC75190738AAD1B8A1954BC776A3B6278D0C96841F2068361E4E0F6AC
547E97AC9C99F36E65B5A3D05B68480724C5B5B5B5B2E2535C7DC58A1526AE6E30A4006639AA0DE8
EC18F229B3968B1F09F42792EBE4856EC6324C3DB3B9A686A19BD9DD4D8AFB69C176FABFC8A742BF
7C4346A04AA45063C688525CDDAC1D83C1903544DB0A9603089498E2E3C68DA383C3ECF09280ECBE
32F4853BED52B58A65742C62DC607040EFC3A9E39FD2C34BFF146D662B9F5DBAC044C7450DF7545C
F6D5D553784FD39F06B93E9EF8F3A7F814B51F45FEC2C40D4A11B250720B05F7542A9AE1E6B23E42
09E9CF180C8622415C3EB7F730EC6D2C37926C8BF5A2B6B6B6682BBF1B52C31FFA4275A3DABA478C
55190C843C0578289C3C448998356BB944C19B4B6784D6ADC5944867C4B03AE7323C70A488A54B24
21CE8933CBFA68336A29024FF79A96565D25810F3E99F290DD5683C19009181C0EDB7BCA94298C75
11B1F4CD9B37C70FB389A574A00DC5FAFA7AAD2A8C3B8B77B20E7D3118CA0CACBEB762C50A7A1F24
631D13201E1FBD596D73602942C4D59938833910183BFA3229D69CAC24BAEC378A36FBE55755497D
467C9CE27E362A4A1DB8893081C0974503D0CBBCABA961A8BCD57634180CC382C10C98316036D4D6
3DA2CD09114BB7C9A4D421A12F4C8EE32DA6381558D5A9E39F9A7EA3A1928191CFEC18B7FA5EDD23
78DF9E8E9286BE7158CB1C3713F811FE5CBEEAA9406784FC29158471BCF8A728A32AEBA38D939283
5B32E0F4193CEFA2BDCF30511157B7FB6B3018860583C399C0ABC5D29D645B9B494A1D22AE2E11E3
4C28E026AD1C6637DA5051E0869254DFC3D371C9A8EFE035C5D22D38B69C009B39B8D26B4DCDD3CF
2C0F2C59E5D47AA6FF82B59E2D76BACCC05CB99EDEC38B16D4F96B713ADE6A83C160D010B174EAC2
C90442B1F414B387CD2AA508C6AEE076330210B6012358445CDD6EABA1D2C00D2598C7CE869255DF
2B63F0A653B24FEBF6C090E64D17016DC6C37B4746F81437A3C68D1B47C1408BDF284BE0D6631830
0942148F9904413D93219B130C064322C8A7409D1805C1C5026636FE94944C4319C099FC991180BB
AC235866CEBB75C58A15C6AA0C9503D9508221AD379430193EFDCC72DB8E2E6F30161A46B2764638
85A5985C436D1FC761A1C5FD6C9C941F707331034866A544CB33B5DC823A8A10B8291704FD94C4C7
0DA50E56EBA070A8537B85DA563651942B445C1DB71EE681DCFA2953A6AC6969D5A97336060C6509
FA2028968E194F17F3158F6DE88D34E416FED90C131D051E1D6704C5D5C57D89E1217A4D3892F181
26FF58F6D0E2EA323C40BAF127164A93752A3618A5328402293EB562C50A981330AAA558617D7D3D
C58BFC1F09BDD986DC8209055A5C9DEA461457B73B6E2863381B4A1CFF7A4329F4161AF2878442C0
114F25A6414C7D52378482159809393CB83E727AA458BA693F9637F46DC56C8039410F0FA6D159A8
7CB1C128952114B05011B7E624D88FE26F78DFCC89CA0123C6C1A3ABABA7E8343ABC6365C80C6509
BDA1246612050A9CEA7B662C5508C4194171F50B62215E94A110D388EE4BE153363C2A07B8E9D433
91F13076B417D44171750BE9291218A532141854B3F102C8EB1ED1B9D84CB685757DEAB825DB560A
C430C07D7FFA99E592A67DF9555522AE6EE924863203EC1F66C7C07E76742F75F53D43D9C3F13461
AE63A5574DA3846131BFD88CE74A83A6CF1205CA5181D903A305AC8AAA8FA137D56094CA5048904F
75AE7DC99139A257428BA51B2A0D3DBD871905AA59766D6DAD1657375BC250EA60EE0CA63B9D3C25
1B4A9642589990DB8D6910BC491647F9C13BDA3F65286F24BBCBA78E7FBAEB6017E60A1D0188D983
E6936995840EA354868281EAD9528C434779AD6969C552E2DF66B199A1FCE0BFA722144C1940B02A
2662E3F7D8D1D1D412B3250CA50E5D2DA266E25459EF30C26123BDF9E69BA1B7D05000A498C7E8BE
64B17BC7286285322B416500B827833944649D60442D5A5087B9C5827CC285512A43C1E0174B6735
228A1785DE3C4331E0C49F3FA5B83AC814670346BC38E2EA6654184A0B7A4349A7C954574FC1AC68
6A03150ED26D9ACAA2A5AF7FF0CEE557559155D1956903A65C91FA86522C97D6941657A79E09C5D5
6D488405A3548602004B00EEF59A96568AFB5D10CBC5C69FD4894DE6B918B2F5A2C2C0D0D037DF7C
13AB0307C905911D5A589ED4B9B2BA9686D202C5286802CD9E3D5BD4B045B3CB34582A048ECA9F7E
4D4937D67A1643084385EAFAF20EC5D57124187A465F67281BC86D3D78F020A714AE9222F0E56CD1
18C2C2486EC1086F9FDDFD3206B58C988BADEBBACA8E4AE82D34141B445C1D4C8A0306EBC5942953
A88766D385A184C00910768ECE13C44C085B486F28D9A8AE1038371A7FF61DE8A6EC800C0FAC8FAC
F58C9183DF78EDA867E3F89EDEC336662A1C60E218095819F5F090AA76965A653094194E1DF772B1
A915CC47FEDB7F7731F7521C8D1A7BEA0D434ADD48983877F5E9B1621EAE89AB1B4A02DE308E149F
62E283983DCC8BB10DA50A84CBA722E1A09E1EC5C4A9BA982F4913E3E199613A73DEADDC59E2020A
7A8E4F617935B3B93221B79BE1C4B5B5B5B25F8D172059583AAD6495C1503660300323B8B4583A5E
9BE6A7211912E2614E9FC138E1F6BE84BEB0707CE7DA9748BB426FB0C1900C74B6B2E61AAD65D8C3
5E25EBDA5AAB645DC910411E0C83E5AB9EE2FCC6FC62D025B027CC6FFA6000EFE820401C4C566503
A962A1B71F3D5593DA5A2D220A920E438B5548426FAAC1601821B8B7C6C75C5CD278CC1F5BF618F7
662D5BCA9002B25E60456049689D852785E3876CFC188A0C34724E1DFF946214980065039929813A
26C75081E0B4265AFA98D9A4DE3DDEC1F09025522637395E1267181C486784E8BCD9645809F0DF65
D6FA0487E25892E181F9076CCB6A3B1A0CA50BAD862D62E91744B482998B6DDB26868C40E969AC0E
3A715BA4D2CC3A35141BB4BAA96C28D1AD60EAA6950C29E9CBFD46ADA5CF60ADC0E28CC2AD601E33
88540B56E03C4CCA73BEC55069C0B4F3F433CB75CE260B889B068EC1508AE04CCEDD92D9B367EBDC
017BAE0D59406C09E6E13A4E4F29E8632684A148C0E42911A3D0D5223080FD45856CE85614287FDD
B9F6A599F36EA5903EDDEE60461833A045A937881CF56C114465EE95A7F3968612A0A15C81BB4FB6
2EAC0A038CA9556B5A5A312FD96EB6C15042607002F894C4F472BDA0505B3A621466601802C15099
9EDEC32C1C4F4B5532FD31BAACC4A121746094323A0B368C6C14535385166FE82D348405D1D2C7FA
88E1A12B0AF9F71B53541501FA0E7453B0428F3158D1FE0A8FB69E561478BB19D48139476A3B6295
D43E501B1506439143722467CF9E5D33716AA66214F68C1BD2014D56C636700B4E12BA6163F85D00
06436120C95314A310E7145E538C420E0BBDA986024382FDB8DF08E356C42860EB62C5C4B0C94848
9FAB2D663C963973224BB9DA9A7A4F25830BA516ACC078C35CC4A00EDBDB31188A1C2246A1EB2348
79569BDB0D3904631BC0D6038B7158951F432141DB156372C3860DD46413296C46739994B181E2B7
5C1F75C01EC5251C1337C568D1FF2247C319744008865F544FF2F4191B75950C0A5678FAFC91FA35
9894383C304751063059E16983C1102218CCB0A6A595B12EE29F62EE80F129430EE12FC6C1C14681
23BA444DFFC4906F38739A648F5E7E5595F63E605694A16802A7150BAE8F8E5609964B8C99ECB44A
F410E2D893321352F6514712DA90AB28E8DB2D8215B2FD88E181A94962446D6C180C4502AD418405
42D25B38A533173BF4461ACA15187E8CB3922AD2A27FD5B9F6258E3D5B2F0C7985943297001BEE06
B394B969175738A85522226C54B766841E288F1E1E598F136E69E26C92BEA7BFC2E4FA2B1C22D7AF
33EF28874246EFDFF331180C05866441325E579C53F84DB9009BC90DF983B643B80B77F95555B25E
24E85F996BC09053E81145154A7A1FA8DE7641ACC2A69532AF7078E1A0914ABE9E52BA0A51C64CB5
68415D8A02ACD9CD5452545AC742E37BF18EA556552CE48E630070E747B61F999781B94BF2F80C06
43E1A10BB032D84F17ECB6C22B864242A8FD8B2FBE88D541C62165635954DA6C09433EC03C058C3A
6DC462D4B1565AE8CD33141ECE560F272518AE129227AE819C4F4A0955AB6A6AB412606DDD239D6B
5FB23159C970D426F59485216A369BC11022A8ECC7F86DAD65CD4803CB79348482CD9B376BFDAB4B
467D87AC2AC586B0C1900598F7FDF433CBB5ED8A1733E7DD8A89D102692A1C180052E95E2717633A
627E53FEDC9742E5580B584430B8345B557403C6006599E952E7F8A4631DFF9242243677190C8501
83FD1860A0CD89DADA5A09F6B3E7D1501868713F8C3D3A0EB80547FD2BEEC25136D686A56184E066
2FEBB42628FB4D9C2A25A76D98551A12761155B09F147A6661B2CD9B37D3553492B4A9C0D7CE9F2C
4BE45445C71065FD3E1344AD58C842896180F94A0701E235C5FCD3291E6A301846089A0AA78E7F0A
939542C122088027316BE522832157E0FC8F71E80401B2CA2177E164BDB0C5C290113860602D6302
C48CE78C2E0BF62B6FA4395D600CF8634161AC5229BD30738E7C0BBE9141801249327674424D619B
032B164CDC6010A0DE18A748A9D9720643BED17FFA0C773628D6AA230ADA3B5EB1C9D95054D8BC79
B323595C5D3D0506C6860D1B2C08D0903E646663B01F6B0089F7E19251DFA9ADADDD75B0CB0655A5
C159F2383CBCF4FF48997B191EAC24128AF0238300592B4D42BC48F128A08D35DDB9165BC72B0A54
02C442496F3B85222FBFAA8AF262A6566A30E40A4E80016B0071E75FD60BBCE0CEBFCCCC06439180
8356AA1C6A696B6ED206DAC0B6821802112D2D3D71EAD8D197C99E3FA5786093E8236D08952B92DD
592AFB319153E7D6D1394521FDD467C82B24CA4B3BCE44EA2DAC56194287C47F62882E5FF51446AF
0C0F7ADE31E3E93C0E3FEFB6916330A486FF1911951898133AD88FA6A939880D45085D0E983905B0
84254C8BEE2A47742B456282A1322186042CD2450BEA606348E614CBF86A291E1B301502FF8DA636
359528C42825DDCE792C68A6C34CC6305A02B359379263986AFFD99DDC50360075625D33CDBB315A
74F29DC1601821B8370BF644F920D97C930D2E83A1F801F6545BF788CE296029EA352DAD96FF6270
A0ED87BE03DD18243A730A43085647D43B6F7CAAB28131D0B9F625CC2D34442518BEB6B6D65F95CC
192D3951A8C80812E5A54508D95AD1CD188A44F887DEB186824117AEC208A1EC982C94978CFA8EB3
796E739DC1900558184E6ABEEB4CDBA6A626D6B8743E62CF9AA138C1545C980D7A30737300B631C5
00E5C8D05B6B28069C3AFE29754D7509697AE7F1BE66E236668A0A85B91D9239555D3D450F0FCC30
1B366CD04A1429DA939DCB29FD0FEAE3B566052C6489F2A22C2AAE82D955A925D66DA89737748A87
0C690C0FEA553284D5C680C19011F463E5C8A43348C00A5B184A028E9CB5D412D2953719DEC0EC2A
733A5432E277FFF4996816DEC4A9E05312EC072B74F9AAA72CD4B98430920739F0B34C9B0258732A
41453F9239F5E69B6FFA354573E59F1AF9A54903B0BED3712FCE08CC8A12CB5AB028689B698B10CC
AEE276BA102BD1353D75DC5895C1902EB05888F2AA2653A933A74AFD112BF5F61B1C04A604F24534
56C7A9D8128965F582C62DE8A5B2110F758E599BF881B53973DEAD3AD4D9668CE2445EEF0B4F4E32
A29D535248B7009B8D39A463682DEBA1489EE9059162C4981B3113E6BB330DC58F0D1B368065CB0E
A4247D1466A81B0CA50E589B4F3FB31C0F91535363D1823A4DA66C33DF50A29011DBD37BD8535C49
DC37C0C867F998D0DB692818B4463AAC65EE26E9A0172A5F79BE099BEE2A1574DC80685086826E1D
89895ABEEAA9D28D1C46CB31E9E12A74F8A296ADB0E5BEA2E0DCEEA8D2A95A28BFFD7717E34FCA56
58E5088321109857B937CBAD7B262752110866863F12C0602875C04866B28C8E6EE5986F6A6A3262
550910B381A26D5A8487D63245DB6C02AC40E8B47D27FF88D3057E63F6D0D5C34B0E32FE199AA265
2BA84BC09950FBEEED41A81C8837939BED74576160C03EC4EF2953A690771BB1325438F4AC78EA78
746F56C74131C164C3860DA5BBF96630A4038CF037DF7C932A5832FEA982D5B9F625D3032C6F605A
83C508BB5134FDA4D825AC8561B3F50D6509BD518F31807900F303CD48D96FC4FC807F259B1F4A6E
B9640235F5002F503F781670A5B404D20C8A2EB96B37A403EE3B3DB6EC31CC8DDA5064FE9DF06EBB
FB868A059E11463268AF2E560D6AFA95F4E69BC1903E6838C16C9839EF566D4B8058D5D7D76FDEBC
F9D4F14F876CB1282F50D194B924AE0D59F7881681D4B0315021E09CC0C80D96B617190ABCA3375B
B2181223944FCF5A2730F57FF942E2559C874292A92DDBB4022183843BF08B16D4E93051A9C2667B
5086CA049E0B9653770460A9292D9A7E16EE62287BE8F01EDAD853A64CD1E6049E11D66DB1F08692
86CE9992DAAC921423715CBA40CF50B646ACA17401CAB0EB6097142393023D42A67A7A0F478F2CBB
E2E0B2E20B9D946BA738AAB824FC25B752280219CA06BCD19821376CD88065D14FACFCA180360C0C
E5072D77C6303F874C8D1B378E057A5877C0C894A13221FAB19A5831C2810130CCB1B2A7A3144101
0AE64C0993BA20924027D9D6C5938C1F7A032A0D5264415776BE20965584679F64AAEC59B653C18A
418F0EB132978481D3A9D6CEBD20A6E7E3547B1C2AEBE7C55039D0E60137161C71606EC2637AD49E
A9E269B6C15048E8E7054F04CC09C76305DB9B013016165B42C0D2CFF46A16DAD31BEF3006F07E8A
D2BD3617953D58868C7518B55E349F777AA8393CCA753004D6CFD2DBB07449E86787312D8B16D439
15FD929DD9507E9048510979D2C303532B1EA8352DAD6996BD36188A19BA6CE5AE835D8C72D17A3E
1744EABC73AF895E5A1BED06C350E49189BE88E55388D297DE8863594FAB265FB4105319D6A04E17
9518152EF769C6A8D85D2E33E0864A003C1E676778C03E24D77616C7F27351A51EF372BDE82B86CB
3A72EBACC9853EE4BE4439F58CC18F64F7971E5EA6A63A8F12E65E513809BDFD064376A039016B70
E6BC5B9DCD25A69A72E7CD76DA0D8640381E5E985E14197636B1750C8C9913C500162BC7D4575BF7
88C3A49C8A2A76BFCA1281AE16FD5F0C0F46BFEB0A8C7A6B5D3C2FF91E21D99D3FAC718BEFA59E36
251075B897702B702E9A16A6625139D04F1CF52B9C5AD8B20F491D5DEDB732188A199262CF19CF3F
A4FDE69F8D6D8361588839E1DF88134B8C99B9B61797AB0ECFE0E0D3677A7A0F63B17E6CD963B83B
8E9DCC1B84D5DC0952B2A9AFCC1078433D6614F357E209F5AF8CA40333E7DD1A55B4CB7FD8674E4E
9BE64972FB5D4EAD2E4C86FE3D5BC6D3B2A615136A02B79BECE92B3F38FB90BAD4A9333C66CF9E8D
B9DA36F60DC5060C5DBA5C317A172DA8630083B35ED4D4D44890928C5E9BD00C86ECC08D0B4A5838
CF9A17E730712AAC77092AB3072D57E05C8725181D8BFEC784864EAEAFAF0787E28EA8FF5EE05F22
E2676EC4F2860C0FD657E2F0C0C207A30ECF29CC7EADDD77810A492293721ED5F45D541931F4118A
0416D5006617513F93C6B393A62D7415FDCF1D27DC113C8CACE7259E2CDBE22863907A6FD8B001F3
B07FB994C7100C0B077078705B123FB67A1AF2078C4C6D4E60068345070E059BC1096E91818A512A
BA2BA6E06730E404CE469C760AFBCD09BC892794E6049E591C4C73829B1BF648A606673C2F84AFB6
164415131DFA13169A7FBA93490FFFC591589DD9DBC36689DA2D284548828F08D8A6333C68DEE330
ED401179DB7406C9480851AE14D7753257166D4EF3E02C5C75CE7C48B344EB6AF247882D6F160E83
A5ED2F5E6028033823E7D4F14F771DECD29BFFFEE1A16772191E126060FE2C43D6E0E68FCC4EF8E1
62C1F5C23F14F13EEDB7C7963D26F12D7A048E64B634180C29C098C01491457C42F57A8163F050E3
81D559F00607E282D77BE0D27B586D1DC62ADE289BE52A0192DAE3B1AAD8F08832EBD8F0803D468A
4D0E255542529F39F501197199F4BF6BD817C30AEBE5DBED9534B4523F6EA7CF700798DBBFB061E8
2BE4F62F9994C40A720384DA05F8A03DAAA58E7446BB44D193616178E021BD29F223862E87078B42
8A6CA00D0FC348C04989461A193DAB4278E6C4C4A9187EB367CFAEAD7B0407B05C8E66F136F60C86
9C6358438B8B059F5C98703027603030F4C8D945976C6E53CA1A165C7FD9A507233F12199273679F
DD88E287738F6478D00BCCE1D1D37B980BA29F626774F2146F6671CC502239CA62B065F4C191F867
FD9B1223216E9C15253293B7898FB3CE4730941C463E6132DB914FB1FCC8E36CCE29430EE1A40F88
2D9162B72D30C020EB3D3783C1901D24A088469DB35800B0FA2C5C3C9D6E1CF96125D7C925D7E022
E9A5614944A6BC20B5BB7324B17FF9B8FC749A9AC366A43E673A1B50B9ED901C76A021C48E2D86B1
61A858D8F03338B0F1103ACAEF16FCFF014722F5ED
}
\caption{Extraction of frame mask feature between two signals. (With only the time-shifted windows shown)}
\label{FIG:W2W_MASKS}
\end{figure}

\subsection{Experimental Settings}
\label{sec:evaluationSetup}
For experimental evaluation, we extracted two sets of English vowels, /iy/ and /u/\footnote{We use these phonetic symbols as in the database's documents.}, from the TIMIT database \cite{timit}.
The vowels were from $390$ speakers.
For each speaker, there were $3$ samples of /iy/ as well as $3$ samples of /u/ included.
The signals were down-sampled at $8000$ Hz.
Fundamental frequency was obtained with the method proposed in \cite{Huang2013,Huang2017} and assumed known throughout the evaluation. 

We chose from the $390$ speakers a reference speaker whose fundamental frequency was about the average of all speakers'.
For the NSGT, we used Hann window with support interval of $20$ms length.
The time hop-size $\mcnt{a}$ was set to $4$ms.
For the pitch-dependent frequency hop-size, i.e., (\ref{EQ:BF0}), we set $\mcnt{q}=75$ according to pilot tests.
For $\mcnt{p}$, we used an average value of the first formant frequency ($F1$) as anchor frequency and the average $\bar{f}_{0}$ of a speaker as reference and fix $\mcnt{p}=\left\lfloor\slfrac{F1}{\bar{f}_{0}}\right\rceil$ for the speaker.
We used $F1=280$ Hz and $F1=310$ Hz for /iy/ and /u/, respectively \cite{ladefoged_course_2011}.
For (\ref{EQ:HAMASK_SOLU}), we empirically set $\boldsymbol{\sigma}^{\mathrm{Ref}}=\mathbf{1}$ (all-ones) and  $\mu=10^{-7}$.
Part of the routines in the LTFAT toolbox \cite{soendxxl10,ltfatnote030} were used to implement the NSGT.

For each vowel type, the frame masks for an input speaker were computed from $3\times 3$ pairs of signals \footnote{As there were also $3$ samples from the reference speaker.}.
To obtain a variety of masks, for a signal pair we computed the frame masks as illustrated in Fig. \ref{FIG:W2W_MASKS}. 
Hence, $C^{\mathrm{A}}$ and $C^{\mathrm{B}}$ in (\ref{EQ:HAMASK_SOLU}) were one-columnwise for the feature extraction.
The obtained mask vectors were used as speaker feature vectors.
We employed fully connected deep neural network (DNN) for the evaluation.
The feature vectors were divided in the following way for training and testing.
For each speaker, $\slfrac{2}{3}$ of the speaker's masks were randomly selected as training data, and the rest $\slfrac{1}{3}$ were used for testing. 
The DNN structure  was set as  $1200-1024-1024-1024-390$. For DNN training, the following settings were used \cite{HintonSalakhutdinov2006b, Hinton2010Guide, Xu2014}. The number of epoch for RBM pre-training was $50$, with learning rate set as $10^{-5}$. The number of epochs for DNN fine-tuning was $25$, where in the first $5$ epochs only the parameters of the output layer were adjusted. The mini-batch size was set to $100$.

\subsection{Results}
\begin{figure}[!t]
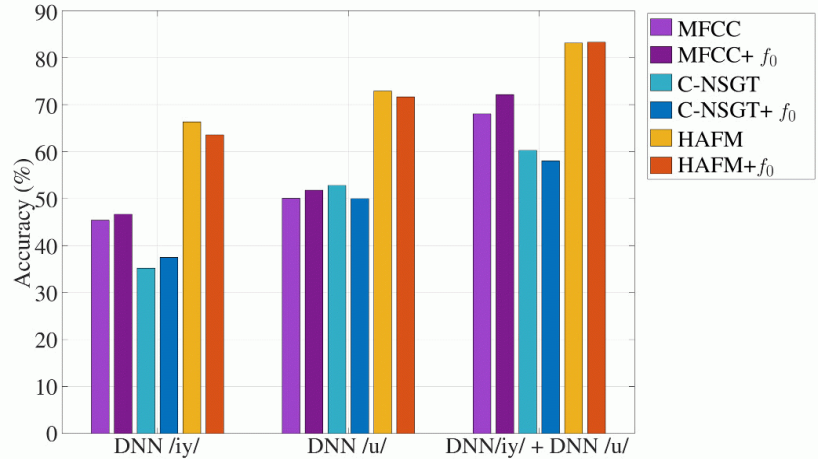

\vspace{13em}
\hspace{5em}\pdfinlimg{0.33}{0.33}{1052}{599}{
789CECFD7F8C5DE7751E8C028D12BB9F59DFD8993A464219D7B6CCEBA6A253BB43D84887EC053516
14693C484C9182E8A92B998412C76C1B9883E002FA81DAE94D4D57EAFDEC86922510452033B83173
AABAD680D790C74A51304C508E027C06A344325340A588DC2BF11FC3AAFFB9838BBBCE59E7ACB3F6
5AEBFDB1F73967EF73F65EC4A3A33DFBEC73E6CC9E77F67E9FF759CFB376FF7FBBBB0E87C3E17038
1C0E87C3E170381C0E87C3E170381C0E87C3E170381C0E87C3E170381C0E87C3E170381C0E87C3E1
70381C0E87C3E170381C0E87C3E170381C0E87C3E170381C0E87C3E170381C0E87C3E170381C0E47
8D78EB7FBDB5D5DBDEDA7AFEEA5FFE55F2E0EBAF5FDFFE5EFFE0577EF84AE39FDCE170381C0E87C3
E170348B9B376F6E6C6C2C2F2F3F73EE99338F9F591EFCFB933FBD641F7CE38DFB8F1E3FB0FFE093
67CF9EFFE6B756560EADADADC13B002569FC0771381C0E87C3E170381CF503E8C0BE7DFB8017F09D
C01AF6DEBAB7D7EB8983AFBF7E1D0E06C6C17702AD809DC0351AFF591C0E87C3E170381C0E479D40
6501D8043002A14A00D1004EB1B4B42498021C0CFBB7BFB7CDDF61E7CA0EECBCEBE0E1C67F2287C3
E170381C0E87C35133804A7CF83D1F06EE805FF2FA252C708247DA035402B8031010FD3EB0139EDA
DA7ABEF19FC8E170381C0E87C3E170D489CDCDD3400756560EE9A79E3C7B16680280880616449DFA
AD53FA60D8A90BA81C0E87C3E170381C0E47EB81854FABABABFAA9ADADE7F70DFE91FA8022C5534F
3CAD0F06028252853BB51D0E87C3E170381C8E4E01EDD5A615E2F2E5979044005FD81D9926E0CBF3
DFFC963EB8D7EB21A780C39C56381C0E87C3E170381CDD017AAE01B4871801B003D4291E79F831F8
12A804B2862B2F6EEBF7210272FE996F88A76EDEBC7974FDC8E947BFF4C8EF9C82B71A826F3BA607
F88536FE191C8EA9E3F77EF7CB2B2B87E0B1F14FE2704C083E8C716037FE915A8CCDCDD3F488279C
CF4F7C09D4E19822E00F0D2E68A82F88A75EF9E12B4813F06FF0EC57BF821403E8033F0CFF245FDD
79199F455143C0DCE9980580FA5D7FFD7AE31FC3E1983ACE3C7EA6F1CFE0704C1D8F0C56ED1CF540
700A87C3312D001DB8FA977FF58E3DEF042EB0B1B1413B9126FCC99F5E5A5A5AA26227B8A1A34E61
76D92602A22F8F376FDE34CBA51CB3C033E79E8113DEF8C77038A60E9F0C385A09E7147EB61D8ED6
00FEC4902CF0BFB5DEB3E761E781FD0749C278EA89A7C931A1DF048806173538608A0B13DDC67FCC
8EE0FC33DFB8F9E31F35FE311C8EA9C3750A472BE164B94EE03C27BFE4E9D59D971FFEF257E077E4
70D4804746B591F4A5B98C3FE7809B35BA2A5098800DD823A29CCE7FF35BC81A44ED13821CDC9A3E
B84E51275CA770B415BFE7532F471BE12BE775A2EC65042BB7DD76E1680A8BEB1D00B2B0FDBD6D60
E5F8E5F2F2327004EA46015C091987D9D8EECA8BDB4840A8C536C1758A3A01A7DAAF7E8E56C23985
A395704E31CF67DB6FA98EFA4143AE356BF2C01150B920D9057E46640D2641A054285D78032FD461
508E19A1AF53DC78A3F18FE1704C17701939F3F819BFB93BDA87CDCDD38D7F86EEA0824EE1971D47
5368CD9ABC102910D820CFBC00A229C3ECC7DD9A73B210D8FEDEB65F001DAD84D6401D8E16A0D7EB
35FE19BA83B23A85730A478358449D82FE5E6803A8844910801A849A6E23DD30EBBE16F19C2C2EDC
4CE1682B7C6C3B5A095796EB44599DC26B9F1C0D62D1D7E4E16F07F39D969797CDBF23786A696949
DCDCE14BD138AF4DE764B1E0F32E475BE163DBD14AF8C0AE136539C5E367BF21E64253A4181D642B
1DFC91273C5D0B3A7FC65FF4A9DF3A059461636323F47BDFFE5EDF67F1C089937C27BCCA74672FFA
395944F81FACA3ADF0A997A395F0815D274A738AAF3F8E1B68E97AE4774E0136374F4F0599F9D893
DCD66B9E12C4BF9DCF4FCAA25FE7B3807E649CF6032FB8EBE0E1ADADE7E3BF77E00EC03B1E79F8B1
EBAF5F879F1736E0855BBDEDD839F1DAA7BAE0B727475BE163DBD14AF8C0AE139378B48F1D3871EE
F095DEDD57FB58FD3FFEE33D3B804936563E70EFD427E17338ABC76FAACBEC1D39A76EE1D6E48FAE
1FD9D8D8803F9CCCCE1AD8681BA8D3A9DF3A05483A98BCF6A94EF8EDC9D156F8D876B4123EB0EB44
354E81939C7BF73FD8670487FE0C980502A8013DE66C8847E0148D9F9029C2C9C2D4E1F3678D45E4
598B0BBF3D39DA0A1FDB8E56C207769DA8E0D1DE1DCD968FAD9E004E4184E2F7EFBC84F8DAE11768
0F6EE83DB8211E57F6FEDAEE8CA7E293BCB974CEDE78636BEBF95EAFB7D5DBBEFA977FC5DF391938
F9277F7AA937F877E5C5EDEBAF5FA7FDAFEEBCECE33F82D6CC9FA738C817B41E6C41E10B058EB6C2
6F3D8E56C207769DA8ECA7D81D708A470EF7B07289D487FF78E8CFE84B0E73A7C0EAADBF667ED327
CF9E5D593974CB4FDD42E83D7B3EE7D3BE63CF3BF9AB30BD137EE47DFBF6F1FD11C0917A1601EFB0
B4B4840760200F6E2F2F2FC3530F9C3809AFA283798D135086D5D5553C189B26E32784C78D8D0D98
2DC386878447B088DE81C967A1C9FABD76F0AC8580DF9E1C6D858F6D472BE103BB4E4CE8A7401211
A109581C0587A15481CC8276F6B7471BFDDAA7BDBF16993EC1C0A0A97E285793A3D7EB716A209E85
69183D0BEF8C7E58DAD8B9B273E6F133FA8530E1A73714937FF892A88AFE76F073DD7FF4383EBBB9
799A0F72780AF610FDF1FE2C1178ED939F93C6CF76E39FC1E198057C6C3B5A091FD87562124E71EF
A18DB8EE007C21CE3832750A02A90380E47AFE81FD07E96011CE0900EE90A427BD67CFF367E167C7
97ACADAD855E82C4013E27DF09678CB48C3FFFC14BE60B81C538A748C2D7E4FD9C340BBF3D39DA0A
1FDB8E56C207769D989053442A9A508CE0860BEEBC105C030F4B720A98999316A0DB1073E0147D73
F3344903E200523DE292C73BF6BC1337FEE44F2F090D2224A9604513DF43EC26CE835004317B253B
E8B7E6DE01E39C2C5A3DD8E2C26F4F8EB6C2C7B6A395F0815D2726CF92E50C82D881A012F4659F4D
8CB806EEA43D399C02A6EB4F3DF134A90FAFFCF095D0911B1B1BCBBF721BCCB5F0C853BF754A1C10
E2148229DCB8F61A6ED03705B6A20FE3B87CF925388C0E207563757535797A812839A788C0D7E435
BCF6A9E6B3DDF867703866011FDB8E56C207769D98BCF6694822468462B831A87A2A7C4954029F62
07D0469C53C0F75D5A5A82E9131084780D12F285EDEF6DD37CBE824E416C627760BE40BF83364AE8
0FB93B2020B487EAB5926150FDAE04DFFC96FEA80EFE5BF3F9B380F3AC3AE1B727475BE163DBD14A
F8C0AE136539C5F9B35FA789F1C95F3949BE094E28B44E517058B09A28C135229C02BF29EA14DCAC
6D8E16CC65DA651A418453204DD0B37DEE89A0D4A9471E7E4C7C2413C447A8628AB38C387CFC47E0
F3670DAF07ABF96C37FE191C8E59C0C7B6A395F0815D27A6E2D1D60C22DF9ACDF9488E9F024B8336
363678519398DEC37E9C7952B8537EED13E2D59D97390B205EB0B5F57CA96850AAD48A0B1CDE533B
13EE1DD0709E5527FCF6E4682B7C6C3B5A091FD875625A1E6D5EE9643A2942CA05B763E4708A338F
9F818D577EF80ACDF3C53C1CB509DC264E11D129B09E0A002F44A0B39B7CD65C1641338506FF0C7C
FB811327731CE58E4C78ED939F93C6CF76E39FC1E198057C6C3B5A091FD875625A1EED1095488814
C438069464F5DDF7C6BF3BE914BBAC1E49FC08700CECC10F19E31437DEC0A7803B2C17FF9103028F
BC71EDB51C57B889B5B5B564FCAC231FBE26EFE7A459F8EDC9D156F8D876B4123EB0EBC4B46A9F84
33224D228A87E5EB144F3DF1346E93558107B7621CABCE5C2A5BFB74E6F133F4B65C1309E9142678
9FBB9CD0274712EE1DB0CF89D783D578B61BFF0C0EC72CE063DBD14AF8C0AE13D3E11464CD2E69A3
10DEED9C2C59AC7DA22F71C64EF3CCB5B5B58D8D0DFE692BE73E914E013F2FF753E823B50982F650
ED137E17ED9BC00DB75164C2D7E435BCF6A9E6B3DDF867703866011FDB8E56C207769D989A9FE29E
1DEC34215A54E8C7D097E7F27ADE61ED137E865EAFC767EC58A4C4CB93623AC5A8F629C4294810D9
1D74BED3B94F39A00619F9B94F8E087CFEACE13CAB4EF8EDC9D156F8D876B4123EB0EBC454748AB1
D670E8CFA847B6E1D72ECA19FCE07C4EC1A7FABBACFBC3D6D6F30F9C3829ACD0156A9F4CD1813A62
9856EB88CAC0FDDD7FF2A7971AFF752F3A7CFEACE1F560359FEDC63F83C3310BF8D876B4123EB0EB
44594E0133BA9B3FFE116E93471BD50AF444C036A709F814C919B481DB7824ED2C5BFBB4CB58C3F2
F2B29EB7E7E43E85740A0E8C96455C7FFD3AEDCFA9595A5D5DC517F2A22C4735B87740C379569DF0
DB93A3ADF0B1ED68257C60D78909758A3E8FE0D94D8A4D98EE09FE144792532C2D2D51EE138166FB
C038CC5C599353904BC2EC1CA18D0F1833F58E3DEFBCEBE061F194067010A21E57FFF2AF7262A3E8
DD7ABD5EE3A3626EE1B54F7E4E1A3FDB8D7F06876316F0B1ED68257C60D7890939C5EFDF79E96B87
5FE07E0A7255FCFEA72E857C16B0C7B45A2439054CE9B5A301DB49509F3BF1D3856A9F28CD89F7CB
4E7E77E1AA30690590023892D49CDD418A147D2F2E7308C0BB6921C6214E91CF9FFD9C3408BF3D39
DA0A1FDB8E56C207769D988453DCB9BCFAC4AF6C3FFAB13F9C16E2FD29B05AE9C0FE83623F290EFA
25E483D051AE14459BEF9E860F4049532B2B876E5C7B4D1C003C05F6EFBD75AFE61AD450BB9F5265
55EFA0A4A2551887F815B877C038275E0F56E3D96EFC33381CB3808F6D472BE103BB4E54E014B801
D3E6CB975FEA59FFB6B69E8FECD4CFF2A7CC6FBA736507BE2FC9044010601225DA558B1F04DE8DF4
0BC443C71FC4A917BC1BCCF0C9DC8D0401F644140471C6E893007D38BA7E04980B7C247C43781F53
BC809DF0FEE4ADC06F0A1F69636303585288A438F469F4357901AF7DAAF96C37FE191C8E59C0C7B6
A395F0815D2726D129A6DE5521F486AFEEBCBCFDBDED3FF9D34B00D8002E038FFC6098AE8BD76EF5
8687014F8157E13660776070C077BBF2E2F6D67F7B0137E031D9239B7F0B7887471E7EECFEA3C7D7
D6D6E0F1CCE367F0CD93670667C54043E085C0471EFCC2E781E9D098F7461571F8FC59C379569DF0
DB93A3ADF0B1ED68257C60D789CA3AC5EC109F90278F99E41BCD7A4AEF9461F213E8F36701AF07AB
F96C37FE191C8E59C0C7B6A395F0815D272A64C9CEE7C4987FAA39FC843944C991847B07349C67D5
09BF3D39DA0A1FDB8E56C207769DA8A053B4E9173423BDC3312378ED939F93C6CF76E39FC1E19805
7C6C3B5A091FD875623E758A09BFC54C3FE174DFDCF948D9D3E5F3673F270DC26F4F8EB6C2C7B6A3
95F0815D27E6935308E4170E657EB65034D3FCFC5C0E13EE1DB0CF89D783D578B61BFF0C0EC72CE0
63DBD14AF8C0AE1393E43E391C35C3D7E435BCF6A9E6B3DDF867703866011FDB8E56622106B639AF
7EE489AF3CF5C4D3671E3F530DF072787C73F727A1F79F05E69353CC1B6D99B7CFD359F8FC59C379
569D5888DB93C351013EB61DADC4AC0736BCFFC35FFE0ADC8571FEFFF8D7E1BF33B00D1B08DC8F8F
3085EE1F73F61BF0258779133FFEC1BF77EDD37BAFDDBB0478EDF82F5EBDEF7DF0087BE2DBFD970C
B6E1F1FCAFECB9F2E2769D33D839CC923551ADC669F2D7567E79D9A05AA72D9967D5E7CF025E0F56
F3D96EFC33381CB3808F6D472B31EB81FDCA0F5FF97FFEDFFEDEFFE7C23EC4CD6FDF0678EBBFECC7
6DDA83A09DFD8DEFECFD7FFFBFFE113CC2F6E6A76FD193C05FFD85B7F5A901708A4FEFBDFAEBEFC9
7CEC6FDCF7BEFEF67A9F53F0D66935A053B94FB5CDDB2B7F23671671B87740C379569D58DCAB9FC3
11878F6D472B31EB817DF52FFFEAF7BFB8A7CF112EBE7F88EFEC7DEBB9E18609601CFD8D8BEF4766
011BA73EF536F1B670673FB9EFE781170C99C28059A06651F8523D0EA58AF5BDC8292E5F7EA9CEB3
BD283AC5B4505674F049FE5CC16B9FFC9C347EB61BFF0C0EC72CE063DBD14AD4C3298634E1DBB7A1
EE80DC41881405B562402586A2C673EF3F71F7CFE2BBF139E7C6FBF6FCCDFACFF5D901AA0F233102
5983B9313CF2D8521FF7BDEF1B1FFFD9C5D2291A99724F58BCE4585CF89ABC9F9366E1F32E475BE1
63DBD14AD4A653F4B50994219E7B3FEA14B8473C720983D8C7E6A76FD1EFFCABBFF0B67E1513F005
A14120706771A34F2BE0CBF5A1A5E2FCAFECF9933FBD54E7D92EC52960F6A22B4F6AEBD730CF5462
9E1B6AB409EE1DB0CF89D783D578B61BFF0C0EC72CE063DBD14AD4C029CE7FF1FF142A730A01D9C4
5B2FF45906FA29F87BE29C70E37D7B9026208832F4BF3CB6441BF488A4836AA5FE66FDE7BEF1F19F
BDF2E2769D67BBAC4E71E6F133F423AFECFDB5DBFFCEED53C4DEB7FF9FF9F9DC9DFBC9367C3C18AE
65EDD8889B837FFA8573FE23377EC27D4D5EC06B9F6A3EDB8D7F06876316F0B1ED6825EAE114C811
D07CDD572206455063E097B47350EFC40F30758A61ED137A2848A4A02F5192200CACD95CAD00A201
9C62CEFD14307BB9F9E31FED0E6677F7EE7FF0D42FFDBBCDDBCE3CFA8BFF7E2AF8F07B3E6C7ED3F3
DFFCD6CACAA1D5C13FDCD8B9B293F369373636EE3A78188E5F5B5B5B5E5E868DDD41F9166CE07EFA
27BEA47F0F9C38A927F9F0DD61FF3BF6BCF3969FBA0501DBFBF6ED839309A36BABB7BDF7D6BDE6E7
41A1E7C0FE83F4427CEDFD478F6F6D3DFFCA0F5F81A76A16AA160B3E7FD6709E55277CDEE5682B7C
6C3B5A891A729FFA7E8A81DCF0D60B016182B109B18164447BB477479C62A841105988273ED19703
96019C629E6B9F768BFD29D63EB2362D3681F8F83BEE887C6B98ABD33C1CD9411CC0CEF8BC5D8C2B
A01BF42CFC50009898D1C6230F3F06E4059E02A6402F411101D807BEEAD46F9D226A031B9B9BA7E9
0DF9AB08C026E8D93E351B7C1E2C5C813DF4DA5EAF57CFAF1E7E96A79E781A3EF602CD483B357FCE
54ACBC1EAC4EF8BCCBD156F8D876B4123579B49F7B3F992386948110902A701B058B18A718F82924
83E001B3DCA03D3A1231E77E8ADD22A7B877FF8375720AC0D2D212CDBD6F5C7B2D7EF0D1F5237430
30083D0CF0295353C09F11E6DB9C1DC077C497C0C7304729EC4476000788B7BAFFE8717C2D158FE9
5F44CD9C023E2AD6E1037B2A3B0C9AC2227A07CA3634A9E01BEA0ECF6A1C3EEF72B4153EB61DAD44
6D9C82CA9924B930AB9E38DD08D73E0D8B9DD61959606CA2C020D836518FC5D5291EBFF5991A3805
CCFFA9E248D304318AE0189AC93FF2F063FC59541C229C820E237670F3C61B2479D0B3E6F78503E0
55FCD953BF750A5F0B673B3263445A31F5B45EF33BE2F7DA1DF0267D7EE6168B52FB24CE79856B5A
3EB3589473D20EF8BCCBD156F8D876B41275D63E51EE9341288A1E0A1E12054FC56B9F72EB9D4612
063DD50E9D82F845A98D4C4E81936164169179174C9561624FE54630AB97C36CC411229C6277E01F
C1EF4226085C278F7C6BFCA6F425FC4273BE11023E738453001D868F3DC91F087DEC7DFBF6C14F04
1BD75FBFBE403D475AB0266F1AF3433F6CCEEFBA05E76481E0F32E475BE163DBD14AD4AF53682376
E129A15C0CB6E33A454195209D427C6949188BC529D63E720C88C0631FAC4FA7C029F7DADA1ACED2
81389887A10681CE8820A7180819C9A93EFEB0C40B48A4D88DD20ACE29C82B9133F13BF3F899C86F
04DF672ABFF7C8D99B672CAE7760FB7BDB2B2B870047D78FC090585D5D853DA1C430182A700C1C79
FFD1E3E8BEE1CF1AE764D1EAC116173EEF72B4153EB61DAD44AD7E8A8B8A3B28A7F6B8F689F18E4C
3F059727C65FDE6B0B190BE7D13E76E044CD7E0A985FC10778E587AF9051C29C652195C04F5B9953
702705D5503D70E2A479B0F818BD67CFE3C6F5D7AFC73FAA78397CAA4898B02035D500DF0B29129F
882E4A80ED62ADC9D35985E107C309C62D3D0523134537FD2AA0CCF0148C1CFC1236E04BD819FA2E
5EFB54277CDEE5682B7C6C3B5A89DA6A9F42254FC28E6D064025FC14219142B9B6F9610BC729A698
FB8432474EEDD399C7CFC007585E5EC689FA534F3CAD0F5B5A5AC24578D229F49A3C4CD5B0800A39
859854A3D2415F122F882C08F377A06DF87864EB4EBE2AB4877F8CC93905FED27910D6A2108ADDC5
9C3FC3988131463493B0B1B101FBE17224AC375C98E0EF1051E516EE9C2C2E7CDEE5682B7C6C3B5A
8966729F8463424B15234E91D6294C0621140AEB00D8B958B54FDA4F51A1020A5EB279EBEFE5FB29
B0F87FFB7BDBBA1809814FE187249D42CFC7B84E71F3C61B37AEBD06B33BC49517B7D13D410713A7
C8EC8B412077F6CACAA1C97F5913720A38270F1D7F10A6B2588EF5C08993B07DF6AB5F99DDE89A3A
1671FE8CBA035C76C47E640A9C14C320843D5C38A3BF35DCCF958E853E278B0B9F7739DA0A1FDB8E
56A2619D22D4A282729F32740A23DC895B27784D142717EB8BA753444844A62FBB54EDD387DFF361
3214C314CBD40E60F64E915039B54F21D0EC9D226401E6A48E43ACF9539E6DA476251F93EB1468FB
5D5A5A5A5E5EC646DEF8811745AA5844EF000C542071BAA40D4D133C5B18B38B61CCE83741266896
DE2DE239595CF8BCCBD156F8D876B4120DF82944D59365D62E1090B0477BA85384EB9DF47ECE3BEA
E7143A64357EBCF6684F225294E51458FB84DBBC85DCD88F304873A235E148ED53C44F71FDF5EBAB
ABABA453704F445CA7D0F373B293DF75F070FE490E615AB54F7842F4C7E8F57AD8F56FFB7BDB5319
5D53C722D63EC1AF1E4DD97C276A46B09F9FEA7D837F26CD3FFBD5AFC053A15E2A0B94DCB5E85814
F6ED709485730A472BD154EE938C81D23E0B76588853BC76FC17852421D262CD16DBFD970C5E35EB
DA27DD570BED09F9EF10E1143578B4A9F60941BD2AE8A46D6C6CF02AA358ED532A4B9638055734CA
4EB6A95BF781FD0705E3889FF6ADADE7E125CBCBCB38CF840D2CC7821F19636001B81F1E931559FC
9BC2C16608D5D1F523741AE1D97FF3DBFF7A76E3B03216B1CE070621FE12F9207CE587AF2C2D2DF1
1E2BB007150D537400BA87CFEA1A2A18C9AE53D406E7148EB6C23985A395A8BF3FC57043B7A510B9
4F19B54FA853D8EDB345902C2B88AA21F729742B2CDBEC4CFB29304BB6B22F9B6FE4700A5E7F4EBD
A79147A0B19A4FFB27C97DE2D54AC429CA2E08D307289B014B2A8C599425F6C098C99FEA008BE452
0E02280CFF79E1DD78F4D0FC6041B3645757579156DC7FF4389C5B38F9DA73BDD5DBC663E077A1DF
E1F2E597905368AF379C13785BF85B20C01F35FC96AFBCB8FDFA9B7F7BD3FF4DF51FDC3E9AFE08FE
CFFFCDE49F8F6DFFD7CA7FB31ED870AB153A853456A87615B28FF6A0E75DBF347DD0800CFFC1CDFD
577FE16DD8415BF79E30FA56F0F4A7D1314F7FF4ED30A38077EB4F06D89B4FFE0F3E2DCE34E091E6
1ED40D217F461AF75364963F715F76E5DA2704773AC0CF2208C2249C02813F2C953095B545C0EC91
3E216CE79F67ECF40D63001D10F81BC4FCA8E153837F304E323BA3111E3AFEA0263830E9C5C92A7D
4298F1EA93D6381648A7108214D10A9499802388E3315D1660AA4EAFEEBC8C9CE2F1AF3FAEBF91EB
14B5A1D4DF9AC3B140F0B1ED6825663DB0837E8A30B9E09C2299FB8494C1AC710A854111B9F8C37F
FAAE66FD14BB297E915FFB946FCDAE5CFBB4CB8295D00DCD279C38FF2CE5A70815266DF5B6ABC90D
BBCC4BCE4F7526B9108755FB000238AD15DF42977561C5CE84DF6BEAB8B9807E0A3AC90F9C3849B4
420F4820CBF814D007FD0E7FFE8397C86DA1738F17F19C2C287CDEE5682B7C6C3B5A89590FEC42ED
93D9F38E140AC62CC6C1B3F1DCA775DB31218264CD0300B3F668EBA9EC24B54FF5FB29B08FB6182D
3AA98950CDA36D8253832423B8C95A3FF47ABD7C3E12171DF4CF5836B509C50E98DC8A6E1ABA1A8A
FA06CE15166EFE4CE7F9FAEBD7914AA0DCA0F39D88535CFA0BE30A406E0BEDE05E509EB5A0F07997
A3ADF0B1ED68256EDE7863A6EF6F78B49951C2346EF39D718F36E914A26BB65DF8C44BA46AF1686B
4CAB3F45BE1811513472740AFD81C907AD9FA2967346ED53CAA3AD874D661513BC33CCFCF9141D53
A46E8976A9C037841F810704699D62C2DC276C412E781986E58A929BDEB3E7E79053DC5CA8DC54FA
F5C11F350C33344ABCBAF33226381DD87F90FFAEE1E7C2FD66ED135A30E000DDE471E178D642C3E7
5D8EB6C2C7B6A395A8B5F6497BB103714F858D819F42BF33EFA36D18B4551894345C2C487F0ADA5E
FBC8DA631F4C93083CC07C043CF29EC71F657DB4E34BEEA6A301E7C9B78CFADCEDB2B9DC230F3F86
4FF1801D04C61F95AA26A2467B11B3363A3874B53CB0094ABE3547384CEC6192196F8D3739A7C08F
2766AD780245EB0D945726F95EB3C022AEC9A3BD1A592D8D4CB257903F084617720A335B8C9ED50E
EE9B8BE95B5F50F8BCCBD156F8D876B412B5D53EC9F0D870EDD3B08F36A31EB1DCA7FBDE17897BD2
6C82E73E2D844E41BF20E0149BB79D215E10071E231EC9A98D7B7EE9A73F6E7E539889C16F8DFAC7
C1AC52500FDEE78E5E0273391E9104B332FAE4F06EA41DA0C021328E42D406DE617979995E081345
987BF79E3D7FFE9BDF4267077C125EF8244E1DBD100E832FE1853045C4A2979C5029F4684FF2EB7E
E0C449CD14905308A281C153750EC51C2CE29AFC5D070F631F6D3D68B9B7E5E68D376259B2CF9E47
4EA1BB2E2EE239595CF8BCCBD156F8D876B412F57BB479966CB0694551B088EB14F12614B2D51DA3
1BF3AF53E0EC659886F491B553BFF4EF901744C408CE2C44F02C7F169EFAF07B3EACFDA7BB83D938
CCD8613E0CC00DE017BC460EE883986BC10178306073F334BEF6FEA3C7770705514040F0297AC387
8E3F08E420F4538B4F0504045E05E4024D16D8961AF6C4C357F14DE004C2674312012F8477804F95
59CF839FB6ECAF987F780C1D1207A0B5445883E793532CE29A3CFC964D710AFED29126D0484696A1
5DD8BB83412BCCF5FC9C38A7A80D3EEF72B4153EB61DAD446D9CC2C892CDA87D4AEB14C78229B276
8B6D963DBB103A056DDFB9BCBA7AEBAF1D7DF767E1916365AFBDAD219E054EC1BF57667B38F32553
41E5772BFBC2DA1A6969D78999FB84AD9927D4446681855B93C74E1F7026F1B2267ED1482AE98A87
61C8E4DDE60703F184A7CC4485853B270B0D9F7739DA0A1FDB8E56A2D6DCA72289E0254F729BF7BC
CBF053045364958D82F7C85B084E51CF95676EA7E553F9249163922F2FFB93DE1CF47CD92DFA26C4
9B1CD87F50CC4B37374F3F74FCC1C64FA6FE59166EFE8CFAC29517B7C57EA41BD4607D77940D053B
F5AF78A868587F7A8BE55B5F74F8BCCBD156F8D876B412F5EB1476370A737F499D42963959F54E0B
E7D116139E37777F52C3E7346BA26AC6ACE59259FC5098138B554C8F3CFCD85D070F9BDF1126A53C
83687710B1650610358B455C93870F0C8C40670B6CF5B6750F47F87B8483510DA4DF0EB6C30BFDA9
2E22CF5A5CF8BCCBD156F8D876B4120DF4BC63814E9C68905431D429CAFB2982EE099D0D35D8987F
4EA1A32C6BC0AC57FEA7FEC2B2EF3CBB6F845E0974B5C3BC34E4F8806779CDFF9517B705C598132C
E89AFCA9DF3A85D14FF48B8633BCB4B4F4E0173EAF0F7EE0C449788AAE0398430B3B5B764E16143E
EF72B4153EB61DAD4403B94FDA9DCD158A323A45BFE7DD3199F29464130B94FB04B3D39B3FFE116E
CF4FB951352CFAE7CF44AFD74357BBE86AA7D18FB17AF63CFC8A4F3FFAA5C63FB689C55D93DFB9B2
03BF85E5D13FD83633637707C31218C7DADA1AD0BABB0E1EBEFFE8719D1FDB8E73B288F07997A3AD
F0B1ED68259AEA4F11CC922DA553F070A75484EC7863D47D7BFE750A5DFBE470D48645AC7DAAF663
EAEDD0DF5D47CEC99CC0E75D8EB6C2C7B6A39568CC4F216A9F5095F82FFBCD2CD9984E61F6B30B5B
2A16AE8F765303C3E158C42C593F276D82CFBB1C6D858F6D472B516BEE13F353A46B9F5889545CA7
30FB530403A018CB709DC2E188C0D7E4FD9C340B9F7739DA0A1FDB8E56A27E9D625C0425E409CE23
B2750AB31B856E4561CA190BC1299A1A180E877B07FC9C347EB61BFF0C0EC72CE063DBD14AD4EFA7
18BB278AB94F661FEDA44EA1439FB2648BC1010B51FBE43A85A329F89ABC9F9366E1F32E475BE163
DBD14A3495FB64A63CE9DAA7723A45B88FB6E9E35E084ED1D4C070383C37D5CF49E367BBF1CFE070
CC023EB61DAD4403FD2932FB68E7E91476E89369D32E8A1A0BDAF3CEE1A80D5EE7E3E7A4F1B3DDF8
67703866011FDB8E56A2C9DCA740EFEC523A85EE4F11EC7C37DA58A09E77AE53381A84D7F9F83969
163EEF72B4153EB61DAD44ADB94FDC97CDBB6617D58A723A8555EF94EE7CE77E0A872305CF4DF573
D2F8D96EFC33381CB3808F6D472BD1B04EA15D15AA235E4E7F0AEDD48EC816C32F17A4F6E9CDDD9F
20ADE83D7BBEF7ECF6D633E7CE7FF35B30AB812F61A3C497CF7C43ECF9838B5B7342586AFB18C98E
660E71BA7C4DDECF4983F07997A3ADF0B1ED68251AF453486345353F85E5C88EF7D1FE9BF59FC3ED
85E014BBA309F0273EF7C047FFE0BBB77FE7D2ED7FF47D09B1D33C461DF0DE4FDE97F319F4F43B7F
663EA3A97BFEDB3A779804EE1DF073D2F8D96EFC33381CB3808F6D472BD154EE132F798A64C9E6E4
3E89AE13B122A8E2C6FCD73E3D7E7658650173E3DB3EBDB1FA9D1DA003CB172E2D3FC7706104B1E7
B90046C7FFDD8377D73FDE4298EEE45FBC9BF9E64E37724EA3CF9FFD9C34089F7739DA0A1FDB8E56
A2FEFE1466078A905F3BBF3F85A012C10028F6B8103A054D7D3FF1B90780537CF40FBEFBB1FFBC03
007690BF616EBFE7D78FD3378A4CB0E9A99B37DED8FEDEF6D6D6F370D2802AF2975C79713BF93E91
77AE30BD87710BDF143ECFAB3B2FD3CECB975FCA792D7C78F829E0B53B5776E087E23FE0F5D7AFC7
3FAAF9B171A37D24C57353FD9C347EB61BFF0C0EC72CE063DBD14A34E8A718EF290A1393F4D116ED
2A226AC542E814CF9C7B66CC29EEDB002270FB772E212940F4650BB5CD778EF75C608F839D7BEE38
92390D066AB3B4B474CB4FDD02D87BEB5EDC7EC79E77AEAC1C82A736374FEFDBB78F1F6FCEC0B77A
DBCBCBCBF826007893D0917018BC391D09DB0F1D7F908E019CFADC517A13DC78E0C449A009B011F9
298044DC75F030BD277C66FC2EF0F8C8C38FF59E3D0F1BF0489F843E4026E0FD5B462BBCCEC7CF49
E367BBF1CFE070CC023EB61DAD44ADB94FC24FC10885590A45B68B747F8A70F45344AD98734E0113
D4A79E789A730A24141C1FFFD64EE61E7C042A411B5CA708E1F2E59768F62E9667812300BF20A2B1
9BE7BCB8F2E2364DC257575723DF1A09089005B11FB800EC3FF55BA7680F4CE6E9930087D51F001E
E17BE1011B1B1B62CC03A1A08F44D3C59B37DEC03D6B6B6BBD5E0FBE05007E7DF4C9E1CCA06A0394
0AB949FB96AFBDCEC7CF49B3F07997A3ADF0B1ED68259AD22942D68931B9605F26FA53042C157611
D4FA22E53E719D62F9EEA3CBDFDDE1224502614B05BE49D2A30D93649A57F3FD9C3BE05C1D45874C
001DA0393CCCE7E3DF5D70049CD503DDD0C723D7A02A2CFE6949CE10EFC64729F2026A08025C18BE
04CAC07F5E3A219CD1E0019C8FB4069E9BEAE7A4F1B3DDF867703866011FDB8E56A2613F855501A5
6B9F72729FCCAA27FDC8EBA0E65CA7D82DF6BCC3DAA74C4241478A0DFE0E714E410A85A86BD2C022
A2FC1FEA811327796953A86408F6C3B3E229240E5ABC409813FBDB3F302CB8DAB9B293FC79CF3C7E
86C6ADFEA17ABD1E955A89A7363636DAD7A0D0D7E4FD9C340B9F7739DA0A1FDB8E56A2A9DC27CD26
4C7B4552A78835CE4E39B8E75FA72878B4EFDB209D821E6FFF4EBF96894C16B807150ADC430790AB
8268458453F0B57D6E823681B3F1FC1FEAD46F9D3ABA7EE49F9DBC9F688539088102E8B7C5E35756
0E99EFBCB9795A9C61AA568239BFF801F5CBD7D6D64837811F4A8811BB4CA7D09C62ABB70D7CC4FD
14AD879F939ACF76E39FC1E198057C6C3B5A8926FB535806EDB23A45A271769147E0618BA553683F
05B75D0B1736DA253883A003683F518F08A7A005F9A4488128CB29B0980A458750E9145215314B27
DF04CCE177153580C12688000922A1AA270E180C67BFFA15FA52138408A7D06801BFF035793F27CD
C2E75D8EB6C2C7B6A39598073F4541A160D6EC4C9D22D238DB1029E8B045F053E8DAA7B14E71A1A0
53702A01DB42CBC0B827DC4F6A458453ACAC1CC2C9F3E6E6E99CCF8911AC99B368989023A780E369
CE4F960D7A1353A7E05E0C51D6AEBF3BF00E4A798A1C968F48EDD3E46F3E87F0DC543F278D9FEDC6
3F83C3310BF8D876B412B5E63E597E8A02D1F82FFB4BFB292C0345B0B936E31D8085E014224B565B
2A0AC54E1754A8EC8582B481B42259FB84F370788459F4D44F02F0142A5E02CA491C81BC0C085E52
C5F3A3787CEBEAEAAAD951028F2702A203A672FA716894D2295A00AFF3F173D2F8D96EFC33381CB3
808F6D472B3127FD29904DD859B2C9DCA751C493EDCB567550C43B16A2F689B6D14FC18B9748A110
F481540CAA7DE274838E09710A3E6F8FFB9ACDD7268F81A93E9FE4D3441DC07F1D219BC68D6BAF89
C6107056CDEF4B6A0B154465EA08A1C3923A45FEB75808789D8F9F9366E1F32E475BE163DBD14A34
EFA7E05F2A4E91ECA31DECA05D2C7C32FDDA0BC129427E0AA95014F7A054C10B9F86C7B0CE77214E
8131AAF9360481E4A49AFC140498A26BBF36D53E996F78D7B13B39AD78C79E77722F39BE84FC1A91
D37EFE996F10F5A0B70A51860EEA149E9BEAE7A4D9B3DDF867703866011FDB8E56A2A9DC27FAD2DE
3FF253E4F7D14ED43B1569C5A2D73E8D9508AD503C67EC27EAC1DD16214EC18580B29C2253A7109C
6297CDFFB17DDE6E469C14900E7A1542546A1DD87F30D315027C84DE24F243758D53F89ABC9F9366
E1F32E475BE163DBD14A34DC9FC2746D2BC691CC7D8A77D0164FD1F6FC730A3E7BC1DA27F24A546C
78C7F6C7FD14E62C3D099871018B0456028F046179208FB678A1F06B9B1E6DF394725A014C849EA2
DED9820298C447301A1339B54F6D827B07FC9C347EB61BFF0C0EC72CE063DBD14A34E8A710E14E66
EE538E4E61F28860CF3BDA585F802CD9F367BFFEFA9B7F8BDBA8537CFC5B3BA8564C8E08A7A0E614
F91F1827EADC1C411081B49B9BA7D14F21E6F6BCE6EACCE367E05713E714F4726EF4E6A400BE11EE
34FB6E8BEF9E939D1BEAA3DD56F89ABC9F9366E1F32E475BE163DBD14AD49AFBC4FC14E31A2753AD
C8F653442409E9D1C683EF5DFA9BF59FC363E65FA778F2ECD937777F82DBCB771F5DF9EE5F70E775
4177D01BA19D299D02664D64703067E3FC48B1B1FDBD6DE00B6B6B6BAB837F47D78FC0B698813F74
FC4153A780C7DEB3E7891DC0CF2E3885164DE8FB52D76F000C39DC89AC2454D124508A537447A7F0
DC543F27CD9EEDC63F83C3310BF8D876B4124DEA1422FD89F388F27E8A908D8287C772B56221729F
9E7AE269DA468FB6B0694F823D771C11DF8EA6E8DC5F30F51162FA2908DCAFCDB9007CB6FB8F1EC7
56772680FEE8F028125CF899D4A0A82B5DFBC4E58C50ED539BB29E38BCCEC7CF49E367BBF1CFE070
CC02F33FB6DB7A5F73CC140DFA290A254FF4A5D5653B9EFB64063D85F670D231FF3AC5F9B35FC7BF
6B78FCC4E71EF8F8B776303F16DDD613E23DBF7E1CBF4B3C881566F239BF65732099EF8C7E0AFE94
388CD881C87D82B3471DB7F53B937B827F92EDEF15DADE452E92499D025EDB353F85D7F9F8396916
F33FEF7238AAC1C7B6A395682AF7C96845C1CB9F8A2A46AC3F85D58122D293A2BFB13ED42CE65FA7
78FCEC37681AFCA1B5CF2C7F7768854072516D83B6B54E41C06F4AA669EA4667CEC971DE9EBFA6B1
B1B17160FFC1F831F4ADF94E2C3D0A7919F6DEBA978AB5F887810FAFF9023135DAA3750AFD1351ED
D343C71FAC73D83405CF4DF573D2F8D96EFC33381CB3C0ACC7F6545406972A1C65D1547F0A4E316C
8F76B64E116C6F570C8F2D78B4074FCDBF4E813DEFF0EFFA139F7BE0F63FFA3EE1A37FF05DBD9DB3
41D8F7C9CF684384181B14D67A74FD88886F8297DCB8F6DADADADAD2D2128DA2E425080EC07AA4C8
01BB2CCF963F4515593A1B96D29FCC267D68CD40B5820743119E39F78CF08F983F0899BE575757BB
70B1F535793F27CDC23985A3AD98F5D87EEA89A74F7DEEE8C6C6C6C9CFDD71EAB3F7D263CE063CE2
0B7DFDC451168DF929AC6D3C66AC538C02A0B2FA53247B670B43F722E814BC3FC5D9AF7EE5F4A35F
FA37BFFDAF1FF99D53413CFC587ACF08F06E399F0126D2E44AD87BEBDE878E3F78EAB74EC1050777
C2FB67FE2C40137ABD1E9554015BD9EA6D0B9EC281854662274EFBE15B033B801302C7F49E3D8FEF
097B225DBFE13402FD2166013F029C5B981CD24F079F27F4DB81D76E7F6F9BF40EE235F0CD6FDE78
A3CEF15333DC3BE0E7A4F1B3DDF86770386681598F6DB8C7FD7F5FB80DF0D60B7BF1916FC02C4B3C
F5D67FD9CF9FC5A78095347EA21C8B855A739F427E0A61A6289FFB644B123AFAA978E442E814F3B0
1E8E936AB846619A134639C19EB223015E82D8DA7A1E001B94D114D205C41E32320057C54C5AF807
04E7FC1F9FCFFC418008A0A1E3AE8387E111B681DA244F321C831F3BF4F95B095F93F773D22C9C53
38DA8A7A380530051E7D438F34E3127BC68F17FBA4E3D467EF6DFC443926819EDBCC7A4AD970EE93
902A02CF96EEA36D6DD30665C92E964ED12CC4C7A8F6A92AFF2CCD7EF7095FBBB8F0DC543F278D9F
EDC63F83C3310BD4C329E4FC2A54616EED7FEB85BD273F77C7ACCF4337EFADF5A3B6F3DCB89F4270
8DB160C1BE8CF9293E6D943F45E4897196EC22E43EA19FA2F59887AB8A330E0DAFF3F173D2F8D96E
FC33381CB340DD9C82E7F38BFAF362743FE917C029423A850E3971D480FFF9D65B376FBC71E3DA6B
D75FBF8E8F00D8633E72C0C1AFBFF9B7F85A6A7936233496FB24B68B4E6D71702CF7E9D8924122CC
6227DA1EED719D62CE61FEEC5D3E21F59F7F9F3FFB396910CE291C6D454DB54FBA3E241CB959C8DB
BC18E41493DF82A75278D0417CF83D1FFEE44FFF3AE0E3EFB8031E0FFC9D83998FB801587DF7BD14
E3392334D89F22D2F64E70EA74EE538A4D18B9B2AE53381C51786EAA9F93C6CF76E39FC1E198056A
D2292205E7A1C711B49FC227FFCDE2D881138FBCE7F1C76F7D66F3D6DF830D006C3CFA8BFF3E0438
E0B10FFE7B381E37E0F1377F7633DEFF777234E5A7089AB2D15254F46897EBA3ADD944C0CABD109C
A2837FC59184DB0E9E8D667F11BE26EFE7A44138A770B415CDF829E26CA2A865BCF5C25EDD89092E
8030A97BE587AF10FEFB5FFD0F78143BE3C0BC44BF9B97C5DA47D688530065800D240E114E818721
FB78FC03E7FA5F7EFDF1997EC85A739F2E1A9DB2B54821B364F3729F247D0884CAF2972C44ED539D
1FAF71842E328D5F7C1AFF008DC0BD037E4E1A3FDB8D7F06876316A89353C84C27FEC833FCC57EAB
F669EB997327F7FDFCEFFED24F4F827FB477B9F1F3BF88B873791539C56FEEFB2AEA14A83E20B9D0
001281EC030E436963F3EF7F79D1394524F7C9562EAC2CD97C9D829B2644A8AC56345CA7703822E8
DA9A7CCEDF1A1CE3B94FB5C13985A3AD68C0A31DAD74D24F9959B270F5DBBA7B4F61AE659697AFF7
1FC54EDAD8F8D8DEC6CFFF2262ED236B429B08B109AE538C29C6E0E0D6700AC34F1118CFBAE75D3C
F7A92040088A6155462D68CFBB99A21DCCA51D3FC5FC601173537BBD1EDC4C1F38717273F3343C3E
74FC41DCC09D00B822D1C13860E067DCB76F1F365EB9FD03CBF11F195ED235F5B04138A770B415F5
708AC2BC2B6CCA366DDAA64E2138052DD8C6527146499BB473E323B7367EFE1711C02904410821F4
6C0B748A50EE13CF8C2D281425758A7EEE93922A6C1E51F4712F44CFBBF367BF1E3117CC08F33033
9F87CFD085CF1CC722D6F91CD87F70DFE0DFDE5BF7EE53FF60A738FEE8FA11D84F4403AE57F025EC
0CBD7FD7B49B66E19CC2D156D4A653C8DCFE9C3DDF19F6A7B039C59D632AD19F80ADA77CAC24528C
F6B84E510D9C53446C1471E562D13945A83F85A953A041BB949F6238B61577D00C420EFBF5C5F068
CF3A4C78A698913F22FEF2F6CDED9BC2C2CD9FB7BFB70DAC016EA6671E3F03803F1FDA803F3D784A
580E1F79F83160104277801F1976EA1EEE08CF7DAA13CE291C6D4533B54F5AB3D09DEF462BBD7D4E
F1B9A3E26D81536CDFF3AEE1A46BB0A2DB2F410F948B88FA73DA769DA21A8853C4458A385A93251B
F453683385C8928DEB14C558A750D093990D35FFB54F5CA7D86D74B65C7F8F9B69455867BE705198
489D9F73E1E6CFF71F3D1EFA8B46BAC1EB9AAEBF7E7D6969492B1700D809E05552FCFC2F16CF5A68
38A770B4150DE43E452C15A2F83C5FA708F7148E84E4B84E510D6B1F39F6D8071394C1A41BB4B305
3A4521F72964080A0DF8BC3EDA5CA18830083DF8E79F536092F04CE7F391B79DF534BEF26B73E8C6
547EAE45211A33C262CD9FE1D34656601E387172DFBE7DBC9210458AB5B5B55DF58BDED8D8084915
8BE831595C38A770B41575FA29648D93DEB0427272FC1458FB246945719B9C17F4A5738A6AE0B54F
955D156DD529F490D6E2458E4E51D0D454078A885F7B21FA5300A3BCF9E31FED0EE63C771D3CBCF4
B14F4E11BF74703DF363E08C0B86CA9517B7B7BFB77DF9F24BF06BE5D330D8230E9E35E0BBC087C1
4F627E0CFDF9691B782EFE203B5776F8F8BF71ED35782A791EBA8345F45368E06F8DE803011D16E6
9F24EC44F385D921C53DDAB5C13985A3AD68B23F45A8EA894A44A2B54F26A7306B9C74308EE73E4D
025EFB94E3A7308F59749D22DF4F5130078D2C15499D4298AF4D5FB6D9B462216A9F7ACF9E87390C
4E6C3EB4F699B73D7CF196D397A685F7DCBE92F9316062B9B4B474CB4FDD02C07211DC5E5D5D852B
0C5CBB6002A65F25E6635BBD6D380C5F0878C79E77868EE487E177BCFFE8717E007C477A0A373637
4FC3FBC346E4A7001E01B34AFAEEF053E0CBE1F191871F8397F3DA18F848FC33D0ABF44E02BC43CB
48C762E91411007F8441C5EBB8803CC2EFBDBFD3121DE0EF0E39051CA645B1769C938580730A475B
D1989F2254F5A4E3FD033AC5F7EF1895881C934550F6FAAD98A1B99FA22AD63E722CC73111112FDA
54FBA4FD1446A38A808128DD9F22E2095232DC82F6A7D8F7C9CFDCF22FFFEB90117C717B72FCCC6D
1F1706703D25BE7CF9259ABDF77A3DFED4D6D6F334A5373985F9CE30B1A749385092C84B969797E1
988D8D0DB11F19CD03274E9A9F2454034F6C025EC8C73C3C057C843E124D176FDE7803F7DC75EC4E
9861C2B70000F5C09DF06E304D0520A542AE71FE8FCFD7338AE0031F5D3F023FC88D6BAFCDF41B2D
7A9D0F8D3AF81DC1B0E1BF7764B8B01306A47E215C1650C510631EDF13FE8A75DF58DE19D61FA7F5
08BFB279F818FEE88F537F9CF515E3C12F7C1E6B9F72529ECC721178F9C9CFDDC1DFF6ADC18A4A5F
A7E054427BB4C3C521BCF689BFED3CFC46E6FC7177A053909F02BB690BFAA0BFE4FE0B6C9077E6F1
3333FDA8D75FBF3ED3F3F0EACECBA69F42D63815A39F8447FBD4A7DEA6DFB9D0475B7B82AC9064DD
9F02A68538A4A7F853D30FAE271E93720A2414C02C262414F83E1FF94CFCBBC37C12A7D02B2B8776
8BD32ADA065E00079846D710486500C02C3DF2B3C301A218094E6088C2A0BA01BF50B11F3E6D9C71
EC0E982F1E40652DB8474C29E984704683DF82F39199E2C0FE83580F09673E49E526443B6A9F76AD
C227F845236B30EBE576AEEC60142D1A9A0AE7E4C61B7D516CC03161B00137C10D185A3801764C0B
703EE16CD359F5D3EB68136E5C7B0D37A638B0F95B9DFADCD160CFBB108A07BFF5C2DEDF387AB7F8
16434E3148CEA43676667F0ABBE3F07DEFE39CC2510AA45354089245BAB1F9F7BF8C9C628A100398
06F6D4DF19F7C0DD59D43E0DC945D1852D463BD53E21CBC6DA27FECEBB034EF137EB3FC705B58481
E2DE619032BC0ABF7CFAA36FE79C62BA184E3906730F9C78FCC99F5E0A05548660700A2405482B26
790444390570229C3F235F788BF1083D618B143269C0849C26F9260B40A0A221DE0D898398D2D3F7
A5893D7F155552C1508C7C2AFC7644FA608AC87F28448853EC0E12876A28B3875104DFFDFAEBD7F1
D38A92B0A9E3AD05AFF3C16180894FE20741C704FC334705304AE4147A1160D1CFC9622179317138
1614F5D43E0D17693565E0D0894F033345D04F71E792F05398DDEE8CB49C11BF107E0AFF338F83CE
8FE879B779EBEF2579843806758A997EDA5A739F069CC22411D29D5D64D3BCF6894E6FA1F629DA38
DB201783B13DD3DA27FD67F2D6A06AA2D49BF0696A81534C05514E41DE81F8547C775028824686CC
2B035CEB6032BCB1B141B4421423E1067C5FED8F88EB2640D9C419863F1F7C49A48B19617575F591
871FC377BB7CF9254D1C229C0266FBF0ADCD3300F3FFF37F7C7E2A974D386F74AA51619C29162E4B
D6048C37E0B0E242F7D4134F63ED93295D39A79813B89FC2D156D4ECA748173E89E04D38D2F253C0
D54FF6A7E035E7D18273EF79372190532053985F8FF6A0A86F76A0DA27DD81C2287F52033BEED11E
EA1456E253845C2C50966C533A055C37B848111B3F830B63DC1C2D00D73A2405A42098DF05AD1C62
27D9C3610EAF5F02F3437873BE87684BA8EA89030643FC1714E114029C411CD87F10085A3C512AFE
0EFCC787772BF53E93A01DF3671866DCB98327167E95B01FB8864999E1AA859C422FE9C080F7DCA7
DAE09CC2D156D4CA29720A9FC4AC2CC029A44E11B1AF9ABD862D9DC29109EA4FD1B75A17FD147A23
A465B44FA7D0329C6E7E271C43B19E77C796AE298376845C70D4ECD1AEA653D89C62723F459453C0
9C3FE977E093DE5266E1CDCDD334C7A339FFDADA9A98459B9C827B31E2D3DD7EF9596FEC079FCA6F
90388568C76C9E13029ACDA7222B60C2D5547E961CB4C04F81854F9A055C79711B1BDB995700D889
9C425357F8152FB46F7DB1E09CC2D156D454FB34A86232539E38C4926FBAF6092BC9477DB44DE2A0
6373887438A7A8065DFB14572B5AA953E82CD9421396903FA8B81DE414441C2CC54DD73B09F3C542
EA144807A6A153FCCC6D1F0F7D5F2A7C32E58009CB781E3871920CB3401CE87B09FA2C6A9FDE1A59
E0797C2B709FC8002602A203A6E8DD323F332D6E87748AC85B4DC829E89D9164E924A2D961D1758A
B706A15EC00EF4C9873DE8A7307F409E252B9E729DA24E38A770B41575F6BC138BB7913E77E4638D
E914BC3FC5C073ADE3FAEDDAA7D14CCC394535A04E915FFB84594FADCF923587B1942A46B422A7F6
2954B36758848A3BEBCF928D2CFB9B98954E31C0DB3E76BFF94DF9BC5D940C8566CEA558064CF3EE
3A7898FF8CDAAF0D6F8826717344110DC10D3CABFA33E0641E200AA2F221DE13A69A99B54FFA6394
AD7DE2000A7660FF4134B6E34C58384A668485CE92C5DF1D9EAB907E14EA7907F416CFB3F9B68B7B
4E160ECE291C6D4503B54FD1B827D9206C90251BD7298C3EDAE6EC8BFBB531F7C9FB535442A6475B
F7A72814442D38A7307ADE0586B1CC95655F86748A7E8253844DB01A3FB3086A21758AD9FB292858
35D386501670AD13C19E3045D77E6DB3F6898019B604342C982151C9C45AA21EF456214337E914A5
480A16924D2E08A2AB7DEABF8E0816BDF6090B9C741E2C02A39FCCDFF543C71F84A7CC61B3E8E764
B1E09CC2D156D4C929422E6C59F554EC1D16F353DCCB6ACE55970A3EE912F331DE9FC25116BCE71D
EA14C23D91E3DA5E743F05710AA3E79D295514C77FA28F3656F40522CB6201B30396BD783DEF6AF1
53DCB8F61A4DB077AEEC4C1E5824DE81D73E1168FE4F8BC3142715023611E07440546AE5700A7AAB
1CE7C5F93F6E46A74098066D34E9C00FB8B9797AEAD78A45AF7DC25677215E0C2C0F2D15E6A9163D
F25A734E160BCE291C6D45FDB54F37ADF427738D77686B8D730AD382ADD66FCD6958A738C5149372
29F729334BD65631165CA72878B48B262043AD08F4918FE814C8298CFC01214C6872B1A03AC5F4FC
14118F7668969E339C80925C7FFD3AFCDE69438C31AD53EC0EFEEEA8A209978EE33A053FA59C56F0
181FD23272288060342692B94FD47D3B07F96E6B1C03FCFBD2A880A92F6DC3A99E6E4DD4A267C9C2
6F13085DE400E0623ADC09DBE185FE545DA7A813CE291C6D453359B256D2A6B1DE3BD808D53E7DFF
8E3D9A53D8A5E67A7517FD14DDA87D9A904DE897739D0280B94F1166C19F251746CB750A4D2258C3
BBB44E6131E560511F17291AF253CC8B4E91E214D4938E3EB018DEA13F16721C70C00C8D1F6F720A
1C2AF412F8C1437E0A0DE011C447382980493BEE0C4D2CF9A7BA25233BB7D7EBC539057028F82470
F6F011DE8DCE24DF8FFF4AFD69E3D910B359384BA20AEBC0FE83531CD20BBD26BFFDBDED083518B7
B9D9D8E0E94F98221B6926B8D0E764E1E09CC2D156D4C629121DC12C8F363E9BD4298441BB30F50A
CCC75A9FFB34D3FE7DF93DEF223EEE45D7290C3F852A761AD207D1F0918DF048966CC84FA14BF8F4
310BA953CCDE4FB13B2ADDBF4575974B022672ABA37FF05ADC103370F8520731ED0EFE1869D28E31
50825344228FB0500A415546D86CDAAC68D27FF8393A057DBC4C3F057E176034402574F450A98B0F
3AD92F5F7E4974BE107E611860F0BB2BF52B8B60A1D7E461FC0075CD3104C1AF158E84E17AD7C1C3
B0114FD65A68DFFAC2C13985A3ADA859A7308C1516BF281490E4D73E85B3710AF331F268B7975320
7ACF9EDFDA7A1EF10717B7683BF7CBDEF6D633E7F4DB724EC1F982EE4921D8041ED00E9D42E73EE9
EA26C1A3B94E81E422A853581D5562A1B264261A6C2C30A798869F2292258BABE2C2341D425962CE
FB53E877E07E6DC1058EAE1FE1C150E26D6182ADCB9FC87091CCFFCCE21413E43E55F668E34F6A1A
B491658801936C5358EA5B2F2EA7D88D8E4CB3CD7DCE7B2E34CF5A3838A770B41535FB298C5956A8
266AE4DD8EF4A778EDF82F6A9D429739E9473CACDDB54F702B59BDF5D7BE76F885277E65FBD18FFD
E17F3CF467B041C0FDF12F1F39DC3B76E0C45F5FFA6BF1CEC42972ECD8212DA3553AC560C416FA53
E8562C459D020F8BEB14F678B66AFCC4970B59FB343D3F45284B164166041E8C139A7769D34404E4
D10EBD1BB103318BC65E03E687816DF89CF812AE0800E5BF6514391BFF5439B54FF97DB439A6E2D1
C676097C0F46FEFEF90F0A9C023FE124DF88A3956BF293D7B83AA7A80DCE291C6D4503B54F9642A1
A761B4A89BE8A3CD2D12EB56D2663890B3F53AC581BF73F0F73F7509D8441FF7EC9C3B7C85007BE2
5F027A775FFDCD7D5FA50543BA67713F45A9DAA736E914DA4FD1DFB8F87EA3026AB45DD02906FB23
3A45BA96AFD8F06EE13DDAB5F829768B5D2ACC41481F0CE6ED30698FAFC3D3C1B071FFD1E371E7EC
2E6BBAC777A24C10AA3B829700B4D0F0C8C38F855C15DA4F11D729A89546A4DE5E63F23EDAF8BB10
4406237F456D0F566755FE46022D5E93AFCC2C16DDB7BE58704EE1682BEAAE7DE2D3AD6FAB9A28AC
3C173B039C62FB9E77E992723B0CC7EAA6DDFADAA7BE4EF1EE7B9123F456FF0FE20BBF7FE725BDAD
3790599CFAA57FB7F3E235F1CEBCF6A99436D1269DA290FBA4FA5324BF4CEA1494FB64DBB1AD2C02
1AF90BA953D4E2A7D81DFC5D00C8627C74FD885E69873DB01F8E29358AF03D77A31DF4C8AF4D7B76
99895B9F46345FE839369D467C16BE35AF8C225051539CEC50C694E9070961724E81F4810F86DD51
E42FFDBCF8D4743985AFC9FB396916CE291C6D45237E8A42124EA8F29C7BB4233DEF3EBD57A76EDA
EDB3C58A6E07729F8853105F203D02B749BCA0FDB8D17FF69E1DE41497FE42CE4E45ED13E63E45D8
C423EF795CF7D16E9F4E115128421EED5C9D22833BF3F13FFF3A05CC5E0A9C02E800953F4D887FF9
5F7FE6B68FBFB9FB93E467A0F4242C0D7AE0C4C9471E7EECA1E30F2235C80C444586B2D5DBC60670
A8086C6D3D8FEA86492E74190F2ED71FD87F10250998635F79711BDEE4AE8387B1C0C9E40BF45AAA
E6824F0ED75B38B7300387CF8F3F08FC6891DFCEF6F7B679BF6FFCC16167CE9F0F96724DC229E0A3
C23BBCBAF332DF69EB1403723495B1BAEB6BF28173E29CA2CEB33DD3F717F593BB1397C6391C99A8
D34F61CFB544976135FBAAD29F22B0B42BF6B45BA7D81D700AD21D90412065C00DFA526378408A53
E4D43E091F776B740AEDA78829145C9B63CF66F929E2BE6C7640BFABC582F829CE9FFD3ADDE33EB4
F619CE08FACC6292C72F6E2F7DEC93714EC16FAF3075073681D938F0F8CF4EDE5FF6D4C148E0F906
009893876279F05BEB5E751B1B1BF814D007B8662E2F2FC387595B5BCB2FFB8723EF3F7A1C63A900
F04301B3E0A559E6AB7A837FF899E9F3C3043E87295CBEFC12FF1615003FA9195DC53DDAF8FED3F5
53F89ABC9F9366E13A85A3AD68A4F6C90CC991FED6D19E7C8FF6B889B64EDA643A051DD0054EC169
02270B26A1E0B24554A7907DB44308A542B540A730729F8AE3D6942A44284132F729DEDBCE64D30B
E1A738FBD5AFD05CB45F427378F3967F726A5AF899F7DE8EEF5C6DBA9BFFAA488D539D277F2A1F12
0597DA3E3F7E8B03FB0F9AD556B7A8A68430C0925E957CF89ABC794EDAE75B9F5BD4A953381C75A2
D99E77C9DAA7B44E21F29DA2EBBAE350D92EE914DC4CC179042F85D2142347A708E53EC5F3A0D0A3
DD2A9DE2A2C11A3465D6B54FC9FE14BA02CA1EDBC53D73AE53BCB9FB135EFBE4B7BF5920E7ACC68F
99D1EFE5C6B5D7683EAFB90302880666E4D26780FB4866EF8CCC93E39C42C07956CD67BB9E6FE457
5747CDA8ADF649CCB5326B9F127E8AC06AAD19B3891B85FE141DF05308CF351925921550993A453C
E8298245D729B49F625C046552E362ED53299D2214242BB9C6FADE45F1538819639D68F6269B59DB
3C6B85A5C193899D35B060EC1D7BDE691E83ADA2F91E3872C2D05A0E5F93D7709E5527EAE114BE6E
E3A81F735BFB34AE1209E814DFBF638FE8A33DAE7D327DAC225AB663B54FA1D858A3F0894847B69F
A2AC60D1029DA290FB540C1CD00C627C40793F45A699821F39E73AC5EEC04F516745B1BB14F5D988
EF991D802CE068810DE1DAE053A08D8D8DADDE367E79E6F1336507581CBE266F9E13F7ADD779B667
FD2DDEDCFD091534C2237C47FC321F6EFA705440AD9C225C6A1E912DF23DDA46ABBB40B42CBEAA0B
9C82EB14823B0849C2241A494E81B54C996C82F7A758744E11CC7D526459B7CFA68D78EE9318AE41
33850A4C5E089D822690377FFCA3597CA43959B42F456726F9CC351B222601DC14CC085F01B8C8C3
B88283AFBCB83DF5DF94730A3F270DA286E9FA6DEFBEE5BE83EF3E79E71E78FCEDD5B7E3063CEA0D
FA523C75D7F21EF8D39BF38B8963DED04CEE93390113519C39B54F81DCA76083B0A253BB0B9CA290
E3A44D13E1C2A7717F0A956389B54F481048A720669159FED49ADAA7B19F420FE0620C142717E9FE
14A62F3B459917B13F856377EE5940CBE06BF2E639714E51E7D99EF5B7001E51689C94820C30FCF6
6D5B5F5EDAFEDE76E3E7CAB15868A0F629E4A7D0BDB6533A8591FB14E111C535DE8ED43E91475B38
B243E245EFEEAB9C6524750AE20E7DE921DB4CD1029DC2C87D52903D1C55B15FA9FE14865728D09F
622138459D1FCFE1E0F035793F27CDA2064E71E2EE9FA51BCD902F14176F8D85AF22B9704EE1A880
FA7BDED9666D5DF23458D44D7BB4ADDC277B0EC6577DBBE7D1E6BEECCCC227202313F6A7E88A4EA1
C6AD6010482EAAF92972C845E180F505AB7D72386A86AFC99BE7C47DEB759EED597F8B9377EEB979
F1FD069BB0A65B3A60C43985A31A9AC97DD2039B1675D5C80FE914DBF7BC2BB4546B5859ADB2F32E
E8142659302509B34B45D9DC2761A6A09E778F7D503A2C165DA788F4D136EAA0D498CFCC7D2A0CE0
226BD6F90334BC178253D4F9F11C0E0E5F93D7709E55F3D99EF5B7E8EB1417DF2F2DABA31BD0784A
A6D675E94BE01466D4B3C311C13CD43ED9CD29464F25FD14E3DAA7889595B5BAEB3F7BAC43B54F42
8C10854F3CEEA9AC475B1BB173640BAC8F5A749DA290FB64FA294CFA5CA63F8569C1362BA0C4985F
88DA27D7291C4DC1D7E4359C67D5899AFD14A456E0BD49579EF37B96D73E392641633DEF4C254E6C
94CA7D6211FD7A0E26C23687B94F1DA87D8AE81411708F76059D42BB270C2DA3DD3A855AFF29DB9F
C20C1FC8DA5890DCA73A3F9EC3C1E16BF2E63971DF7A9D677BD6DF82FB290C07AB9A74E95074E714
8E0A68AAE79D9C6599CA454AA7E87BB48500A1B237F901E38D2EE914853EDA4C8C10C2041E26EC15
D5FA686B3CF6C1B679B40D3F852531CB019FEDA7C021AADB4F18A66C611A5A779DC2E188C1D7E4FD
9C348BDAFC14B1947E9DE45F241DCE291C15D048CF3BB9664B4E0A73E447748A681F6DA350A4380D
EB08A730CDD7C258113A2CA73F45D25861EA14ADA97D0AAEFF44FD1439FD2944ED93991C6BDA2E16
8253D4F9F11C0E0E5F9337CF89738A3ACFF6ACBF05FA29F05EC3DD1346ED53F1CE355CFBBAF87EE7
148E0AA8BFE79D4D1FC2E24548A7A03EDAFD09D57DEFC3AAA7786EBF98A1758153246B9CB44DBB94
475BF7D1CEE9A9DD369D22DE9F62D46985B66F66F4A70885C7B6A6F6C9750A4753F035793F27CDA2
363F852DA3EB0A73F75338A684667ADEC56B9F38594EFA29905044DA0A874365BBC029B8473B6EC7
3629C62CB2645BDE473B70A136748ABCDCA7D0780ED53EC1E3FCEB14307B214EB17365E7EA8B5B97
2FBF342DBCFE83ED3A7F76C7C2C1D7E4CD73E2BEF53ACFF6ACBF859DFB64D53E193AC5A8F6C9739F
1C65D164ED9339F5FAF66D62DE95EEA3CD2AC9B9B7C20EF667F68A2E708A7C53366713391E6DAA6B
127DB433D19ADAA7849F42508CEC3EDAC10433EDAD18F18EBF59FFB945D129905320ADF8E79FFA65
B8799DFFBD3DFD5BD83470D7F29EE98A206D9254DAF4B34C72129C530838CFAAF96CCFFA5BA09FA2
203D145572391F53F72CD7291C153047B94F91DAA7CC3EDAE176C33CFD89F6748153A0479B37D1D6
11B2A198D9497ADEC5EDDB1DD1296C8A4107C4FD1456E389A0B782F7735C90FE1434B9FDDC917F04
7FE3FDAB04DC012FF6FFDEF33734E0A91377FF6CB59F62FEE7DB914F38FF1F7E7EE06BF21ACEB3EA
446DB94FB20945A0F89C3FEB7E0AC72468ACE75DB8D38A51FB14EFA33D986EFDCDFACF195E6CDD32
AC633A85305F73D6209CDA7A7F58A7B03DDA268F809DD8F38E739036718A447F0A114490AD5360EE
930E1688782B16A83FC5F9B35FC739303C3EF2D9D5FE4983F3230C5655819C82DE9F0F891BD75E7B
E587AF20AEBF7E1D1E610F6CD031B08DC7F04700A92A21245DE7FCDBE17B8A6FC13F1B6CC3B77B75
E765FE39F907887C18FC31E9BB006A98C02C167C4DDE3C27EE5BAFF36CCFFA5B8CFB5304D66C6D79
C2FD148EC93047B54F812E1525FA689BF52161E5A20B9CC22C6D42E542130DB13FECA7303CDA5804
9559F8D4024E91CE7DD29AB22A618DE73E99F57B412705AB009C7F9DE2A9279E163A45419DCFC4E8
78E17F3F79E71EFA469C595CBEFCD2C6C6C6D2D2D22D3F750BE1E8FA115E33DC7BF6FCDADA1A3F00
8E7FE8F8833B5776C47B72F47ABD77EC79270CB9C8541F8E81EFC5DF19012FD43BF7EDDB07A4E0C1
2F7CFEC0FE83B47365E550CEB985C3E8257B6FDD0B1F1EC66A9D8361FEE16BF27E4E9A45CDFD290A
8BBAA15059CB4FE19CC2511675730A359E63D3B0B27DB403A14F22FA89E6635DE014438FB60A77CA
4452A7D05442AB153A4BB6059C62E6FD298A0A8519351062CDF3AF53C0ECE5E68F7F84DBA7D69787
B54C9A97C511380C3945647A4F536E6019E6017027A563E23F087E179CF96F6E9E4EFEE0C05F88AA
E86761D0BEBAF3323C055C80DE7F7979993E0CB283C88F0607706EE20551267C4DDE3C27CE29EA3C
DBB3FE16B23F456415972DC8F08BB0730A47053493FB649588F4BFBC582CEAC38DCC3EDAB0B11EEE
0B26E84627750A619728D441DD7DD5DC1FD7294AF5BCA3835B56FB14EB4F217AAFA8B5A09CFE14A1
D43263788F58F3FCEB148F9FFD064D77B1F6A97F8920D1011713067C0D3742DB7A0360FA29F8EC9A
A481D02081FDB4CE1F7FABDDE2343E3487A7FDF17746C0EF8E3F0B97507AFF8D8D8DF889E542CCBE
7DFBEA1C030B045F93F773D22CEAD329B4986B918B4C3F050C12B8DCC14EB846E12322B97DE5C5ED
CB975FC23DB071F3C61B8DFF0A1C33C25CD43EE99E77BC0230EED15E2FE814B17997AA4EEF02A7E0
7DB451ADD0466C7270EB12A94C8F763EB9688D4E51C87D8AF4A78834AD48E53E09892D143BA079F4
FCEB144F3DF1346DFFF34FFD32FC8DF7A58A1C6D22898B36A7E0A0597770F0DC78C39CF99B94E1A1
E30F124949CECA72380540700A2E55C46DDA70C099C7CF38A748FE167CFEACCF89FBD6EB3CDBB3FE
16433FC5A8F649CCAF8C52A83C3FC5C6A15FFEFE1D7B103007CB7CE41B80D38F7EA9F15F81634698
8BDCA710DDC8A97DCA0FF02F320E78DCF8C8AD8D9FFF99827BB475A76CED9E289525CBFB4D44FCDA
68B5685F966CF5FE14ECA99CFE1442A1888C6A52EBE69F53F0DCA77F71E41F1B27CA3A9F99955153
E01479337F041CB673652752D154F99D1170095D5D5DDDD8D8C017462AACE0B780B60EE71471F89A
BC86F3AC9ACFF6ACBF05F6A7208210695441EA70E18040EDD3C93B575F3BFE8B70AFE9DF6EB03264
F0A837E84BB97DEF92738A16635E729FC24D2B62B54FC2972D36D60B3521720276DFFBBAC029C634
810B10775FD56CC26C8417D7294CBE1022112DD3294AF829ACEB768E4E610A13F144597C76FE6B9F
04A718173D2A9FBBBD04C10CDA858D81E899CF297436140D9ECC993F50633806DEE1811327F125F1
2E51C977D6160FB884AEADADC1784B5A3C8050C02FE2FAEBD79D53C4E16BF21ACEB3EA44DD7E0A6B
FD36B8C61BF55320A72824F39B18A4850CB7D7F7F6633959E6FF230F3FD6F8AFC03123CC45ED5348
868BD63E7DFF8E518C3F8DD575CB49613508EB54ED53C84FC1D9846E7897A95354AB7D6A814E91C8
7D327365C508CFE94F116ADD18760C5D5B842C59A953BC103881396A85D240CB720A73F064720A78
2B9C8991AB82D299CC374F8A08C00BC41E602BABABABBB2CD0C9FCDB814B22D20DF824588BE59C22
045F9337CF89FBD6EB3CDBB3FE16A853DC1C953389703C630E266A9F02FD298053F4EF35C7FAB79B
E16318785782C3FAFD588965DCBBF4F097BFD2F8AFC03123D4CA2922A14F6C9951B08F741FED40DC
93E9601D1E39A0215DE0145AA488B30971644E7F8AFC0859A41E2DF36827FA5398AA5C667F8A80A5
229E280BFB1755A788F008B1CE20AC85C5E0B80AB54FF46170239353603C14BD965C0FBCE1851E96
9177C65428BE07DE7C73F334720A4AA382CFAF090B3008B8D8C2C68D6BAFB94E1187AFC9FB396916
B5F9296C193D3E071B1D039C42ABAE439D6250E931CC12A18D634B7C0F3CE21E3A80240CD7295A8C
C6729F54244EA129182B142991FB946FA9E8929FC2EC6DD7BBFB6AA47D36F10BE0143C991F11AF7D
EA884E91CE7D0AF5A76087E5E43E09CA9C4C945D483FC50BB60F45C792D829EB7C7F198F36CCD2E1
A6493126575EDC8647DC99C32956560EE1341E41737E4A678A5455BD63CF3BE1B580FB8F1E874778
095000D417C44BB0F609B7E1F3E0CB7BCF9EE7C75CBEFC523FC66A90A6E2B54F49F89ABC794E9C53
D479B667FD2D50A7C00BA32C3E57CB5FBA2E3751FBB43E2C6A92654EBCF0091F114831469331F753
B418CDD73EE90802117716D629B0AEAF50FBA42D159A4D74A9E75DA48FB6102634F59869EDD3A2EB
1485DC273DB99DB83F452CBB2CDC816551FD14C94AA7F0868EA44E720A8A69828DA5C13FDC400900
BF4C720A9403B824013F51BAAA6A942805DFE2E8FA11C0DAE01F308BD5D555DC2F5E42B54FBB83BE
1E265FC037A191E99C220E5F93F773D22C6AF3530415F3D045B598257BE5C56DF5B6439D226DA928
B20CBA49C1CBBDF6A9C5682CF7C99A74997423E1D10EE53E6936A1AA47BAC029327BDB856CDA156A
9F78E21339B21FFB606B3DDAB1FE14A69FA252EE5384478831BF1059B2693F4564C30CE6A5B33A8D
2CD9DD515FBC08A7802B9B6E6C0DE7015F183A21F1DA27543AC4CECDCDD3C417761921A22544E429
3B577686855B23DAE29C22045F9337CF89FBD6EB3CDBB3FE16DCA36DEA14E342112B3FE466B2F609
4D13DC5871ACB093B4897E05D4A707D5BCA35EAE5EFBD4623496FB14EA08AF2663219D62FB9E77E9
3AF360A5939590D3054EC13DDA21A9A27C966CACF629540DD5D69E7789FE146239BD7CEE93CD8815
B958ECDCA780F410298232962046FB2B78B485AC90E3A7C05857B8E762F5146EF47ABD783A53F29D
854E011F0CB364F9A916667038E0C0FE837480EB1449F89ABC86F3AC9ACFF6ACBF85E9A7D06C425F
3F872F89D63EF53DD73AE8A9B8318EC4B9B7EFD11E120DF753B41D4DD63E999E56E5E34EFA297846
59DAA9CD867D1738455C9530E39E32748A72B54FADEC4F9193FB64079AE5E914A68122CB4FB168B9
4F9F3BF28FFAEB69709B635786B19D8A6F98254FECB0C859E548562861F3B8C8CC1F6B9060260F07
EC2BFE231D41581E32DF59DFC4B99F0241B405882D7E095C869E258F36271A0E8E6EAEC947BA25EE
3ACFAA1735FB29EC599658EC12ABBBA9FE14851B50F8C65458195B1F2E8839A768319ACF7D8A8CF0
52B94F2ADC29584042B94F9DF16893AB028983B04EE8B8273A66F2DC27936EB44DA708AB6F923B67
EB14A817DB3C22500745A37DFE3905CC5E688673F2737718EB69D530E81B7BF2CE3DFA3BF209154D
FB43C724D5046025A11F999CDAA64C10EAD02D3E03FFB40F9C38899C829EDADC3C4D6670606742DA
8878B4E3B3CAEEA0CB6BF291FC64F7ADD7869AFD14BAF39D601926E388EB14A1BB9270ADEAB5AF6B
DE9FA2D56820F789D3611D92A3067944A7A0BABE8248116A4E81F9666C32D6059D42D73E25A50AE2
20D56A9FB49F02798720178BAE53E4F4D12EA8CC96609193FB24B36455D619A7D5D706591CF35FFB
F4D4134FD30CF99F7FEA97E112313C93A3C58410F831C38ACAC1067F16CF6A64FE2CFC14657BDE5D
79B1CF1AF400A3F7A1F7D70DECAAF5D1163A05891D682A177F4A5EFB944417D6E4CBF2C72E9C93F9
414D3A85553E2AD761E886252666E1FE14E4D13672CEAD0E4A05C79FEB146D47F3B94FBA4444E73E
95ECA36DACE58E06361FEA5DE0143CC429C426884A68DE318BDAA776F8290AB94FA1FE14A600C7AE
F0597E8A904E21E409F6E59C730A98BDF49EDDBEF9E31FE197C3DAA7C1094193353135BD41A72EF2
A5D629C4FC2A3F9D49CFFCE1253057E7C281064CF2F1E5820BEC3211219F533C74FC41EEA7C06F7A
74FD48C8B881B54FEFD8F34EE71421B46F4D7E7205AACBDA4DFDA8CD4FC1AB3EE404CC2CD36537A9
44ED53281EC7ACD41D28EFEEA7E8021ACB7D8AAFEB66D43E511FED61539575C92F4255E8F4651738
85D02942ADB443AE8AE9F6BCEB904E61496F056DEEDB89FE14665153F0023ED826DFDCFCD73E3D79
F6EC9BBB3FD91DCC85FEC5917F4C52661A7A9D4D1D90F468539A6B68904466FEF07B87FD70FD89CF
E268C22FFADFE5730A7A7F2026DA194162C4E6E669F313BA4E113FB78B3E7F861F0100533EA09C30
423636366048886360B0C11F263C0B9C146671FA80969D9305426D3A055E308D901CBD9C3BBAB466
7AB48315E6A228D7D2293C4BB6C56826F7C97459F29E7779B54FD7A8696338C3DF1EE783C3BAC029
4C9142D007B3291EEEAF50FB948945D72966EDA710354E71768C7B16C84F418C923805FC994F0B1B
877E597F471EF144137E9D9488A09E77A465D0234CEF31B835F2D3C16164D9E012C3EEE0C225DE39
F4390978B01ECF2B2B87CCFDC0D74C09C3CD1484455F93875FE5534F3C0D630CF802FAF43560182C
2D2DC161F825FCBCF065E4A7EEA66FBD29D49C256BDF8CCCC9589E475BF829422522E33BD4FA304B
D6FD14ADC762D73E89ACA7F562BD53AABAAF0B9C42F7D1E64A84B66FE765C956E979C78F6953ED53
E9FE14ECBA9DA953E8216DD00AC62FE6BFF6E9FC33DFA029EE835FF83C4CBC61863C2D7C66F5FF4A
DF887F5F20027039A2F936E2E8FA11CE2C7ACF9EBFFFE87171CC03274EC2AFFBCA8BDBD4C61AA673
9B9BA7CD3106B338F871885300969797E1D701DF05DE99EF07C0E7D19E0BC4AB3B2FC3991107732E
03EF29080B4C236192293E3CBC2ABE40DD412CF49A3CFC3631612C2296C13086B12A342C98C8C14E
338E6C77F179D662A13E9DC2E4115AF32D1647DD4CD53EC50270C23150C3E999D73EB51A75738A40
DC53488C8BE73EC93EDA7A8A154A21403F4537729F4C01225E013549ED534ECFBB36D53EA5FB5398
3BB3FB53D817EA4871D4E2D43ECD74E51C6BAB3291FC242D58E46FC18F302D2CEE9A3C125B2014B0
1149428603E0305177073FF5D2D2123C1512C89C53D486FA7BDE85440A7DB7A21C8C647F8A029B08
ACE5CABE155EFBD47634D8473BA2BEDD143DEF4C9DE2EE3D8566D9EBC1955B31F23BE5D1363945A6
99826A9FC43DA85AED131D09FC02B64D4E31C569CF7CF5A7285EABD39C8277B5338BA02CDD8D5466
D429EA9C4312A7C8FCA6BC3F85C3513316714D1EFE5EE09A03A400C842FCC3C3A51588836EF2BE3B
F0E6C0CBCD1580F6F9D6E719757BB4D5CDC858CB157E8A41EE93D1477B75B950F86496E346740AF7
68B71DF5700A739136C426C47688538CFB68C3701D95EA150636776D0B123D5037BAA053F0DC27A2
0CBACC49481854FBF49BFBBE3A79ED93790CAF7D9AC5F452ACD14D1D15FC14B2F62992254BFD29A2
89B25A89C39DDFF8F8CF862A6A6684B2F708E014757E3C87836341D7E4B1E4498789F19F6B7760B4
81C3B4797F77F0770A4F99CD101791672D2E6A5851E11EEDE1222D6B817753E49C67F7BCFB8DA377
637D8879EBD1FBCD4530D7295A8C1A38059082F1BCCBECB412E879374C897C61EF43C71F146F0B57
BFAD3B978602DCB125AE5318D57D01122D748AF62D9CC6FB68E754434D25F749D43E9959B2533CF9
F856376FBC31D3739BA353E81256FE8710D129860B41E16E77915520786DFDB54F8253247F9BAE53
381AC4C2ADC9C31FCBEFFDEE979153848205E847C302A75EAFA7FFC4E0A7C6B228B3BB8A738ADA50
7F1F6DB9D21591D4533DEF0A854F4501225413220A484E3FFAA5C67F058E19A1269DE2BFEC076A10
3753C8B9D96867C84F31D42958ED9339EF32CD44F8E5917FF8F38D9FFF9962E5DD47901A68A7F639
D5E14E6B16E8D1FEF31F1456BCE1D6433A45B2E75DA8328A3805BFEBBDBAF372EFD9F31570FE8FCF
C30D1436E09EF8FA0FB68703BB2E4E11EB4F91BA74076B9FD4CA8F51B36A0EECC10173EED1DE759D
C2D1281671FE8C84022388E1E2037F41F047073F05DCC1F9551418071E697AB1618A886F228809BC
C3E27A4C161135ACA8F4FD14A66B35E4A7B06A9FB27ADE05126539BFE8AF9261E0FFBD7D1BACD73E
B518B5F929425285C926B8B128E6A7209A7CAC10FA148FDC24B5A20B7E0AE214A25FB6992E9B9DFB
B4F668B4AE2927F749FB291EFCC2E7D72E5F5BFEEECEF28B574BE1E3DFDAE9BFEABB3BF0F2B5CF9E
C2777BFDCDBF9DE9B90DD53E9934596B19F1DCA771FA77446E0B65C9CED2A31DBA1556E014AE5338
9AC2C2D5F9C0D40EADD96B6B6BF71F3DBEB272E8E8FA91E5E56524080F9C38497F4D7067C49D66DC
315CB5F07D38A9A79CE4BB0E1E86B7825B363C02363636601BB8C9F5D7AF3BA60B6085376FBC0180
EDE1E3CD9B537CFFDD11A720CA102CCA555F92B7023805FCF6E14E8AEFD9FFC0376F6E1CFA6539C5
0A79FD841FF0D812F10B1857D3FD791DCD82C630FC838B4CFF7F833DD3FF46376FC2A5891A5AE9F5
5B834D146BFC50A7F88DA377FFCFB7DEA20F096F0B97C43FFCA7EFE249027A0226A8B47C6AA053F4
D767E86D6773121AFCFDA24ED167100312615A27349B20BA0107FCE6BEAF6EFDB7176E5C7B8D9F25
E014A29629C926F831F0DACDBFFF65E014F06EF0CEF43B053AB0FCDCA5DBFFE8FB1FFBCF3B80E50B
976EFFCE25E00BF008DBB4219E8247D8031B00D8B8EBD89D38B0E1A23DE5F3597C84BBBC59FB14BB
74AB010F9C82FFF5E192E3AFFEC2DBFAE272A0C35D2879808FEDA73FFAF6ADADE76731AAE1E3C194
06A71C705FC059076CBB9FC2B14058389D820A9FE0AF8FEF477F04FCA3F2603C1288C3AB3B2FEBF7
214EA1170148A770B25F036ACA7D0A54E4CA3B91D53B299EFB94E9F5E3F32EDC80D702DC4FD162D4
EAD1D6532F934DB06C1C6C62256A9FFAF9F683DAA761962C0CD7FBDE37AEF1536D1C8DA2BECEF4A7
209AC08509B3DEC9142FE2FD292AF4BC437E61EA14F05B5EFEEE804D3C77A94F2EBE73093722DB08
78495FAAB87069E5BB7FF1E0173EBF3B92F2677A6E8DDA27BDF8A3D9C4686C27FA53985D1A4D4F9C
2A919AA94E1182EB148E05C2C2D5F9C06D34E483A0A7904460E813FC8BEB14FA0F16FE1E9DE9D786
DAB264E93664B4D28E5689246B9FF4CA2D9F74C92F8B113AEEA768316AAA7D8A8C61550A25167B23
3DEF048F48860F8829591738C5D70EBF702ED0473B2751369E251BAF7DE2FD2974ED93EE79D7E714
2F5E1D8A1403A690BFD1AF7D1A3CFEB393F7D733B00B9C8295A14AAD59B728BD381EFF39FD29CC51
2D15676E946BDA4F9143167CF6E268100B57FB7474FD08992904E042843401B542F8CB424EC193DF
E84F1278073EAB9774164EBB5968D4C629E8F664C7F85B4B61D4CC22BF8FB6C123548B8AAB18C5E9
3DEFDA8E3AFB53E8912CD3CC423DEF227DB471B8DEF7BE609588492E066DF2BA93254B1A44244556
4B15D49F42BC2DE714118F76281BCACC7DC2A1C2758AB2E8BFF0BB3B1B1B1BC3813D638F36F9290A
DCE1E2FB6F9A45AA4A564E700A6D0BB2B60B34996DCFBF4EF1D4134FBB4EE1680A0B377F7EE0C449
E402E6B33C63163DDAC032CC34E99D2B3B4840B44CB37059580B8D9A7BDE65E53E093F455CA7082C
73059776BD3F4567500FA728E86E58FEA186B7542B6884C739C568F4EA7CB3C244CB726D7741A7C0
1A273453986A0535A428B089BBAFA2AD3B59FB841C81D48AFC5228A3F669C029C60421EF114BA186
82C57777E0E65BCFC036FA685BACA1A05088DAA732FD29E4B6D9A56274499F7F4E813A456B68456B
7E908E60E1E6CF9B9BA73121563FD50FE25B5BA32677245BF47A3D7D30300E7C162E5FFA7D168B67
2D346ACB9225E941EAE656C588D02CF2758A50C1B91930E29CA2DDA8B98FB639C58AD43EC5758A61
8F7818C0039DC2CCEDD7366D2A3BEF02A73005089D226B4A15F9B94FA660C1B58CC73E9851FBC474
0AB4634736884A14ACDC5CA7A8BF8FB662C7628DA8E0838BEA14638F7628C44CAF0ED101EB9E255B
98E4FB84DF21B070F3674C73025A615ED69053907D1B750AF34F0C2BA3CC1AAA85F3982C346AEBA3
5D8825D44D8785D72FBBF6C9BE3DA9804DE9635D1F7EE99CA2C5A89B53987D28CCDA27EEA7B0FA53
689D22641AE2ABB87CFC77A1F6C9EC4C11B25194F56867E6C70ABA11AF7D228E90502846FC828E94
3A455DB54FA28D5D6254E765C99A54428E6AB52E447F02F3AF53FCEFCFFE01CCEB10757ECE6A887C
48CAE19CFAF79AAB33A33FCC5C7DBCB258383FC58D6BAF699704617575156DD7F84B41D7B66E144B
4FD145529C93C5D26E161A35F7BC938522E1C9185FF89A44A7D01EEDABC7C634C439458B515BED93
CD174CDF5031F12CA4537CFF0E56768E3A851AD526711E6E74A33F0529115A9288B82A38A7D0F921
A453F0DAA71C3691A353148A9A2E58F2C4857171148F81927E8ABA6A9FEC2C599323AB05A2581FED
D1908E47F699237CFE3905F7533CF885CFFF8B23FFF87347FED1FD478FFFF34FFDF26F1CBD7B92ED
A3EB471EF9ECAA39E3C5D85E989EBDF2C35730C4181E01BC9509851BD33178587238992DC6C47787
F7C45C7A4A51C6EF8209F0987E8C9F10BF1DEEA12333E9067E23FA4901B3FBBD2F28B358389D6277
441C36374FEBA744438A57775EC60227F123C3E381FD07979696909888DFDD229E93C545ADB94FD6
FD48DC984A67C9866574B3FCA9A05338A768351AA87D0AF58257ABB8C9DC2753560B15F5E9D2F42E
708A3153A08AA670786CB647FB18E70EC9FE14A5FC14D89F624C132E0CE90375A3E01E0A7C76DCD2
A2463F45224B36946F56EC679AAF53D845AA56D9EAB5A6739F72809C0267350F7DE21FDCFCEC6DFD
BFDFF5BDE8FEABFC881B21FD11CE09308E77EC79E72D3F750B3E02D6D6D6381D800B0B4CDEF02904
1C0944D54CE624F47A3D38F2FAA0D554085BBD6DF85EFC9DE9FDF54E9810C2007BE8F8833039A49D
58309FC4F2F232BD27BC1CD8D68C68C5ACA5C0D96111EB7C8829082E00A31AF6C3C0E63BE197AE9B
6563E33C5A75D1E7C439456DA8D9A31D0A9235E5753AB26CEE9329A3D3F46CD898F898D73EB51C4D
D63E85D48A8B852FD3B54F83102723F7C90A25E06CBA0B9C62E8D15664413AB5EFBEAA0D17397E8A
CCDC27A16BE4E43E49B562A44AF0C227DCE0B54FD447BBFEDCA750634783358F0E2BE5A7080D729D
B031FF3AC59367CFBEB9FB13DCBEEFE0BB890E4C057002E3EBE7344B37F3367707B32F3A26E7C739
B0FF201C990C5D876F41EF0CD37EF300F848C28D0B1C813E4C9CB6EC0E7416CE4DF48F3615C0E8BA
FD03CBD8702DF991E6108DCC9F81D9C1B0C754D8A5C13F98E1C3C881EB1E8C8A9C1B31BC9CE803FD
4E8102C39B68D16179F08FEFD77BC43959389EB5B868C64FA1EE4705D9A2AA9F42DC7D42D32DD729
3A8206729F2275E6FCD98CDC27D1CF51CFAF84299B27E4744AA708F5A710854FBA0E2ACE2922F461
92DC27EC8BCD6B9CC68285B2519066811EED66729FE8B27C5129C82C3059EBCB119DC258F34975D0
5E209D026674377FFC23DC7EE813FFE0DAC05A2E3C7D44A30A1B91FDA37738FEC1BF87EF1C9A3ED1
327E647E45EBFCF11F04DE0146423E01C177860F1079675C4F1E8F878D0D7AFFD022333F1826AB24
76CCE2778DB485F811FCD9CEE2BBCC1435D7F9EC5CD981C9FCBEE83FF8AD6D6E9E4E1234F8BB5E59
3904C3030E062025091DFCD0F107B11505005E02D7C608BB749DA24ED4ECA7E0B51F46BAAC4934CA
F6A708341AE61B5476EE7DB45B8CE66B9FCC7A2791FB14D22914410ED53BF1789CAEE914A44AF0AA
A72CD7762A4B761244FC14C4260A668AE70A6C82F6504D54F37E8A78ED53912CC7758A6BF716EBFA
8AF165C6CEE29EF9D72960F642D39B8D43BF4C7F9839C55DB1FD8337214E1102718AD001F99C6277
308DA7373CFBD5E1BDB21A5BA157C10C9376F6FF2E9854919460600239534EB1B6B6861F1E7EE9F0
83D73CD2A682DAFCC8F0CB3ABA7E044E17962D011D802F3706FF60860F802FE17CC27E621639137B
78DB9B37DE809F224781EA9B6B322E86EEA7A81335FB29A45AC117BB547F8A6A7E8A90C3C2F05378
ED53AB5133A7306D14E3CE7766664E40A7D8BEE75D4681536048CB794867748A90353B2E4F100DC9
CC928DF7D1D65F466A9FC6C1B04CA7089929701B0F0EE53E61C1C0CFDCF6F13EDE7BFBD2073E31DC
CE031EBF7CB764B5869FA24C5FA1849F623D6608B25D424CAD9B7F9D027E299C538C9B2805A4F344
C557317B219F534CAE53EC0EA6F13B5776328F2FF5CE089A79E20B4D972E02EED4F0A3D1B748720A
38FEA1E30FE6DF02F074C13BF3EAFDF90CAA4AFE2035CC9FB15504F005142092E707E66FC035B439
A2C24F57E100AF7DAA1335E914BA1B456045D72CD02DAB538832DDC2BC6BE4A7F0DAA7D6A399DA27
936228EE9CE9A790538E7539A40BABA06CD8778153C479843056882EDBA5FA5344D804F6A7289B25
6B8A143C4896D30DAA7DD23AC5FFFEEC1F2C7DF6F76FF997FFF5962F6EF71F4F5FEA0337721E01BF
F19FF67DF233BBC57B71A8E79D79E9B623FB523A45CC46A1A7DC231A021B0BA95344D8C4BADA8832
8E24A7A0EA20BE93FF7293337F3AF8CCE367505380993FBE44DF853992EFFCEACECB620FCC485757
576152AA2BACC4E410450AB2541CD87F307E1E50FE28658800AA8EDF65D6236AA6A8A1CEE7F2E597
FA2EE9DEB6F99B8AE3FEA3C7E1375EF95B5763795EFB54276AF353048943B8FE9CD6BEB2740AF366
24A65EA303BCF6A90B68A4F6C9ACE8334238F3729FC8A38D6CC22CE7D34BA0D73AD39F827BB44355
4F66730AE4145696EC319E16CBFB6867163ED99CE27347B5475B473F097982448ADBFFE8FB5AA7C0
05492005C811DEF6F0C5FE06900BE417991BA72FBDF793F7D1E7C43BB5EE79271B55C48D42713F85
0E310B5CB48944F361BF583AC5439FF80770932242949024AC4411511896CF29426D17F2D504782B
ACA281F19093CE44D420F4CEBCEA69F877F15BA770864915504F3DF1B47E61EFD9F34837E8932475
8A0A9C022EBC49DE34FF98B54E01D7CC24A1139F47EC813F10EA6117396C81CE8983A3369D026F34
661F6D634AC66F6465FD1456F3BB90B7C2758A16A3FEDC27A3BA29DA233EA2538CFB68E7155A170E
B8EF7D5DE0142683D00E0B1E0095EFD1269D22D244DBCC868A78B475730A11FD44FC8277A98864C9
F639C5677B634E81FC022943264E5FDA7B6843DCD0FFFC072F099DC26EB6225873716C47740A830B
875391C5CE05D5296CEE102F82B2560FA6E0A718ACC62739059A94E94B6013F82ACC6E0D5598E877
A623B7B69ED79F8A38453C330A1804DA75E9C32739057EE04C4E81D9BFF02D908B99FD0A17A5FC69
A6753E701232537FE300AECA3F64E6B9ADDC41DE7BDED5899AB364258AAE55BB7AE4E2FBCBF929A2
0B41430EE27E8A0EA0999E77A68355F38B68CFBBA1479BEB14F12A91AE66C9028380C750F4136713
74587E966C961D5BA918892CD90B4CA4280A165AA720E3B6AC7D1AF929CE9FFD3A708A3E9BF897FF
755CD4441A44CEE317B78153884F9BE84F614A72EABA1DD729048F084ACC8A71CC3FA730FC14451F
7A096F853A51214EC11DD054A704270A001BB88D8F38B78F730A78B7BB0E1E86413B6604BDE19CDF
EC533C1C96234E01BC6073F3F443C71F8483D1D740328478091C404BD6D4AE424C895FDD7999D801
8C4C244DD3D52930E196D88A68C7BC286C827E0BB35B93875F4DA9BB6A8405D439F5729DA24ED4A7
53B0A9546415577BFDE2B54FE9FE148A5F781FED8EA0A9DCA72059366B9F2C9DA2D0475BEB11D1C2
27A4211DE11449533667137959B2B2E75DA8F31DAF8FCACF92E58D270471306BA2223AC59BBB3F39
FFCC37FADAC480200C758A81F4201FC31BF02AE414A69F229860662EFE64EB149A269BF4C198752F
42ED93D42954B7CA2089580F57438D4E57BE4E011B046C19801BBC675CE84DB08A494CC8F313A5E0
48745EC3BFD5C13F540D5083C093838FA453EC8E4A8F7056CF4723BE09FF60F99C22A7231E7D2F38
5ED7E42066DAB07BBAA873FEBC286CCBFD14359FED597F8BA19F6254FB54B80D856BCE336B9F0A6B
B69A3E5897E5B14EE19CA2D56830F749CCB2C6E94F825C64F6D18E646C6ABAD1318F36F753149ADF
85D9C4B9581FED824EC1F942B2E15DD2A34D9EEB3E8FA006D9CF29B3B6502846B94FDAA3DDE7149F
ED190C22C74F31FA92FB2910218F76A43F458129A7740AEC37AAD7820A8B42EB8A620CF6CF3FA710
3AC59843996C22DC92C32C0FABE6D11648720A18AEBAC405CE03BE307442E24E8D78ED13FF60B7B0
867DF89E7FFE83E197E4D1E69C02CE36CCF961D00270031E517780A1025FBEBAF332ED0FF58BC46F
A47F34987800D1D8B9B2035385482CD5FCC0EB7C345CA7A813B5F5A7C86213C545DD0A7DB4057D30
6662C5DC27F768B718CDE63E19FA0517E94AF6D136450AC93B3A9925AB5983C92644038B647F0AA0
09E89528DBFC2E9E254BB14E9C4D8C458A626514118A5096EC58A7404221F84204ACFC49D73E699D
C26CB0620FF21C9D420768F021ADBCCC5731556370F59EFFDA27A1538C17BE144BB2970B2C8186BE
9CC4A3BD3B320E2439053C0B53682C9ABAF2E2366E50D1946979D8CD706AE8170A4E41B4E5AE8387
710FF63EA30348A7E03BE9FBE620C40B7ABD1E3C0B3F23DF096484732BE417758EBD0A6870FE8CBD
02E10CC3E081F3393F5DC83D4BB6E6B33DEB6F51F053841DD9A19CF338A748D6A09A95517489769D
A2C56832F789AFE286CDDAF97DB46D91623D40A53BA95304239EAC9DF91EED52842251FB74A14016
8493C2EC7FC7FB68DB3A85E61479360AEED1169FB6E0A7508356AFF9889AD5B44E21AA9BC28F9A35
2F964E81B94F264BB2D58AA8533B8753842A94E8237135C17422F78DFF235B01FCC38E66F8486F0E
33C6F83B9B9F4DCCD8F1EF626D6D8DEFA1C93F105BFC92CF06CDDC27F8BECBCBCB30F95F1EFD836D
FCA8403D96D93FD8CF199FF82DC3F1AFBFF9B77C27BC842E11D4C002FFFAE6B6ECA7C13A1FA087F0
0BBD39F887FC0B4E3870527E02BB764E3A88DA740A2A67C23B8ECD264C9D62923EDAD1455DAF7D6A
379ACC7D4AC62667EA14D642AE5CF354EB991DD42932D9046F5A51B98F765CB988F5A750D6897169
53D14CC1E3A1A88FB6CE7D1AD73E95056316393A8521B499639E0DF87C9D223E9D163B174EA73084
98387708F38B52B54F957BDEC17EF891CD97533A93B633C0F166966C7C2609435A742BD8DC3C8D6F
020C1ACE24491BF83E5CA788A7002103121C21F261E063DCA29A7AD069A417C2241938E3EC06DBE4
98A94E214E20FF72E7CA0ED143DE45022E7D115F7F0BCE8943A0363F45A4F649DEA42AFB29CC489C
8072E17E8AD6631E729F8CCE14797E0A91251B61CAB25EA2C33A8519212BAA9E32FD14481CC8A35D
AA0E2AA4530CFD11CF5D126C42D82838C5209D021E736B9F4AE914514E91F053B0E15DAA8F7624D7
484FB0B97F19B6174BA7C0DAA72C013D6AD9A6FD93FB29E29C025943A46A85DE5F37B0ABD0471BFE
2E4463657A13007C2F31819F5D7F0AF85EC24272F9F24B9A9ADD7FF438C985F389DAEA7C04A7EB3D
7B3E349BDADA7AFEA1E30F367C4EDC6352E3D99EF5B7409D22968113A87D1A6E47B364133A854ECF
18F929E04B78F9E947BFD4F8AFC031233499FB1429F32BD3479BF7BCB3CB9F4C2F6737FA5360962C
10041D0085FB43554F397E8A478BFD29E26C22D3A3AD5509AD5C709102D94442A7204E514DA788D6
3E899E7746EF6C4BB348F6D1160EB8F8A81697F445D729F45FB141B22CDA35153FC56EAA331D4CC5
513808BD9C2C0FFAB00A9C02A69ABAAB32B00CA215F25B8CAC13D3EDA38D9F1CA6C4FC7B610E9538
72C236D03560A6753E91FC2BA06070CE43CFC2B0A9F9CF96C3758A3A519B9F42D43E19B5E58154FF
9BD5FC14A104BF624E94EB142D4663B94F21A9428DEDB44EA1A335536512D73A93FB4454A210F734
E85861520C215524FD14F16E7793FB29884DF09E77859D17582FBC64ED53599D22DC9FA2A053D0D8
BEF87E83328BF51FF654B28FB65901155B0E5A403F8559FB5451B6C8AB7DA2D978E8EA17E114D809
22BECA8D1541E8561033F6CC6E7A853FB7B5353D11253102FBDC899199F92D929C82D3872B2F6E6B
8336FC1ECDA0AAC8CC791E30D3F933FC2DECDBB76FE7CACEAEC53AE197027CD01C78F03B6DB002CA
FD14359FED597F0BD9F32E1EF7F46D55A05EDE4F216662FCF6D43F9E729FDCA3DD6ACC51EE935AC5
4DEB14386E473A45429B288A14F06517748A78678AC8CE1C4E9153E684D9508F7D50ED4FE53E91FF
9A8B1466BB8A44CFBBCA3A45D4A36DF829C24B40A29363599D227EC5D693EAF9E714C1DCA7088F50
3D2B856C41C7C43905653A618CAA790CC537692DE3C0FE83FC85A15ED2248560FD3C1D460D264C95
C42CC5476EC22FD4B81FE6EDF014EDA7D7E23C5F147799924A7E7F0AC0C35FFE8A2620671E3FA339
05FC0D0A8964DECCDAB3AEF359593904B4027EF59AAFC100403B7FEFD9F3E229A0ABA2C8AD4EB84E
5127EAF65398594F56C8F9CD3C8FB6EDA748258AD04B9C53B418CDE73ED1600E985B433A85EE7997
6613C503BAA35384DCD9F13D613F45DAA32D18C7B0DE69E4BF80EDA49F62689AE0594F170A1E6DDA
19E979B76B7AB467D4473BB210A44974193F45E88A2D66D7B473FE6B9F74EE930E3C4FFBFEAC2E78
113F05CCD9C8DD4CB8FFE871BEF68E65EDE21818994021AFBCB84D6DAC61320F3FB279F1844B137A
990930C3BC7CF925B83BC3645BBC33DC5B71415B00DB49FC9BDFFED7FA63D031F09EC2DD00DF9AD7
440DBFC5EF9CD2AC014F3EF22398F7E6CCF96186ACE903E65F197FC81DD62976075E6CEC8D0E0306
869C48C182FD181706C413CE15FCA9DEB8F61AFC66E1B7D9609DB967C9D67CB667FD2D44EE53614E
95F46BA76A9F72FD1456C1AA738A76635E729FCC9D797E8AC2DC236AA91074A30B9C42173EC53D14
E78ADD2B32B364856021BE34EBA322B94F3CD9A9204328E336291725B2642B4915A5758A9087A24C
7F8AE415DB2C5B853F81C5D2298053A06E4818FF8CEA4BDA1939ECC83FFCF9C6FFF4DA079800EB55
74B4AB8B9D40D3EE3A76E7EEFCC913841AEA7C808261CC17104CEDA3873915260F8B7F402E5A7C4E
1CFC6CCFFA5B88DA273B8D70B402C6EF4D65739FE4CA4F60A5D76B9F3A82E6739F022D2AC61BA93E
DAC1DA27330F8ABDA40BB54FE4D1368509B2696B3611F5684FDA9F22A253700621ED1217648A2C2F
8E8AF5BCE33A457991225EFB444B40B6475BCB16997DB423AB3D6681103332CFBF4EC1192550FBAF
7EE27FFBC37FFAAEA73FFA76F8E408A045FC4BDC43DBF8943886BEFCC47BDE2ED24DE77672BB4030
5D2468EB10A77765E590180FF376FE6BA8F381AB19153EC1C40CCE09900852C4906E6C6C6C2C2F2F
C37E00500F2E42B5F29C3808B5710A2208BA1B45A28FF664FD2984E88C7E0AAF7DEA029ACC7D8AE8
6EA2E75D40A7103DEF4C6D4213E74EE914DA8BCDB983E017BAB376A4F689E738952A85C2E3833AC5
05A6445C901E8A428B8A7896EC54748A32B54F41C1825FC3ABF5D1CE1BDBB47FCE750A98BDC0DF2F
B6ABEEFFA67EFC23D880DF17EE990A76E76F1EBB40A05307BF56FC83C20EDAE6C14B4B4BA27C0BF7
CCF3F9AFA1CE074E9D28397BF2ECD97DFBF601B930ABDD1A8767C9D67CB667FD2DD04F617BFACCCA
90321EED903C2196C2E4535EFBD4013496FB14CE80E22BBAC3DAA7904E11E99365B6BA63A14FC09A
BBA053988EEC78E193E014FA0E98AC7DD2AA84F06847740A5222CC6E771406C59DDAA57BDE956D51
51D5A36D382CB2FB68D3404DDB85947379CE3905E0A9279E9EDD95679E67B38B02388737AEBD86FD
2F7607D1B5FA0F1601FB79638557775EC697CC33665DE7836DECCC86839B9BA723D14F0DC2758A3A
D18C9FC234AE2A1F6B85FE1459497DC7465D2ABCF6A9D598B7DC275D3112D229B6EF799798741546
35792BCC0958C7748A58130AD5ED8E36923DEF901D089D427B2BCAFA29744B3BA15CE8A7861BA1DA
A759FA2984FDEDA6687217720FA5748A9887C2B40B31C631FFB54FBCBCDC29C0DC0266BFC011B6B6
9E17667081B5B5359C7B5C7FFDFAF2F2F2CD1FFFA8F14F1E470DF36794240EEC3F088FC02348B340
1D0D3805302F9859CD0FB3703F45CD677BD6DF42D43EC9B52FB32EB7B851CE4F1188D7708F76D7D0
64EE130F3413E482578CA4FC14C6B80DF4A428983ABBC129221A4452B0C8EF7997D93B3B47A7D0A6
0999FE14A018A853243CDAD3F353E89E77A19227D3A39DA3531474E47078ACDE3FFF3AC5F9B35FAF
6D36E59C651200A1088D257E628162C318E8F57A8D7FE01CCCBACE47C83A285B00B9E0D1B270C656
575781B5CD49DA92EB1475A2A9FE14A10AA8B259B2869F22102A82FB0BFD29EE5D7AF8CB5F69FC57
E098119ACC7D325777CBE73EF13EDA9185DC71F179F73CDA6894180B10775F8DB009DE5F7B2A1EED
0AFD2944573B1E212B28061D90DBF36E4A7E8A824E71D132BE294F1C55F4E5EA14345CCD22D55043
EDF5C5D029DEDCFD096ED733E7776651EA2CB5FB74CD74FE0C6FBE6FDF3EDD9902F61FD87F509492
020BC3C4A7CB975F6AF69C7B966CCD677BD6DFE2D4A7DE66B309B31CD75A0ACBD229749EB9D5110C
6B9F68A7EB142D46F3B94F4A8093B3B2641FED545948E8CB2EE814661F6D59E374F75533182A47A7
2885CCFE14A2D59D90270A955117A44D3B51FB54C14C919DFB9418C6BA663547A74806805B57EFF9
D72944B4A6C35127665AE7036CE2AE838743CF02ADD074035B6FAFAC1C6A30FAC96B9F6A3EDBB3FE
16BCF649A43FD93726A5A757D0296CF59C0AD13D4BB603683EF7C9AC1B61A33DD1479BEB14263B0E
AFEE768153647AB1231EED904E8104C1D429A8142A64A6E8F7BF8BE814CC914DCDB2C7F4E182B251
E4EB1465158A8CDC27A3F629D0A2A25CEE53BC7974A4F66911FC144F3DF1344C9F107026F5E3AB3B
2F4FB2CDDF39F42DFC31F311CF279DD8E49937371A7CC40D8ECB975F9AE99A7CBF4D362B03E30204
FC61527DA6005CBB9696966831A46678ED539D68AC3F45A0FC49DF9BE29C22C7942D17C158ED9373
8A16A3C9DC27EEA430958BC153F0F2B27DB445AD486867173885D99F4237AA302DDB393A85A00C48
314C1E2190F0533C77491738F116783C4E764C43429C4278B44B8914393AC54555B6676A708C71A4
FB531C0BD7EF59B638BE5234FF9C627730B982A9DDCE951DD8C06D0ED84F1B30022BECE77BC49B3B
F201670F4F203FE1FC948A6D3A58FF8E9A02F01A78843176E5C56DF848304D9A759A2B4C99F6EDDB
179AA24772B16E5C7BAD29D7B667C9D67CB667FD2D78EE5342A128DEB6727ADEE9DCA7905AC12B78
E9FEE57E8A16A3C9DCA770D26666ED933180AD855CFE2C1FF91DE114719B76846554EEA39DE3D14E
FA2990325054ACD626C66DEF463A05AF7DA2028319E9147696EC2800D910294416415E1FED48BC46
B0727511FA6813DA5DB7EFE82CA83536CCCAF84DF6E68D37DEB1E79DBBC5913F0F7F05AE53D489DA
FA53C8D2DC009B28EBD1CE0A8F15B54FDE47BB1B6832F7C92A7F1A8F6DEA1A5FAD8FB64534C6B3AF
CE78B483466C65D32639836F27FB533C5AA9E75D84530CFD14AAC049B7BA2BE81417B2758AB22245
2AF749F4D10E2E0415077CA93EDA21EB84F9E5D545E8A32D2651F330A77238A688EBAF5F275AB1F7
D6BD771D3C0C7F234F9E3D7B60FFC1B5B5B5C63F9E86FB296A3EDBB3FE16A853E08DC64894550C42
2FF996D529328906BCFCF4A35F6AFC57E098119ACC7D0AD73B65E63EC93EDAE65A6EB81CBD3B3A05
C5C69A7A040643998245BCF6C99427B899A242ED93DD2CBBD8058F6F90413B96255B4AA7B028468E
4E612C04698A51BE8F76D05BA12D4283DCB385F0683B1C0DA2060E8BDFE2E8FA1120149C5C989150
357FB6D0F7754E511B6AF35304AB70493A57C551393DEF2249E67AA37087723F45DBD17CEE932A08
91D3B3A44E41557CEB561D94EA4C418F5DE01439854FD53CDA33AA7D42C5415BB0858D026BA2B848
91D5F32E27FA294FA7307ADE59ACD9B874E7E91476DC7760A7D8764EE1706834325DBF71EDB5D38F
7E09C8C5DADADAE6E6E9C807F32CD9EEA0A9FE14A1D957598FB699FBA48371E859EC4F41FB9D53B4
184DE63E59235C4CC0923A85CC923575B79193427CD9054EC1FB537036210A9F38BFE0DB3959B299
B54F25748A0B52AA1059B2C24C817423ABE75D4E7F8A3C9DA250FB54F46887AC43A5748AE4900EAE
0B2D42EE93C3E1982B78ED53CD677BD6DF82FB29A4A155DFAD4A7AB4B3FA53E8D211D7293A800673
9F826A85988965FB2974B1537099B7C31E6DEE9B884B1570645CA70845C5C6FD14C32FC37E8AA448
A11F13B94FBCF6A9AC4891D229827DB44DDF50D5FE14B260D5AAF4A3B523D7291C0E472978ED539D
A8D94F315EB30D15E87EDB98A495D329C265BADADFEA9C62AE00BFE5ADADE7AFBCB80D1BFDEDFFF6
023CE297998FE72E5F25B1B5A9DC275B7D53DB593A45AA96CF48C8E99847DB1426625DF046D24652
A788D43ED14EECA34DEC23DEF36E4C1046054E66D693D02C0C3F8559FB14D729C24423ABF6C97452
B0D15EB63F85E01169A63C7A89EB140E4753802B03DC584B75240F1D73F9F24BB57D6CCF92AD134D
FA29C40257A8F6A942EE53C85BE13A45D3885C85E0A9F77EF23E2C118119D7EA770A7D87731E01B7
7DF1FF7EF5C52D7CC3866B9F44C1B939DA233A058E5BD229D615AD580F4CBD061BDDD129883E843C
DAB65F3B234B56D73E657A2B2259B24425C6814E966B5B38B84BF4BC2B2552A4729F429AB2C19DC5
08FFF66D99FD29F4E53A9469E63A85C3D12CE0CA73FFD1E3BB13FB2376AEECC025B1B68FED3A459D
A849A7D06DB2C5C25728092AE5D10EF6A730FB0B8FB264DD4F319FD873C7118AC781D91756926702
0F064E410B20F5E73EC5C6B3D813D52964CFBB40E893DC4F83BC1BB94F39AA44A4086A2AB54FA853
E47BB4B95461C63DE1E017BDF0447F8A2A7E8ACA3AC5C5715E9FD1AE5451097A369DFB6419DFF4EA
9020CBAE53381C0D626BEBF9A3EB4726798757775EDEB76FDF6E8DC66DF753D489DAFC14861EA197
768BB54F391EED12B94FBA55AB738A39C37B7EFDF847FFE0BB30955AFDCE0EEA0E34F5CAC46DBFFB
244D399AC97DB286B418EDE3B9597ECF3B33E88993E5910CD79DDC27A15364C63DE131A85358FD29
B26A9F04EF10C7846A9F28CA89040B1DF734DE7321DC9F82D73EE5F8294AEA14BAF629D2D55194F6
D146A23F85F04DA4AA9EE825AE53381CCD02FE22969797E92A540A7077434251275CA7A813B5E914
78A3319B0EF3FA10F3E65559A7304AD0D70BB54FDE47BB665035A6B946F1770FDE4DDA04720A5ED7
14D918CEBE9EBBB4F7B1FF5033A730EA9DB87B82673DA94291889F427AB445B4AC59D1D7318F36D7
29845AC14B9E4C9F45D2A31DCF7DE22A863826A953489122EAA7E0FD292AEA1471E37685DAA740F9
1395ADE6EA1491CEA40187057CE93A85C3D138CE3C7E66EFAD7B373636AEFEE55FE51C7FE3DA6B4F
9E3DBBB4B4B4BABA5AFFA7F52CD99ACFF6ACBF859D251B522BB4477B023F05EF46C1F7D09DCB758A
CA98857089B54FFDF2A73FFA3E6FFB958F0F3DF21FB0F6A9368F76301B599108DE477B5CFB14D029
863DEF32E29E4C06DD054E8159B2C9161542BC4879B463B54F997E8A48962CF9296C6D229AFB94D5
F3AE9448915DFBA44984D19922DB4F114B8B0D6C78EE93C33157D8DA7AFEC0FE83C02C002B2B87E0
1207AC61ABB78D412BBD5EEF9973CFC0DFCE5D070F0395C0BE784D4DB7BCF6A9E6B33DEB6F31D429
06F71A234BB6C82FF8B3A4BC57CF7DD2B3327408BA47BB3C3889D8B9B273E3DA6BAFEEBC0CD38F4C
BCF2C357F011F0DFFFEA7FFCF90F8CD887F77EF2BE8FFEC1776106B5FA9D1D3EA122BB44F271EF63
FF81864A33B94FC42070D235A20F6629544CA7287AB4CD3E2CC141DE0D9D820CD7A14ED971D96292
DCA710D7C0DCA7884E51686367F5CED66A058EED829FA254EE53799D22274BD618CF93F5A730ED15
C3152136E05DA77038E6077065E3DDB4C53F7A6A757515EEFB4D7D48AF7DAA13F5F9298A45E6B959
B2F97DB4AD187F31F512C640E714D50054E2F6BF73FBD1777FF6F83B4F54034C08E11DB4D2D1D729
46E54CB7FFD1F7F9D42B130DD73E59EE09634AC66B9F2C9D62FB9E7751C9933DD10A4DC94655525D
E014634922EAD136658B6AB94F3988700ACA7142C4E5090A3D33FA689BB54F95448A78ED13ADEAE8
66F1863057F46827FB6847C6B61D3EE07E0A87632EF1E4D9B36B6B6BCBCBCBC423F011FB6BC7D944
3C0A722A1F0FDE073E61E367A923A8AD3F0535B08B67C99AC99C15FB688B12747623F3DAA74970F5
2FFF6AED7FFB4CD9B916CDB870E3C0DF39A8DF9938852CFC78EED2783216D918D43E5D797138545E
7FF36F677A1E623DEFACEABEC2726EAA3F05F926F80036E3718C54A8FBDED785FE14438FB67052DC
7D557C192A8E0A78B4ABF4BCCBAC7DC2B18DE3593B29CC0028DE473B2B4B365FA1C8C87D0A053DE9
A15EB98F76D08E1D18DEAE53381C0B8452BC00D8075C932F5F7E091FE142047FEC708BD46F02D740
B8CC1E5D3F02B405FE42AFBF7E3DFE19DC4F511B9AF15398E99A2A3F84EE53A5FC1489B273AF7D9A
1893700A9AB0253905D59CF39958FC11750AE2144DF5BCB39770C5382FD947DBAC710A09169DD229
904198C2043CA57B6DA73CDA557ADE65EA14C3214D4EED92FD29D23DEF4A291465729F62BADB047D
B4636C62DD12325CA77038DA080C6F21E7C581FD07A9806A6BEB7971F033E79E8123E90F13DDDF91
EA26EF7957276AD32948434F2EE76AFDA29A4E61163EE106DDBC3CF7A90238A7C874AD6A443805FA
29B0F689CFACE2F5E77858E3B94FB1E19D5DFB34EEA39D1341509C7475A78F76C8A39DE8A37DF7D5
1C8FB659FB141AED9959B22248968FDEC8F04EE43E993A458E429191FB14F4688B48E4927E8A483D
6AACF6C9FD140EC7C2222958C0651318C4CACAA1E5C13FDA1687C1FD1118C4230F3FC6DF10BEDC7B
EBDE5EAF17FAD6EEA7A80DB5F929C4925728F09C6BE865750A7B39D7BC7FB94E310126D12968F695
53FBC4AB4462EBBA6C0E56BF4E1132AE2694B864ED93E83DB16E2B71FC185E25D5059D426B109C4D
68E582129F72FC14F93DEF040789E81485BEF017C6439AB7AB10921CF753D8B94FDC4F5156A4C8AE
7D8A8CF0C29E3C3F85C98BEDD8016591739DC2E16819901D0053B871EDB5E4C168D610DD31E04B4C
A032998B738A3A519B4E81371A5D289273DB2AD19F42192844097AA1F6C9FD14953079ED530EA710
0C42F417E60BB9DCEB5ABF4EA1EDABC954FFA44E21FB6847B2648BFD29E8802E708A647BBBB8A562
F2DC27F3A908A7E0EC38244C08CDC2F653983DEF78EED3643A8551FB5464C4B659BB8C4E2107B02E
820AF4BC739DC2E1987F94B5573F79F62CAD9944007F8CA85FE86FB1BABA7AFB0796B546BCEB9CA2
5ED4E6A7482FE1B2CE77542895E414114F9FAC83628F9490E39CA2022A700A3DFB8AD73E7177B658
D4155C83DCAC7864831E6DCD8843CD1C293639E1D1264B855AAD35067951AAE802A7E01E6DAE4A8C
CB9C983091CD29AAF4BCCBAC7DA2A16BF883D8A816DDB413FD2974EE532954EB79A76B564BFA2904
8F085DB1F597AE53381CF38C0A614DF012EC5EB1736527DE9EFBAE8387E1C8CDCDD3FA297839FA2F
CCF7F7DCA7DA50B347DB36B432E95CD63EC146344BD6EC4F612A17A695DB39450594E214A165DEB8
4E21FD1481455D3D016BD04F614B15A17291321EEDBEB1C29A5FD9F5E79DF153C4B489019B0805C9
463DDAE57ADEE963123AC5854BD4EACE70063D67443FE5FA2950A7C8572832729FC84F5110E3B445
68E2DCA748CE862882729DC2E168197ACF9E172D2DB67ADB66D6133E6B8A0E70DF1C9645A9FBBEEB
1475A2293FC578CD965A83856B4572FD142AFD466BE854FBD47F748F7625989C82A65B7AC33C6C72
3F8599BD392FB94FA687A228C6C5758A501FEDF1201775E9C5D1DE119D422811A19E77868F3BE8A7
28DDF34EE43E45740A61CD36CD41A2966F6CE84EE63E55435EED5330009C362ECAB5A09CDCA70883
B04D16EBEEA77038DA06B8B2614B6ED1324F34B6D8DA7A1E59834E8202C014319413E53A459D9856
579108B89F822787D8F655D63E3BC7A39DD029D40A58FFF8919F025EEE3A451CE6F0109CA254F453
BC3FC57B3F791FF5B9E359B29A44D8FDC22E1438453CB37A7224729F32D48AA04E71F79EF15CEB98
61AC08159C77AAF629A4506879423C8BE55275F6BC7BE0C4C961CF3BD6CCCE20C54CB02888717FF4
7DAE535051DFF9B35F0F66C9663E9EBE047F74E2D3C63DDAE608E7E103B811D42974CB95B04D5BD7
F8A14E51C36D8BE09CC2E1A80770EF5B5B5BE30DB8F94DFC9973CFE04E7355E1F2E597F055401FC4
F5E1E68D3760A607734800BC166608B8019C05336C1D5304FCCA66FAFEBB453F851D78CE96BCCC25
5FE014A88515DE767539166F6E16EBE23AD8689206FCE2F4A35F6AFC5730A3D3DECF64FEE6B780FE
2F2F2FC32387DE2370E7F22A6DFFB393F7BFB9FB13FEE6AFEEBCFCC99FFEF5B2732DAD53880FBC3B
D029FA554FACE79DB45414F7E859D96DBFFBE4D6D6F3F0B3C37BC29504DF1CAE1E700D81FD04FA12
AF33B441DBFCC85EAF071B30D1DA1DC568231EFCC2E70D8F76B2F29CCFC45ED8FBD0F107C5DBC295
7398FBF4E9A19A56306287FA68938A81B94F83DAA7C6C7E18CC63660E5DD47E2A14FF1CE77C82970
768AA305DF76C2DC27DC73E6F133E2339FFADCD17EED136B0D9FEE4FC19EC5DAA7B5CF9EC2DF295D
B4FB9C22D2F3AE4CEE13FFB43B57764259B2418ECC09C500A853885F1C700A9E909CA956F03F81A7
3FFA76F8939CC5A0826180170A044C3CE0AF1E4E8559BCED70384CBC5592EFE3F1FC55707D5B5D5D
45FA001BB41FC3668138E85EA5BB839510E4147A11E02DAF7DAA1135E53E85C3CC23AA7A5696ECBA
6213C5189C986BBBED7E0A20EC4FFCCA36ADD0E2226DF2F1F73FD52F26A1F5DE63074E88AB04EA14
8F7DB07A738A48ED93D629489530C8C5689B0ED8FBAFFEADF653ACADADDDF2E5ABFD855CC097AFBE
EDE18BB831DC63018EE91F363A06B69F7AE269711E50A7C01226595B1E6013E3923FD879F1FDF0F2
939FBB439C84B19F02C7F67DEFD341B2A1E2731AE142A7287B9D9F7FACBEFBDEAF1D7E41F8268C1A
A722B9E0CA05EA14E2CC54CE7DC2ED78962C95F3C53BAD08C102FF22E0D1C87D123A45053F45B4F6
493065BB8AAF78ADA6C3363F7D8B1E75A853A4F334422B42B57BB4E147709DC2E188002E6EB831DD
BBCCD1F5234010969696A8F00038C5ED1F5806E280EB7B02AFFCF09510A7809940EFD9F38D9FA88E
A0363FC578212BAF89369FA4453885CD26989942B30C4E37DAEDA7006E0E9C0209427CEA15292001
4E21DE966A9F721672CB728A8281A218244B31384695C868EA75DBEF3EA96B9F7EE9E03A967984B2
71862C836FE397837A12DCA3AF54467F8A701D94592212AA7DDABEE75D059D625D96F38951AD3D17
DDF4688BF427C1260A5F0E748A3FFFC14BE26D337BDE4588469C538C6BF9B44861C91663721DF268
CF32F789740AE1C80E0E727640B0F649F980F4555AA7100CAFEAEB9E25EB70CC17D6D6D660FE6FCE
F32B03E9094A15BC5F36EA14AFEEBCAC5F423A85B6B3C105D3758ADA50C30226EA1478A389195AE9
0E25B264C33A058F3A0FB109B1F6359CA775A3E71D728A61E4E6A8D557924D200DC1837B775F5DFB
588153C080C9CF7D8A18B7F3B364794E8E9E74F1081DD8B3F7B1FF404385A65ECB771F2DD0049EE4
4F0BBCFC592A4D878DDFF84FC829FE1F8F3E223E2DEF4F9116DD4454725EEE13D9B1E375E68509D8
40DAE802A7C8A4C9A654417E0AA55394AE7DE21EEDA04EF1B9A3CB2F5EA5B11A69B6523AF749E814
25439F4AFB29CCF51F1AD26C4FB28F76A896CF1CDE34F26BE614AE53381C7100A7C0522598D2C3B4
3F1E06CB21AEBDFACBCDCDD33C3976ABB78DDF25E4A7400E023750FD8D9C53D4869AB36435C62422
54FB14CD92ED4FB7C296554D34C6B94F83D8FF76730AAA7D32D3FB4DC03CAD5FFBC48E119C6277B0
8C69728A9C555C9A89E567C9D2044CCCC17853307A0A38C55F5FFA6B7C37D229C69C82B309DAD61B
30E3629C025F18E214268F2874A3E0C9664CA44BF7A75857A3D7AA339745E90374C1A3CDFB537071
0D472FD7DA4C7211D0294AE73E6984729F728409932F07FB5350ED93600A39C8EC79C7B264EDD801
6219C5EB7630F749A72217AFDBA65D88C6BFE73E391C73059A446D6D3D0FFC02E6FCCBCBCB687A9A
F09D7BBD1EF6ADC02F61E281ACC12C64C2834D15C3758A3A519B9F42DE7D5800942ED615B7AD127D
B4436C42DF9E3AD0479B740A984AF51F07732A9C89998FFDDE61A3F661F0429C954DB7F629334BD6
9C7D99714FE2F8984EC1C982B9A24BCF5A73B364ED53A4784F16F825758ABBF710FF35466FA44A64
1450D0059D821732C1A08DC816824D9C8B65C996CB7D429D02FF10908344729F882673492266AF78
2E43A7E01E6D1AB7992205F36873E89E778900282D3A87758AC2153BDA99C2BC92D79FFBD4EE7B84
C33175C02D6C7575756969096E9193D4446DF5B68123F0251AAC6ED217D8DD512A141CA09FC2BC9A
C64F4B47509F9F6254FB64A8156281B78C475BC61246D9C4359E25DB819E77A8530C1B0D0F40EBB7
380D43AB056DE0B35F3BFC021D06F334AD53E8DAA70A6BB955748A0BCC4071A1E0B02014FA538C44
D831A7280B3453C01CEC37FE93B9F82CFD1445252238014BF929B8477B38BCD7832BB7668D5F8774
8A2259106A85E9B6C8C992CDEC796713E76C3F05B59FA0472952987E8A3AFB535C54A5AA2A9D4FB8
2DF0221FC992E5E54FB65F3BD460C8FB53381CF30ACDF429A9A95F13557EAA09AF12BA035C03E10D
E94AC8B1B9791A0E2631577C30D7296A434D3AC545D53E69E4D78EC813391E6DA33F454611146DB7
9B5390477B482B944D355107F5A97E4FB11C4E510193E814BC2E9D370B13FD296C3F45F94917BE30
A453187E8A908955CDCA229C82FA395EBDEF7D18BF1929089113B06EF829421A44884AE4E914E56A
9F50A7E0C7243DDA218582D3671E3B00A31D378CFE14A28F7685FE14D13EDAB2D249D9DF64152BDB
08D53EF5AFD883BAD371D440201ED90C83728FB6C331FFE8275D7FF35BBC8D1D6CAFAC1CBA7CF925
7E4CE4E5F0B8BABA0A2FE1FB77AEECE07BEA97445A5778ED539DA8DB4F215C7EFA9624F2369359B2
E1652EC1260AFB3BE6D1360DDAE365DBD17EAA8FC2F55E7C61A4F629B3CC49476EC639C57039F7C2
255D28423A0585C78A80D904A788143EE9CA10B254A46A9F82F94E917291C197F0F2B89F426A7081
AEF17ACC7756A788443F093963C2DA271CC942CB8867C986140AB33905FE0994F0684F4FA7D0B94F
06412E2E07717D395EFB645C9CB35B3AD69F25BBEB9CC2E18882CF21B1DB0BFCC92C2D2DC10C1FBB
6303358039031C76F6AB5F018E007BF8740E68021CAF5B4EF47A3D3852B4D2DE1D65CC72BF066CA0
419B37B3E0709DA24ED4A653E8E5ACE4BD898C81A5FC14F24E64D58790A5A2DD59B2BCF6896A9968
A265EA145CCE8083ABE914F1A4CD52B54F215B2B2914A2463DE8A7A02CD9520A451EA748AA129A4D
0C6DDA019DE2FB77EC21C5C19E5FADEF0DB109B33F45FBA0758A1083E066ED716554A5DAA7E45371
9D226E9D10594F38C2B9471B7BDEED8ADA27D1F36EAAFD294883A34BB1A15098D91A519DC2AE6B8A
D06436DA5DA77038E60AB4300BD7A5074E9C042270FB0796A9631D92022E496C6D3D0F4FADADADE1
4E6C9C0DFF802CE02D1BF6C34D1CF6684281CFC253CBCBCB7C277C093B43C287FB29EA446D7E0ABC
2B15B264F51DAAA8B0D391399CC250CCD7F7EA1B16E53E75A7F6495086C8977AA7C92978EE537ED6
53A9DAA7F1F2AC95B749224581745C187AB4833A45B549D708C9DA27CD26ECC84D469C73B2648726
20B18A1BD2E346E3BF0B9CC2D429CE995DB3ADDCB30AB54F99833C96FB64756CD4DB245290A85122
4BB624726A9FCC2EF07C9B9237689C47740AAD3EE87521B36CF59AE73E391C7306200E4F9E3D7BD7
C1C3A84DE0BF8D8D8DC8DC1229039008D8863F67600458D184EF005FC21F5D3C8701D50AF8BED8B4
E2FEA3C771BFF92AAF7DAA13B5E9144179422FE7AA2662F1DA272321C40A98ED60ED13EA14A44D50
E8534ECD39B5C9AB5CFB642EED56EB4FC1275D9A65F025DF12B54FA5D48A6C9DC2CE7712CD29687F
3C4B96740A3DD1D29D598A53AF2E700A5DE014E99A9DEDA72897FB24E846BCF629A7773617298841
07B364CDDCA792FD294AD53E45E4894C9D223874438F8DF6A7D8754EE170444142038A05B8F41167
04D8690218C4F5D7AFD39170D9C9EF6D412FA1C4F808BCF6A94ED4E6A7200D3DB89CAB9D80997DB4
C301FEA1D042DA6E77ED13E914BCA289AFE286B27126AC7DCA2914497BB4599913CDBB92A9FE414E
A16B9F9293AEE2F1714E617772542399BB8A22B54FE4D1E6E54CF14997F82BE802A7C0F119E2C811
9F4554A728D7F3AE944E916C4841C54E3CE8CCD029CCDCA7B23A4546ED9359B0A7950B632D28DA47
DBCE2E4BB65FF13EDA0EC7FC813845FE0A2DC637EDBD752FB76CC7116F9017D9B98BB54FCF7CA3F1
13D51134D9F32E94B7A902FF4BD53E19B724AB1CBD3B3A054DB148A7B04B9E026BBC9378B423B3B2
7C9D42D48418BD86D997D3D4298A1EED741F6D3586ED8A113EC2233A45D110640EEFE08A6E37FA53
08D1CDCE770AB0E6B0477B0A3DEF227E8AF8E815E46258E317E979A73DDAE58BFA72FC14419D22D4
052FD2F3CEA20C59A3DA3DDA0EC7FC0138855E458903AE874843EAF984AE53D4899AFD144678ACD6
2684473BDA47DB2C350FD59C8B4778F9E947BFD4F8AF6076E0B94FE32E1519F52164DF36739FCC3E
DA658D15E5FA531463638D12F4D163DA4F913FF52A22B7E75DA844442DEA46748AED7BDE250B9C02
13307B39B71B3A45B0AEA938A4AB79B44BD53E6566C962D0932D525C188B14C2BB9DDBF36E7A3A85
51FB6456F4E99E776C7FBE4E2133A0A214C33985C33157A8B6305BAD634535B4C64F5167AFCF49CE
F6ACBF05CF7D320CAD01D35FA64E114AD74CC46F7646A7F8DAE11738A1080913A253186FA89DD429
4A2DE466F6D196FEEBE712ABBB693F05D63295D229B2739F22252211E522A1538CFC14C6605E2FB0
6673C07781530C45B7809F42944509A23161CF3BDAC0FE149C83E4F829428285261A14395BD02942
3DEFCA8B14B9B54F716DA2D8AB1437627E0AD3EC663670141E6DAF7D72381607717F446D33E4B6EA
14F34931EAAF7D4A2CE77E5B3526CECB7DCA8DC419E53EF51FBB91FB843A0579AE43919BA67E3195
9E77621537B7F6A9E8A7A0E6C266213A45FDA77BDE89F613D95245A8F629334BD6F456A4739F74CD
5EB8E6BC6BFD29423A050DF54841D4B47ADE69C4B26469489B39C9AC1DBCB06C4F59A7288EF0FC2C
599362849AB344740ABCF69A6B4186B1A2C8415CA77038E60D37AEBDB6B272082E50BC67043C3EF5
C4D370E9C3ED0627C0EDC892C5DE1FD37AAB999EED599F0AE414E38473DED58EF27046DBFCB695D3
9F2266A3D00BBCC5DB5317740AE1A748569BE7D43ED5DC477BEC5DD549FEAC4484FA5364D53E95F7
5394CB920D0970C42F523AC5902F8F748A61447FBC9323D58D74A38F367AB4E3F421449CA755FB64
3A86821E6D9EE3A4E8836884477F05C33F8429FA29523A05D53E190B4119B54F699DC26A4561CBCA
6AB1C8750A8763AE009723CC80DD7BEBDE57775EE64FC1DC75E7CA0EE63BE957D5F9095BA953CC27
6AF353C8DB935EC20DC5F897F55304BA26159EED46ED93F65384748A714F6D56FB34CC7DFAC831F1
B6A69F22A950E4D73E692A31962DF8A31229B0F629D6F3AE541FEDECDCA7744E32038DFF844E110F
19D021C9BC1CBD1B3A45447733ADD9E2B0C9739F4C0F51D24F4154828F5B33545664C9DAB94FA2E7
5D599D22AFE75DA9DAA7B44E21AED266430AF362EE3A85C3316720C3352D7A08C05F10D9B11B512B
DA5AFB34E13999D13B37E3A708F717D675B9F1DAA7B83C61AC7D0D9477FAB2DD9C22A95398519CB8
B3599D827BB475A2ACEC52C1B6D3B54FE5275DB9B54FDF8EF28B6FAB4E1665748AE4C0E6B5221DF1
5370592D5EC2A7E906EA143B5776C4DB66D63E71C6A1B58C904EC1ABF8CCAC274E96C74E8AD197E9
9E77E59307D2B54F17D512509C5FE4E914FAA21DE11143DEE159B20EC7FC01F842FC6FE4951FBEB2
F7D6BD0DFE1DB5264B16CEE45F5FFAEB9D17AF5DBEFC1200AE8497FEE2126C841E39E066873B6F5C
7B6DD6677BD6E741D43ED97ABA756F2A95256B168104675F239DA283FD29F846885F50ED53A69F22
19FAC43DAD399C827BB4A57D554B15CC6D919B259B33EF2AD39FC2403C2A39D34FA154898887823F
DB054E21740ACD26427EED1C9DC2E40B7A786B2D23CB4FF19DB13328DEFF0E5F92E5A7284B963372
9F620479B2DCA79003CED82EEE744EE170CC15969797E307C0F50A38C581FDC61DBF1EB446A778E4
A147366F3BB3F9F7BF4CB727BC01D16D486FE85BD5C6C686B0BD4C170DF4A7B0D235CD2F49792FE7
A7D08B5D7C7BE4D1C65B5517748A9048115ACBE5CD2CCCDAA7FCFE14919DF91EEDA45491A553500F
BB9CE8A7323A45C1BE1A59BCB50A45D2B94F4C5F0B0EE96EEB1486FA76F7D57835148CEA6AB94F26
1EFBE0785467E63EF122A848AE2CD729727BDE951429D2B54F17DF2F288318CCBAF629A95398AB40
B18BB6D73E391CF38A959543F10376AEEC00A7585A5A6AEA13B6C64FF1D0F107E1768349833C7210
D7BEF86304F71F3D3EEBB33DEBF30037177BB536B0783B4D3F45608197C4F476738AB19F2290B719
59DA2DDB9FA22CF23DDA622616725594EB4F9133EF2AAB53041A6A27B264C37DB4616C1B332B9DE7
DFD5FE14BA8FB61EC0916AA8FA739F882390F4166A81C72933EA14E9DAA76A3A45B4F68986ABC87D
92E103BC3EAA6C7F8A54D9AA18F6AE53381C7305E014F16924CC840FEC3FD820A7688D4E01F711E2
11159AB1E2AB5AC02978ED93AE800A4DB7F846964E1192CE7550219336DACD29B03F85A87D0A4915
857917E53EDDB353C14F9133DAF3758A31B35051397A2636691F6DAB3824D34F516A488F6B9FF2FA
53080BB6B469C361A39C9CEEE814A28F761205C6D150EED358A4C87052D091B959B265458A54ED53
21D3ECA21CDE418939A33F8519216BD20AF194EB140EC75CA1F7ECF9A3EB47F81E5E51F3D4134FA3
83FB9F9DBCBFA94FD86096EC748B8BE03EC22757FCF614997489BB1850BC599FED599FD5607F8AF0
D26EA1F6A98C9F22F438BE2BADEF1DFB29DA5EFB54C87D52DA44C8BE4A93AECCDAA77C33C5A3D93D
EF441F6D5E7C1E6A5181CF26729FF2458A6C9D0246690E9B18C726734B517EEED37AB1C02910E04F
BCA30B9C22A790CF7452E4F829327BDE95D229244D28E60C842C15E4A748F7BC2B459633729F8269
B1C5F17CB398AD51AA3F4528C4CC7612B947DBE1983FACADAD016B803F4CB83451378A57775E5E5D
5DDD7BEB5EE414705569EAE32D6EED93A024701F49AEDFC65774DBA153F0DC279161AE7365F5BD29
57A708B7A2201231BC6D7535F7C99C5951CA93380093FFA7D2F3CE5CD7CDF76817A658C5A0272155
94F068CF40A790029CD5535B528F00A7F8FE1D7BC4426EBC5CA4B3B54F63FB8F22C8BDBBAF9AAD2B
A692FB24E846A69F42F4CB1631C866395F099D229F2C1791EBA7E08399F7187ACEBAB67FFBB674EE
9335A46D1B05BBA43BA77038E610771D3C4C5D2A60031E894DC046B353FAC5AD7DD29CE2F10F9CAB
36F56A13A7C0FE14B23598CED5D4919BA3EDB23A85088F155E579EFBD46E4E61F6A7301DACE6326F
8853CCDA4F21ACD966D76C733BABE75D24FAA9924E61D435E952734D992BD73E59932E5D7CDE859E
77118E9C041C3695DA2713A1DA27D1EACEF65394EDA32D3CDAD3D02974ED93CE0037E973AE4E51AC
568DF108B5D36B9F1C8EF9C49367CF2E2F2F1395C07F6B6B6BD4F0AEA956DA8D64C9CEE287D53A45
059C5C3F35EBB33DEB738B3A05DE68E4BD295AFB9493252BA2CEB583D52817A1DCA7B667C9C6FB53
98714F7CD2F5B5C32F4C4BA7A04A92CCDA279E7E63246D5ACD29F025E9DCA7F293AEB84E21ECAB86
3681839CB58C1F8FFCCC2CD9CCEEF0DDD329504D8B0F6CBE9318C754729FF01832CD918411D729B8
9FC2C81928E60F8C8F67B54F748F963A45799122B3F6C9E808AF5784CAF7A718C665845684B447DB
6B9F1C8E7905553D5DBEFCD2D6D6F358ECD4148F101FAC669DE2D4E78EC2D4FDE4AF9C3CBA7EA42C
8EAD9E80C92D3CD24C86CE21FA297256BA2285E82DD029D04F6114935B54C2E87957CD4F61364EE2
5CE35857FC14E702FD29426C42EA14E5B36435839844A7D04D284C5705F28B12B54F19F6D5723A45
884D04AAA1E23A459F2F339D424FB76C7E814FDDF7BE2EE814C2A31DF707E9013F49EE9330CAF579
C4604FA2F6498D6DA3904F45254FAA53A48CDBA56A9F0C5E1CE82B9493FB949BAAE159B20EC7C202
A6C4CD4EB4EAF453E0FC1FD8C4577E6D07233771218B96B3921BF0127821BC9C5F79F06D1FFCC2E7
4BD957DBCD2992FD29860BB942A74865C99A7E0AB31A44DB2B5A5FFB94EC4F61666FD208AFE0A7C8
CF372BE5D1B66D14A213F160F655224B3673D23549EE93D026CA64C906B5096BA2A58FE9824E1161
C4A186777C7FFD3DEFC490D6AA84B6575089143C56F45394D7298C9E77AADEC9700965FB2962EB3F
A9262CAE53381C0B07B87635F87754BF4E019C02A74F913AF3485D2ECEBE68CAC1758AB2C284460B
6A9FB89F621C9233B8EFD815B965B26493FD29C47A97A01EEDE614C9FE14F1819DE9A7A896939CEF
D136F336F9642C2B4B166B9F4C33C5643A85CC7D4AAA15EC98904EB17DCFBB24235E576C22D06818
B7BBC029861E6DABF7448E6031C59E77624FD24FC17BDE451A38D2082FD1F3AE3C654EE814D6FA8F
DCA3FB0A65F6D1D6894F29B2EC3A85C33187B871ED35F8C384791A606BEBF93E7AFD0DF8F2C9B367
F7EDDBD76CCFBB9AB364EF3DB451964AE8D99739E578FC03E768A54BDF7A9268874EC1FD14050D42
CFC18A93B19C2CD978C179E83E35BC9D7540A7F8DAE11750A4085109214F502ACEB9096A9F420C9A
CFD3F2FB53C47CAC452B772C4B166B9F4C1E31259D22BFF6294BA7287AB43577886C7450A7308589
10C588728A19F6BCE3514E3C64C0ACE5E3CDEF64966CA8E75D599122BFF6498F649D0155AA8F764E
EC80C5385CA77038E60D6B6B6B3CE5E9C0FE83B84D1BF8AF418F76CD3AC5B1D51393700A84B93655
61FDB67D9C22E4A748AC7DE579B4854E617BB4C57D8ACDBEDAEDD136758ADEDD57D17C9D1CD21366
C9C62763F97DB48301508A74C43CDA213FC5C43A859DFB1466139C3E27739F62DC215C16D51D9D22
348C4DD310DF8017FEE6BEAF4EABE75D6696ECB0F6C96A4E21BA378ACE77B93DEF4A91E58CDC27FB
5AADAEE1825C94F253248A548B84DA3DDA0EC7BC6173F334128A959543441F804D2C2F2FD397F054
EFD9F34D7DC2FA6B9F50A7D0F5E439DB74F3D2B353DEF3AECB7DB4FB3A85D58A2218E65FAD8FB655
702ED7C43A96253BCE7D2A2EE4623FBBF160BEFBAAA9BE4D31F7A914A7E054C2EC02C677F2767831
3F8599FB9437E98AEB1489DC2731D48B633BA253480D6E3DE5CBA6313FD8DF298FB6D98742EF2918
2B2AE53E4DD8F34EF45BB18BFAF4501F0DEF44ED53798522BFF6C918E47AFD27BF3F4578C187530C
73C07BED93C3315700CA00F72CD220484E2500E9D057DA3A517F962C700AAC5F2A254CE0DD8A5EA5
750AF468574B3827B4C64F61AFD6862B466EB20CA8B27E8A848CDE999E77F13EDA7A39D74C973D76
20DD9F229F3B57E8A3CD4B446213B00BA9DC27CD26A6AB53C4D984B5D81BD329783761CB3741D4D8
8CCAE9824EC109425CAA302FEF93D43EC5AD16213F0567C11C218E4CDB257ADEE593E58CDCA78223
FBE2FBEDA1AE9680B2FA685B9D29F4301E6F8FBA5AB84EE170CC158EAE1FE15FC2AC43CFD6F6DEBA
B7C14FD8944E115AE08A6C70C46B9F04B9C89C83B549A7C03B8E34B49A6DC2AAF929020D85EDA5DD
914FB0DD9C62ECA748F9B2CD03A6DB475B2CFCE67BB4455908DF23448A7259B253D22962B94FA124
A8547F8A421FEDFBDE17B26387DC43DDA97DD2DC2134983566D4F32EDE9F8286B4A86EA2C16CEA71
25B2644B8914A9DA273B4B56D73E89C8BE32FD2986D7ED68DB3BF7683B1C730BBA2E1104CBC03D0D
CEB51AF7531065C042F4E18635E3C2D91A7E19F2684F2252B48353D859B2A277B6D5477BB8279A25
1BD22922F5E7FC46D66E4EC1750A5AAA157E8A905431CE7DAAEAD136ADD9399C4278B4850B5B34CE
E62245CCA34DB54FF993AEA9E43E05CAFC862D2AE27E0A24149FDE0B833C61A95015E95DE0145CA4
88F4B9D37B30E26F5AB94F7A6928A45388E863CD208448C1C935AF7D0AF6BCABA053E4F5BC0B2D01
492AC1C676293F45305243E974AE53381C7385E5E565FEE55BFFEB2DF893E9F57AF4E5EEC0C47D60
BF71C7AF078DFB29C67C8105E6E023D95D692646742347A788A812A1CE4A2DE014439D6270D33127
60A680CE6F52A5FC14C64D4A718D8ED43EF1FE148509986A7E27AB9E78EE53799D22C73D94EFD1E6
F3AE984831E0201F7A24AA53949D7795D1296C3621BA54084E9DECA36D8E64AB4BC578FF808674C4
4F91E9D12E93255BBAE79DD635423A05EFA01DECDB688914B93A4559912255FB34CE92B5A8B1B1CD
C67689FE14919227D556DB3985C331578049D4EAEA2AFC99C0550E66EF88A5A525B891DDBCF10660
73F3F4BE7DFBDEB1E79D4D7DC2C6B364C7AC61D42C8CF308D19298266026A7E059B2D5D01E3F4571
8DCB56288A7DB4736A9F74EE93B1CD76F68F3FB6441EED0EE63EE9E996592812F16887FA53940A22
C8D729F2458A44EE53D9CA90E2C155729F149B10D3B0FC3EDA660C8ECE22A0D95717740A6D91D0DC
A177F755B3F95DB5DAA79C11FEC87B1E8FF829780481C81FE0FD56385F1E3EF5DD9DB5CF9E1203DB
D629CA9829AAD53ED9CA45199DC2ACD9B32906A715EB5EFBE470CC17E07244F94E74E97BEA89A7B1
27053D05BCA3C14FD870ED93620D7CBF5ECB45FD425F79441F6D935C74C54F31EA79672EE486740A
3AB2B44E61B56A358DDB9DD02902FD29F41E5E8E8E0DE2A7D29FC21CF065FB682377A0E04DEDA448
F82944EE53BE9362C2DC27734FCA4F31EEA36DE5C406A75E1DEB79672A1471B2CC2FE053AC7D9283
3C94254B154D8A2F9B2CA3907E96EC795716656A9FC420D72B4285BCBEC153993A85345388E827F5
94EB140EC7BCE1951FBE0294617979F9F2E59768275CAF0EEC3F487D2BF85335A391DAA771B57931
6F901EF91EF3CEE5FD294230FC1456C0A6716FCAA87D32758A6065885ADD6D37A7409D82F4B5D05C
CB283B1F1C3F5D8F76BE4E11EAA33DCECC619DEFC41CAC9C477BBA3A45269B180DF2727DB4290659
C7E328C6D191DC27EED12EACF658457D62D8C3C09E62CF3B7E4CDCA31D0A4936DBC4D353593DEF68
78672A14656A9F86B94F214B8528F6CBF6539875AAF1675DA770381605C035E0CFA7E6F9BC462359
B27A994B548C8410F75350ED93B83DE55789B4A0F609750A5AEF3242CE2DA31F5F0DABA05308EE30
FE727D9425DB8DDC275DFB1459CB1561FE93730A5AECD5033EB33F85E963A5473D078BF5D1164C21
D3493195DCA790533BAF8F361167732DD7CC83EA02A7885F9923FAC5B98CFE148850E52A8D6718DE
8F7D5046CB866A9F38952006415EEC02CB784E0EF87259B2396479829E7776E64099DC27A3724F2B
71AAC790F7BC7338E60D7039BA7CF925124FE710F5EB1458FBA46DAA91199758FE326B9F5CA740A0
9F626CF7131330CBDFC705F7527E8A60B1939925DB0D9DE29C2AE78BC86D822F9BB54FBA3F45844D
54D329F8B2AD4CC2617E6D91345BBAE75DCEA42BA553C4729F420D5932FC1432BE2C52C227ACDC9D
C97DC2F1C943F928D629A71A6A929E77A1B11DEFA3CD7B6473FA608A149C6E489DC2AC7DCAD42904
923DEF2E868BFAF8C0BE283945AC3F455C44D6A9E06CBF730A8763AE10B14B5023BC6651BF9F42EB
14DC8E9DB5F675E8CF6C9D6264A3A8D6FCAE1D9C02750ABCF504FB5328B3F678ED2B9A251BEB4F21
D804DFD999FE14434E1156DCB8899593E519F5D1CEF153C83EDAC5355B4131B86BBB5CED53F6A4AB
62ED93D62F8A4DE4E33A457F6CAF173CDAB1F001B1F0DB194E91D423221D2BA658FB2410C99235FC
14AC908F1364DA59A5E75D0E59CEE87927FA68739DC2A4C97C7F6E7F0A367483976EEF4FE170CC2B
D031417F2673C22338E6254B36238D90EF8C7BB4CD6955640FA105B54FE8A7D002BA11FD54AC7DA2
8D89FA5384B2643BA3535023151CCF4422E259B2638F768A5394AAE8A305DE527DB40B1DC1ACEEC3
34494B708A4A93AECCDAA758EE13E711A30661699D424DB762E103C583BBC02984FA90B3FEC329F3
E43DEF50CB78EC8352C2C8F153F031CC550911308B7F05597E8ACC8ABE0C9D42D73ED911B2E2629E
AD5384C673C25B31D8764EE170CC158EAE1F112D2A4CC07DAD29BA517F96ACEE79178ACA31665F79
1E6D73C94B7C694AED2DD029B8473BD89FC22A11A1FB54E5DC2791B1A96F58EDE614A28F764E7361
4E2BA6E2A708D18D7C8FB6260E85B956D17691A87D4A9A29B44291D2292AE63EE1088FE73E453B36
9A5F928AD1054E31F468472FD121A23161CFBB487F8A08A710554FC3C0014D90557F8A58EE13E914
D52873C9DA27630C0BE29CA153C8411BE0C5C6F0763F85C3316700A6009CE2951FBE123FECA1E30F
36F5091BA97DA2DCA773969342D007FD25BC768AFD29F80D6BD6BF88A6FC1446D97948A798A03F85
110935D8D97FD57DEF6B3DA730758A883C812A062DE44EE8A7881F90DF9F82CFBE38B3306BA212FD
29220BB6299DA24AEE93E9D41E0DF2884EF1FD3BF6708F366C632B6D6DD0364847677ADEE9EB70A4
C64F974255AB7D4A2E04E5F829228103A65A91AB5394152952B54F919081D8A53B5FA708F8E02242
86EB140EC75CE1FAEBD7E18AB4B6B60693B41BD75E836DDCD37FBCF1063EC22571D6CBE3113455FB
24EE3EDC4C116113B4021CF768BB9F82EE41E6CDA870AB123DEF2AF5D1162BB7DC49D19D9E77C24F
11D129CCDAA76A7E8ACCFAF3FC3EDADAB56D1E503A4B36D3493179EE93E9D4CEECA36DF2E2482241
C7FC14C81184B18288435C869BBCF60981B54F1CF97E0AA347BC305650AD54D24F11D729C24D5812
3DEF4C4D5957F4B1EB76DA4F61B18664FE80F7A77038E610CBCBCBD484826FE036ED699053349525
ABFBD99953AFD053218F769C3244E4757CAA657E8A42E0B929A68B9E77298F764CA708CBE8B4DD1D
9D821272749E3F9F89E93F81630772394572B40F5771071BF91E6D43A40807FB57CF7D4AADEE56C9
7D520521825C94C87D8AE66D76D34FC195652EAE454635C78C7ADE95EA4F61F78857195059B94FF9
457D293F8551FB6432E280471BB763B94FC7D263DBCE49769DC2E198333C79F6ACC926F84EF877D7
B13B936F95345C5473643495251BE70EBA6264B844F6A9619261C8A36DF6A7084DB75AAB53E81EAC
81F227EDF5ABA253146D14A1CAA8D6730AAE53849832153B09C122DFA39DAF4DE4EB14B28FF67345
2AA1BEA475DD587F8A781FEDF0A4AB62ED53B8E03C47A7E88F6D56FBA43B8599657E9DF2534484B6
D090E617F3A9E43E99BCA3547F0A1135606640C5740A9E253B259DC2A87D0A544005AFE179B94FB6
D666453F0D0F703F85C3316780CB117007F833D9DA7AFEF2E597E0FE0B7FA1F008B8F2E236ECE9F5
7A7001DCD8D8C8640453B772379E259B63F1CBD4292A2C7309B4C64F315EC8527A84C8C3913A45D9
DC272DA3EBD95737B264F3FB53E8CA9009FD1449C651C2A3CD5CD805CD822FEA8E1E137E8A087D48
2DF04E94FBA4675FC9DC27AB1A24AD5674C94FC1FB6887484450711E708A9D2B3BE26DE3B54FE648
7EEC83E3FD65FB5370D384D9A2A25C966C599122E0A7D0B54F461041E05A4DFB73FB68EB983EAB2C
CA750A87636E7174FD086E44E8001D5316498A919CC436E2A7C8F4680B313D9E258B1E6DBEE45596
50B449A7C01B0D2F1431CA9F94808E8F993A45A8FE5CCEC78E8D5FD26E4E813AC5984D04A65E2677
9E4AEE53645DB794475B4CC644F9132F444FD73E99D14F118562F2DC27ABEF705CA7D8BEE75DC1F2
8F803C21EA46BAA353E470E473561DD4E4B54F788CB8B0473885F053088522B4417DB4D33DEF4A91
E5323A85C926E405BC7CEE534298B0AC43AE53381CF306BD38B3ABB840A93F5B78ED2B3F7C052630
91F784037AF0EFD9F3183915A11E4D65C992E3CF54CF43A5509959B2E662175FF50AB56D6D01A7E0
59B21242A1284EC626EAA31DE89D34ECE2DAA53EDAA2D94A7C0E260A45E0C87C3F8539E30A2DF3E6
7BB435B310594FFCB14A7F8ABCB49C29E43E093345461F6D23F7C9542E46D66CA4CCDDD229541721
73D9476C84FB534CDAF3AE547F0A9EF524A50A56E057A2E75D499122EDA788D73E85C945BE4E1122
C8E686EB140EC79C434CEFE1CBA79E78BA6C45D3BE7DFB969696CCA76E5C7B0D9E5D5939045C06B0
B6B6B6F7D6BDB4D2A2D178EDD339152412E717787C48A7107E8A526A453B38C550A7E02BB4BA13AB
997C3E7AB6949F4266E0688FF620F709A75EEDE614D89F2222BD990A051D8F53AFA9E8149A5F647A
B4757EAC58E6154BBBA5739F924E8AC9739FCC3D95729FECB996D5BDC2758A04779E20F729DEB4B4
547F0A3186CDE8A7DC2CD9E9E91485DA27E6D18E476A88AB7A3CF7C93450C46307AE797F0A87635E
11A70C40014ABDDB03274EA2C55B3F055727D8BFBABACA7722AD08F5C868BCF649C810A66EAE4B76
B54E11EAA3DD354E31F453146B9F228BB722C63FCE29CC3EDA990E8B0EE9142C48D63456F0E10DDB
787CA69F22DE473BF42790EFD1A6055BD1264C881438432B97FB149F6E4D31F729B027A25398757D
417F907AAA0B9C22D2F34EEC3405E80AB54F999DE2B3FA535C2890652D4F88F4B374CFBB0A2245C0
4F11F16873494236D7CED629703DC7A85935EBFA8A0E38D7291C8EB9C25D070F2F2D2D61DC93D858
1AFCC33D7A0A1742EFD9F3CBCBCB219D6265E510EC17D5563027819DA176DEF5D73EA14E31F65394
59D125D83A4531F7A96C38613B38C558A708148768034561E1375CFB24E2FA63B7A4D18CABCF41BA
DA9F22327A354D2E95FB944F25929CA2CF2630CA69C42978873BD1264C06E6243DDAF9332E3860B4
909BA87D8A2A1166125452A7E8FB29D6D9B2ED7AB4B9B05EDDED8C4EC1FB5344967DF4059C740AB1
BC060C9AFA4D64E63EC1F1B2C64F738ACF1DE53A85D95D25D44ABBBA4E91439933FB53A4740A18C9
622D289EFB646B70A1CE14830D6416A8534C3D19268276AF3B391C13626D6D6D5FF8DF81FD073151
962E5F1AFCCF19B6E178986320191147F67A3D933B602F6FF82EC047F4FBC305535B33668A63AB27
28E586EE38395482EA43427DB4E9BE831BA23B9258E06DB39F4208100C91CC73EEA7D8DA7A5EBD2D
AB7D0A6813F61CECDEAE78B42BF4D1CEAC7DFAE44FFF7A65E90D1F439C02E75DF008E4424CC30AE9
B21764393A4A18D3D4290883695BC24F114F91559CBAFFD4C5F7C3CB4F7EEE0EFD5BDBBA733C7431
54369338D30147FEE1CF373EFC668A95771FC1BABEFC651F216400A7F8F31FBC24DE96D73EE52F04
65D63EE9016CDA82F878E65DB649A7B8FEFA75DC801B657F70863A3956D5295EDD7959D43EC5E489
D17ED17B28D74F610D669326235CA77038E60AC02956560E016580C92A00368EAE1FC147F8727575
150E809B9A9E210B20B380E3B7BFB7BD7365C7D429E0AD6EFFC0325C4EF5CB6127BC44D444211AD1
29B0D223E7F6246C80386183BB5B48A7A07B0DDD9EF20BA26AE014352CF8F09E77A1009C589BB070
CF3B9943284A4434AD602656780A666BEDD629C659B2E13EDAA2E9F0B889FCE02F22EED136A75BC9
262CF88710E714639D42B4B7533E0BBEA89BF66897996B099DC25C34A0DC27C9944329B2C5611FCF
7D82F1D95F9E3DB61429F033A2FE314BB61B3AC578B85A4A9C2E79E21968218F3666C3E6B0093DD4
4359B240079020EBCC58DE6045881485A6DBC9FE149A4A54D52912FD29846011E014219D22D6A534
ECDD163AC52C4654E8568877767AB64E89C4E1987FF0C613E65F07DC3A9332015ED39E7AE2695CE6
7D75E7652C9DE2C7A08401C4012636FA1BC14EB46008D5031F9BD12946DDEBE2AB5B7A27DEB9F494
036E01993DEF1AE414B5E63E718F76C4A05D2D4B96F471714B120B5F58CD3BAACE3DFDE897EA1C69
35C3EC4FD1BBFBAAD947BB4031B84EA1FC14A2F629EEA77834100095AC7DA2FE14BABCDC0EF6BF10
CD9215224552A1A08D54ED5361EA25B88399223B9A7701A108FA29EEDE331EBAD8F0CEAA391F8F7C
D198B8339C425F96E3254F7C9CC7FD14D8F03D14A941431D6B9FC4B326FD5CF9EE5F882ABEB1D6A6
3B5314DB3B724E01F7471AD8439D2244877390D9475B0D6F7E3D177F02B811D429226D56428D2A18
FEF09FBEAB1E9D826626F087EF3CC2E1082147C883A93EDC97437F47B81F2E3B64E5469D42D43E21
D10815386D6D3D8FCFEAB962537E8A1C5542DFB9BE76F8854C9D228936D73EE9F010BE5A6B4EC9D8
C257969F22DC3B49508C6EF6A7E0F32E142022332E3C1EF972C44F51D61F443AC5A3657ADE0925C2
AE3F1FCDCA12B54FA5740A3C7E64AC4867C96ABB840E2218E917C3AE16A1DCA7BB87CBB9A17C27AD
D089FDDDA97D22BE9C340D4DD8473B7431D78B4526A780515AB00845FB68F322A8984E71F6EB7D4E
81B4A26A04410EA7A032D4A0604169E1A30B7B8853904E615BDEC23E0BD229F2CD9E15A0A73D5EFB
E470C411270BBB032D636D6D2DFE2607F61F842B0F6E03A7D03A052A11F0CF5C55B87CF9257C56D3
07B86062982D2DC8E0F6ECD60A22B94FA1BB528E4EC13DDABA3552059D621667A0563F05EF791792
2ABE5D0CE784ED70ED13EA1439B7A48247FBBEF7797F8AE48C0B1E71CE266A9F6010C21FFED1777F
B6EC60161333C129706C8B9E777C7EC5176F6509FA8531E3B8ED779F4CF4A7307BDE998F79B94FFD
51AA28732C098AED31750AF8AD0D7BDEE97997D9264C33E8A24E31D38B67CDA01F847BB4A94E2F54
F53416E9D8FE80477BECA7A876D1865799F70220052BDFFD0B5DB31762197480EE7927FD144541AD
B24E41A722D6F38E5DA8439551892CD951519FE9030A7A2B46DBA853CC683CE3DBC275834A26E071
73F374E3C3DEE15838F03FD2D5D5D577EC7967E418F82BE3974DE41442A7005E804A04510F8E5777
5E466943DFA9E12F1A180D7C0BC423BF730A2EC8B0B1D5DB86A76EDE7803AEA8537B1CFC5BFBD889
D0826D642D972FF9024E3FFA25FA78F8CE700B78949952AB4DC0E054E0DBC2037E6CBE3DF9E32B3F
7CA5F0E6D33EC3BBA453846E46453F85F1D440A7E83D7B9EBF337CCC93ABCB91C5AEA0DD8FADE502
BF8071F5FA9B7F3BCD11D5DCA318D8F0087FA47D9D62243A845671B11A8AB6517AA33D6B1F39067F
F838AAF19D2F5F7EE9933FFDEB99DED5D061BFF4D31FC7BF3EFA85C277214ED19F7DA99E14DCB84A
3A058917C3DAA77FF56FB7B69E87DF29BC215C79FADFE0C73FFAA583EBA5FD14AAF6092E447486F1
FD1F3AFEA088BB311A3B46BAE05DEC738ADF387A377CC2F1DFDD20A162EBCE61933B6AA7122935D7
9C1A3DDAFD3519362AA67FF16C6890C3C9871F8DB26449591317EAA45FFB37F77D75EBBFBD303E45
782F501EEDF8A55B1F035FC2DF1DBF5CC30786A132EE3411651066E73BD429D63E7B0ADE16DE0D2E
DAF881CF7EF52B76CFBB927E8AFE081C7D547804569EAC7D32D4B76FDFC66935708AC27D6AF007FE
ABBFF0B6A14E1168016F7B85982AF7F447DF0E7FE0F451A735AE6E5C7B0D3E21CC34FECD6FFF6B9C
6FE023804F515AC3CD1D8E690117FF69E59F4B00B8017F41C80542EFB0FDBDED9595437C8F59FB04
EF7360FF411D248B80DB3D7E176180DA6DD04F11E84F11B2FE71C13D94FBC42B7269438BE9A1A957
9B3CDAA1DC2763D225748ACCDA27331B474FB7D647B54F1DF05360ED13E9146248C748F40C7ADE89
3D66EDD37B3F79DFB82F98AA3617654EA45050867FDFA3FDE5AF53DD66CC4F51CA5211AD7D32C39D
A49342B089D1B3493FC5D804A4966DC5944CCCC1BAE3A74882D7AF8EC5B88CDAA70AA57D982B6B7A
EBD0A32DCC14427413755005A7B699258BB54F93E81479B54F4191A26811E21B713F85CD88C36D1C
69E7F95FD9034C7076E34ADF10DBAD653B1C136275751527F3D48A82007B901AC023956E6AF4BB60
170B66CCDC2798CFE0DB6A9D02CB2738A7E0E8FB299EF9469DE744F6A70817E5F6EEBE6AB6D88697
98538E524B5E8D708A5AFD14A2F6C9B4AF5A11E8391E6D913D988816E9467F0AD3A39DA9C425FB53
F0C19CAC39D748FA29C6DDBED4FA2D2717BC07712CF7A9829FA288DC3EDAF1B55C310D0BF7BC1B0E
DA91F93A2EC0698AB1F1915B1B1F7E33C5B8E79DE2C89989B2118F3689CBBAF6498C70CD3B22B54F
A6233B68D6664C3937F7A9AA4EA1FFC04D3F8529228BAA3FDA13C97DD2616589846476BCF7D17638
E60A47D78FF08614B77F6019E7F6B0413B57560E8596AFE1E55BBD6DB1F3F2E597B89F025FDB7BF6
3CBE1B3C4B47D2DB220DC15428F16EF5F7D13EB67A426810F1E9967836D447DB3DDA88B14EA16F46
56F9131DC0FB5324FB688714733360842663ED5E838AEB14C891CDD10E07C01FC2D70EBF60720AD1
473BC288CD218DE43A99251B2A0E1115507CF695E879A71772F3265D713F45306053B389A29031D4
294C8FF69D81C6D901D63C6E4C3C5A01769DC21CDE7C844FA85344AEEDC13EDACF492F765CA11892
E591477BAC53987DB467A053187E0AEE9B18953C89EB764E7F0A2D4CD8E94F6A75C8FB53381C7385
B5C13FF83341C09C0A26C3B0018FB8FDEACECB2142016C0258004C54E825F8269B9BA79120D07E38
F8CF7FF012B295F37F6CE43EC114119FD5D7079809D4CC29B4473BB4726B56ED466A9FCA4AE72DE6
144410E472AE366BD31EBC49C53DDA5644BF492578ED13DDD7DACD29F2758A73AACC8FA65E91FE14
F9C358CFD3223A0541B700E34BBBBA9264CAFD298AE552219DA2D06B38C08E43B9B2499D22489023
6D5906FBBBC029844E51601383EB7924DC2C9925FB681971598CF0904E213408A3964F59B64BE814
7A784FA58FB6AA7D12F94E46D9EAC84FA17F6B58FB64B0094596E5381F1DE03A85C331579824C400
0360910B509514FF877BC858813B4D73045C0CF1483DA70546333F59B2E674CB382050FBF4F807CE
F1FE14DDE414BFBDFAF62C9DA2E8A1E0FB239CC25ED72AD6E51A06C06EE43EC19FDED70EBF403102
216AAC8737F5BCABE6A78858B309499D6258466EADDC7236C1CBD1D359B2DA2E91745564E814F672
6E600F9F9845748AA106376860175CB60DACF15EEB547F8AE2D08D27F865F829D6F8284D5EB4457F
8AE1D53E43A7109165324E7654CE373ED2F453709DA2924851B63F45A8D8895FB7133A054BEA338B
5423DE0AF45338A77038E607AFFCF095E431219D022E65F072B8E00800D7408270E3DA6BB8078F5F
5B5B83FD9CC5F0E4A803FB0F0AAF377D97FA750A537D30A50ABA738997E82BCF835FF8BC58E3EAA6
9FE2C4DD3F6BFB29744461F156454796AB7D4A792B78B948BB3985E8A39DAC15E1E319E766656B9F
922139393A85D1D5EE3959E354A8806253B5DCDAA71C146B9F927E0A411CCC8210E1724DEB146155
C274B6D2975DE014213F45E6BA50BCF62954B997BC6847740AEDC836B509EEA4E01EEDB14EC16B9F
221EED89FB53D0F5B9E0A7285EBA439EB8A04EB16E5309B34855F8B8DD4FE170CC27346B803DFD58
516B86998C067A75E765747C8BE37BBD1E708AE5E565FD12601370BCA94734EBA7306F40DA64A18F
2C5BFBA46DADEDF653144A44D40D886ACE0D9D22DFA31D2A3ED7B7ADCEE8149A53F0E45873C00FEB
464AD63E25D944A64E6178B42FC8F55BDE0B8FB71E2E57FB94AD50C4750A51FB148B1D50A523493F
05E91414FD14736AB32FBBE0D136750ACD8ECD8B798E4E11A97D8A5CB4633A051BC9482EF4D8E636
0AFA2B88E914118F76559D42D73E055566ABF629A953C4A9842D5B8CC6BFEB140EC7BC01EE59665A
2C500398EA732183D84184562019D159B208D22FF83B5C7FFD3AEC3CB0DF9854EC36A753887B50C4
5E612E8879ED5308058FB6F65304584666ED13D729629921EB2C96F3D812668F7424F76958FB142A
02B1248C096B9F72B846159D827928965997612E55A46B9FF2158A6C9D2254E0649B86D853219DE2
FB77EC11799B09F58D9207462FE9884E61466A44FAB0F0AB7AD2A3FD6846ED931EE4597E0AD6A531
E4A7182B14913EDA718FF6F4748AC8A896F6B7B23A459EEE461BAE53381C7305F87BC4793EA6BC0A
B200172BD8AFE76F716030ACC89245004F81A768810501F364D869F6ADD86D42A730FA68B3395832
84D33DDA71F49B1F8DEE4AE3455D759F924423A33F05E714C62A6EB82B6BA7729F6894C6C7B308C9
89E81499B94F71C4758AD5EFECDCFE47DFD72EEC4262BFCE984DD63E659A29CAE81466687FA8C5B0
D42C223A45D8A36D16870802D2114E111397D97846369157FBB446D7DE0AFD2932750AE1090A253E
11A1909CC2AC7DAA2452C43985BC5CAB0A55B3B550DA4F11087D8A174179ED93C33187C0F9FC5D07
0F3FF5C4D3B847D00A98CFC301D80139097C2D5C829606FFCC632E5F7EE9C0FE83402BB0FBE743C7
1F0452132214BBD89FA25E8F36D63EF1552F7BF196F54E1ACFD346B3B57896ECFCF6A7F8F18F667D
7AFB3A457CE536947C5EA6F6C9987DC533CF07DBEDE61499B94F7CC6455462C29E77C9A19EAC7DD2
4D87359B20DE41B3AFB49F2267DE5552A7885583A8CE149C77A4FB53043CDA8508022B0FAA0B9C22
B4CE6332084D96933A45F2A28DBC837BB4E33A8528DB13F60A996F5614ECE0A92C9D225FA1A894FB
A45BDD55D029C6D1C77161C21ADB5EFBE470CC1560E61FB769DFB8F61A700AB8064EF7FBC29C10E6
6F009A1C469CE0CDD63E85C488C801700B0B79B44DF53CB347583D7DB467DD4A5BD43ED90BB9FC9E
25FA5354D32954ED93B126D6764E813A051729C4600E0DEC42EDD3478E89B7D59CA26CDE669C5390
E8205ADA89EC4D5D2555AEF6292E5294F453186D8545933B51EC97D2298239C92AE28CAFE2D2C15D
E0143A4BD648300B18B773B2641F55B54FE6751B8E011E817B86DB519D42A622173B56687B85E1D1
8EF829CA229AFB243CDA862417D6E0E2B94F526E8BAE02F1239D53381C7385BB0E1EA66D7332899C
42BB2D3231F904751E6A9F226C42DCA4323DDAF3EBA718C9E893FFFAF0B5DA8333ECA33DB8D18C73
9F84986EDDA472729FE2F284BE3DF539C8B1F197EDE614633F455EEE9310E02AE43EE56B16991E6D
BE4E3BB4B55AF66D9A7D4DDA9F22DCF32EAD5384D2FB0583CEF0538C750AEEBF5E8F896EE3913F90
36BAC0294A2D04C9EB76CAA39D59FB641E13F75388CEEFA2ADB674550C94B8DC9E7765458AC16342
A7103AB21ACCC3546495251BD3298A6B3EF62A90B0C2AD0FFF04BCF6C9E1982B24C9024CDED07031
EBE5EB10FAB54FCF7CA3CEEF48B54FBAA829B2D8553820DC9F227FDED548EE530D3A05F6A7306F4F
B2EAC952D533B364CD0AA8E0ECAB1B3DEFB03F058A0E717F909E718DB36433748AB29A45964E61B5
18167B0A252217863A050D15C9294A89140A493F85CE1C30BDDB99B94F38B643DED5607232F5A7E8
40EE138E4F7EDD0EB1099E69467B26AF7D0AB18C884E61A6228FDD13CF1952456ECFBB59EA14C18B
73D83494D39F42D8DFB42AA747BBEB140EC75C01E6A8D43FC29C4CAEAEAE62AFBAA97C3BF12D72A6
AF8DE814FC8E1382E9B628EBD12E7B873AB97E6AA63F3BF7684F482E343DC12FB13F4562C6C577F2
491A6CC7FB68874BCDE33551F0082F3FFDE897EA1C693543E814A6E226C63C77B64E9EFB1419EDF9
1E6D9196C39773F9BC2BA15360EDD3043A45A4F6498FE4189B605FE6FB29E4FC6ADDE8C3C2477817
748A481FED90CA9CAF5320CAE63EC56B9F7894934E36D3FDE2E9D9ACDCA738592E93FB64F829740C
B2AE65CDD629EC946F933E2BF6E13A85C33157B87CF925E00BA2E2857074FD08128AFC76DB9913D1
FCF96A237E0A51FB24EAA0CCCAA8219BF8D4A598473B9525DBB89F023DDAF0DBB971EDB55B7EEA96
77EC79273E9605BCEA67DE7B3B3CEA6F31AC7D1AD5E59A6B5FE3E273AB40B79C9F22BCB45B988975
C64F712E2FC74CE86E65739F32C9321D16E71411E2202271443149098F7652A4284A1B591E6D8B29
47F207D23A8599F514B2B2B22E155DE0149C20E8951F9329733DAE42ED53E6200F7AB42FC42AF7A4
48F19CF453D8B54F66CFBB29E914B1DC273DBCB3750A33BECC9427E4D8769DC2E1983F005F401776
BFC9DD8D37AEBF7E1D36E00F076D14CD163EED36D7F34ED43E85AA9E42A6BF9C3EDAE2F6848B5AF3
C02900FFFDAFFEC72D9FFE37C37B13AEE8D2D26EFEE33F39F53FDF2AB82A7685475BB389A271951F
46C6C02C9D422DE7DAF3315EA3DBF6DAA7427F8A680C8EB9903B95FE147C30F3D15EB63F855DEFA4
52A1629CA2AC4291A753181ED5D0748BD3E7A44E11E8796717F5710F11FA29BA51FBC42FD7660BA1
6103476BF003A7D0F18333AD7DE2A4D8EEBDA27B55D0C139B54FE5458A78EE534C5C360393993FAE
541F6D4D25429A85730A87630E019738A20F6203FEC583A1668DFAB36465CFBBD1226D8E925EAA8F
F61C66C90205C0F93FFCD281110C6F37482BCA02EE6E1FF94CC8A34D0421D8F34EADE5E678B493B9
4F415BEBBDFDA5E076738AB19FC2EAA31DCF1FA8ECA7D0E3DC5CF2CDCF9215019BF13CFF12B54F49
1B4585DCA7309BB093FC033A85EC79A7275A662B16C6AFBBA35364860F884B7AB5DC273DC875966C
52A7D072DB38F149F92950B38BF5D1367BDEE5209C255BD0292E06553639E08BD7ED6AFD294217F0
E101EED17638E615FD0286B5B5E5E565A2122B2B87E6E1CFA7D9DA27533AD7B32F7DD8A2F7D1BEFE
FAF55B0E6F8E6F4FA853E0AD2767035FF591CFE86F21758A6FCB72919BC5D0FEC9FB53D8132D5228
8E2D51F648BB3905D729220BB621BE5C21F7299F356766C9EA5996D974981675D3B54FF90A4585FE
14D6E2ADCD9DF3FC1466C953A87484EFEF02A7D07DB40BA498D53BE900A849729FF420177DB4831E
6D213D843C14172C312EBFE75DA64231428E4E2102936DB5A2647F8A205F0EB82DBCF6C9E1987F44
6A9C9A2A7F6AAAF62939D7E2B922D5748AD09D8896BCF43133F7688F6A9FFA3A05728AB2CD583998
4E41408F36DD9B12B94F42A74879B44D9D42DC86B8AD55DCC8DACD29787F8AC890E613335E34F2B5
C32F64D63E9515E0923A85EDD17ECE58C82DE1A7E0633B2FEB2947A730729FC2A5E654F834FC32D5
479B7BB44DD66C7AB43BA25398E9DF917A27F1875021F7498F7371D14E7AB4A94505EF4661DA2878
4272899E77F90A45A64E41639B99B583FA72661F6D334F2330C88566E13A85C3318748528906FD14
F567C9EA9E77A122F310E9307BDEE93EDA65675F75EA14634E5161D56B34377BDBC7EED7234764C9
0617B802FD586F463DDA76786C3836E41ACF92ED8E4E51A63FC5B951F95FE5DAA71CBA91F6685395
48D1ACAA1B0DE37EDC4EF7BCD33C224FA7A898FB24D844F180884E217ADE6569708C41778153E45C
9C23CF4E31F749B08C10A7A0062B63C52DA4BB51FE00EB4F9195255B52A428DB473B34C24BF5D18E
943F194B437C8D68DD750A87631E01B72D9CA4F1EA7798BE5EBEFC52E39FADA92C597E03E22B603C
5D93C7C98A3B94D629227DB4E788534C51A708D43EF12CD974EE135BFBCAF16807FD14E1B6773419
6B7D96EC50A708F4A708795A89564CABE79D492EF26B9FA85C44B71BE655E8786489DAA7E4389F30
F7299A999CCC7D0A1587D87621BEE4DB999E77DCA36D8AC826B9206DEE37F77D75F29E7726427E0A
320A91AC660A13A21BCB9059E7F4BCABA05358B54F91FE1439A57D499D42CB6A8220EBA730030DE0
3A85C3316F38B0FF209AB2B517FB91DF3935EB496C12F5FB2944ED53246C33A4539C0BF829F8846A
CEFD14059DA2424BD62F066B9F7896AC59552EBF54C52465FD146931BD4B3A85D99C2289B147BB52
EE53489EE8AFE20E9ECAF168D3742BB48ACB439FD23DEF747F8A9C815D2DF7293ADDA267433AC5F6
3DEF2A94ED15B970610997EFA762BF6EE43EE5EBC8822F9F8BF92966DBF38E72066C3F0517DD58F0
EC34739FCAEA14D6455B67F471A69CA353844C7021DF106D3BA77038E60A3C33D63C0026B16B6B6B
0D7EC246740AD18F552FE42611EA4F9139D79A239D620298B54FDCA33DD629CC5B92B9310D9D624C
37461EEDEEF82944ED93C880D27A1C2DFCE6FB292A8CF0CA3A850EC6E16524E9DAA74C91422DFC4E
94FBA43BB024750AC188ADE273FE9478B6233A85699DD0D76DB3C1D024B54F34C831F7498CF960EE
53B1EA2914F4A4DB5864E53E95552832FC14E2124D0B3E76B15FD9DCA7481F6D25C3D186D73E391C
7385E5E565A00CBD5E2F7400FCC102E980FB5A53968A39C992D55121B24A6430072BDB47DBBC3135
E9D1E63AC5EA6385E95659A92250FB64FB29CCFAF349FA53E488E96AD5ABDD9C02750A7348275677
99AE91D3F32ED9BA11C7363F2CDFA3CD2B4642A14F84747F8A509CECD4739F4C2705375394E9A3CD
75375BA4E0CD1CBB54FBA48B9A344DD68755CB7D4AF2E5B847DB68A2AD1437B3214B56EE13B70B65
8A1481DA27EDA7A04BB1BE801B65AB39B94FA158B322C5D0957ECE291C8EB942489E20C08D1838C5
5D070F37F5091BC992D5C421DE7498400DC59259B2A504F486758A6A2D59BF18CB7D92AC41549BF3
FB5199DA27A153C4C47435136B3DA718F7D1B6748A6409FA747BDEE5EB1466B193AE3CE7360A7270
27FC1413649A55C97D52D32D296724FB680712726C0D8E3DDB054EA175B7480594562EA652FB642E
04C5FC14C5E6EFA13EDA0527455CA7101EEDA9EA1485DC27451C42B54FF9B94FB16AD5E231EEA770
38E6132B2B87E207C0F5043805A0A94F380F59B211A94227A287B264E37DB4333945AD3A05AF7DAA
E4A788D43E15A4076B0266DCA74637A9727E8A8898CE6B9F3AA3530C696F86AB625A3DEF721675F3
6B9F645B8A0B051231E417A3E9596EED53CEA8C6E327CC7D0AEDC9E9A3CD740AC99D434D87477BBA
C029861E6D2B33565F9FC5C04EEA14A43B24C7B9EE7997F45344829E844891F653708F76799122B7
8F76884758B54F15FC14896A5554960703DB3985C3315748EA143037BEFD03CBC0299AAC7D6A344B
96C4717396155AF88A7BB4F3E75A024D66C9964524F72950DD14AC36E7BCA3AC9F2254AFCBEE5F74
58BB3985E8A39D29BD11AD9E6EEE53299D623C952A9AB5B94241B4821F56A5E75D7C5177842A3DEF
02FC22D99FC2E8A31D57E2F898C7DCA7CE78B445EE53BE5451B9E71DBF8CEBA23ED893E5A7784EEA
6EBCD88913EA21770EE53E65EA14E1D2BE6A7DB4A5BE5C46A7885F9CCDB228FAD26B9F1C8EB9025C
EE227F2324523468D36EDEA35D2AC93FDAF34EF4D12E9BFE04C73F74FC419EF78B8FFFF3ADB7E0B1
02609605F7A337777F42EF096F851BB3CB7D423F859C509956D662EE53393F45A41FAB2E41FF7487
729FB84EA1A75862AE25F6C34BB49F22D2F3CEE4CBFC599AA7A5758AE7944E612A14C58A912A3DEF
E243BA5AEE53984D70729DDF473B34860B8C83192E3AA253F0E863AD50C42FDD956B9F928B42397E
0ADE6F85473F9914234BA7880FEC30B2740AF3A2CD567E2AF4A708B186C81A91EB140EC71C62DFBE
7D7089A3EB12CD549F3C7B16D8048A145B5BCF37F5F19ACA922DDCA132129F687D2CD2F32E8735C4
9F3DBA7E44BC2DDC47DEF6B1FBDFFBC9FBFEEEC1BB731EE196015335D8804700EE7CCFED2BF87B37
748A096ACECDDA27D429F04663168A04A372704FD93EDAA1A6C33A4BB6033A05F753689D42185AC7
9E8BD1C2EF847E8AC8626FD93EDABC0F859883E1766EED53FEF0AE96FB24D40AB34B45869FC2F068
A7D84407739F70B80AB3B62E4FA59DFCC25EA1F62947628EE81442A4E0BC9886B7A8892AE1A788CB
7095750A451F0CC54DA86F65740A419613AB40CE291C8EF9035C31B03F053CC294F5FEA3C7979797
2960161947837DB49BEA791757CCE386D6A04EC1DA67EBDB5352B3103A05E0CDDD9FC09463E5DF9F
C71AF2E16304CF151FBF3BBC9D7DE2BE0D7C4323F729791B2A99FBC4B36483054E5665484E966C96
295B65C9762AF709750A33753338EC070424E4A7A8A1F649947F149A0EAB56779C6294EE7997942A
26CC7D0AF48B4FEA14A28FB6DC582F6A13DCBBDD199D820FD7B888AC11F6684FA1E75D4E7F0AD90E
3EE0DD4EE73E719DA2BC4891D9475B046BC83D5AA7F8F66D397E8A64EC80DEF0DA278763DE806530
771D3C4CCC82FFD373E39AD14896AC0830D761E6A175B048CFBB9047DB6413A15B98F668C344F143
CFF4F8ED06E760FA913F8B0AFBED7FF47D2CCD7DEF27EFA3B38D1B76ED535984729F78C16D789665
67C996CC7DCA522BBAD1F38EEB14BC3E245E0485074CD2F32EB29C4B4FE5D73E89755D2D52F069D8
34739F4AF6BC8B392994452857A7A091BC6EB30929CF75C94F91D947DB942D1AE979C7CB99785378
83383F57601FD24F514AA7482D04E5F6A728B6B48B0C72DC88E73E05AD13E6757B7D3CCE5DA77038
E616A808C0956A6D6D6D73F3F4566F9BF637F8A91AAC7D0AAD6E89BB9279E78AF7A788DC9E228205
EA14E23702F7918FFEC177FB04E13B72C196833F8B3726C0EA7776703E76DBA73744EDD3F5D7AF1B
B54FD3C87D1AFA2974ED136310B6713BC3A35D2A42A4B0F358FB6B9FE2FD29727292CBE63E25D944
8E9F82F7D1D63C42178DF0A7CAF5BCCBEB4C11D72964399FA94DA8148261157A76966CBCD2490FF5
EEE81411912272B93E3759CFBBD0B09FA83F854A94A5E13D919F22859C3EDA66A8ACBD34343A26D6
9F223068632AF38832BB4EE170CC3342F4A19BB54FDA4F91BC31C5FD14E4D18EAF6B456E55DC4F81
BF14B88FF4758A516993D4D09F935522FD032E5CC255B2E104ECBB3BC029F03D6BCA7DD23DEF94A1
55DEA7F26A9F4C9D2244340A59B29DD129C60D29CAE43EC1DF42A88F76A99E77A1A7CAF5D12ED684
18FDB5473331E0143454D23DEFA6AA5364D53E8935DEEC9E77C93A73714017380577C0998B3F91B5
A06A3DEF3211C97DD2BCD8887B5223BC44CFBBD9E814A1CAD5825A5146A70832E2480F53EF79E770
2C2C661D611A412359B2A2F6293FC6BF944E91BFA84BD0B54F5BCF9C034E21DA0713F87E9D5848B3
AF0FAD7D86CE366E181EED0AD14F56ED93F053D8B727736917EF53518F76843E445846E7FC14011E
21B833F95851D768B0E71DF7688B155D592B529C8C95A87D2A19E65F31F7E96261B4F3DAA7A19F22
AE5304C236CD9D7CA3239C2279598EA872156A9F74BE19F00ED19F225EFB445285E9C5D6520555B1
C286AD53F02CD9F22245A69F829362D9B75151E6A44EA1139223E4A270555F778FB6C331D7D03340
94093A95252B6A9F22F204CFC611AB5EB64E11600DB80896E3D1D6E40E750A7E33E2998454285258
FE529C62AC534CB58F7630F749F4BC2BDE9B64730A91255BC64F519871452AA0463AC5C35FFE4A53
E3BC06607F8A73C5DA2731B9D26BB9E78447BBD13EDA62BAC58337B97241E552257ADE25158AA24E
11A97DD2C57BB6B742B18F44CF3B18ABA453AC172A40820C7AB4DD054E915CF30929CBE7323CDA88
52D1DF699DE24261B5C796DE14832ED1F3AE944231497F0A3DC8CBF829CC651F73BBB03178CA750A
87635100B3C157775E46A7F6C6C646833DEF1ACF7D0A755FD50724FB688BFE142671A005AE1C4E51
D0292E1456B470AE255671459008EC5CFDCE4E4CA7A82C52046A9FD04F91E8A0AD17C1CAFA29ACFE
14C6A26E976A9F0AFD29CA745DA1A997D99F2227F729C2A6939C42D43EC56DDABC4424E6A710B94F
653A53E4EA14F1DA2751E017D52964CF3B2BCF5FB389E17EF468778053708F76E8BA1D2217D3AA7D
D23DEF1E0DE73E15ACD9172477084915C357E5F4BC2BA5508C90DF9F2264C7167D854AE81496FA66
0E6F5A3B729DC2E1984308BE005FF69E3DCFD39F1AAC7D6AB6E79DB96C9B44E5FE1449E8DA27D229
F4029708E1E40ED6B178314B9D2298FB44B54F667848EA86554EA7108B5DD68218CDD3DACD29447F
0A4D8723EBBA95739FE4FCAA582E52D6A33D1CE7CA8BAD99321E90DBF3AEBC4E319DDC2792E4F2FC
14C1E481E208C7C001CE3EBA90FB94498D0BAADCDD57697BF2DC277C8A6B19E8D18EEB149C02DB71
4FC50E8FD49F22CBA35D7645A84C7F0AB3D98A18D2393A85C98583F204BF92BB4EE170CC37E0EA04
D32A91280B5FD29248231FA9E62C59AC7D121CC15CE032D6BB527DB4F9EDA92CA188E814A225AB8C
EE37AB9E2E8CAB4A6C8F36F6A728AB50A4729FD04F41E54CDA5811522E8677B40A7E8A7014A1D035
DACD29227DB433997266EE53D9BCCD2C9D826D101D366D14FC0FA144CFBBA4936216B94FC584FFFC
9E77E65C2B681AEA8C4EC1E563BA3873E2C0D984E8CF3249ED93F053084523EDA7784EAE0509C559
5CC64BF4BC2B9B011EF553245A3772858E0FF564EED3312355C0AE7A62529DEB140EC73C03AE1BAB
ABAB824D2C2F2F3F70E2245C009BF568CF61ED5364BD0B5156A7D0D3B0CCDA27E1A71095213CBA5F
300BAA8C823D75E63ED93DEF22B54FAA8F5295FE1466E8D3C0E5D79DDA27B38F36F70469550E0736
05FE57AE7D9A8453088FF6505CB33CDA9C7A6479B44B2914B3C87D12B422BBE75DA8D2C93EA01B59
B2BAF6495FC64971D3C754A87D4A56F43D1AF5680BF784B84473B5824B72CBF1DCA759EB14A69F42
B5BDAB92FBA4D944F8EA4D3B5DA77038E60A6FFDAFB7E09ECB4904D08ADB3FB00CFCE2F2E597E8B0
AED53E090D42D007E1CB36BF0CE914493F45FC56A57F11BD67CF03A728CCB802FD8565D7A4517114
4CD812B54F656F4CD1DC27EEA7182FEA9A15B926E3A8D49FC23400D27A57A7729F882398F102A1A9
17BEA4498FB695254B354EF61A6FD24F11CF7D8A620AB94FC5995842A7089958ADD9972E8BEA02A7
302FD7A145213DECA7D2F3CEF45C846A9F78D5D3587D33158AA2752856FBC4758A4A8D4A73FA53C8
5E8DBA4EB58C4E91700685AFE457BDE79DC33137800BD19367CF726102B69153C024411C5CF3AC5E
7CCE9AB36475ED9359E364DEB6681A96DF473BF3C68488F829F8BC6BACA75F907330919923FD1453
EDA39DCC7D1A57E1B29995E011851B592A4B36A6530462CF3B95FB44FD29923ED6496A9FF251B68F
368E583E74E34EEDEAB94FA9CE7753C87DD2B3AF00A7D8BEE75DDA0A247DAC91AE61DDF053089DA2
5496EC14739FF2B364C77C5999860C85E282FC43C8F5539411294C4E61F4D10ED73E0DBB3796D129
F470CD52DF062AB3EB140E47E3B8FEFA75B81CA11E4184828489A5A5A557775E6EFC43129AF56827
A984B83DC5758A09D772D3B94F451736A70FE3D997550A15D329F233FCF3729F78ED93E8A3AD2509
73EA55D14F61969D8F6A9FBAD3473BD29F2232DAABF5BC4B6E94F2684B76ACE65ADA5294F068C707
B0C0CC729F0A7E8A3C8FB6987185FCDA1DD42974401F1FC9A1D5A1667ADEB1451EDC1011DFC24941
A33AABE75D59846B9F82FD29223459A918F93A856D11B23A6BBB9FC2E1681657FFF2AF5089E0A609
B8BED1DD1600CFEE5CD969FCA3121AF75350CFAFB8A4CEB7937DB4B5473B47C2B0FD14DFFC1679B4
79A69336560872C1ED8133F1537C31D6F32ED8473B9C735EC14F614FBA02A122ADAF7DC2FE142852
E8F4E3788CFF58A7A8BDF64964C9EA2059232A87F5D1CEAD7D2AA9534C21F749F18B7C8F76529818
6F0CBA5A7444A72815D3272EE953AC7DA26330F729E2D116FD83ECB6146CFF58CEC8AF7DCA8ED440
E4E81405312EB4165426F729A8BE450266BD3F85C3D128CE7EF52B48220EEC3F88D60973AEBEB4B4
34579C621EFC1499DA44059D42DFA1E25985719DC22CC735B30A79EE53C14F31BA3D5D7FFDBA51FB
348DDC27F453E862A7605FB03259B289752DCBFDD7353F45BE4E718E850F8CB364A75AFB94C3296E
577DB44DD62C4638D68D94A87D4A3A29A69BFB641547A5FB6887CA42D44AAFA821E9824E91ECA34D
8CC31CFF156A9FB4FA564AA730FB0709B22CD58A914EC16B9F5E7FF36F874325E4D1CE512832750A
1ABD17DF6F5363DDC93127F7293A808D82A8D105DC750A87A359C0D50C99C5D6D6F35C9E2000A7D0
7E8A06D178966CA44097FBB5877E8A4F05FD14D8473BA4538496BCD23AC5C04F41332EBDD2A5B30A
45415489DAA75208D43E15FC141665B8A95B694FA25384FB2869676BBB39C538F729A38FB69E7785
748A49729F72FC14B77F2738AA8500A70B48AAE8141935E7393A45A2F6C95AE04D78B4994ED1E717
7CE8AE4B8A21067617384564FD27B41644637EA6B54F319D42ADF9680385B0C50D458D69E914EAA2
5DB18FF677F6F2EB76D9DC27ED77934550661C81EB140E4773E02BC670F1595939B4F7D6BD705DA2
6A4C3CC0D4299A6AA2BD3B07B54F91C5DBC8AA57A88FF684CBB9DAA34D3A852955888956E1AE94EC
4F31611FEDBCDAA74202A1E89764FA294AE63E85B6F9617D2602B3B5EEE814561FEDB830776E361E
ED1C3F05AF7D223D4264E688F14CD3B0B49F2253A4509828F74977A9A0DC27CB4F21FB68A70236F5
626F173845A93EDAD93A45B9DA272D5BC4B364E9A26D88149645686C2C0AF5D12EA55358D7EDACFE
1461CA6C308EC1F08EE91445B35B4898D04B4370DD769DC2E1981FC0B5E2E8FA118C7BA2E458F8D2
3DDAA277527E480E1D10EF4F917F7B4AEB14033F855173AE139F2CE542FA29A6A853447BDE91475B
DE8FC2F7263A78729DC2B87375CA4FC186B1C846D6EBBAA86B443CDA919E77910D81FCDAA78FB16E
C2DA28242666C02968A818B54F13E814D5739F8A2BBAB23F45D44F910CD8D4F521AE53F055A0F27D
B44BE43EA196B179DB1962CA787CA8F6896882E1A7D0DE8AE2C84FEB143983795A3A852217B2AF50
9E9F424BC976B6C6FA789C3BA77038E60162A6F7C8EF9C3AB0FFE03BF6BC13EE5FA64E412B21F5A3
A92C59EEF8D31950C986DAF1FE1491A5AD88C81EF15388FE147C994B54EDF27B131E66FA29EC2CD9
69E43E9DFAD4DBCCE996BDF16D657ACDEEA3AD8BCF8D45DD6EFA29061C4177FE8A53667C765A3DEF
04B9C8CF7D42EE2CE8038D6A4E31627E0AEA7927FA684730DDDC273101CBCB7D32FB68739A6CD8B7
072FE98E475BF7D18E2FFE64728A78CF3BDE475B2B71219DC26C07AF850961AF30729FCC2CD9AA2D
1D73748A8847DB28ED1B6CC4739FF805D9ECC362BA87AE79CF3B87A36944AA98609E86F6EDD5D555
61A9A0CB57231FB8F1DAA7885411FA3247A7A8300DCBEDA37D615C31A27B5588DCA7ACDAA76A08D7
3E8D17B25451AE44198FB6EC4F11E9088675BCBC8F7637FC14F1011C998F359CFB545CAD15DDEE68
C6C52DDB89DCA7B20AC5D4739FD49EB8473BC123C2C9661DD129CC3EDA3A09CAECF618F65354A97D
12943998253B6A75A79DDAA147EEA7B06B9FB84E5156A4F8A25DFB54D02968DC5E944A8471DD6657
F8447F0A251C078D6F457EE19CC2E1985BE0F4EFF537FFF6E8FA116C81077F3E702386F91BEC69EA
5335E2A710FD294C376B7C51D7D62926F353C4739F444E8858B9E54F89DCA7A0477BF5316321B78C
5411A97DA2F5AE424B56057EB7CAF1689B4D28E2F521FDC7F54ED43EE5E43E15CAA246D9FEA46B34
D2F3EE76D6477BCC1DB469E8B902D1C0C744EE53A9BABEE9E63E891482A89F62AC53B08A261CB7BA
50446F74C44F212ED79A209B06ED29D63E6924B36479A0B748E73337D23A05F7689752282AF4A7B0
5264C703BE6C1F6D2BD3CC2C851A1FB0EEB54F0EC7C26073F3F481FD07B199C5DADA5ACE4B600279
E3DA6B9986EEEBAF5F07240FAE5FA7A0DAA7C8BA96BE67F10998D99F023DDA3CF7A902C530FA68B3
FE1414A129966D8584216AA266D59F22A3E75D812F0436A49FA2427F0AAB1BC5B834F7D868A1AC23
7DB4C3FD29E25EA1C9739FF88017C524F93A45A8C89CCAA2783D49BAF6A92C5F9E6EEE5351AAAB90
252B2B9DAC80826EEA14C9CC010E18D8BFB9EFABF5F7BC2B847B5BA97DFC925EB05D709D22D447BB
A44211E214461F6D31BCB90CA7330AF2FB53243BC2AB61EF3A85C3319F08CDED1F79F831E01474F9
0ABD04A6FDC03E9080C0E3CACA2133A8165F056C05DF13FE61D3BDC807AB3F4B966A9F846EAE275D
46DEE6284BB6541FED1C33455CA728F829D4B22DAFC8E58B63D49FE2436BC3C9BFE1D1FE62F5FE14
66ED13F6A720C3B5AD9BEBB52FD229F2FD1481684D519ACBF77441A718F6700CD3076E68E593AEA9
78B443C8F7688BF2729A71093F053E56EF7997C244B94FD6ACAC9C9F625DADD9DE2BB58CAEF9294C
912234B6C5E57DEAB54F853F87804E31F607F15833D56048F75E913A8559FB14D229529439A73F85
797D0E918BACFE1461415990887152DF60BFEB140EC7BCC16413B093F6633C54E4F8B5B535201130
DD8529D9EAEA2AF6BF0066216805BE109FE56F7260FF4140E8E3CD43ED93796F0A8917B1FE141F38
C76F4FA5EF4D293F85B6639B65BAE5FC14D3CE7D1AF7A730E39E089A718C6E5265758AA048411122
5DF25388DA2718C94816926BB94D65C90A8FB621523C572888E2E33FDDF32EBFB40F8F9F3CF74978
2BB2FD14B2274524515659593BA253E855A0B8B8CC37A6D2F30EC6F3631F2C91255BB80E5F084B15
567F8AB44E5156A4F862C24F61843B05D685CAF6A7D0576659E6E77DB41D8E16C1FC3BC2E9223C25
84869D2B3B2858D00C8D26960F9C3809FB7BBD1E3F1EBE849D21CB46E3B54FE69D2872ABAAA05398
F421B3E75DC14F51EC9A243236A54EC16E61893EDA65EB43A84A2455FBC48BCFED86DA5AA728DB9F
42CFBED6D93D6BB087EE50EDE614A8539C2BF6BCC32F936939633F45ED3DEF444EB258CB8D1BB7CB
79B4E3284EC926AD7D52543AA153E0E81DE9141136718D37081B1DD3114E618EDBD025BAB073363D
EFE27E0A51CE17B1514CD34F31A14E11E97957B451D0D2508E9FA254F9131FE1CE291C8EB9C2E4CD
EC4C2E80DDBA850BE3D59D97B13E4A7F06142FA8470647FD59B222F7A9545D2E21A4534CE2A788E8
14308FE25E09D198B5D09E55E914B3EB4F91D3F32E52F82434F4CA7E0AC3004877289EFBD4F6DAA7
711FEDBCBE60624A36159D82E66638F833750AA1B269D14D1AB747F524256A9F72B291BF38D5DC27
9DEA1FD02974CF3BDBD01A366B77815324AFC971C7D0546A9F4CDE11EC4F512CE7CB6F4E11CB7D8A
FB299248F6A7609762C34C112217C9DCA79C3A559D25EBB54F0EC77CA32CC5803F677C8978E1534F
3CAD5D1870213DB0FFA09E1503D05811726D3458FB84732AB102A65986EEA954AD8F76F280481F6D
91196B243E094B607E7F8A6A3A45A0F609FD1452370FC53DF1C5AE927DB4ED252F335776E4666D37
A728F4D1B6CC41919918456EEAFE14999C22BED29BE9D136448AA2542144BA74ED53BE42519C954D
33F7895A5494E979272D15666FC7EEE9147CB886A464B18185AC53E97967B28CA49F4247F39922C5
789CFFD1F7737BDE95BD80E7F5A72864C9AAAC6F435FCEC97D8A876904CAA29C53381CAD418480A0
B35BD4382D2D2DC14E3DD9DE1DACA0A205433F55BF9F026B9F7256BA42B54F70779BB0F6A9844E31
C87D124B5E46DC938A2BE46551B3D329F4E91DFB2998F460943C05167B4BE914B6655B6D74A4F649
F4D136E39E0AC6558AD0214E71CF4E4ECFBB7C6B760EA7B85DF5D11E4FC394659B97A6276A9FB84E
919FE43FC5DC273513CBF7689BB54FB666811EED0E708AB848C1DDD9C685BD52ED131FE758E6249E
CAA97D129441CBCA229A80FC14593DEF4A89140394EDA36DAA15D5FC1486A01C60135747A97D5EFB
E4707401401036374FF33D705D429385201A88ADDE369645E916DE4DF5BCD3D6899CE614581F0233
B7521EEDFC3E4A119D42AFD986140A91FB94E5D1AE10FD74FA52A43F45A17D52A4E449F4D18EF6A7
B075F370136D1D0075FAD12F35FEB7333BC4758A18531E109048EDD3277FFAD72BD3E408A778EF27
EFE3029C9E5C853CADF42C700A1A2A098F7652A7A0E3075F96A87D0A681322F429A15390E880A57A
8A47189A05F356B43EF769E5DD47425103C02674CBF873CA6A11AA7D7AEC83E3B19A53FB04C70BF6
61720AC126B870AC99B2B8B607B364CF7E7D163A45244B566E84648B6FDF16E414EB161116DC21A0
3837E2A768F7BA93C3318778E0C4C927CF9E153BD1880DAC41CF067707355468A910B1B13023850B
66EFD9F3757EFE63AB2722B94F725D57553D21B44E01A765C29E778F9AFD2958EED3B219F4C4DA67
D301213F050FFB2AD43E55522B829CC20C33B796700D9D2255FBA46985C8C011B730CA7D8297B7BB
3F45C1A34D6A455E9E3FD53E4DD8473B246184740A9EEC14B45184BE04B2FCBB4F66F5A7C81129CA
7AB4D9442BC826B44E61728ABBF7F4C7367A7FA8EAA9580412F268E3CEEEE814441FCCF14C2E2111
4A803A855ED1E2B54FC0142A84F5C1ABCC7B41214B3664A6D001B3F91E6DC67F4B5CBD4F5F02222F
3EAD91251B522884413BC7A3AD1ADB19D770B5674EFC1493DB511D0E470470F139B0FF205083A5A5
25212EA0C302388569C446FB361C006484FE4EC9A9B1B6B6F6C8C38F01E0CF9936605609D7D29B37
DE98E2E3EB6FFE2D3CDE7B6823E45A8DB4611577A8477EE7D4CD1FFF08DF165BFB094E4166EDD0BC
CB94DA8FAE1F419E459F16CED8FFE5EBDFD29E3E332147B4BAA365B10FAD7D06DF167E83F89961A2
78CB3F39355C9EA5D957A9FE1483DCA7FE07C5F370E30D380FBBBC8F768047D893AEA2471B68263F
09F0E14FAE2EEBDA725D582EB806F611C303601BA685F8FB9AFAD06AE0917E90D13FF8ABE11EED88
5F55B771A43D6B1F39864385CE12CCC450A78853E6C850474E31FC9483718259D6C029FA554F454E
A1E39E44B908DFDEFBAFFEEDD6D6F3FD1188637B30544AF4D1D626EED1DF42FF0F7CF0696104E220
843FF0829F229019AB07361E86B94F0F1D7F103F2D9D61F803DFBA73A93F50A93FA3AAE5937E0A45
378EFCC39F1F5E37061F782EC6E7541E073F0EFC681F7FC71D5A7A2885DFDCF7D5ADFFF6025DA9F0
5700A35D88CBA1414E176D9125BBF9F7BF0C9C02DF90DE167ECBC429346B08259B89E8A78D8D0D1A
D8F8CE67BFFA1559FB5456A718D43ED15F220E1598B48BDA2723BB4CD167A14403A7E88F40766EE1
CB5FFD85B715729FB28BA068603FFDD1B7F7FFC0A73DAA71506D6E9EC6F906CD3A7006B2EB6CC2E1
98316E5C7B6D79797965E5D03EF68F973FC1751539054C56F5CBE1AA28E26709371BEA7947DDEB34
9B88B7550AF5D1464E5161A58BDFB64259B2C2E567663D11D1D045BC319D8216BB4A015F12CA7D52
77A5488448A99E77E4A7B0D50AB5B37FFC7DEFC35B5847740A51FB441B9C41846A9FE23A45A4E03C
B28A8B7F11C9FE1449CA2C37068FB7FDEE9376ED53661F6DFD6C4AA728049A8564383DEC473AC5C9
CFDD21DE76ECA7209D42C910493FC5439FF8078D0FBF9962F5DDF7D23ACFD70EBF20D871883567D4
3E15748AD06559982908580715D229B4F12D74DDE697EBA14EF1E2D52C9DA2ACCA5CA1F6C9AA56D5
47866A9F8657EC4074869D04C506FCEC6A9F225D809D50381C7502EEB6C8299696965EDD79197792
4E11E714FA4E0D7FBFBA926AA6A0DA275AF58A249F8B141178C4FDA6475B78FACA4EC050A7106FCB
6B9F28C93FA45068DE216A9FF0F604E77C76B94F059D22D48D427928F8CEB89FC22EBED5865661DF
1EF08B76FB2944966C9C1A8BA917D729C4DB92479B0FE07C8B10E914FA03F3DAA7615A32EF141F88
DCE473B344EE53299DA2B8FC7BF6AB927E22A740B981AFD0EA18FF9BAAA7767F75F7E2FBE1E5A1DA
A782D6B66E8B6E917ABF23FFF0E71B1F7E3385C87DD29A45A84815F7473CDAC24F91C926381331FD
142BDFFD0BA15314CC14D6D2D0F0300C2B48FA2926D029F41FB859FB640E6C2EBDF1FD214E210A56
1345508A38CFBAF649DFBFDCA3ED70D40FF24790EEF0CCB9679053E89AD55DE6E0D6F4A1119D8277
01336F4321B705AEE526739FB4473B796F0AE914C0293EFA07DFE5B94FDABB4AB72A717BA2033EB4
F6193ADBB831EE79572DF769F0AA601F6D3EDDE2F5E4EC3664585C472FC9C99235633665FA93C87D
EA4696ECB962CF3BD34F318ED944E2CC3DDA93F929428867C96205142F0589E814E3FE14913EDAA4
5344A29FC4782E560046FC1485A957381B470A19013F05FCD686B54F9FB6FC14AC0E4470643E19EB
8E9F0287ABE0C5C95A5618D8C829C4F58AFA53F4A9C1E82A1DE1CBFD636E3B230E08E53E0D233284
0BFB42A1CC89EB74A45388DC27AC29DDD51EED29E914AFEEBC6CE63EF17C6FA363290E7E8C0DFFF6
6D912C59F35AADCBF94C1EDDB847DB350B87A31E3C74FC41DE1D9B58C6D6D6F3FA609822869E6DAA
F649DC7AE20E565DC71BD729AAC1CC7D024E51F0535C9045B9BCC27CBC1436E020F8E5CA77FF229D
FB5452A148EB14A9547351F8C48F8CD73E998B5DE63A182EF9F6E7699F1EF6BCEB50ED53864861C8
7013F4A778346CD086795ABAF68956681571961523CCD9BAF7CB5F1F738A516BB0ACFE14623C5305
604AA73017696D19CEAA7D82CBA6FC03FFE6B7B6EF7957810EAF4B761CF3537446A7485637455806
7CF99BFBBE1AD229924DB4E317F650CF3B528A75D99E300A71597938C84D8FB6D029CC908169F4A7
18B26653AD30F7476B9F22011A21B58206FCEC740AE1EB24B47BDDC9E1985B003BE0ADB4E1BA874A
844910B67ADBA862D0920BA1FEDAA761CFBB4F5DE2779CC8BAAE58E0C567ED2CD9C9FC148F467BDE
89D25CD1AE428814E259CD29C63A8529A34757BAD2B54F6CE5566E14437278ED534E1FEDF16DC80A
1ED4FAC570ED7734496BF7FD62AC5304FC41BA77F678FC0FFE042AF7D14EB2E964962CD53EF13918
17260A22C548CE0866C9C6FB68EB999842A8E79DDD9F4215998B24287476C7FA53F0A23ED3222448
341DD399FE14A2F629448D43391B997DB4E3C3B85A1F6DE40B661721BD1DCB7D32B3644B09CD56EE
93EE4F11149439591E6DE3753B98251B18C6216DA220CFCDD24F1182D73E391C8D00A507B878D29E
BB0E1E365DD8BB230BC681FDC6BCA2AF533CF38D3A3FF9B1D51354E911D2D393303905BFD16099AE
5EB98D0BEBA19E774669AE0A0C313339D3FD294A2D7395D529B43C21DA5214B371F23DDA11072B9F
7AA147BB239C62ECA788F6A7082DF356AB7DCA0C832AEBD136DB52082ADDCF7DAAD0473BD5ED0E11
F1680763FC035F8ED974BC8F36B2096C7BA7577103D523784CEBFB53904E9193CE67CA16D53CDA39
D273BC8FB6CE408EE4988D1336B89F6224C0153CDA39110425750AB3E79D412EAC9AD51C9DC258F0
49F10BEFA3ED707404BD5E4F74A3C066D9A45C70C0CE508BEDA67ADE45A65866FB6C7E7B0AF929C8
EE570D91DC275AF2320BCEE574ABB8A8BBFA9D9DB19F622A7DB4A33A05F6D1C61B4D610266192B4A
79B4837E0A33C99F3D76CD4FC13972EFEEAB39FD29B09363294E9164137C9216F1682FB30C7F336C
D3D4E3B2FC14665D5F885CE0246D34554BEB1466A619C913968128A2530CF9F2805060815FACF25C
4DC33AA453584C99AF0EF186DAE3C11FE414419D22225BF0911FEFA3CD2FCE850AD5A22D88F38EA1
6097D347BBC2A5BB64EE93DCA3063C21E8A7F834AB3E8D5A27F498779DC2E1680D92D6A48D8D8D95
9543E225C032969696F46BB12C8AAE8D1CB0734E3885B85BC5B33773FC1415EC15919E77667ABFA8
12E1D5E65C709F894E31E84FA14FEF6FAFBEDD58EC0ADC9EE47A57B48F76214BB668C4B6950B5646
D21D9D22B978ABE933EA1A913EDA82535418D8E5748AD10837B2372F8CD39263B94FA2E54AB25644
ED2CD1475BD9286EEA48A878CFBB3B97AEF18AA6007136D778873A45073845E8421DB9867324758A
9C3EDA02483A82B54FC52C3EB33BBC0CD6607E0A23F74964C996BA6887739F74ED53213039700D17
AEED481FED645B0A83658C5AC9BB4EE170B40937AEBD061C6173F3B4A009285260E32ABE1FD39FC4
CC0DFE48E1E09069A27E9DE2D8EA09BD6C9BDF6BF85C46ED5335634544A7E0DE55EEFB1351E7DA12
B82CB264B54E5179C9EB8B767F0ACC7DC2DB905928224372D4C26F7EEDD338D629DE47697D6F7774
0A115916F25608574524F789B264F387B12E2389E73E71014E446E72D6AC6768256A9F42DE0AA165
A4740AA33F45B1E6DCE61A299D42F005BB2C64BDA853B09774A1F689EB14E3E032A64A98B232E714
561FED63C491237DB4896860CF3BFE544CA7B8A0C8725483A34BBAD42978ED53A4E75D559D42D73E
05C7B3201A6C78C7FA689B2DE00301B3AE53381C2D06DCEF30AF094801FCAD610F4AB86062BB67F3
25306D83837BCF9E47C681EC2332976B3CF7C9A412A152289A95D93AC507CE25A944353FC5B0F689
DFA42E18D32D51A31BEA4FB11BAA7D2A295544FC1446C99339DDB2AC1639B54F629D96DBFA38B918
F7BC1B1CD66E4E01B4FD6B875FE0AC01CB9904CBD07B68672359B2E386145AAA50929C58E6CDAD7D
8A28145C77CBD329120C2250FB14D129BE7FC71EBD5A1B2A17D1E2C5D52E65C9469263E3F25CBE4E
212AA092633BA4537C4C74D0BE10D3E0C4151B36727BDE957C2CDDF32E6F2D28A25350EDD3B80C95
5FA5AD5428F753381CED034E17B77ADB2B2B87969797575757EF3A78585F3FF5AC12AE811B1B1BCB
837F0F1D7F50673D8983EBAF7DE2D32A2AFF88DF92705D17833A4D9DE2C12F7C3E5EFB94142FE23A
85B069EBFE147AED0BAB4456BF63708A42EE5335046A9F863AC577869120F6329755944B47E6EA14
56BAA67D7B1AE9145DC9922DCEBB38CBD02A066DE7D73EE593E51C8F368D6159CB67D928C4FE74ED
93601349A65CD64F61320855FB845F26758AA1477BBD20C005B3714653B2EEF829E8221C52E20C01
7A947596E3D1C6EB73A445050A7099FD29B8D94DE6CA06748AE578EED36C740A5DFB24ABFBC2639E
0E8BE914AA61902E5E153BE93AEF3A85C3B1E8101C2114E35CF67D22873552FB24729F4237A3D0BA
6EA99E779908F5BCE3B54F227BB0E09B604443585963B54F15FA53D04BACDAA7A19F22DC74554EBD
78966CCA4F114A38D7EAF9B85CE4D8F825ADD729423DEF4C25AED0E731ECD116B54F39A6215E2B92
A353F47BDE8D5208ECAC279D7516D729FE655E1F6DCD3B4653B5A49F42138790EE96D429C6B54F18
FA34687B176713A4CA7547A718AF0259B57CF12FF33DDA652FDD91DA275EADC747AF760915748A0B
AA3F8599FB54EDA25DAA3F45F88A5D4AA708260C98CDB5EBEDA3ADE19CC2E16804D3EA2F597F96AC
E9D136A75BE6BA6EAC3FC507CE6572874703D185B1FE1422E2E94250A4108BBD89DAA7CA4B5ED1DC
27EA9A142F0E31676865FD14BA68446E77C94F21E65D91E12DD078ED53287B530C75BEC09BE84F51
49A188EB1438B0655193EEA35D3EF749173819732DEEB66044A30B9C22E78A7D6E64B2D0233CB3F6
29DEFCCE44A8F649F8290C29D9BA687F6CD44A3B51FB54F5A29DA353D82D1D03F1E0499DA2708936
FDDAE161EF3A85C3E1288546B2644DE9A1144AE914995922719D42D73589052E1EF4C4BFCCCA92AD
2052046A9F0C3F05F51156613872BD2BD59F22B8721BB83179EE139F5F99D9FEA4D64D8553845A12
E77BB44DAD2DB4CC5BCE4F115728CAEB1425B6933A4592179BFD2946E5521DF26847C48880D01CE5
14639D22A79576298F366F27A4B589B89F225DFB5449A4C8E414864261F6A760633B47A7304675A8
93A3FB291C0E4725D4EFA7A0DA277DEB09361A56FD29423AC5247DB493FD29F4CDC8ACC8A5E67754
C19B95255B01A1DCA78B630D3D96705E9C80E5F4D1367BDE19B245276B9F44EE1396336572E4B147
5BF92932739F426C2247A73083713835A6E99608832A9DFB146EE098EFD196C3382E5560C78AA84E
B17DCFBB7461B9511962E62777C64FC105B8CC6E777C7847740AE40BA12CD9C81ED43572FC14DA49
51C24F91EC79974F9CF36A9FEC2A3E35A433750A2E52E8351FBBF669F42AE7140E87A3141AEC79A7
B94364992BA953243DDA13EA143A6953C49B0FD7BB06ECA35C7F0A5AD19D46EE13FA2944C2B95179
AEF243726A9FEC6955C8BEDAB12C5991FB6496850879C2F06867E814994DE1ABE914A6AD7579D40E
4C14FEE57AB473148AEC2CD9F4D44B6F97EA4F214ACDCD00FFEEE53E69079C18C6F4ACD98125A953
3C3AEA3711B9809BAC39A73F45C44921B439F253AC7DF69418D852A7287FC54EEB146C794774A090
75AA65739FE20DEF7465D4E0255EFBE470384AA1912CD978ED93AECED54F55E84F216662FAF66472
8ADEB3E7C9A35D48230C5827F8BD89666889FE14D510C97D0AF7D1367A814DD09F422CDE4AD90280
77B4D17DADDD9C2294FB648E6A8D09739FE2C8F768CBF1ACFA6BF3D66055FA689B2245059D22541C
A2C7F9684FBA3F05F3688BC16CFA583BA5534C58AD5A2A4B362EBA099611D129882384B29EF44622
F74978B4CB5FB473B264FB1B1755E720B39DD0E81A9ECE7D2A5E998317EDA296E19CC2E1709442E3
B54F115522320DABDC9FA29A4E21BC1263555D67E3148B464259B27D4EB1FA58617235A5DC27F453
E87A72492E5040D737ACF27E0AE39E25F67446A7D0B94FC08831D02954764EEBBD53E97917522EF2
3DDA7C3C179C14560C54BAF6296EA928AF5348592DA250E83AA8B04E817CD9A80949792B3AEBD1CE
692AC4758D9C2CD988181141A43F052F6A8AC813BCF06F286DF03EDAA19E77954C70393DEFC4A55B
AF05697DB9741FEDD0459BB169D8F6DA2787C3510A0DD63EC58B43CCBB55A64E9139D1CAE11422F7
49949DEB007F3EE3A2DBD98CFC14C99E77C1FE14A3D52D7362564DA78854860C6F52DDE84F311CA8
C5FE14217FEBAC758ACAFD29B8CF4207E6D0504FE81471FA104548A79043DACA77922C23E5A7187A
B4994E615789A83EDAF4D8054E618EE1D0153B5BA73856EA126D0FF2904EC1C6B0765284A48A59E9
14A3315FAEE79D5A0B0A71E79CDCA790DC16BA867BED93C3E1288B46B264A9F6C96CA29473ABB275
0A56DD64DAFD2AEB1491867734EF12F54EFC8E96A87DAABAE465D63E997E8AC2EC2B706FA2F8D9D2
FD29CCCE14EAFED5119D42A7DFC4EBA0CAE63E6933855928220AFCD21E6DE5A7A06D5182CE877AC2
4F11E10ED16C9C08A708D63EA949D7D8CA9AF253987DB463635B89175DE014393A72C85891D3F32E
D29982F75B79EC83050212AF7D2A64805BD76DBE5FF4A718EB14A19E77E5CD14E99E774AA13087F4
44FD29021A9CBE68C3A3EB140E87A3149AEA7917B93789787321A987729FD0A33DA3DC276191D04B
5BA2E19D48268CE914A5FA820984739F68B5566E848C1595FA680B13ABB86D0DEBD28F2D61554977
FC14348FCA77554C9EFB145FEFCDD429B857623CE08B6DB5B9C922D89F42E43EE5881479FD296C47
B6562854DD488E9FC2A8E88BAEE2D29E8E64C9E228E5A57D9952458E47DB2C798AAF0BC5739F4490
AC58F69197F1884E11E97997AF5064E43ED97DB49512273672729F82A66C35E645519FEB140E87A3
14EAF753447ADE45C2CF878AC6A72E6144677D7DB4BFF92DF2538442CECD5228FC32E4A7B8FEFA75
59FB545EAA88D43E15A407533137D7BBF2FB6847EF53620F1DD96E4E11CF7DD249FE949983FB1BEC
7927FA68DB62DC85B1FA465CA37A7F8AAA3A85EC79A7D58AF0042CBF8F369980C66EA05037ED8ED5
3EE5286EE642505CA7C889900DF18E909FC2E8AE7261ACBB19D7F30B8C4D87748A0A7DB48BC8A97D
22C9D82411526ECEC97D2A5EA563CAB27BB41D0EC70468CA4F115FDA8AB8B3FBAB6487FE2CE4D14E
AEDF72193D5FA7A019972ECA35AD7FBC5F58A2F669329D429FDEBE4E11CA2134ABCDCB64C90A9D22
92FE34FC7294FBD41D3F8519B9998306739FB8477BF939C98B75300E3DA6B364CD08820C9DA27AED
93E96CCDF068E7D43E19A5E9031AD2054E91BC32470EC8CC92CD417ECF3B9E5D66EACBE39DEC11AE
DB593DEFF2FB53A4740A5DFB146A756756FAA5750ADEFC3DD2E48E6DF79565EF79E77038CAA3FE2C
59B3F6C9BC25716BF670DEF5A94B919E7742A798859F226EA3D08C23AB3F45559DC2AC7D423F85A8
24D7B32F4E2EC4B2586E7F0AA19B9B3939C595DED6EB14E4D1B61772592AB200AA6FD5729FF838AF
5EFBC4648864360EB1E9123DEF32C7F904B94F3CCACC1CEDE92C5956F5112F14294CCC805374A3F6
E91C0B1F88745AD157F573038FF6CE951DF1B666966CCE853A47A7880DE90B6A5B982F726A9FCA8B
14157BDE29BE5C4EA7280E60535316239CAEEACE291C0E47293452FBC4FB532417BECCD9576D59B2
C24F21D26FB84EA1B767EDA7306B9F50A720B210BF4F193DEFCAF82912E94FEB43A241B94FEDE614
C24F41C355178D08B28C05EA156A9FE23DEFB81297DF475B3365D33D845F96A87D2A39CE73739F8A
DED5900637CE7DCACB92354BCD8DB2F3D1E0779D22C8A3479ADD241EED4755E640AE4E71E112375F
6B0F85DE437E8AB447BB94429191FBA4B364A5DC3C999F22AEB8191BF72E79ED93C3E12885F9C992
D5F1E6E754AF6D3CD8AC7DE27DB467E7A7E0EBBA991D9456BFB3F3A1B56191520DB94F3C4B5642D7
3E594B6165FB5304934344FFBBB6730AD34FC1977623533254DF4AE53E9542BE475B32E541D59F70
0FD19725B264330776E5DC27AB0F8B51FB94D9F32E3AAAC508EF48ED93D029B412112F678DD43E11
5F302FDA62C03FF641C92C227E0A992D100E7DA20C70EA4F9195255B4AA408E81459B94FC5E15D36
F7493B80CC212D39B5EB140E87A3241AC9924DDE80C402AF4EF2CFEF4F61666F9A4B5E397E8AD8ED
49F92968296C46FD2982B94FBC2004EF3ED1DA27444E96ACD99F421007E36ED5313F85A877C2A226
73848BD13E79EE5368F0E77BB47991B9DE230AFC12B54FC24C11C724B94F221B473C4BD14F499D22
9E871CE2D1DDC87D2AACEA84C33468D80B252EBF8F76E9411EEE4FB1FCDC25E1D48EE46910B3963A
45A8E75D059DA24CED93D1E7CE549C537DB443BE6C71DD16431AB79D53381C8E52683C4B5607C926
55F588471B57BAE8F6544AADC8F253A48489849F82EB1413F7D1366B9F867E0A6E9408D53E59FD58
73FB6847967359B42C7AB46967EB750ADD479B9676B539486C4F98FBA4B9F3A3D9FD29B847BBC088
75D9F9730521A384477B4A3A85CC7DD26CC29A8FA5739F700C934EB15E987A4554B94EE53E696A6C
B209CC9BCDD3298C3EDAA510AF7D2A286BF1B5A0512414F5A748EB1465458A2FE6D63E09839B1192
CCD5B7927DB48D351F254F50C286D73E391C8E5298AB2CD9249BA02F4BF5D19E84538CFB6817176F
238D93C86D410B5F099D223F4244ACEB86729F2EAA0811220ECCC7CA6F4C99B54F864E11EEA94D36
0A9A7DB59B53987DB4F3ED424D65C9728FB6081CB0C932DB5F3D4B36A55354CE7DB2077CB48FB6EC
79172F1151B3B24ED53E9DABD4A234DCF3EE18BFFC8A25A0B85DE8D1547F0A5ADBE1033BAE564CCD
4F11A6CC3959B2F1DED9265F4EFB29AC712B958B7549345CA770381CA5D0889FC2F46887AA734512
14569E87FA68739D622A9C02FD145819A2BD7EC6448BAA4432FB53949527F83CCDAA7D2AF8298A6B
5946CCA6D0D6F3FB53441B6A17B26459ED53BB3985D9475B7BB4CD9C9C52B54FA62411512E923A85
F0689BC60A232427A253E8DCA77C97D017B36A9F926CC258D14DF5A728516A5E8C87EA02A7185F99
AD8614FACB821E37B3DAA708A7E0957B917E2B22D3CCF053447ADE951429D23DEF4CFB9BA5C7D1E5
3AA953F4D3BCCD2B76C0BECD5539D7291C0E4729349525CB957431E98A35A7C8F368CF48A7A01997
A800D133AEA09FC2F468574530F74937638D9B58B99F223FF72960591575B9C3DAA7C19E8EF82904
9B3029F34C750A11FA94AF5324CBF9B8B1028E2FE7D18EA33819CBCD7D3209B2552E92D3479B7BB4
8D7AA740AA7FA7748AD01250E8029ECA7D2AF4D12EBB1094A353E826DA218582FE0A623A05718A6A
3A455ECF3B2D4F44729F927E0A3992353B768FB6C3E19812E6A1F6294222E896C48F29DB9F227F5D
37E2A7E045E6429810F7291E3C9BEE4F51D94F11A87D423F051284C204CC92D179ED534ECF3B5B9E
08D38AAEE9143CF749B7CF36D66F470ED6487F8A7C3F45A422BD541FEDFE4E16F764A7100C0E48F8
294A8D6D7CC984B94FCA5BC1C9457ECF3B3107E393315392EB02A71022852928472846D93EDA21C5
ED51E5188A78B4692D082FDDB63C516CAEBDACFB5398B54F15448A804E11C9928D259B15CB56F373
9FF405DC162906576FE7140E87A3141AAF7D3A77D8E8A3746EE4F50BB18F9AFB53F0AA27AD9EF309
18518C127E8ACA08E53EE92C59EBDE64169094F253D8966D4137A88F76DB3DDAC93EDA2677C6BF05
2217C70E54E97997F4B7E67BB471EA65F6AAE0DDEEF0A974ED53BEA5A2786495DC27ED152ACECD62
3A45BC3824D2DBB14B3A8599EC1D1AEA852F33FA689BB54F39D273284B5684F5C5FD41B42791FB24
6A9FF2158A8CDC27528A65100111641AE18A44E7E73E8925A0C81AD135EF4FE170384AA2FE2C59AA
7D2A943315E572333984DFC572748AB2D3B0CC3EDA76F6A02A44A7E9D9ACFA53046A9FD04F41F7A6
4C7231AE7DCACC7D8AEAE9E33D3CF7A9033A85EEA30DD4D8E011C570332CE7C3B1AD3945A83F4568
BA65EECFEC4F81E51F85323F2BFA8904B812B54F994E8AA9E43E156B9FC695E7019D62FB9E777173
4450B0B0C293B15CAA0B59B28555A0509B957037EDFA739FC4C0369BAD883E2C740D4FEB149C3267
8A145FCCAD7DEA6F5C2C74A030D4B732FD29E282B271D11EED719DC2E1709442833DEF42CBB9E62D
09E766C435421EEDCA2205E2E4FA29F1B6BD67CF8BFE14630FC585F17C4C04E694E84F51B98F76D9
9E77A91615391EED849F22A45C74AC3F45643C879236719C4F37F78913E7B4479B6AF6347756EDE3
69904F27F7C99A9255CC7DB292A0726A9FB487C29E809931FEDDD02922D767D169C5382CAFF6A99A
CA1CD22904592022AC6B56754641DA4F313D9D2256FBA40D14812CD9FC3EDA760594A953B847DBE1
709444237E0A1D86A397B9CCA5B048EE137AB4457F8A9C25DC4C9D825671B58762BCD8C5A944B23F
85A87DAA205558B54FDC4F61F6B64B6C67FA2942F7261D9E33A87DEA2FE77640A7D07DB4B9A73564
AC4091E25CD5DAA71C94D02904710ED067DC999BFB949FE43F79EE93B92759FB54F4680F897068D2
2516788F75A2E79DF668477A6A6BBA019C62E7CA8E78DB64ED53E8428D1B718F366FE3F8B1516BA1
E5E78C45217E498767A59F22D4F3AEBC4851AAE79D91F534AA7D2AA553F4AFC0A158B38099C23DDA
0E87A31A9AEA79979FF5C417BEB0A617666EA5FA5364EA17C9DCA7500EA1AED7E5E4A250FB14EA4F
5149AD48E63E995D2AB8C5CFE8795756A788066FE246BF437197FC14F93D29F8426E53FD290C8FF6
88388BAC275DFE54AEE7DD34748ACC723EDB3D94ECA39DACE8D3A9509DF15344A4E4108FE698B0F6
09AFE170CC631F944786740A7D8926258ED7AF0ACA4CB54FA5758ABC85A0844E614A12913D393A85
D93628B4E17E0A87C33101EACF9215B94F91F52E6ED31EAF8F453DDA7CC9ABECBC2BD29F4276A628
F2085E88CB7BDED13D2BDD9FA2AC54F1C5DCDA2773EDCBDE3312D3CBF6A730AB73B99FA23BB94FBA
8FB6F65388114EC439D49F22C92992F966F065298FB66842416BBC72E497EA79971429A695FB249C
DAA345DDB4479B67C9462A46AC8CD92E700A1C9FBC722F874D900057A1F629772D289CFB24AA9B96
9F93635BBBB6C9A39DCE7D2A2B5204748A50EDD370C1C74A4B9EA69F222C317BEE93C3E1288B666B
9F508310AAFAB9B04DFBF73F15F3534CB8969BA95348739F2A38FFD87F96215133EA4F11E97947B5
4F46CA13DBE6F7267A49599D225910455FB69B53987DB443C5217A8487748A48ED5372D24507E4F7
A7E04550624AA6477E899E77994BBB2394CB7D126C42EF89EA14B28FF6BA35C50A2DF60E1A8A75A7
F629245524F754CB7D12409D42EC09D63E31EF0FBF569BBA1BB7CB499DC2CC7DD2C33B2F3039A7F6
697CF556DA44815CB03CF044EE5358650B656EE04EE7140E87A3149AAA7DCA2C7C0A158A98CB9894
251BB93DD14E536AD71E6DF453E85842324D48F59C5959233AC514FA53046A9FD04F21758A88D18F
2F8BA56A9F44C94766AD48473885D9475B8CED482E41A64E512A88A084475B4DBA8841987550593A
45BE4891AD5318319BD644CBF46E27FD14E65CCB1CC9C25BD1059D22242B27A50A1CF333CA7D7A34
E2D1BE60AFF044148AB16017D22978ED535991E28BB9B54F465BD2B0BE9CECA35DD64CE11E6D87C3
5119F567C9EA9E77A1052EDD4D29E9D18EAFD9264B76233A45A85392C89515ADF1304B36D19FA26C
C7617A9555FB64F829CC8614010DBD824E114B7F22BB6BF7729F225D2A7465C8B8F6A9763F85D029
F48CCB4C94CDF25324A75B619631859E773A212795FB54E8A36D1A8502BDE3BBC029785378832C17
E391F59ECAB94FA2C39DDE13CB92BD306E78A7CD6E3C88400EF8909F4264C9965D0E8AD63EC5EA54
03F1C8B42753A730D9B1BE5CD3C1AE53381C8E52682A4BD694CB236A8598A195F2686722E2A7D0E9
9AC249A19DDA5C764FD73E55EA4F9153FB64182B9453BB50BE5BB63F453CFD89F7A768BB473B99FB
1457DF423A45D9DCA7B2FD2984475BCCB844CA59AE9F42F7BCCB0CC9996AEE139F89C5750AA38F76
B859182FF3EBB84E71CE8A47E67B527E8A12B54FFAA978EE13770085E883502828F70936ECDC27AD
53945C084AF4BCBB1810E034D728A3534444B7782DABEB140E87A314EAF753246B9FC63D29D44A6F
661F6DE1D19E3CF76938E3BA20AB44F45AAEEE79B71CEF4F5159A788F6BC33566B53B54FB43D497F
0A91FE74ADD89FA2DD9C22D947DB942A68A8031FA9AC53847A3866D63E718FB65CB0B5CACEE9B144
CFBBA44231A3DCA70C3FC558A76095E498251BAA36779DC2E4179175A169D53E09DE11E314233BB6
20178262E895A2897ADE4D90FB14EA46AA47B5D429A27E0A1D2F108C81123BDDA3ED70384AA2119D
82F7A7E0B7A158863F563D8D3CDA993A05928BCCBB55567F0AAB4D926E78C705F7829F82EB14AB8F
55B92BA5729FE0E612D12322B54F391EED9C25AF485D7ABB3945A88F3692053DC27127F5913F17E8
4FA13945B29576299D42A4688AF0012D55F0C78AB94F118562C2DCA7509BF80C9D8267C9C6CBF9F4
7CAC0B9C820FE0526C0287FA54729FCC2B79C84F212FD4EA12ADFBB0D8FD294AF5BC2BAF531859B2
5C898BB4716497F4587F0AAB784FB38921FB5867CCDAFB53381C8E9298872CD99C855C81501F6D53
A730EF53991E6D9EFB242A9A969F1B73077157E23A45BA3F4565A46A9F7405945128228A49CAF829
723CDA5DCB7DD2F138C9FC81FE246D344F9BB0E75D886564F6D12E58B00341677C9E5625F729AFFF
5DB9DCA7D090C6DCA74A7E0A3ED1B2B9339B987524F7C9148E437B0ABEA169D43E21F2739FC6ED47
9942311EC017244DA6312FFD1466CFBB9CC15C32F7896A9F6C5F361BF9A5FC147684ACB5F2A3F77B
ED93C3E1288546B264F51D2773BD8B9ED5579E1C8FF6243A85A8BFBDBDD8325BD4E806FD1491FE14
53CA7D8AF5A7D053AF6FAB495AB64E61660FC63DDAEDE614F97E0A6D681DEA14F7ECE4E814F92891
25AB048BA048C1BECCED7997939033ADDC27C126D801E99E779633285270DE299D82C87228543674
E9C6FD0DE43E153B688BEA26B3966F2C6D24758AAA05ABB91EED400CB81D419093FB14498ED595AB
239BB6EB140EC7A243CF1267FDED1AF753446E49A2F62992FBC46B9FC49257660BBC78EE53C1B57A
41AD7D158BCF79EE539647BB928C9ECC7DB26F4C5A9BE035BA518FB6E9A7B0E75DA3A274EC798731
FEEDE614661FED50A795029518D53E55EB79176113997E0AE1D1D6294F869FE242869FC22411793A
C5946B9F708447740A9C688D748AD0BA2E1FDE7C7AD6054E11D2DD429A321735AAF5BCD3D458E43E
C53DDA22D09BAF05056307463A45C1A31DEA7957A96035A7F68937D4B66DDAEA1A5E2AF7C918DEE6
D8768FB6C3D116546016D5C848E359B2210F05BF25D136EAEF5F3BFC42BC3F450E4C4F6B48A7D0F7
265101650689CCBA3F8559FB84FD2964D4B92E14299685F063B2748A406179C7750ADD473B34FB2A
D88806B54F483D92B54F5C6E0B6D6BE47BB4ED6C1C35C2716789DAA7CC909C09739F42B139E4A7C8
E979471EED787108FB5BE802A7109D49435769330CFC5CD59E77497139EED1E6021C1FC0A17864BB
8F7666CFBB38CAF4BC3378046B6F77537BB473729FA286A09000E73A85C3D162CC42C268DEA3CD26
6039CDEF72FA68670A13A5748AB100A15A5108A39F90300A7E8AA9F6D1366B9F50A7C0BB92EC4F61
AA1513F6A708543D5172C8304BB6337E0AB1969B1CD224C0CD497F8A3129D66A85E5A7A0A192D629
F243922BE73E69498EAFE8A6FC14211261649AD108EF4CED932952E40338C5CE951DF1B633AD7D12
E97CC2ACAD9D14FCBA1DEC4F213CDA65D3BFF37ADE8548B164196CA827740AABE54AAC5A75B46AE4
9CC2E1681F70DE68120ABEB3B24ED148ED1377FC19B656D3E8379877C5B264336270229DEF22FD29
A832442F7989249C71093ABB55A5750A9A8095D429F4E9453F85BE13D9E99ABAF629BB8F7656B579
314BB6DD3DEFCC3EDABA1AC4B0B54EB5E79D295E64F6D196EBBA0179A2849F225FA4987AEE131FE4
797DB4C71EED7543710BE914F0AACE7AB423D76A3EEC27AC7DE2BC23DFA3CD53BEF9A00D39296869
28E6A7307BDE4DA6531438C545830B8B0D293D67EA14F1385955CED7DF58F7DA2787A3CD808B1B10
07BAC4450087E19149A2D154CFBB50F9538E54712EECD1E6B94F652337E33A055FC832166F959FE2
63FF796CB548F4A7A886819FA24A7F8A80B7A2824E11AC36A7907F3E25EB864E21FA68477ADE893D
93F7BCD37C39C74F21ADD9C571AED372781854A2F669029D620AB94F8A71C43DDA0906A10805AF00
EC824EC1577E92B97CF42740AF9A56EE93BE9E47B2640B35AB91E8EFA2AB2296FBA47BDE9519D599
B54F21139CB90494D4294CEB8431D4554DD455EFA3ED70B40BA4503CF5C4D3FBF6EDDB7BEBDEDB3F
B00C1BABABAB11B20033B70FBFE7C3303D06C0C19B9BA777C32A462359B2FC5E732E5C7F6E2E7FE5
D43E55433CF74978B4C5822DDFCF6373A49FC2AC7DAA2652046A9FB89F625C2832A012A65421FB53
54D02922A9FE5DAA7D32739FE29A05FF32BFE7DD74FB53088FB66E135658E6E51BC9FE14990A4519
9D222BF749AB15499DA2481622E1667CA3533A45F2E27CCE32591026CC7D12990388441FED62D593
EEE4A817886E37FB684FA85364F7BC133A85BD2E14102F62FD298E19FEEB60C953B159BCEB140E47
FBB0B27268DFE0DF81FD07F78DFE2D2D2DDDB8F69A3812669878304D356103BE849DA1376FB0F689
D304215584348B78EED3E31F38C797BC4ADD9EE23A058F3A0F45E20C8F194DC0B89FC2D62978CFBB
6A08E73E09FA60CFC18A39E715729F0A65BA8166ACE3DAA7CEE43E15126FC26D1C0566E1A7C8AC7D
E21E6D994510CD80CAAD7D2ABB9C3B61EDD34563BD37A2536CDFF32E121D126C6234D7E2EBBD9DD2
29F478167A5C689C4FA5E79D89889F82EB11FCFA1CAAE81B53E6CC9E77254775BCF68906B3345614
4DD9B2EB4A52A7485927ECFD8347E7140E47CBF0D0F1073923B8F9E31FADADAD11C51007C3B574EF
AD7B85EED0EBF5E060781FF3FD1BF4688BB04171632A248A14F3FCE198CC3EDAA5EE5070C0D1F523
E26D797F0A9DDB2F566E0B39E7EC18DDF36EDC9F82DF7ACAF7D10ED53E9972B9510A652D7C853885
30FAE14D2792F824BFEC869F420CD43895A02CD91C3F45642197573DF1719EAC7DEA2B14C5189C98
694875D32E57FB945CCE656951494E9193FB44878DFB53C4FD14C796883504650BABCF63173885BE
56873438835F4CAFE79D38DED429800E904E219972A86C55741DCAE94F9139BC4BE9147C6C737159
05230F35E88BC367E37DB46DA61C922DD6C72AB3D73E391C6DC22B3F7C451307C0C6C606D28AADAD
E769E7AB3B2F03A100E8E361271CAC6337769BF353085F76A68D225EFB14EAA34D779FE44D4AF7D1
EE3D7B7EAC5330E95C371716D3307A49224BB61A0677A860EE93F0F199E5E5AA1A2AA78FB65CDA32
9D7D4AA1E848ED93D9471BE8301FD8C494F5800FF929669AFBC43D14A11291887B28374BB614591E
6CA4750AD15638E4ADC8F153DCBD87EAF4224D28E47E66A9E802A708E91411D64C7B60604FB1E79D
BCBC076A9F88236B852254B9CA3DDA869FE2ECD7879C82B3897C9122A053187E8AC0B24FB093E9B7
6F0BE9147D6559D73E296FC5FFBFBDB78B91FCBAEEC41E4245DA5D65236967B5C6662444B434D66A
C5CD9A6842824D6A03796468A9E12016BF60796298A2A095AC8962900323803844247B3716232530
ED9128621E161217312713C69608465028250818DA00477E30180736980761483890F81244D8A746
905375AA4E9DFFF9BAE7FEABAAABA7FEA771D0A8AEAFEEAEBA75EF3DF7F7E54DE385535455ED53C1
E78B770D54300B9D387102DA84AF5DBA4457FEC6839F826B6826E4853D8826F61CEE484F61AC4716
F1492F550DEE53FA8CCBAB58A3CD375D662005351A62E5DA708EF6B2A0A7D02F2FEA294443E1F2CF
B59E22A7D196D6373E5164917937BF7EBF7B8AA49EC2DC956D554F01776BF614BA5F30C7B9B83026
F32E3ECE5D56D053F07ED9555258FEC9F0F008A7A09EE2AC35920500478DF3DC2AAAF4141A5C1635
DAF7A939C8E31C6D3BC331BC9CF27DF2BA060FA118E5FB1404388ACB2E4EB1C4883D4D90AD835BDE
5A384555D53E15CAABCD429C023F8078588DD7E809F670BEDB4189B7BEE9E8F51486463BE72242FB
31A82D69B4354E31C8BC33CF6FCDB3AF8C9E826BB4476CBD1CEE13E214B8E248337FE50125F51421
F709710A4FE517ECC1E8A6FDEE29CC1CED6EDF27D553E47D9F82A623D0532C4A4959BDEC157E7D1B
A748B6CC23700AB5AD72D10AEA2C429C6271A23BE4EC0D103727A8623A7A8A41AE90C35315F044AB
A7E8E03E21C14F78C9DA38C527EF459CC248A3708C35E883D0E829849E228F502CAFCFE82904BE6C
F6CE03A294DF530C1470A1C3860955144E515575A35760E8C46FC20EE2EAD5ABF8236C3C90E06482
0E57BFF914DE0AD397B869273DC565A5D1D6111531AA6EE7682BEE93DE5CC56B5613A7B08D079D48
D603DFF769244EA1B84FFAE5B5BD6495B1B9ED251BE653B86E848E21E1EAEC774AF914975B3E66DE
7E2CA3D1CE6882CC2BDFF7860FE83FF8673E72FF8AFB34D45318BA219EBD92C9A7D83E4E1118E348
17D9584F81DC274BB56AF711426A71CF8929E0143CA59403137C26F7A6EE663E45BE161A8A25AE01
97238DB670FF6E89296CEE13F51482FB140C6F6FEA0EB94F0369B689262FAFD41115AE465B4DD45A
52E12117A5A7A8AADA9B8A7325E0D613274E409B40777BF6EAF3D865C00CA01F0B57224E01CD85B8
E9E8B94FF79D7ED0CBD1F6284F6233B625EE53A0D1F6F414F691D795418EB6C97D1AA3A7D02B54AC
D1FE633B7458F61A3DDCA7C8DC29D87D2D97B329E01462DF75F5CE9791D414109F686F365A4FD11C
F94DDF27D14AF0D1ABCDCD689396D553E4718AAECC3BB1FBC2F6015D71CC46C3C129BEFF4B6FA6FE
D78024CCEC95E1A1EE14700A73C43631E575B84F7A54A3E682EB296C2FD94FDE7BF083970507D59C
A8F9F8A7E95D6AB4BD7C8AB87D70C06568E4F507DCC62984988206369FAE9773B8CB7DF24DBF3357
224E312E4E775C554F5155B5BDF23ECBB00E4283409F3EB81BB19BF8A9023D9C500CBD15879D003C
F675F5B5BD6984EB29BCBD566CFA64FA3EC112E069B4F33D85C77D32994E9A16C2BFD36A75FADBD7
84EF13BCB632473B8F50B0232FEA2930E2107FC5F9BBDEF8BA7223F4100A43A31D729F5C00DD239C
2FDDD111A778F8E21777FE99DA5E7939DA6613A10D934D9C02DED6B8A70872E1933D8518B45CB5AD
110AC115C19EE22787FFF6D0CBA7C8F7CBB9CCBB413E85EE2F9C5E63855338997703A0EDACD329FB
A411D1531CE51EEC688A739F9A8A21E14BC07B0AF1CA8CF07D7AF467076A6E13A740DF2741D873E7
6A065B503E05C729AEBF7A7D31542E3D3EC8A73067661FA168729F5C3D8539CE935EB24B8CD8307A
72A8501CC278E2036FD16BC1A6CADC75544F5155B5D9CAAC47A74F9F8606815F03F32AF6147F7DED
2FF5FD61D67AEFDBDF0B77D007C5F041864F31DC8117EC785F7FEDC7B45F5DFF3BFF71A1A7B82B3A
E38ACFBE605D232109BD5CB08EAC035260514F41CF0C3DC5CF3DFEB409A307075F014E81CF3CC029
4668FD8638051F3303EE93C90F11723F2B47FBD967BF239E79C17DE20A56BFADD0FC28D268F3C170
435FA01FE932F614B4E922E8216971267A0A7A5AAFA7E84A1CA69E82FFFDE425CB33EF8807D5248D
A0469BB61CB4F532BC6433FDB2C22978C80E7EC0673885D935E065E6DB6F73FF963D057F119EBCFC
E422F30E87EB7D27645F6C8D64228DE087027B0A6F90ECC177CFF7298095F93530B03F73EA31EC29
F8A4DDE5FBE4B5CFD853F0BF16710A91D5286669D37680E314A2A780675E709F62D3A7185C5E729F
E8758082255BF414BA95182883965A397E13F61462D25EE829EE7F6760F4343808524EE0D85398D3
DDFA17601184FF1DBED3C6035EEA40435A5555B58D82CF38F40E2FBEF8437E25056D6BC5C4E1BCA7
409C421F02FCF4C8BD6431F32EE81ABCCBB36EE2AE17622F597EE415EFB2929977845308B293074F
E0D6CBD55368EE5373DF15548BFB346037890D18F17245E65D885348DF2721A6D0DFB99E62DF35DA
663E85A9D736C038F27D0AB94F79A32731C833BE4F9E68C81453181AED2545A42F47DBB943805318
969BDA1547F5CB8B1C6D4FA3CDF329422E9F79D83B1DEE9337982FB3F03B63566FE92902DFA7E6B0
0FF41434500567CF4399394ED1F67D6A9EFF5895D2682B1842F717E2531068B40364D91EED7487B9
9E42EC34B65D855354551D7199B208E80BB0A730670098B5506DC1BD67B176A5D11E5D1CA7E0B591
CCBB40A3BD6A2586C75CE6068C3A0B4F4F31C8BCEB4228D89197994F31EB29962B4EB4015397BBBC
641B1CDDE18F8BCBD3D053644CCCCCFB6C23479BAAADA7187A0BC42D33D768D35019A9A7109D456F
E69D47E7B3F66341E61D8E6D4911B1B22A5663DEE73EED5F11F7C91CCF81B68286FA38DF276FD2E6
1A6D2F9F427B7A6B90C2832A6CEE532647BB35E033DCA701C1CF44E5F8A86E719F783E45FC5D4FE6
E5FB5455B5DF059B5ED8BAE89DE44B3F781E7B0A93FD88D086C8C8C33A7A9C027B0A0F8F68DE94C1
29C47A943FDDF5700A7EA8E59E7431E79C554A85D0532C7D9F523845B32C9C02F514E2CCD6D65058
183A2C6A99CC3BAF7D1000FAC29F734A3845CC7432A9230867E085FB6E5BCB4BD61BEA014E71C05C
34B5585B6FBD52997779DF27D54DC43885EB251BE2146D3D05760DF3B089819FBFAFA158D1A5EE7F
E7A47C9FE2D6D81CEA9BF27D32BD643D9C224AA670541538BC535EB29E5C281EF34DDF27E23E3D37
98A50D1FB3216735C8D1D6BC26C3A0CFC9BC2BDFA7AAAA3DAE0B171EF63E7130EFC55EB270D38913
27E8C8853FF0E87D9F32CB906086883B98C798B4DC8CD66837710A1E6967EEB8A895E01BB031BE4F
4D0A7A80533CF72E62DEDAF9146C61D2CCF3D84B56B71286CB93E5793E119C820FDAB89B18FC386F
2BC6E114B13D7206A7E03B2EED6FE681141DBE4F396FE40C4E3180DE3C54A2A5A7E0B5F27DA2717B
D6C958E17DC4F02330059C228B235B8D4633473BE63EC55D468053685701175C1E5A3FD95EB266E6
9D07BA399376CAF769487F6A74D0618E766C3560F7CBEC42E1145555FB5AF0598B3F6E28DCE6DB36
DA70A282FBE0E0403F6AD6533CF9C451FE2322F32E76F2177B3012C06E29F3CEC029BEF534F414A8
60D587BAFAF88BA318C4546FEB29E850B7CBD2FF9FFC9A7E798D7C0A41CDB5FA0B7EA18FFB14A0E7
CB1FE9F244F229029CC21BF0014E217A8A3CFAC615434DDF277172AB4D38F9A92FB5D5114ED11CCC
5EE65D1EA7080633F5114BE1F6424F11E468B79A62CDF49B544FC135DA26946C9E08B5F229FA7C9F
10A71052B808A7D013F570BA3675167D5EB24984E2A10EEE9381500CE1099D551AE314FC60C7E4F2
057A8AEA29AAAAF6AF60BBE2F92150EC1D7C18A1713873E68CB8037416F79EBDDB14681FEE8EFB24
4E68CD15CAC0D05BF914DC4BB637AEA299793740222C0B110E67D0915783FBD48B50F0135D5FA38D
8C262F9F82AEECD568DBCB53E0FE343D9C22208478392C688FBC4E3E4513B988710AEEFB3420F55D
B1100A4FA3DD9BA3AD6821199C42EA29149D2F08736C7BC912F7C90B0253BE4FF89029709FCC21ED
292C74EFDCE43E8DF67DF2700A4E46355D9EB4CEA243A3CD8777DEDF2CE43ED1D0957A0A13A160E6
1BAF077A0ADD0B3B039BB7D5747654384555D59ED5CCC2EE5B4F7FFA130FD08FFC56988E60FEA4EB
4FBEE32406E189BBE1F53437F2DA95465B7B3A25B1756C434668B49B47BBB1EF93F0741256218234
2248500DEED33890629EA3AD7B8ADF3AFDA6014E21B6555CB56A9ACAF668B4CDF588F8210332C97D
93D3539097AC378C5767B9F3B6DAD36877E929F4B92E8EF9A6469BC370E6EE4B2014074B2FD9F17A
8AB0CBC8FA3E99E4104D416FE114B37ED933EAF703EF26A8D18EF329449610AF7532EF743791C9D1
166C5501496890E260E8257BE6D7CF8B81DDC0297295C129F408976E03FD7A8ABCF5131FF6855354
55ED596140F6C1C1017EC70BB7DD720714B409274E9C802D076D2C31F94E7C2A11BFD08E4F58C7C1
4BD6F424D4CC10DE7DD838C5CD97398C2ED6A02659B7A1A7581EE49A5B2FB130D17226F5141CA738
FDA8DD4724718A876CEED38377BE4578C9DA51DA8ED62FEE2902E984ED07C5BD64A7815378A3FA72
68B6B9F2921DE5FBB44E8EF662D0C66AD6A18CA2ADD1CEF83E99C3BE2B473B835070542E9979174A
8404EB099BE589F414DE9C6C8214FC02E55388A71D917997C729704E36EC9EF4852B2FF00644E214
9CFBA433EFBABE5B3845CC7D8A3A6506D835710ADE3244EEDF43E950F5145555FB54F07186AEE194
FFF5D13B3E2C1E72E1C2C3D06BA0BF13F40B70C18CBAA33A7A8DB6E43E596B135D3623592F5B1AED
073EF759B371F0AEECCDA71030BAD74488BCB0B69E42ECA9F295E03ED99BAE706D8AB94FAE9E22E4
41D1E5FDEE29A06D87BEC06380045BAFD8F7A937479BEFCA3A34DAC3046D015568A905FEE87AC966
42B435978FDDDAEE299A0885D656F47AC906BAA1E18529F4144D4C3962FD257C9F3271F0A2323885
E9EF64682BAE0CA00A57A3BD059C42739F04BEE6B618EC9AB69E8275CD912EBB34DA5555FB5BB0DE
413D79F949F3027C3713EEAEBF7AFDCC9933B7DD72077413F79EBD1B7ED43B4FAA9D709F061AED61
2E58507CCD8A35DA74E4657613C1224539DA545C4F619A3E99326D8E539CFEB6D153ACF229328621
CE196FE0FBE4AE440ED59C5FCE7BC97A2BD1EA3E439C62BF35DA8853C47C27DD6BF09E2283537439
243733EFB446DB842AF829EEAAFB0834DA9F6FE5687BCDC5B2B308B84FAFA7C5142259BB434F7176
183A2C7E5491C453E829161A6D6BA2F64E7E786DC9F729C02984959927A3D0F3B9D468079977BD50
4513A730F514E6384FE3147653ECB910CC7FA4B3A3C229AAAAA65C41EF103C6427DC276FEBA50FBB
8C346D27F38EB84FDE892E9E6B0577887D9FB43D8EB94271FAAED4536434DA3D38857E7917F914A6
68426FBAD8B23532F34E9F7DE9756A1ADC27D3F709B65B02BC304D65C9D36CCD7C0AAFBA34DAC2F1
498A891851AAA1A7F0D037B3B9E8C52942A8C26CA8839E628153588EFD86A095FF086DC83D27A6A3
D116BE4F30BCB9C2825F2F90B88DF83E995DB3D7530C5C913DC042D99D11F769855398BE4F1924CE
9AB4333885AB1832C7790BA7A074781B6BF3A3584AA35D55B54F35A241089EC47BB69D78C9364F6E
6324BD8953F47613414F413885207E18A2BF2B12A7A0A5AADB4B360955A47D9F5CD15F27F7291345
C15B0CCCBC9B08F789700AE32077A9A4D0BB2FC2293698A3ADB18CBC465B6FBAC460E6578ED45304
C7BC2D9CA2E9FB348027180FAAE125BB4CAFB3A513CE666C821A6D1AB1F1ECADC7FF06B94FA2EFF0
B84FFC54C760F439504503A7E0DCA7117A8AA6465BB70F1667B50FA7509D82664379F845E1145555
93AD713DC84EBC6439F7C93BB98DFB8E64E65DB29B88700AE6FBE411CBB5CA6FB536093DC506718A
24F7492D5266744592FB64E2149A0A22EF33991CEDDFFFF0F7A4E180C22944A331683AFAF514F176
AB83FBA43A08AC01E8A61C0946FA3E25A443A9CCBB8482950FEC6CE69D48160E3663CB6B26D253E0
58E5D14231DF8977CD6BFA3E0563DEEB29B88F131FC64D55C5FBFFE8FBEDCCBB2DE0148646DB4C63
E9C429623F8D2893A5708AAAAA3DAA648FB0269C71F47A0AE23E35A10A01A037718AD8DC2983AA07
7A0ADD3888935BD30BDDD353489C62C4A997C37D5AF93E897C0ACB18479F77E57D9FCC2D96B8F595
8F4FCEF76946055119D9027D133D356DD5369EA34D95D468EB6D183725D0B6C96372B4E3D13EDAF7
C9EB2FD861EF88CC3B7B784F32F3CE1CCF62F61617B8B5EC0633EFE8473C2F0AB84F0BA30C9E94AD
A668AD808B7C9FD6C1299C9EC2C8D1B6A8AAABB9BA534F11F411A6E100FF14144E515555D5553BCC
BCE3E7B41E2DA40BA710CB53F0E3450B6A0F700AED641EE8FB223D85974F310EAAB0B84F984F4130
BA6724EB196F8ED453788E227336EF62EBB5EF1A6DD25370DB019D3BEC95977917E829F2E4F3A446
5BB86EF261CFAFA43B377C9F3C39B67384CB7F6CE7689BCD8585567478C95A1AEDB8CB984E4FD1CC
D1BE1C46E06DC3F7A981535C59F1FAF877318607B4D52B564FB1299CC2E13E0D708A582B344A4F61
8750C424A8E5EC5D384555559559919EE2C8BD6407BE4F39005DA429C5F914BD3BAEA0A710BE4F62
5B45DF3939449C7AA534DA238EBC1CEE13E214AB2D9672209447BB6A91EAD553B8D26CA1D1DE77EE
13CFA7F0CE72451FCDAF4FE2145DA33AA9A7E01A6DE184CC49E6D4262F867A464F2160B80CFA16E2
1491EF93B7E9A2E622D468BBA3DAF73AA3DDD7147A0A7360372DCE5A1AED0EDF27B24DE61F8100A7
10B92AE605ADB3580CF2A4976CE78CDD85537029908D35B3BB25718A289942F7CBE5FB545555D559
C781FBE4F1A0A89B206B913FBCEB050FA7E0F9149984BB644F217C9FE4B6CAB3D964F7840DDB7BCE
FC1ABDDA7821CABC4B96C37D423D85E63E992C1189A18FD253081EC860C7B5F4929D7DDF77EE13E2
14A247463D85D87A998A210FA718A1D1D667BF5D39DA0277E3BB2FC13F5F2B473BC429C6F93E1968
C5523494D268877AD5404034859E226992EC1D13ADC37DE2212C7A626FE7535C199CFCF0895A03D0
0DEED3389C22E43E457A0AB359EECFD1F600E541BF6C51580BA7A8AAAAEAAA1D729FF4FE8AD8E6CD
65ABCBF7297FBA1BE82938E549AF53033B912BFD38C538A8C2F77D3234DAA1A96C92FB641CE45A2B
91BCCF94700A6A90098C0BFC07F8805FDFF7496B5A333885D66873D1C4200E4F9DF1AE95A3ED10CE
F14287EF93E23E191BB090FBF4FCC7DEFA8A97B16291FA065DC69C2E35052FD9668EB6ECA6871604
E3B84FCD793BC02906B603CAEBC9708262A1A5299CA217A448FB3E0DC038CF7C60E8AD11FB3E193E
038EB6424CDD855354555575D54EBC64C765DE5D6EE92988FBC4710A61FD448B546F8EF6EA9C969F
E53A54284E14B9B599A3DD7BA2BB2C93FBC4F514C22467E1B169129F289FA207A7706921C30B7479
BF7B0AEDFBA48515E6EE6B81C1F5EB29F25BAF2E8DB6E63E19DE38F3FB443845EF781E7E04F2BE4F
AEB2B587FBA4794D016C3138E99D8C976C667E16DDF4E58D729FBC0A7C9FE411D0B047D6F3B68D53
0499771BC5290698F273CE1876B84F513E856FA3E1119FCAF7A9AAAA6A5CED2AF34E1C67096B59BD
F5226D20A683F5E214C963DE404FE12553E813307EE72DE668277C9FF451EDC0C0DFE43E3DF7AE3E
3D45CCCEE54BD8BEF7145D39DAE6362C895374456967710A65F7640A2B040B7D7C8EB6A7A4E8F47D
32B87C267891F37D12F9142E098A431593D153708DB61EBD66B3CC6B839977749FD8F7C9D0B85920
853D7B2733EFF220C5B21A3885E23E09FBB2D5C01E0EFEAE7C0AE3474740543845555555571DBD9E
C2CCBCF3A00A3B6BFB437FEA69B4B99E62231A6DE1FBE419E398F63886EFD346F3299ADCA7D5A1AE
295CB5CEC1BA700A9316A2918B5942F104B84F991CED801FB2C1CC3BBDFBCA6BB43DD5AAC9421F99
A39DC02932BE4F3606A74C093238C52247DBEA1D5C46DFC4F4147AC6BEECB822F7F4145BF47D32F1
35E31A95581AE11442A39D1CD59D384583AD6AF5D1A97C8A7437B1F8B17C9FAAAAAA3A6B577A0A93
FEA19D70BCD6C3D36837375771057A0A4D73D2A75E83935E534FB1C17C8A30F34E46B26A96081D82
F568B40539C444D2F9C234EB41EE7F276AB4E1E10F5FFCE2CE07FCF6CACBD1D6DCA7CB4BE7017EB7
63A2D1F6063977F8EFCEA7C89FE5F6FA3E05FC90DE1C6D8653CCFA0B3F146CD04A4F494F117410A6
BB2CF51D6B66DE351AE78C9E42E111A6C202077FA4A7E01AED2E906209C6A572B4C5E46C8EF0344E
1137C5F6202F3D455555D5A83A7A2F59C17D32710A53CACA6B1DEE535F3EC5D0F7C9946993CA95BE
77EB294680140EF709F514D287301937FCEDB1F9148EEE6FE025BBEFDC2733475B0370DCC74C74D0
264E91D153883D98DE8CE535DA42283468319629C3D4806CC6F7C9DA83E5B94FB6D7936AA8039C42
E768377C9F269CA3ADC7B36C961DE2DF9ADC279AB479E6DDC550A3CD09A883EF3E78B11ADB99CCBB
7EF4EDA6663EC59CFB24867184C775E114718EB665095E384555555557ED84FB14E453C4E177A4A7
F034DA1BF79215BE4FB800693B59DC7D69931CD453ACBC64379A4F61729F067A0AA6C23604DAC3F3
AEC52959DAF72976C8A16E6295A3BDEFDC2733479B0F75BCAC9BE558A31DE453880D98D92667700A
5BB87A459EE8D23DB99EC2CEBCFBBC9FA3ED5BF7F3FEBA8FFB6499E10C8E79C7E6684727BAECA629
F41401A6EC4DD7BCBF387AEE9374E45311A542643198BD9B38457E600F4543A97C8A80C5C76F6577
88F4149E20C81BE474FDD9CAA7A8AAAAEAAB63E525EB15D700E29D3338C506F514D248F6199704C5
594F869E62A339DA19EE9394F80DDD424CCBCD5E9C420857B579CE747C9F748E36EDC162600E4638
C6586C564F91E13E094F33739745035E9CF7367C9FB6805348DF270B7A5BD9E60C5B8F8E1C6D6BA3
650EF849719FBCE93A98C337EBFBD49579C787ABB678D2B0C58004D8D45334276D73F037F3299C9E
C288C34BE753CC4E75ACEE38D2562CE7F0C229AAAAAABA6A275EB2E6A19679DEE531786D9CC2C9BC
CB3717819E82E84F7CD3B538B9B59C3639EC6E6BB479E6DD089C22CCBC238D365F9BECD08AA1CF79
DC53D8E6B16A2532B0F509709FCC1C6DF22B6B6EC0F0FAFB6E1BD95304F84513A7101A6DEDF5C49D
A0B88D7F778E76CB156784EF934DE43359E8E91CED24DB7C821A6D4DE4D35428739C6FD0F7892AF6
7D12D2B6065AB124419146DBF67DE23845738AB6AECCE7689BBA6C3DDA9B39DA7AD0C63FF243A1C2
29AAAAAABA6AE75EB2DCC05F60135ECA928953C41AED4C7311E3147CC7E5796C72D15F879E627459
DC27585CEC35485F299C7396578ED153F8B1770B3DC564700AD3D9C96B22F88FA3F5143C65D8AB2E
8DB6C18362839CB65E1BD053788EB2FD997762606B842EE525AB69211EDB7CD8474FA1A71026C931
4955D708EE53D233D9CDA7588E521E4ECA87F48AB03AF4D9D80C4E61D6881C6D16041F709FF239DA
B2230E7C9F4A4F515555D559C7C44B362E111CE6E929C49157FEBC2BE82984EF13273B4959DF332F
68CBD98E7C8A8DFA3E09CF90460CEB7A7A8A68333647DEE9D6FDEE29BC1C6DAF89B8CC487D2B8D76
C2F7299F4F3122475B44B150832CC85178B7AC9EA207A188710A9D79F7BA900B29AB1CFA2C747BC9
B6F82193C229329EB157EF7CD91CF0CD1CEDD19977014E2160354343F1CCAAE9E0C74429DFA77193
76CEF7C94ED076B84F4D9C42B7121EA62CE7F6D25354555575D64EF41442A36D5A3C79675F819E82
34DA4D60C23CDA6DFB3E5D196CB156C6833AF98EF666DBCBA708B94FD42644A1C31AA76865DE2571
0AB98A4DD2F7C963987BC638DBC8A7487AC99ADE02D4230BA2C8EACAA69EA20BA118E21449EE9301
C359DE38CD9E42EA29CECAAEC1D3564C474F11036D0DE33E574FD1C77D42DF27719F18A7D0067DC2
EE49F41A59DFA7B122B8543E85B6901568B21ADB599C22CEAA506D45E1145555555DB52B2F598EA4
8B13B020863899A33D0EA7B818EA29A4EFD390ECE4AD535BCDA76866DED9DC7287974B17F27A0ADB
2DC44F0ADBEF9EC2CCD1F634AD7A57B6917C0AB1DDEAC0294CB1B6DA8C89813D3E47BB8553647C9F
240F448DEA7C8EF602A7B0F821FA50576FCFA68353903E2810659BD6B2EBFB3EE141909ED8DB7A0A
CB46C318E1571867D5C329B897EC28115C578EB66B09CEB84F6D3D85496DB2FA086144F04AE56857
555575D671E03E356D4384EF939BA3BD4D3D85A4945B1BADB69E62A339DA26F709F32988C8644827
86FD85085AEAD253C4CEE7F47D6ABE4FC1487649239BC8A7D0433D8953708DB618E47A2746D7AF95
A33D16A79099779EF38026A5C738058EE1254E913FD79D8E9EA239AA3D85055EBF0EF729500C45BE
4F57ECBE587CE763DBC629BCCCBBAEE9DAE73E699C62600065390F74FB3E0559A58E1354E1145555
55236AE7DC27628934594F7896EBE92950A3CD7D9F84130E9A84F4F6142B3D4562A325700A649598
7A0A8953245359154EA15FDE553E0597F8390885C17D4AE3145ABBDAB0C7D9F79EC2CCD10E8C71F4
D8EECDA74836CBCD9E826BB4C508D7CA563EF847E668272AEFFBA40F755FD78CF410A7D099779E73
ACC7839A424F4139DA3A173E409669861FC17D8AF3EF2EB67C9F385BD54BA910C37BD58378BE4FE3
700A56BD7A0AC3036AC88CCAF83E6970CDBB2C26F6C229AAAAAABAEAE8BD6489FBD40429B8E24FEC
C4D6C9D11E87538895483075F5B2C58DA1B6E2FB34D753C4DCA7819EC2E33B093249AFEF9379CC35
349B9D3DEAFE77EE7D4F61E66807FD85C0E07AF5145DAE9B798D363FDA151221CD1519E3FBD412B1
E2DD3A7C9F0454E191A3127A0A1B77D32C11654A30919E226E1CF82CBDB03B6393F6A632EF049611
E753887658AB2ACCD3A1144ED18B5024700A97FBA4418A1E3D8541DECB9C08CDAF2C9CA2AAAAAAAB
769879179C6EC1EE4B240E8B06C4CBD1E6475E23DA0A4F4FC1CD9DB47308BF956FCC48FDDAE03E8D
03297299773680BE5C980C2FD9244E6171CEF51E0CBB89A9F93E0912482623ACCBF7A9D91AF7E214
820A225039AE15125C912C4E910129D6F37D72118A9C9E22D068C75DC674708A80E094F492BDF6D2
35F1B4EB67DE5D0C35DA3A51481FFB88793BABA7E0D4BE3448312647DB390E5AB999F5E829CCBED8
B3812A3D455555556FED444F21B94FE1DA645E3902A7100E9C59DF27954F21E4D87C33661EF3BAF9
143CF36E5C59DC27D45368198581A70F718A591BD2E23E79388527AC58E5534C80FBE4E56827F75D
EBEB29B440BB37475BD342282F8CEEC089226372B4733845B7EF53603B40DBB074E69DE733606AB4
A1139982EFD342A3AD884FA235BEACC41497232FD93EDF277D581467DED9D64F61AFD1F07D5A13A7
087D9F84B42D100A75E9296CC301EDAD61A118855354555575D5AE32EF32487AD05C783885A7A748
6EC9744F71F59B4F9146BBA9C826BE135DD3D65324F75A69DFA7959E4268FD2CDAB9B4F16F719F32
A26CF3E06B3A38C5E525819C1037AFB91066FEE3708AE420CF6BB48570D53622584A2DBA73B4D338
45B7EF93DE74F1019FD4683BD884BC6649509F384E21E0E6A06B6E6AB4CD431E3DCECDFB78380577
423E50D64F3857AF62EF044E97C1297A418A713885E36926CE85DC9E6239623DAFA7E0C7D2535455
55F5D6D17BC90ADF27EFC88BAF50B431A33B8FCBD11ED153204E21C2EC3C5EEE2A086FB89665B94F
5DF5F00B19EE93A07F68FA93C42C7AF22904DF491CF3D23593F27D8A76568174C8F77DF234DA9BD2
531839DA575CE86DE042D09BA39D0B058B718AC8F7091B071D4EB1BC9BC77D7AFE636F0D28E5B2D1
501E50D3D15360FBC0A570C923A0CB2D3D85C97D4A8EF0404F41EDB08695054811E114A3F514B9CC
3BADA71083DCCE184A6AB49D1CEDC0098A77CD855354555575D56EB94FE622C575163A8C38C029BA
3A08F3C82BC8A7301720BDDDE21411434F11E453F482140FD9DCA7413EC5D0C35C1F73E93CD698FB
94348F5DEDC4D0C9701ADCA7668EB6B7EFC24663537A0A735796D7686B6E3947DFB8E5E6FBFFE8FB
1DBE4F9D35DEF749A315F429C86BB4356C21B8227CEB75CF89E9709F3806E7B50FA625D446B84FE6
341EE753E0D836666CCD5F5D1E0A493D85E7FBD40B523C64739F5CDF275D563E4583FBC4FCCA1A93
B642E5FEAFB37FAF708AAAAAAAAEDA15F729797E6BDED3CBBC13DCA7DEA3DD463E8517C66A5AFA3F
B322A29B7A8AEBAF5E5F3F9FC2E43E9DBFEB8DF6464B1FE40EF514E3F229023185D65CEC774F61FA
3E5DBDF365B852F343743418DE6D83F914BC6BCE6BB4C5601EE014CC27F9D638F32E30923D12DF27
4DF0EBD2687B0C7301C32D7662D3D053E4E767F3A64DF93E095CA39D4F91C8A4E097A10D89708AAE
1CED4E9CC2E53EA941AECF82029C6276AAE358F309CEAA3622A89EA2AAAAAAAB8EDE4B5667DE79A7
5E3A3809CF7261F795C42932B114714F4138851DC3EA3BCA723DC57BCE2C000583FB14640DB70EC1
52BE4F5C3181DD8458A4B8463BAFA7709627497F9A12F749E7682F087BF3712B984EDE7EECBEDBB2
5EB21E24610684E573B48571199EF40A84823A916C8E76A636EBFB24D08A164EB1C8D10E294F9E01
D444F4143C475B63CA4D3FA87532EFE2F2F41482953A18C04E10DE6A4AF7F4149CFBD40F528CCCD1
76BA0C7E658453E811EB78269746BBAAAA6ACDDA9597AC67F1D4042C3C3D05E6689B3885079AE771
0A2DE5D3E75D0243A71CED06F76974CDF329F4CB6BE753988B91B5EF8AB94F039C22F63CB7BA8CFD
EE29CC1CED7848F34F416F3E85EE2682EACAD1E6F12B3A116CA1B388BD64758E7692D1B751DF2729
B58835DA0CA798F51716C3DC1BE153E829F8F4EB8DF0C00F6A1CF7499BF53DFAB3AB5B639CC2D4BE
89BCF8543E85C97D8A710A7F84A7700AD535ACB415F373A13E9C425999D99A38755204DFABA7A8AA
AAEAAAA3D75368EE53B018695B422C2F473B8F9B9B15E9290252AEBADECDA7D8748E76ECFB141145
B46A3BC17DD22A6C8388AB157F4B3DC517BEF47B3B1FF0DB2B33475B0C5DF27AD2DCA7117A8A3CF9
3CAFD1D60E03DC4B76207ACDE753646A78BABB01DF27451DC9E7683794DAE2C234B84F1C80CB7413
FCCEEBFB3E8DC029B4D50029834C9082E3CB299CA217A4C8E0144990620D9CC26E312CF550719FAA
AAAA7A6B27380597AF6AD09C4CCEB5D02FD65378475EFC424085B2718A6F3D3DC8BCBB3244285454
ABB97E357C9FC696C97DC27C0A5BEEA792EF8CBE63548E76D04DBC7CDF84708AA69E22A8643E4597
887584469BEB2684DD9966A1434F4143A5E1FB944428D6F17DE2043FC1861A9BA31D5D60637E0A38
85481432BB09CD7AA20B1BE13E214E31C326D8A720CED1E6718DDA974FA21534CFAFA9A7F0E54219
DF27CF52430FE90C4E9134003789AC855354555575D5B1F5920D0AEE1F6BB4C755ACA730984E5756
FC90C1FA35CCD11EE8294CEED33890C2E13E214E211C6F34D57C95A32DF414A1976C4CC4D59B2E81
53EC774F61E6689B74118DBB6D30477B044E21F8E40384C20AD4A6967933388575C0BB01DF276D33
DBEA295639DA7A8B75F6E42B8222C2DA9029F4145D53B49EB1B7E1FB840744CD7C0A1AE142131408
2B5CDFA78C9E22ACBE7C8AA1E1807D40D4C2293C21B6703033DB8DC229AAAAAABA6A275EB2971D3D
85B7E9126E39B0FB8A35DA193D4516A758FA3E99C472413E373A8B209F621339DA9E469BDA84208C
559C808DC0290430A1DB8AD9FDEF7F277EDFFB9EC2CCD1F64E71390687E8C6683D85094C74E3142A
509BA36F22509B467E36473BAF665DD3F7C932C969729F748EB627A030F1B849E11442A32D067680
C78DE03E79B603729037F514CAE2DB5452A4F2299A3845CBD62C93A34D761946343C5DDF83537820
8539699746BBAAAA6A74EDDC4B562F529976C3D368E7375D498D36E21424B8F69C36B52167369F22
B91875FA3E05F17646D4DDF0C2887C8A8022B2C2292699A39D04E3F29977DEA836197D746597465B
D042787381DFE9E3D091A3BD219CA2C3F7A99FFBE41AE3784A0A1CE1D3D053C413B2A0B3EA5E634B
BE4F014E2179AAEACC477B83DBF9145D7A8AB87239DA763E85636BD68553E8116E4EE38BBB9D2D3D
455555555F1D432F59BE4ED9B0C587FED4D36873DF277D66DBF495D51A6D2F9F422782D934DD675E
30B94FAB7C8AF5700AFDF2723DC56003A6E089115EB2114EE1739F162BD434700AD111A39E228809
C37683C2EFB497ACCEA7C8E4AD889BF21A6D3EAA5D879CE51EAC2347BB0952E4F41429DF27DA800D
33B5BBBD643DC3E4E1DE6C3A3805360E036F81A5E7403C7B6F89FB94C129CCA638E89D17F3FC68DF
A7D64150464FD1EE23C47CFEC7EF4EEA293CB6AA377B174E515555D555C74AA37D7928C7D6B005DD
39D668EBFD5526A5C2C429AE7EF3A9289FE28AC115C14D978D53041AED6DF83EB5D626234729A9A7
300358F5A68B72B427A0D1F672B4A965F030B8D91D36C47DF22AA9D1267693E93F300029622FD9A4
9EC26F31D6F27D12A491A5C5593EF36EE1C0495DC3598704B57CC814708A6628BC573869AFEFFB84
601CF792BD18E6680FF2DF557890399E71C68EF229788E762F48D1C229743E8521CDE673359BC363
DFA7662BE1A115855354555575D5AEBC64B933A106CDBD532F3A25338F31BF72F365DE44248F70E3
9E82F2294CACDCEC3506945D2747DBC8BCEBAF26F7C9268AE81D579AFB64E653B8BA3F1181B7EF5E
B2DAF7890B2BCCF35BD141AF9379670EEF8C9E42D0F602FF014D1D593747BB1FA7887C9F34F769B8
016BE314DE764B095DC5209F024E214E75C45C6DBA2527708A14F729062F3CEE93CC2455542813C8
208DB6ADA710DCA7DE43A1648EB60925875EB28D7C8A0080F3DC654B4F515555C54AEF36BDBBED44
A3AD0FB2F2475EBD3845726DCAE4680B0688DD6274E568D3866A144EA15F5E99A3ADF4143675846D
BD46E75318CBD694BC64758EB6EE8BCDF699808C357D9FC4F0E607BF792F59D10E638F2C1A6ABA8F
EB25FBF95C8EB6DF656CCBF7C9C329B0FF5DE21466E3CC4D72C4689F424F41A0DBB8D97B04F7494F
D734A4A90709700A31A4356155BB82738DB6EDFBC4B94F5D08C5B2B2F914C326C2689C7B700A5B4C
E134CEAB5B4B4F5155B5D705FB7F987F6EBBC5D81EF09A6D42CE9C3935FFBAFDF60FFDF5B5BF0CEE
BC132F59BEB912EB94A1EF53945D2F9F82BC64C751739B39DA01623E7029BC32B8A6A1D11E8D53CC
BD64BD7C0A693FE86581ADA7A7B09174C6339F9497AC99A30DA337DE712DF41473EE53464F116FBA
BCCA6BB4798F2CB65BDC4827A5A7E0C33B4914D988EF9348A958DEC7C32974E69D67FA648BB5EF39
3105EE93EE20BA6CC0A1A7B8F6D235F1B4EB67DE3573B4CD76D8106B5F5903A7C81F04F5E014C641
90093127F414024D6E8A290AA7A8AADAFB829907BA0928E814020CE2C28587E10ECF3EFB1DFCF1C5
177F083F3EF0B9CF7AF73F7A8D36719F067B2A8BF5C425810B72C87203D69BA39DB4848AF3290C99
AA2241F19B76E2FB2473B41DAC5C0658B04D5A97EF9347CA957A8AC9F83E79B4733DC2C5F66C1B7A
8A24F7896BB45387BACB7E79BC9EA2C5861AEFFB1467DE857A0ADBB5CCE2848823DFA9E114A643B2
663D11FAB626F7C9AB18A7D0C73EDAEE499824E368C7C9BCE125CB8771CF41509F9EC204DDACF9BC
C3F7299DE4587A8AAAAABD2C98D34E9F3E0D9F35E80E4EBEE3247CF7EEF9E5AF7C19EE203E95F8C0
AF7FF51BE64376C87D0A2822DE6117B518E3B84F63700A91A3ADA2280C1214DB8F493D85C97DCAB0
44CCCA709FD49197E43B2949606F8EB6B1FB1A2E5513E13E9939DAFA2CD7682E36ADD1D686C9BDF9
14A6378E703F43EED318DFA700A1D8ACEFD3E8CCBBB3B29BF0BA8CE9F83E05D884F683D2D3F8467C
9FB446FBA2AFA730A7657B3C33F590914F6166DEF196B9E720289F4F6158CBF2FEA23747DB71FF36
0E82D87D0AA7A8AADABFA2B368EC0EBC9E02EE861D072DEEFC7A28713DD671E03E05DB2D7125AD68
9E469B739F36D2539839DADA75D370916539DA364EA133EFBAC4140F459977B646DB14FAF564DED9
E490E05097E314FBDE537839DA417F217665E37A0A0AA708B257DA1AEDA1E1C0ECB2E8A39F191CE7
E2FD1B7A8A2E846223BE4F4E37D19B7967BB6B9ADA8AC9E4539050884FDD92C5EA9B6C6CCAF74977
CD01F74924DC99BC3E4D5E8DF229B4EFD3E6708A4113F1DCBB06A3DAA7F9A5700A467F3207B68732
174E5155B5C715F714B061839BCE9C39C3AFC43DE7BD67EF869B2E5C78583F6A87BE4FC176AB197B
97C429BA2415C91C6D132E973A0B7621CB7DEA5F9E4CEED3F9BBDE285722CD30B7CEBBC6E55370EB
2771782B962A78F8C317BFB8F30FD1F6CACCD136F514A637CEA2A7501AED71F914F99E42C013D459
D0F57290B39BC6E468AF875374F83E290D51845358C255ED9623776853C229566CD5849E42D09FC6
719FBC1C6DAAA6469B6BDCF4A190495E35700A534FC147F886F4142EF76969862C4F87D880CFE453
70C8D8F0D318CAB7F1FAC229AAAAF6B8B0A7F034DA28B5308F82B1DD309B919DFB3EE96EC26B2EF8
651BA71866DEF5B618991C6D197E6761133C743B954F31CEF4E9A188FBB4D8476937428B2232C029
92F914DE99ADC98C9A124E21F65DBA9BD02C9135718A784867700AAED13E7866D050902B545FE6DD
E7FD1CED1C4EB101DF27E548E0E114CF7FECAD42C4EAD1CBA99510237C0A3D4570C2C36766CF9460
7B99771EF7896BDCE219DBE0FE357D9FC65606A7B0E97C96442883530C5A860CB8CC867AE1145555
7B5C014E71FDD5EB487C32894C57AF5E452DC66BAFFC48DC74F43805F614B00C897C8AA0B9D00A0B
BD8E708DF606B94F845370BBFE40F48737F1735D33477BC07D1A27A6707C9F663DC572A1B1551562
C1523C28AFA790AB9293A3C47F445C8396ADFDEE29CC1C6DEEE71F109F2E2FB542663EC547DEF02B
7954C2BC9BD75390FBB1BE403B3123F96EF94168FB3E8956C22BBA75D95F9BDC27680AF8462B9F11
8F3FA6F414F7AC221AA5BFD3D04E961FED8A9E22E9E37D0315F414BFFFE1EFE9793869FD040F5C9F
FB84C5F5141E4EF11B0F7E6AE5259B40287897B138116238C5AB3FF99BC550D1DCA78DE3140E4355
7B402DEEBC3C3E8AF229EE0B3D634D31C5FC02CCDE47DF53ECF71A515575AC2AE8296013883D0539
3E99B74273216EDAA19EC28BD28E21F540A32DF2B2631B28BD84C53D8554FC5999DA2B187D7EEADB
E03E614F913FC255DFDF78EBAFEA97177D9F5CEDAADE8331EE13DE33C029EC5D5648D35D7D9F064E
C173B49B47BB4882C26E02F76C41E6DD08B34DFC08C003CD9EE2673E72BF37B039CD8F7719C29CB3
8153883E22C6268677363FE012A7B0108AA8D1F0F41477BE19FB88955999497912D43E8E534C434F
11E3CBDE8510A7E8F37D323F021EF749D86570F858E829B8326E76FD1F7DDFD4687FEDD2A5594FC1
FBDF4DE0149AFBE4596A108595DF019B0B0FA7982539EA7C0A4552F518ADDBE33E797D37AD11FBD7
9857551DB70A7A0A98EE1089D01EE0507F7DED2FB1A780BB899BE093ABAFDC6AA19E82F343CCB861
7379427E08ECBE3CEE139D628DC029A03E75F6BC785AD268DFFA3F3A723F4BF4C7318BC8F7697438
C5430BEE5313A770CD702CE4022F64F4149EEBA6411D619977FB9DA3CDB94F41836C5C33F77DC241
7EE6D607C57B9AE43E991D74ECFB843D85E81444FE976043F1FB4B9C62B9F532708A609C9B7D87CF
7D1A78207B443E9D5231BF333CDCE53E2D5B093CD435294F1E6964A6A7F8D07FB4F3E1B7D50AB84F
3154819D35E929C4D81ECD7DE21A6DA3FDFCE4BD8B9EE28AEA8555268526F5CD2EFCE0E5AC46BBEB
7B4EA36D52FB6CAAEA52D07DFEAE37EAB56095A3DD8C86D7CDC5FC1AC4298E727B5FDCA7AAAA23AB
A0A78079157B0A33E10E76B0D853F00F2C4E14306162341E7C3F3838A00B70CFEBAF5E87076EB660
E3F1CB07A745204566DFC5AF87BA70E161FAF3604E86CBBF7AEF273842E1C9FDE2750AFA1D7C5A78
4E7CF2AF7FF51B3FF7F8D3B43C798C5C618F435BB2DBBFFBE7700D6CDEE01FA717E1B5577E0413F5
4DBF78FE8D5F786E709CDBA5A798739FE0D9E00F86BF169E13BEC37B7AFF1D6FD3A75B1E80EEF93E
C1F6189E905E07F8E361D724BD711CCF73B16CE18FD88F3CF0B9CFE2DFB9F171B593C297082FC0BB
F0F0C52FEA9E4237C8A29BA05B49A30D1F49FE2BA0BFEBE23E0984022FBFEF0D1FA0CF0BFED9AFFE
E46FFED61D7722759C1B3DF1566280535C19F4CB78E57B1EF983ABDF7C8A06215EE8C029348D64D9
894007C187CA6C049E3BF7D3EFADE87CD96699B51EF0F04FDEFD4FC5B40633E7BFF967B39E82E442
B3A35DDD355B29787487BBFFF13F80778DC6037E1E773E3ED7AFC5C7FFF5D76F7FDBDDC856E56A20
314B0B604EA4C67FE6D463CF3EFB1DF191C9739F68F00BEED385BFFF2558AAF8E705D702EC29342B
55CFDE83A99BBCCEBE7BEDCC9933F05434B0E199FFDB8B8FD85EB2E982091F7A0A784E3EB6E103EE
E56847E73FC3797B064FB379037E05FCF8CFFFE11B61302FF0653363C54B2F5D7622DFF8F9373DF5
DF3FB58DC10C7F21EE34F00B3721B7DD7207F7F128A8A2AA6AAB15F414B0F50D708ABF5AF6149A7C
821AEDA3FCF0124E3138A77568E72683D7E53EDD7C59BB977795C17DBAF438F93EE9545652B06AFA
D3A2B9F8A3EF0BEE13BED47FB5264EB1E43E89370E7EFCD42FBF1969E76E1FC1AED75ABF00A77085
AB4DF7A7A5DBFF14B84FAB08B0A5ECBAD9650CF22994EF13E21430B0C5A6CB6B3134AF0FAEF1700A
CD68925428768DD892797A8AF7DD71D6C8D1F6100A4BC40D5BAF4B8F49486B80530C6B804758C7BC
819E027AA2194E71F6A491496185DC9917A6C07DA2EC15DD1DA3F340DC3B434F31DAF7C944DF88E9
1A68B40F986B99E75D265039CCBC737D9FD6C129E61F8A4CE61D59F0456885D268DB38850A7F0F84
42BAB98076BB34DA5555FB5A4DEE137C993D05CC5AD853D0F44BF3CFD1739F84EF93B70C99EB149E
9241E975049600B1DC984B522CB288F51482FEA4572BB14891461B7B0A78A9E965FF2BADA7E85AA1
9C9EE210BD649F7B97B11ED1799736954DEB29CC6528C02C665CA9658E183C7C52DC271368231759
09BD2D5BE9717A8A66525833477B66FDC444D9868F99451AA19E02C6614AA31D98CA722FD9CFFFAF
30D789B18D1AED81972C9DD6F2F6810D6FD16BC0C33FFD8907F4BB36D353E0A0BDEF0469B4E5383F
2BC7361FED13F17D226358314B7B98323F1ADAA0EF9380A1CD9E02C7A72D7C1B0270423A44BE4FD4
535C7FF53A5E58CBF7C9D768DBF9140154C13DC0FFF8DD4D3D45E4ECE453585F9EFBF5ED44A35DF0
4455D5D15446A3AD778350B0EEE3039FBD2A6F9D69B49F7CE228FF8B8546FBAEC1116E5EA38D8FD2
A719172E3CCCD7A32D65DE992E4F447F12F1D9468EB6C8A7584F4F61F614A8A7C08586DA04E32057
79126632EF9016622C5296AA62F1FDBEC57E0C1EBBDF3805E553E896418F6AB13DCBE4688FB0D96C
7AC9F2516AE8B587E1147C6F86F05C478E7613A7A01E648E53789977838ED80BAA5027BA7853A4A7
18F6116D9510BBE72B2A9F62FF3645B7BFED6E0F6E3367693D9FC73845DC327BDDC44527F30E7D9F
06E35975C4FCFC870FEF0503F0BBD7CEFCFA7931B09FBAF4F840A3DD6BFD3DAF94EF5308BA09FA53
269F428F64178C6316045BD5687B55384555D591D5682FD9679FFD0E4669C3FE44DCB4AB7C0A71BA
254E6B8D035EE6B7F9FB1FFE5E9C79A78FBC92C75F064EF1ADA70738C5D2C65F339D0472C157AE95
972CD7688FC629A87C8DB6B7BF927115AACB88718AA09B100C28C3036A1A1AED665C63604730229F
226040F1E6BAC17D52F81A1FDE5CBE4ACD325E6EE75324410A7EFDFC42239FC202E064B3AC07BF9F
A33DD0689F35F65736090AED67A7917927BC64B567329FA86543ED67DE215B35F07DE2231C3A0831
B1C79977DC76400C6F8E2FAF103AF2EBB37C9F644FD18550F4E0147A0C47D604F3CAF83ED9CDB2C3
7DC26B9EF8C05BCC53CAED55F51455554756F07183BE4067DEE1C61233EF4C2213B942E99B76E825
CB972421FDF32ED0D1AEE725DB7B909BC72906A160C398ECC1D1AEE82C629C22D862C52B548C533C
3710A8BA0885D6FAC1E530F3CEF07DF2F3EF16FB34DCA44D269F020736EDACAEDEF9B269EC3F18DE
4BCB022F9F229F79E7250E073805810E034DEB72180BF3810548B1BC75244EE1790EB0FB7838851E
BD718E36A78B34F2294CDAB96992A36C39A7D05334E1093DABF3DA46E65D23477B3857F38C15812F
0BA15C564F91390B5295D15318589B05CCF129DDCDBC5BE6537886B136090AC7F62EB84FD5535455
1D59A1B9938953403D7CF18B70D3993367E81ADA709E3B770E3A8E073EF759FDA85D65DEF1D58740
0A0F9E10F784C79AC7985DED83B9F5D25EB2424FC1494D9EC2E2D6617EAB997977FDD5EB37FDE2F9
6E9C62D856983DC52A9F425370D589AE09676472B45DAF278BAFBB58A4A6A1D1367A87F988F5A00A
18E1038D761AA7E8DA7D35F514DAEB498317B837E37D47D45358E8431BA4583ED0C329782E989DE1
E811459A38C512743047B57DB4FBF1C5866D0A1A6DB35FD0FA20E314683EDA37927907F711660541
4F31989C5513618E70BCF55691A3CD7B8A2EDFA75CE69DC62906A2212BDB518A2C82CCBBF9C46B9F
F9F898325D28EE5355D51E57C07D3A9C4F4DDEAD085268E2D3E12EB84FDCF7895A064D7C0A4EBD4C
8D3665DE5DB468B7999D58464FC13921715005576DBBF91463118A45FD935FD32FEF6F9D7E131D76
0DFC434CFA93BED0D253348FB906BC118E53EC3BF7E96B972EFDFE87BF17B40FA26B16669B1BC129
46F414C4DC13908438DDE54A0A1AF959EE5306A7180EEF144EA1B94FDAEE8CD145029C62A115D2C3
5881171AA79808F749770D81DDB79EDEA1A7D0FE215DBE4FFC2088DFE4F93E71F4813B3E198A6CD6
62183805719F344ED109522473B4317222A9A768E2141A656B9E05D1F7EA29AAAAF6ACF841F4235F
7834E829E09EE73F792FF40E82CB74F5EA557808CD90A28E09F789484D5E2D365D4B9C42AF230F7C
EEB37A0DEA3AD40DF414447C12BA3FEA1AF86197C0290EBC1C6DEA298825D25CA1C48F969E628653
E8CD9579903B3CE9256823C6299AE8B96D4B38BFC37EE31442A38DE18C9AF564D8232F413AD34B16
864A32F38E4CCFF41E2CC62966A64F43A1103FD725BE93F6C9E9E03EC57BAD8786BE4F4D3D850F4F
78DCA7144E21BC642D571C1BA798404F81389A316F33866A3087EF84FBC4F514461A854EBB5BEA29
E0BBCD7D5A13A708B94F76A76CC2CDACA768E214BA4D36276AED345BDCA7AAAAFD2E0CA483AE81BB
929AF7A169102EC0FDE14AEF3977C57D0AE072ADFED3B9C36B729FC6E1146225E2464FE238D7D553
689C620DEBA7404F616A5483AD5712A7889621FF023D6ABF7B8A8C97ACC70CE9E23E8DD00D65B84F
1A71D3643F8EC16D12A7509F020FA7B0B94FDE901EB61E1E4EF1FD5F7AB33B92751C9895C9321DEE
93C0DD6217593ECE37C27D32CBC32904678FF7C27A84F348C7CDE453A4710A434FA1BB89E1443DC8
15EAC229869E030289D30650D5535455ED59C18476EDA56B084F60BF70DB2D77A01CFBC28587CD87
C0A712EE707AFE25B2B375EDCAF7496CABCC3D5890AF6DAE23147817709FE28AF514FCA4CB38CEB5
D2C10EE6CEFF0D9C22B342996B96C57D423D052E34E4251B371103A254885378DC27970735E43EED
774F813885D711C77E50EB709F322D469BFBE41CDE0A95AB002FB2384573AF45B726700A33F6CE34
8F5D89B87338C56CE8DEFF4EF47DB2ADFBF5C66C7EEB14708A60F40633366173714F11F83EB5CF82
1C9C824E75745CA94E6F94A6CA5C4FC1B94F8453F42214BE9EA2CF4BD69CBD633D85E3FE1D4CE334
CF17F7A9AAAAEAB0C720FDE8F329748EB6797E1B5F6F6E39F4E6AAEB4437EFFBC42D44C4B9AE5EA4
4CEED34CA3ADB94F5D058FB2B84FA8A790CB93C31531B84F39DFA7C8336478D235BB300D3D05E214
440EF1B84F7A300FB84F0AA748729FD6C42904A55C0E6925DFBE7599794743A5A1D16E2214435C63
53DCA7C14DB1EF93D26847F20AFAD1CAA7D8BF22EE933798AFDEF9728059F81AED06F78913F92E2E
35DAFC0E4DDF276E2CE0B90DF02320C34BD6C429784C7C12A47828CD7DD2B43D93FBC40E8B9ABE4F
1131D5699FA10AA7A8AAAAD215B418BBC229E2A3ADC04EC4D353E87C8AEEF3AE30475BF040B4DDD3
C133524F41DBB32CF7A933A2A2E1FBC43B05ED6AEE1D7CB5F4140DD693155A3111EE136AB407A92B
8E9B996E2EB001E9C52982265ADC94C429B493FF802532D468477A0ACD7D0AF65AC3835CBC32D668
BF2E64DA4E37B108C8E31A6D07A7C08C78399ECF4AEED3E05676E7E9709FE2799BB718E2A626F7A9
6BA2E63F06DC2742D678D7AC0359447391F27DDA1C4E91F592E561F19CFB341FDEB1EF5300B7D9A7
40CB19BE7A8AAAAAAAAEDA899E423BDE784B159DDFF25B5DDFA79B2F07AD843EF2D2F7D1DC2799A3
AD08BAE42B42075F07CFF87A8A80FB34A22CEE13CFA7E0DC27F37457E668B734DA264E21BA09EFD6
BDE73E114E210D9095EF13BF03751398119FF492ED9554E435DA866DF21526621D0EFBEE7C8AE044
37875318433A361FE03BB100A7509CBD41772CDC9FC468BFFF9D53E829BCD6D8CC5ED1B511EE9398
B41B1AED2B065B95A36FDC9A8F9F02A57C9F02638D105C0EB84F792FD93E9C62D8FF76A1CCC57DAA
AAAAEAAA5D79C91A5B2FD54D042A573DF368DFA75ED8A2E9FBC4B327F879178114B46C897C8A54E6
DD283D85E7FBD4DE777980452E9F22A685AC8C37A7A7A73091B5E68E2BD0531C01F789EB558D28ED
670C9002B7647D39DA99E3DC344E117513C3C35E6C96039CE2F98FBD95DA5E9722A2BC9EF8F513E1
3E89F1AC3571010CB70DDFA718A7E0091403175916DA68C6E1453885CEBC4B221409DFA7A697EC80
01359CB733F9149E98425E60F72F9CA2AAAAAAAB76E2252B561F1EA26D6EC3845D6186FBD4D54D04
3D05E114C21BC4442B044E412B5A239F22634B681D7999DCA7553E85E63B59DB305AC828CC620C4E
E1317289FB34259CA2D93BE8214DF92C1BCCA7487AC91A320AE5E46F82147D39DA5DC7B9193D8529
ACD0DD448F9EC26B9903C2394A30A68C5304AD0487EA46F83E99AEC8C6380F710AA23F69184E5058
E96E91EF93E03E758EEAD1DC27D761238153E8D39E481337BC67E1145555555DB573EE136F228280
30B1428DE03E652AF67DD26C105358C1ED7152F914E3408A877CDF27D52FC88549D09F86087B5E4F
11909D16374D2FF34EB4C357EF7CD9546A9B3BB1AE1C6DB37DF08E79F35EB262E87A2E67D85C746B
B49B033E97A3ED26689B437D799F7C8E36DF5C991B3041859A084E81F3B6E6A67A67415C55B141DF
273EC833F91434C8F52C6D9E0845BE4F42A3DD8B53E4B84FC1C05ECDD5693D85D10E5B0A0B733EAF
9EA2AAAAAAAB76C87D1A9CD3FA1411BE30D1DD8E3E9FC253ADEAD35DDAA4D19DB782533C64739F44
3E854CCDD69747E568278F73A7895348165FCB4B96A4DC7021E3FBB4D97C8A5BAD1C6D497FBA621F
F0766BB49B9977CBEAE63E7927BAFCB0379DA32D18201151644E859A024E91D16807C74147CF7D5A
C538CE2F0C084E57649BCC477B16A7E805291E4A709F9E0BA513E83920A6EEF9F78E1C6DF5A3773A
549977555555BD75F45EB2C47DE287B7263C01DD84498BF2B84F1EE5C933C3C9F41488536843427D
F065EA290E827C8AD38F8E412898F7A6EBFBB45C7406045D93942B30F496976C20E8B3775C93D453
F01E3938BC256C621BDC274117E9CAD13633C2CCB0B06E8D76E22037C6293A7C9FD4688F700A4664
5A68B48324C7A1F3C044700AAF951033B6796B86FB941CD2DC4B36D6685317CCED9EF8F016601C67
49C18594976C7E60277C9F6C8DB6E71F3BFC20B4F3291C62AA2D1A5A9E11154E515555D5553BF192
15DCA7F8D44B7413E893E369B4CD5662233885DE5069DF27D25070DFA714F7692C4EA15FDE859E42
6FBD4C22BA75B94F4F111CED8A7C8A7DF792E539DA9E577F7CB43B82FB64DAC96AF2795BA36D91CC
3542216CA0FA34DAF159EE567D9F429C42E6687B312B9A1FB2BC7E0A3DC542A3CD0630EF209A53FA
3ADC273EC8177D04BBC6C32978236C4A8484E9938B530499779D204532F36EA50CF264DA6A3E6FE4
5384928AA0D7289CA2AAAAAAAB8E5E4F21B84F49E85C201A6BE65378A23F434FC17C9FB4D793A1AD
186A2EDAF9145DC75C49DF27CABC1BE2E383F5681831CCCFBE5238C5706D92D72F0D0C71459B08F7
C9CCD18EF5143C411E21B9757C9F82263AC97DD21E381E4281B53D9C622DDF27BDE9CAE929DCDEC1
EF265E99588E76C6C74CDC0791B811DC273D9E7BBD648DE089A14F72E0FBB4159C229979E7C16D9A
D497C12912920A4F61513845555555571D9FCC3BDB30C4B12BB473B46FBEDC054C74E114028F30F4
7D570627BD8459B4F329C6E5683B99770D3D85B7428DD25334907495A3BDDF3D85CED1D647BBC2EE
89EFBB1083835D9678DA75F229E86E198D36EF8BB5D193A96F8D7A0ABED1EA3ACB0D718A006E6BB2
A1F239DA06E569D85F882E630A3D051FCCDABE4C8F6A71D308DFA74CC51A6D014968229F3E20C2A9
DEE53E797A8A35708AC0F7C96B25B4976CDEF7C99BC04D878DC229AAAAAABA6A275EB2B4DDF2CE6F
8373DDDE1CEDF8E0ABD95370DF279D32ACAD08B5EFD3E96F5F7BCF990549699339DAB06DF37C9FAC
CC3BF3FC96AF5984BC8FCCA7F0CCFC973AD6BDE73E7939DA719A36D5A2A718E5FBD4AC3E2F594BC1
6A32A03A34DAC981BDA6EF9383562C2CCED239DA5A37E19DF44ECD4BD6C3294437A145705BF27DBA
E8739F846202B16613AA588929BA32EF46CCDB39EED30A656658B38162B03BE47D9FA2B320656856
3D4555555557ED8AFB2476530142A1FB8BA4976CDC419876E891EFD392FEA4C57D828E7E947A0A2F
9F62E522ABC1F458F7176AB4033D85DD6E4C49A3EDE568071411DE4423452AE3FBD4C42CF435BD1A
6DAF8910297863B84FFE112EFF7124F7497BE374E1148213C2597C211D7D2238050EEC3881C5C6A0
E73DC5B597AE89A7EDF57DEAD2680B644DA3153C054FE4B06433EFFAF1E50CF7898F6A177DEBC9A7
C8F0F70C226B69B4ABAAAA3A6BE7DC275A7A3C5A88C918E9CAD1CE57904F218CFA4DC0827374F9E9
AEEE29AEBF7ADDE03E75E653B8BE4F4BEED30AA7B016A3C16117BBBE2F9F2236C6999297AC99A38D
7A0A7DA510B78EF37DCA7413999E82987BBA8330797DF41118A3D1DE9EEF13EF26F87E2CA1A7105E
B211F46651D027D253D0910E1FC9668FACE7F04DF93EE9EB5D8DF6D206591A112C276A4D67B5710A
93FBB4399CA29179E774132BFA533347DB32D320ED5BA0A7288D765555556FEDCA4B365E80F42225
6A4D8DB6B76645BE4F437E8838CEE57D84C0291AF914A394148B45CDE23EA19EC264E16A9E79AF46
DBCDA7505256534FB1DF99775E8E368E6D3CE325F29EB8C3CAF729A1A7D84A3E8513A82D8C9EC447
607B38C548DFA765E9FE228553302F598E59C8512D708A69E8293C0C62D02C6B4C7939B047709FCC
1C6D816534F5143A6F454CD4DA182A9579376EEA4E66DE290ADF3A39DAC83BD534A7A632AE34DA55
5555BDB5ABCCBB24686E5E0F5B324FA3DD6C25683D4A729F0CDFA7256EAE65142BE8FC99D5E1989B
4FB1891CED58A31DE92944AFC1AE19934FD18AD59E02F7C9CCD18E7D72C8F7892EACE3FB148CF6A4
465B6BB1A57C7B88CA75E468E755420F8DE23EC55C91244EE1119C5AA37D0A3DC542A31DE6537860
DC38DFA764057A0AD36D60D52F2F1516A271CE66DE8D9ABA1B38854955B59A08D15C647C9F02F29E
4982824F44E1145555555D75F47A0A914F610AFABC6E826ECDE01479757606A788C5145CC13D506A
37F514A3710A87FB94D453689C828481BD7A0ADD4DAC0E78A7E4256BE6686BDF7E3DCE79E6DDA634
DA38E0A971CEE768F36B744C98D00DF569B4E31A7E10D6F27DF294DA014E811BAA254E1177137AFC
4FA1A7C89CFC04C8F2D1FB3E095442EB264C7702EA41BABD64C7E21406F7C989A25833473B682EBC
715E3D4555555557ED5C4F61260E0BAF4282D42956DBC629B6A9A758750D57E499AD0E66EDCBA718
A7A7087D9FCCB0ECC59224F65A8AA0BE563E8529F79B46E69D99A34DDC274F2EC4C544A37D9F9A49
F15D1A6D1978C78922792F59C17D4A2214EBF83E79512CCBFB783885CCBC1363DB04E086B4A829F4
1481F9806E25C404BE59DFA7244E210229CCBE583BF88DCFBC1B9B4F61709FBCD6980FF8743E45C0
6E5A75CD67AD4C96D26857555575D64EBC64C53294A4E6F27BEA7584E7686F434F419B2EBD4209E1
F640A9FDCC0B032FD9EDFB3E713DC52098D5822AA4CF79A79E222099AF6C73EE5B5AE84C03A7E07D
44CC7DE2AE38A3B94FC93E3A9FA3CDF9E7A6B895DFB4199CC2F2F91FEFFBE464DE35700A3393C2EF
2626885334E1B6D8EE6C7BDCA78CEF93B647F60471D453347C9FB68453B0793BC8B9D33379339FC2
682E5A03BB34DA555555BD751CBC64BD98A4A0CC63CC35418A389F421C6DF1335B4D47777D9F384E
71FAD1C156AAFFC8CBF57D629C5B3BDECEF3801AA5A788ED64E9106CEF7B0A3347DB6C932554C1B8
4F19DFA7FCBE6B04F76980B80DA5AC4438A7619FC52932E2A0E1566D2DDF27ED999CCBD13635DA7A
48EB113E859E22E82682A14E3FAEEFFBA4C73F7AC9C6F9145A4661CCE4C3191BBD64FB32EFD2E072
D64BF6B9770DA66B1F62EEC2294C31851ECFA82D2A8D765555556F1D07EE53D03B480C3DCCBCA31C
ED584F21A8E6499CC22696B3E602972482276CDFA74DE314FAE5453D853CDAD25EB2C340257EEB38
9CA249829A9AEF13377AF206B91EDE5D5EB2C9AD5713A710F472B1BF92320AC64EEFC8D1EEC4298E
D8F74964DEA5787D93C429742B4176B23A5C7B756114F74933FA309F424CDD81EF939651AC3256AE
188D068EF694EFD358B95023F3CEB3EC1BEAB57B73B44D8422B07EA2EFD55354555575D54EBC6407
1A6D413EF7532AF8B96EACD11E5D869E62E8FBA4375DD83E68CF1C9EA31DE553902DFF88532F8BFB
8438052E34E6068C9FF11A1AED30F32E269C7B10C64434DAA6EF93D744F083DCD84BB6D7F7C9B4F1
EFCBD11E32A0385744B8FABBBE4F9AFBD40429867A8A31BE4F8E92228553F09C3BDF2A7970FD2473
B4F35085A82D719F628D363FFC19709C1C1FB3D507C1C329CCCCBBBC535F0EA7688FE71E9C42D868
442428E5B351DCA7AAAAAAAEDA9597AC56FC05D45C411A697AC96E434FB162372D1D084D2F11ADA7
686BB4C796C97DE25EB283E5295EA146E45388A56AC874E27A8A29E76837F75D0B20633EAAD7D768
7BD5ABD176418A618BB16E3E85DF628CF77DF2C3E23D9CE2F98FBDD565F1397B2D01584C04A710F9
8C995622A9D1BE98F37D1239DA5881469B73F6382467371A57563885EBFB6466DEE52BA7A730D513
D2153C8D53185A899CA36CF514555555BDB5132F59B1EE98EA3F71AE2BCAD368AF0352C47A0A7D9C
752032582D3DA0D453045EB2A3700AFDF22E700AEA232CDB10F7E02BAFA7302D44FCEF93F27DE200
843992F5A8C6DD5A6FE65D7EC02773B439094438709AB639EBE668FB38C5BABE4F43D74DBC43534F
61A884CCE095FBDF39F07DBAEFC414700A9DA3AD394E8207C59BEBF57D9F08804B66DEC5BD831EDE
83AED9F37DD299776BE75378DCA7C5218F9352D18553C4BAEC008F2BEE5355555557ED444F214CFB
4DEE937DDEB5BC6752A3AD775FC1362CE9FBB4A239590805AE5F1CA768EB29C61E7999DCA7859E42
739FD4D6CB3CEFEAC2290CF7274BCA3A11EE9399A3CD93BF3400C747BE875374719FCC7D57DC5308
EE93E787A3418AC8F7696D9C6203DC2735B6F339DA51E33C559C22C971320B06F6674E3D369AFB24
72B4E97AD468BBDC2736639B5E4FBCDD90C222CFF7699B38050D5D8A0A1A0013E29A7E3D85CD7772
48508838174E515555D55547EF254BDCA701F7A3273EC9C429B6C47D12BE4FB400692F5973EB857A
8A9597ACE63E8D3AECC2727D9F74E65DCB8133C97D922D03DB569952D6997F08BB7EBF7B8A8C9EC2
EC2670D3058F1DCD7D5A13A7D04A5533205E90A010A7189FA3ED75CA214E11F93E7999774BB4AE9D
A31D509E3C46FA9434DA9E5B6CEC227B39C77DF2E6E4F80E4D9C420C6021A6E0D0334DE95BD153F8
1A6DCD7D32E767A383EEF17D32A4702DE7E457CAF7A9AAAAAAB376C87D1232556FBBA5932C3C3DC5
9AFBAE8B96465BF83E69A3270FA1E060FA51FA3E093D858D4DA83E22C97D6AFA3ED974F489E92904
EE1664DE11DB1C2EE4B94F9BED29BC1C6D717EAB7516F0FD3D8F8CCDA768B1A1C6F83E9999777CF7
15E768339C02BD340D45B6C948BF675ADC274176E2A743DE85CB9BE03EE97623C6294C059080273C
59DC404F6166DE6D0EA70832EF3CC7EF913885322EF31C92E9CAC229AAAAAABAEA38709F3C6042F0
46DA3805CBBC1BB1EFB2718AA1EF93588F042F77B08AB1BCB0ADE4683FE4FB3E3D6760E86E8292C2
D37BF329BC1DD7E2479679B7DF5EB25E8E760AB6F035DA4DEE53E0999CC9A7E01A6DA35F1E1A6FF2
013F3E473B402846FB3EC5F845678E76E4FB34D5CCBBFC6036C18B0DFA3E897C8AD8F7891FEF08D5
1B4731385535CAA758474FE1F414039C427BC97AC8723A47DB9CAE63F91B9D1D154E515555D55547
EF252BB84F1A98A0E32F8F0D05D7EB99C7D46867AE897B0AE1FBC4D72393974B1BB3B69E4264DEF5
1F7999DC27D45318303ADF65F104A5B5718AC1C9ADF8F12CC329F65DA36DE668C348465253A37D26
8D7682FB1488833CF39CBE1CED2BAB7E997345F831EFC1E81CED5CFEDD18DF27917997F67D0A72B4
63EAC80AA798404FE1E55398DD8454152532EF32BE4FBA229C4264C15B811426EE1C719F78E65D57
F5709F02C46DE136D08353CC9CF732873FAA8F7EA57C9FAAAAAA3A6BE79977263FC4DB7DD13DE37C
0AADA7481E7F357D9F2435F78A71F6C5B9E85E3E85C17DDAA0EF13719FC836C4F39265176845EBD0
5338A8FA801935192F592F47DBF31FD0A178C91CEDFC416EA6A710DC278EB80D44ACC3518DD7477A
0A2F473B87538CF17D12D73C27EFD391A36D09B4039ADF147A8A606636A950E223B04EE6DDB89E82
9B219BD3B5B632EBCEBCDB443E85C77DB2DD062C1403AAEDFB6405AFB8037B794DE1145555555DB5
133D051D64D1F96D60274BAB15F24302DF27D2688FABA4EF539CC48A0B595F3E45C621C72C8BFB84
7A0A6A13A4B0C23F045B9C7D859977064E11988770D2C87D53C129788E3659FAC707B9713E45DEF7
29D883E573B4054EA1775F7C873626473B87538CF77DB212B49B3885C8D1B6F760670744F4A9719F
783E8547700AFA8E35B94F787D6F3EC560BA3685422A5CFBA02BF3AE1F5CCE709F060E1B3A7B6588
2FE34D7D39DAFA50688932979EA2AAAA6A9DDA55E61D2F4D73D2D88440DE3338C508558591A33DF4
7D8AFD43F03BB7DFECC8A7E83FF5EAC8BC0B572871F0D5A7A770425AE9706C3A1A6DE1FB640EE9C0
2AF99868B417039B8D5E53A08D973BB84FC9AE7923BE4FD6504FE114DE303EEB06B24C07A7681EFB
988072ABA768F83EAD8353085F3E398C5974291FF978652AF36E43F914DA4B96CFCF92BFC77F6413
78C6F7C915538896B932EFAAAAAAC6D6D17BC97ADC271E93A42393440F62E3145BC8BCD3BE4F9A5B
CE0F78F9DA14E829AEBF7A7D5DDFA7793E857E79B99E62855368EE933AE62541F7FABE4F033D05E3
3EEDB7463BCED18EDD92D7F1920D8E767B35DA026E136E397A6FD6974F91A7888CF37DD24A8A743E
85ED251BE886867E5053E82974B3101B3DF109BCA9A748FA3E89EC9586EF937242A6D31E61892CBA
0F89537899771BC2290699771EBBA9F5631EA730BC64FD7EB9B84F5555555DB52B2F59ED10DB75FC
E5E568C7F91471D3D1F47DE2BD8356B09A7E50D97C8AFEF3AE59B57C9F52D45C71F09BCCD10ED9B9
E226DA7AED3D4E2172B4F3A7B8977B32EF9A018E5D7A0AA1D1D616049E6D72D4537CDECAD1CE2014
9BF57D1A76191D7A8AB303CCC2D5B72E2F4CA1A75820C5FE44CD5B69CDEEDBA0EF5312A7F0C25644
208B66407568B4379A4FE1B6C6AA4DD6DCA7D8F7C99EA84D0B82E150AF9EA2AAAAAAAB8E03F729CE
BC33E922F97C8AFC22D5C429A2FDD5D051D6D5536C30473BC17D9287BAEA14774DDFA7C8A2D06249
ED774FE1E56873DF27337B05AE59F93EADC77DE2D844B2A710DC27B32F960492E520CF6AB4378753
6CC9F7698153587DB11ED89AFB37859E42E36E5E07614EE3237C9F9ABD3396A7A7300DC0852048CC
E4AB91CF34DAAFFEE46F1643456BB437815364BC64C51951174E31438A4D51B66A31241E5739DA55
55559D75F45EB2229F82B6555EFBC00FC1623DC5576EBE2C8EBC3C30DDDC7A357D9FB04D10FD858C
211EFA70B6F329469172178B9A9F79A761741BAAD039DAA1463B3ED7327A8DA19E62BFB94F5E8EB6
27AC10D7E7BD64475432477B65FA34EC26348104AF19A3D1CEE114237D9F14E8C6EF10E114385697
3885A69AC77DF4147A0A13A7D0DD8439878FF37DDA084E61E3C80C77D65AA12C4EB1A17C0ACD7D12
835C23147D38059F9C9D3EC286E18AFB545555352C7D8EADEFB05B2FD92E296BECFBB4E6BEEB6298
A3CD3921629725FA0B7E074F4FB1A91C6D2F9F628044F8C7B97C6DCA78C946F914E66A7576997937
259C42E768074C3F6E7DB68EEF93875064718A2142C17D063C9022CABCF334DA3990E2A671BE4FBE
9222C62974E69DE15DE65C83779E428EB6D70B7B035BCCE11BE13EA1EF93B88FEBFBB41CD5045870
A61F150794B99E22E5259BAF1EEED3C209599FFC286C2E85530CA51306E8A68F8396A3BD708AAAAA
2AAA664371786CF414EE8EEBCE97CD15CAC3294C3D4512430F700A3A9B156A3EC3E45CE114ED7C8A
B139DA26F769A0A7F8E30157C408A718ADA7D0FC9020296C1A9977BD39DA64E6CF0D0A46709F3CD0
2D8F53708DF6018BA2102E4F1CA4C06D589F463BB3E9DA88EF9343476FEA29EC280A2BE46E35F2E7
D0C644700AEC7F05C4DC6C2ED0067C7DDF271CE77C624FE6688B903B49F05379F11D5EB29BC8A768
709F725EB2193D85A0A40A90C230D9A87C8AAAAAAA611D439CC2D053581BB0F8CA7C8E36AE3BC935
2BE3FBE43987689C229B4F31AE72DC27FBD8965FBF891CED8C85C844700A81BEA19EC2EC260687BA
7E3E45578EF6089C8267D8E12ECB1DDB6C84E343FA34DAF1F7E1C1EF18DF27337825974F1168B4BD
7E99DF61223D85872CAF32EF864E7D7C9CAFC37D8A31B8209F4283CBEF1FA6668BA9DBF67DF2BC64
474DDAF9CC3B1B65C66B148A11FB3E0D74D9A6479FE9B0517A8AAAAA6957069810B5732F59333BBB
79C6DB9BA33DBAA7B07D9F868B94862A52F914A71F1D9C626DC8F74968B4F5DA24B94FCBEB1777EE
D453B854288EB39F9D44E65D9CA34D889B8D59F81AED7CE65DD068E4B94F820765E6858DF492EDC1
2936E0FBC4AF69E11422F3CEEE2686039B0FF829F4141ED696910B8DF37D6AC2CA314EA18DBE05E8
26841503CF64C67DBAFEEAF5C5501138C5A6F329562238CB2DD6303463D737F229B47FACC22F4C38
A3708AAAAAC9166E2FE13BCC81A74F9FBEED963B4EBEE3E4EDB77FE8B5577E143C6AB739DA822892
373F3FFA1C6D7172CB09E7E2021DFC06F9141BD15398DC27585CE2932E2FEA8EF0F42EDF271BADA0
856C984FB1DF3D8597A3ED6DB744F7711C34DADA5AD3CC0B235A5487463BE33990C029B6E4FBA499
E4A61C5B6ECC88FB34013D054F1D350F82E213A111BE4FCDAEB9914F61250A493C42E071CFB4B84F
5BC02962EE938DB829BFBE763E850FB4D94CBFCAD1AEAAAAFAFF0E61DB06ADC4B3CF7E077F7CF1C5
1F9E3A750AE658EFFEBBE23E99DBAD668DCBD1CE9F7D0539DA5CBEAA8D44C4D12EA7F236F2291E1A
ABA708B94FAB45873B3BE1A68BAE773A8EB5700A05A6BF8CF91413D0683773B403006E9DCCBB6060
D395792F592D9DD0422112B77668B49B08450EA768F83EF16B9E9314740FA778FE636F15F2D5A09B
90EC9189E114A6E977ECB38177809EE2DA4BD7C4D36ED5F749CCCFAB199B45C6F33B0C12F1B8467B
237A0A5FA3ADB94FE620D72833072C52F914013061EAE3CE164E515535C5A2C36AD8694307213E95
F0235CA937E15847CF7DBAEFB607F9A66BE125FBB16B740E167DFFD09F42C1EECBCBBC5BF514EFFE
B2380113A45C7EE7D9C234FF31D668F3E32C71BA25AE5FAC56738F74F84E3D05B478F8662D72B461
A1A1AD146EC3C85133BE667EE51B6FFDD59F1CFE5BF10773EE935E92243F64A8ADC09D18F414D493
B2A73DBDE087F062A160DC06475B8840C1C3F7DB4B16710A3E92D1038AFF688E6A92BE061A6D1ABD
C105319EF14AF80EF59137FC8AFE837FE623F7E328E57A55FE5DFF28AEE43D058D1918ED8B812AF6
54E2474E8E1A1AC97A38053405B33D95A99BF0FA8BE5260D1FEB6AB4CFAE400763248BCDD8BCD0F6
FF15CBF7690407F59817F414DCF4BB3D51B3EF214E71DFA33FEBB6C67C185F648C566A3D166747AA
A780B50035DAC174AD073329B861D2BEFDBB7F4E38051DB0183D058D5B6FAE56B766B84FA9518D9F
82B9740E6EF57A0A4C875F4DD15438DA951A6E71D37D8B11FEC407DEF2ECFFF6BDA31C69FB7DEE54
557503157407D03B9C3871022E100F0ABF9F7CC749B889CE5B78C1AD5FBB74E928FFCE7B3E746E76
723544CFB15958351ADEF7F903E1728053C02A73E1EF7F094A9C68894330E11F72B1954FB150B02E
F9B72468453F737E61D6472CD9B978F983F79FA367C377618653FCE2F901EED055F347B9DCA73F7E
B7BD3C99D949A8F863173C9C62A55DF5E109AE04849A71A5E62B17A2150F5FFCE2CE3F26DB2BEC29
10AAA041CB2FE8F17C79698FEC69B4E1FDCD739F6CA8E2E6CBB06D33710AE8296698DA7C8892465B
E8B51776FDCFBCC0873DDD8D32EFE0EFA48FE480FBD435A4D967C1E53E99033857AE9EE2CE37D356
8A3A05319285BE154F71E9F2DEE314B7BFED6EC2E016F4A77060F3EBF1C7CF9C7AECCFFEE287E269
EFB9E50118A2D85610B81C93A044CB0C0FD16B01B403A7BF6DCFD83492C5D4BDA2B37EF71AF62367
7EFD3C3E1BED7217DC27C7FD38595E4F21CD079E335A63DD65E0A3A004F709D785194E81BD83356F
0F525744E33C674CFDE813FF41E1145555932DC4234E9F3EAD6F3A73E60CB41538378A5DE8D1EB29
EE3BFD20620D83D56709A0C3B6CABB40DF4D9C02D6115868C8BD1C2E88B549ECB8F8F1175D26BC9B
5E28D872FCDCE34FC372830B107D870DD5EAF233ABCBB4079B1D79CD7582B77FF7CF494F012FF5AA
A7F8F085D9D6EB5FFC0F11E7DCFBF1A1859E42BFBC0FDEF9165A6868D121FB41F17D41919AF72074
4FB3A7387FF660C17D5A1E612D2E60CB30BC4C45AE23F8D8FD3E83428D360E5718CF7881087EFC82
7085C2FB6373718F458D58474F81FDB2C77D9AB5C0DF7E0186289ED0F241CE47381FF3FC6EEFFE9D
AFD19603E61FFCC8746BB4F990FEC273F8237CC0C54C75E1C2C3D014606BB062EED1F7397E41E379
F1237D7F6E9179F7A94FFE92781160A3F8FD5F7A338668BFB28CBDA3F1CC07F36ACCAB6B3EFDC17F
24E68DBD29FC7788FB84243D93DAC7897C3A8105710AF1E2987A0A428D6DA6D3FCA6013067719F60
48F3A12BC633BFCCEFB33811FAEE404FF1C86F9FC70B307BC3743D1BA2DE00D6ED869AB74F7D441A
6B68EE533475234EB72C7A94A7A7C0531D3E9EF9C096D3F5720E87797BF689387BF21B3FFF260222
8FA6F67B8DA8AABA81EAB65BEE80323F92D86EC097BE09B6B857BFF9D451FE9DA8A7C04D97588C60
5B85FC107D6171A2BBDCAAD16906CDCF8453686688779CAB75169E9E4240E4B8131317E4773CFE9D
D7DB7F65017F0C7A0AF47DC27DD4A8232FF27DE28BD4229FE2B9770D522AC477EB02B914DA38C5E9
83158C7ECF09DA83E195F2FB1CA1585044989E02DFB53DDB7451098D36D4AC5F98E3713498118FC0
91AC1B6AA87B6E79403CADC02970A0C277BA100C7222967CE0EFFC92F8BC1C624F813B28064CB8E3
F9DB8322EE13F6141CA7805DD3C8F1CC765FE2B8129E1FB94F83807833D84EE8B299015A94797776
4576C2735A1ACCFA02FD4883FCEE7FFC0F763EFCB65AD053E060E6E8835938BC29F99112583E73EA
B126F7891356CDB320CE7D820742C1D39AE74BD053C0D80E266D7302C716031B67EA296828CE7A0A
9EA33D6A6C6732EFB4144E0EE9A55B3876D9A2A7A0CFF882FBF4F1937C66D617B07DE0831C4F81A0
BF809EE2C51725BAB4D5D21FFC9D0FFEAAAA09D6F557AF23C189830EF47984AE016F85ADAC78E00E
708ADB1E444B9CD58AB3A49A0B659F26EED2F9989D79B714FA513731E341B50A1F02052B94C97DFA
F97FFD5D41C16DD777AF91E2EFF4B7AFC12E0BDF0ED8C0604F31D0538C228A402762729FEEBFE36D
72EB15D6E09E01F7E9F4C1424F7136228468C6C8EC4C6CD9564C81FBB4D850CD3B054AA6102473E1
7286869C7F78D70B7041E82990FBF49137FCCA576EBE8C815FC9C18C85620A2813A7F85B77DC8914
911948A16C7032F59E47FE80E31478E1E0CE7B676DF238FAD3F2BBA7D1FEE99FDC622A26DAC4A73F
B9051E2E7A0AFC48CEB84FB889A2FED7296F9CEF37F7095EA595467B0EBD91566876E1AEA51F94D2
50D03530B089FBC4A72CC42988864A3D8539989119452D0622D1424F814FFEC0E73E3B83DEBE7B6D
C4909EE53FCE1950D453CC0032C4AC2F3DBE664F019F0BEC29F88BB0E23EF179FBB94810C7AF41B7
A8F377BD51AF05A8D15E7410B9E2C61A30753FF181B7E84E70AB55DCA7AAAAE350B009C4AE41AB6B
F156B809EE70F5EA557113AEAA47F9A79EB9F541EE252B4E6B1BE5F83EE131263FD72289447CB4CB
6F82BAF7ECDDE2AF859EE2E71E7F9AE8B80891733C9D7E44A62E61E8F41D1629C229BE76E9D2AAA7
F8C5F372DFA5094EDE05E4960C7D9F704DF9ADD36F223D05EEBE8810322087109ECEEF365FA44C8D
F6AF9DFE8FF9A68B5F582D462A9B7571F0B5D468EF37AEBDC8A7688E618BF247FE9C8453D00EE1B5
577E043D45402F8FBFE305EA29F8C6037B0A7E601B8C6ACD8C9A99243FF42F891AC17B8A05F14928
B59B5666B19EE293F792EFD38A52AE06B65934C2FFC5BD77CA0FF8BCA7A0935B5ED2AE5FDBCC2EEF
B3F738058C406A164801941AE14B570D93FB847A0A4E6AE28E19FA3BF5C87837ACAF7FF51BE2AF85
7680CFD27ADC8ACB34C85710C60F5EB6F514DC5523D320AB0B276EFD88F86B3D9C42CCDE7C8493F8
624173FD935B628DB63969EBE1CD87343EF0A95F78F3AEB84F85505455EDB09EBCFC24760DDAAF0F
0AAE44EE9396631FBD467BA6A7981FE1E2011707CD09B3F03C73E02170011E4BE46DAA0B171E5E00
13EFFE327EC7832CEE7E232ED07246577EFA130F88A7BDFACDA76EFF6F9EF2DC423CF7A795E5E6DC
57F683F79F438326A9A78085461889A4CFBB66CC5E8553C08F173E7ED3C2E75C599D534C92718101
161AA780A785CDD8C215A707A458D51CDDF8C2977E6F8F570A787367A39A784DCAD04C5F587C1070
A8DF35DBAADD77DB83E225820FEFE9B7DD4327B762E8EA21CDBFD3012F719FF87B0AAD2E8A58CD51
ED91A0044E81D408CE7D82D1BEF2C0190155CC3F11B69E626EA4F97FFF4FFF341ECC64A44C9BB4C5
439E7B9770D5865F011FF0E73FF6563C95A55095188C5BFDB8BCDBA73FF88FF4C0DE9BA13EC329DE
F19FE0C4CB316531698B0BFC4778E0855FF8C320F30EC730F7CD1083990B2890F584B005CCF31AB3
86A102CDB261EBE48FF3C185F92762A5A798EF7261029F719FC6CED8372D39AEEF3923F514DCF769
7516A45BE3E174CDC16578144CFB264EB11AD869A8828637D4BFF9676F3D62DF27BDB25755551D7D
A162027A0A98A0F4AD7025A218FAF40FB6B8F79EBD1BAE879919BED305F80EF5F5AF7E632317A8BE
F2F857DE7FF3C1674E3D76FE7DFF755CB05878D7C3C33F7ADF2FC353D1DF096DD1C12FBCFB137FF7
4151F7BEEDD7BBEAB65BEE80A782A78527873F1E2EC05EFAA6DF7CE4DD0FFD4BA893FFF97F8517CC
92B7FE675FA66BDEFEFEDBE1692F3DF67BE7CE9D83E784677EE4B7CFFFBBEFFEC04D1FFFDD597DEC
E2A03EFE5BED0BF38267902FEFA5276E7FDF5B2EFFF6DF7BEAA17F1FD62928BA0C17A0BC0BFC3B2C
4FB09EE2BF8F7F2D3C2DBC6B4FFDC29BBFF1F36FFACAAD6F79EC837F1BBE3FF181B704DFB1F83DE1
B10777DE0B2FEC6687D64E2EE82BE1FF82CFD1B99B2F3407362FFC20E05087CBB05382D7195E6DFE
9EC206093A8271E3993E0827DFF41FC208C4BF16DF53F811C6CFC947FF0046A918BAF9710E03F2FC
27EF85A78227847F1F3F8930DAE5900E6A35B6071F848FDEF1617849E12343AFC3FBEE38FBDFFD17
FF5E3086CDC14CDFA1E0E1304BD08B004F0EBFE2CCAF9FFF9DF7BD81062D5CA0416E8E76BC9EBEE3
35FFFC1FBE11369C341DE12BBCF351BAA90BF0AFBDEF0D1FC84CDAC1640E7DF16F5C380FAF399FB4
61B4F3819A1CD278E1CCDFFE35FC0E6B013C1B0D95D95A7070301BD58FFE014CC29951BDBA09EE3F
9FB74F7EE9717892D908B9F4C4993367E0A5803A7FF660761044E3F6F4A362424E4CE0BF7BE2E60F
E25A80AF035CF8DDDFFA2FEFBFE36D7C6CD3D41D8F76BA27CCF01F3D78334DADF43A7CF0ED6F8251
8AC3988FF0F8021FE4176EBE6966D23E5F6AF115DEEC96406C39E0FB7E63D95555374A910AFBAFAF
FDA5BE35E829265E37C491C80DF14756EDB6689068D9FE6EFF9EAAAA63523B1F9347FF07ECFC5FAE
AAAABA11EBA96F3D1D709FA0D1C09EE2CB4EF2DD1E5466F21C31C1EEF069ABAA8EA68EC3683C0E7F
C37ED4F65EC91DBE47353C9A2F4BBD445555559BAA679FFD0EE2146642CD8B2FFED0D368EF711DC1
1CEB91A8C5A1F10DF41F55DD885503A38AD7FAE321F90C5DBF684BE733375C6DFC7538B2B77B1BBF
BAAAAAEA18D66BAFFC085A0628D3C4095A09EC384CB5C5FED5319CE8B67DA0B4A945E118BE74551B
A9F89DCDBFEF35428E61056FCA3676F21B1F03DB9B1E8FF3705D7FEE4D7EA8B7D121565555ED7741
CB70DB2D7798ECA6AF5DBA8438454D1A47561B99C6F7D84CA66A1B650E8F60EBB2F1CD86F7BB6ADC
1E4D6DFB75DEE026F6C88EE2B7FA9A78FFF2CEDBB423F89777FE24555555DBAB0B171E86C641A7B6
417DFA130FC04D648EB7AF156CA8B677B00617F4B2B2A573A1D75EF9D1993367E08DC61FE1C27BDF
FEDEEBAF5EDFC87FF7ECB3DF81273FE250F5AA7125FA82FC00402BE3C3F4AECF1BDBDBEB85FFD5EF
7CE9DEB3779BBAB07CC133C0609ECEBEC5DCD036FFFD60E24AB23ABBA63E73CEE4F5C8171E85770D
8359E1FBC9779CA4BCB935EBF5D77ECC67CE6DBF11FA3F4D7E48F3F7D12703F82BE25FD77C779A7F
007EB2BAFE60FD5E9C3E7DBAFC9DAAAA8E79E1240C45D7D0E77D0AC427F8679FFF9F9F87190FD5E8
F875DB2D77C0F7F7DF7C8097E12698CACC6D155C099BEA9909E1FCE1B0577FF5277F7368AD95FFCB
FFFEC2F9DF3C4FBF059E105E79B8151E8EBF0E21219839CDBFF3C5177F78FE93F7D25F080F37ADBA
CC42BC0973CFE1979E38716253AD22FCFDD08DEEFD20B9D10BDEF4AF7FF51B34787060C3178D5BF8
BAFDF60FF178113EC8E1E13074E9E122AC810A86EE53DF7A1A9E939EF0D9ABCFC3F3C0C33FFD8907
E87ADAEF897D0534B9FCE11FBDE3C357AF5ECDEF3DF051EBBC4A3086BDD3953D2B98B5E0A5863788
A61D3D1E6090C08C61BEFEF07082B061EE82327F0B4EADB00FA49133CBD979EDC7F0703E1AE1977A
FB58981B6166A6BF10FE66737AC4E7C127C17330B8FF6BAFFC68FD170AFE4D782A9C39B7572FFDE0
793CD9130B10FF82D7019BA643D517401F0DDB757A2BBDE31D7841F05DA3278435051E0EEB083D1C
BEF0336BBCE9AFFD18161D1C18F05BE06322D28882C24FD6B973E7D67C2F6815ABAAAA3ACE05DB4B
F8B40A2136CC2D3075EC3D4871B89CA271ED837F19D3B8A8603EA46511963FEA2CF863A170158087
C37D82DF85E6BD34ED8BF6ADB92FC259BD57320F7FFFAC5B59FEE5B86DDB08B2007F3F3C336C4E76
FE2656350BF600B837E30786585FFECA97B1D1087650703D6D3883951D8784DE67F2873FF5E413E2
21FC32F4BCDE3ED3FB8DD0B0E3E9F43AAF0F7C36E1497416FCBE164C08F052E39989B8890E3AE026
D8A59B0F872D2EED63E36502DA437D5A4246E5A7ACF30DFEEE2F1E3E9FBEF4A8C051FDC86F9FA7BF
BC3909E78BFFEA6D17CEEDF0EB84733BFC765CA0F14B2C4F54D84CE13350F7A15F557465D1C702DE
C3C50BEE3D3CAEC5AAB79ED38B58C5AAAAAA8E73E10915FD0833096EA177FE871D599D3B77EE5418
FF4753AE793E43491F7804E4FD169893611DD713237F78709E03BF7AF67007CB300BEE0C7F334CC8
DB78D160998027DF63ABE17D2ADAC579038CB62EE668C16601EFF0DEB7BFF7CFFEC2DEDB1CCE7762
7AD7C11FEE7DCAE8E1BDDD01C228A6795DBECCDDF57E17BE23DE7FFDD4B79EA6B7CC3C4B81778A66
ADAF5DBAE4FD16D82D9B58068730E287072B11BEF5DE4E7B9DBAFEEA75F8F77503BEA52210C16BD8
09C2D3FC1F8484320753D8B3EBC619BB727CECFB6F765F6DEFE171C1BB0F8F5AF3BD300F43AAAAAA
8E6DC1B616B60A67E65FE6C4B5DF15F71487CBCD39CEBA7AF7022F175F22BD6DF6B597AE99B32BBC
F8FCE15EC8203CBC5732BF59C858FC6A78D1E2FD61D5F1A9664F01F5C86F9FF7FA62DC4E201705BF
3CD5364C207AFC634F0137D13ED6E3D8C37D7A279F536B5B49E08BB3263DE3862B6AF1BC3BE04E15
DF327D12727AFE454FA267457C476030985D2AB424443AD50031BD9BDEC3E95FA0BF7FB342982326
DB504F11B4574457D30B04BE53748760FB0D9F627D2C860FA7773378C1CD87078529576B121E9084
16BC32555555C7AAA6234BF42AEE29F0F5A115168FDDF88B06933C4C9BFCB0C83CCC8109D6EB29E0
0F401862C4C38382D5C14446CCEA12C916F1E9C6AA4C4F71B8DCB790BF341F003090AEBF7A1DEE80
CFE36D3C60A368AEFEF0F0D75EF911ED5B604B693E1CFEBC2EE40BB6B2EFBFF98090917153D93856
E18D5E014E418D1EBE32A72C8612BC83B0E5BE70E161F13CE22DB8F4D8EF995B5C1809F0707A7E6C
5BF4DB0763C9DB21BFF483E751ECE6FD835D8341DC791CD906D79111149D664F81026A3AD7120427
5C7AE83DA55341FD0AC0C750C33A8853D0F3C302E76185F0F02E2704F82C9F9A6B61D6798F8AF854
5575A3547513584D9C028B7644E2FC0A9B82C3258B899FECF16358988DFFCE9BFFAE7E5A98FF71C9
2612942933F41E6E16FCC6D75FFBB17766C5E7E7A4778728A42E7B904AD571AB644F814997F8C5AF
479C028D05E80EE64E1E46B21E15C8DEC70F173D1CF72D62E0414BD2D55378C4A7AEF14CE4CF49CD
8731F749DF4DCC483019E25E9144BE26490936C91CF6A257189A539C45CD87D3DD829E02DF7A716C
CE8D8CCC32BD8FC47D906CE3B5BD41FDEABD9F18B7FB8597A289531CB20542E85C505772B8ECB316
0753579FD7FFB2D914504B029F23FA789A8D80D992042F759EF8E4BD65C81028E25355D50D5A935A
55A9A8A710E73F6201826916E75B410E419C022F737E88F82DB0A7322758EA290EE7AB92B740C3C3
F33DC5E1F28C882F52B056E2F3737F1B0EAFC02FE5A7B5B038E22A064B83E8B670F5C44DE6CEDFBE
AA66257B8AC3E59E9F5A002CEC297057C9FB0E4D0EF19A028439C4C3F5BE05C655574F81FF14FD8F
F4F1A1610F577EFDABDFC0BBC1DFA0B791A8960A6450FB5AC99E828436E27D819E825E64D166F297
D76C0AE00ED492E887F382F7CEDB4FF2B71EEBEA379FC2E7A1F607262884DED0B0E8A9279FE01B7E
54A36B6D3E926D82D375AF70F2C771DE5519EED3E1521A2FAC1A0F5953C09F0A0F01C433984D017F
F8A5C77E8FDE0EADD7EEEA2938A5103DA6709D852F785BE93FA2254643EAF028ECB68AF854557543
546D08B1BC9E42146DCCA835C01790700A2C9C366192141B15988D3D9C822FA6E42508CFC9DF204F
8EE115ACC57C9686BD01FAD9C2F30B2603CCF662AAA7C2CDA46828E039CDAEA7EAD856A6A7C0C146
7B4858C7C509277D3AB82FA5181B664F01CF03239F767AA4FF85EFC215B98BFB84C7AA62238ACFCC
473239DE9867C8DAF16922B322BE7DCD9E02F6E1B8F11646BB1FBDE3C3B4EB464F727C36B1FDF380
06E43E8987EBCE45C01CF406C164C8DF7AD8155FB8F0300A99096E16C67A66DB081D87DEAF8E26DB
8829375F499CE27019532B2075C229B07039C375443CDCEB29F8C36906D07CB6644F41CBE229C5E3
457EAFF837D1EEC35445A1FBD68836ADAAAAAA6A5795E13EE174875E8562E180C9936F6CD033442F
911E4E010FE758367F38DFD775E9293CDF12A46FE98D253210749BF0ECD5E70F7EE1DDFACA223EDD
5895C729885F2138D5E2D85313E9B13CA0413C9C675E641E6E16FE0D82FD62EECD70D7449B16BE75
99A0E31396F9F6E94D1D9902A1788AEEC081864386119C625434B833BC1166EA073C1B8744614AA1
8773728ED792E0F188E0BC117547E0A7444915BB7D9824750B80C4A7715E79A83ADF869E820AFB1D
B157A7A680FE6BF289129D94D91460C73DFAE15E21A55050D1CC7F93285BE2A5831FB7E75B585555
B5C19AC8595CB208936DBA18E1CE442C1C8F7CE151B155E38422DAF3784D01E73ED1C3B9CD14BE59
7FF6173FCCF71478F0A57107FC4FB5E29214E87C9986DF0B6B0AEC1CB4E39358FAAB8E79E57B0A8A
24E33D05E73E51611FFAFE9B0FF8A20FCF6F369B9A89413635FCE1B0FFCCF7AAA7186248853D82EE
294E31EE13159E904FCDF1092BC97D823D24DE531C3870A0018B04D7FCC039C029C47B84B663F8BB
9A0FD7C4A7C3A5CB10B1ECF4FDC5D082A14E6F3DFF8DE6CC992964848E38544FE2143F5DC68C9E1A
5A0A08A0E190B9379F1CDA7DC77A0A2A44A2F18BBF68F99E02DF0BFDC9F25A7EF3E3595177555555
3762114ED1CCA7E60B314D805C4F41C505D7D880045EB2FAE1A886A0051AF9A879CF4CF4EDD4AB1B
327E4D733FDCE3F1832994D69AE621A6A0B2EAD8D6089C426CC0744FC1F5DA64BF13E829043DBBEB
E1BA70AF4B7967547AD342B978FAE3A303B9A63398E9C58FEFF6D20F16A70DE2B8780634A8E84C12
5C93C0D96B0A38754A3F9C0CE5CC879B9CB7C3E520370D2E28548E5F09EDA756F7231090E90BE0B7
C00C09F7C47AFDB51FE3DF0F2393AE81EF70B7E6A0CA739FC82704710AF224D4BD21A1361CCB838F
A1FE97794F417F2AF592E2E1C99E023F59FA2D4646968953888F279E6815F1A9AAAAEA86AB244E01
6B04D9BDF2EBCDA6E05009AE331A6D5E24F746F75A78F84DFFCE4D997F072163D3E8D5C329669CF3
AB8BCD03BD08B0D2E92D2802E5457CBAB10ADE530CCBF69291A9884DCD8F0785AC86F3F10850C3FD
83298820DB2871BDF9F0E4D042F68B76CB47C685DEB49C9AA7F511BC88FF02DED98B69DEEF12D230
AF48FC22F6F0665370A804D79EC85AC31C874BA720FC750893A19E42BC35F8D6C37B2A1E4E3D85B6
DAC029F1D450BC7F6A18B372B83C9FCF906DD08F88466FF005FF51B34D6E66DED13F82AC24D1E0E3
F0F69E16E5CFB8338F35DAE2D540C872A19E9E67ADE67B0A64087BA34EBC20849273F49F884F53FB
60565555DDE865EA29682A230E367186850DA61044F0C24D0B6EE3839E82AFD77C0A255E2BFC85B0
5026B94FB894D0EAC99FD0EB29B0F077D1AD309FD39E8D9E043B9D6D64D7566DAFF238051E248AE3
4193FB844547ACF89000A7D00F9F7DA6AE5EA587C32E02F69F5F79FC2B99FF48B7F6FC7AFC1B68D0
929EE2F5FFF7FFA17BE2076A828E4F5801F789CF183035E12426F23BCCA6E0702807833B40993DC5
EDB77FC86C49849A4CE014F887796F3D7A4D78561BD82C53A8374CDA3ADB021D9F32C427F84B1EF8
DC67E1C541EF0BBC80D3357CC4E81ABCB5197DC2718A600B8DC0F129C54333B94F78818ECB4E2D03
2C742FA6B94FE245A387277B0AECD74C1D0DBE05A2E53761C4754868555555553B2C9A7891FBA467
75BC865A0F4149D57A0A2A3A1F3B353F22EEC229C4C3E13EC99E02894F78B2240AE67998D2BDBF16
9736D42DE2411CFFF7B1BC05BDEA3817F514E6428F05EF326E06C8E447F83E79240468B1E948D614
59072D097F38EEC79A3805DAFE9C74F279696FC6AFA48350AED1265F9A691E84522B17DF4DB71EF8
72794DC1E1D2C301CFB7E13DCAF414FC2DC0EC1BFCA5B021170F474E8EF9D6FFD95FFC30E829C85D
0A77F826450A0F4C62F7BFA0B6ADD1F6EE16340587F383293C28C015045E257107A41E790F474E2C
3CC37FFAA95F3DA912CFCD0AA2EE1087125304FC01276EFEA0E82990E555C4A7AAAAAA1BAE602783
CB4DA03BC66D8CB92B135EB2543843F21C2273EAF6A853585CEE9DE92962AF0CA4630587D5B81980
3607A67D7D8E87C4A720BBB6EA7856DC5308104A7F1078689DF9706C63F1B1BA29F0B84F54C4123F
358718E2D8B24327EF0CCB3CEF5DE114ACD1E6C4A70916FCEF98DA10DC27D8EBC23B2E700ADEAFF1
806CD30CD68339B048D4734AA57622F1C9DCD97A7A0A1A09A877805FFDD7D7FE52534391DA6A5246
93B5ED9E02571078D7C4FF15F7143F6501D9E6A7061FEE7DE250F7C43D439AFF0E360EE613E29388
B77EF5F15CBE74B88A8D881DACAAAAAADA797DFA130F04075C874339EAA1023202A0010B21758F62
EAB524FC0EC943C5C339D6C0F9E4A66553F0D7E21113FEB5FA8C088D4D9A4AF6AAE35674481BBCF5
E4F8946F0A383F903E20E6BE28EE2968DB734A71ADCDA24FA2779349D8E61B277C4102D466EF2BE0
3E61914A57B38F0E9D64075EF028DCC49AE71B339CE2C927CC07E2AFA0F656F414C15B1F68B4B1E8
5C8852BC7961A4E93AF16AD8B36CC9F7C9EBF70F97C8CEA19AEDB5E0DA7CB829F1E6457EC2199C02
5F648F52187C3C4F0EEDC2925341555555D571AB60BA3E9C4F92B45F32CF5E60A3EE85BD5261DB62
F614B0B1A1877B8B02F602199CC2737CE2FF69DCC2E0F221760214CF418464F30FAE3A9E15A06C3C
EA4ED312B06093660AA2791114A25B12542E780FC73F80BA9E26831AFF176F0CE3BE4E6C69F04ADE
D7085F9AA90D63EE356ADE0136FC38E37988E7C944420D3E83C7AB6FC29DD892F0A915315FAF2F86
AD2F1A110470334E8026AB53383EC543423814E18505E9D4C1298227840F1DFEE5D853E855009AA0
932A9662F5667DEBE96069C067C3B326DD14C0ADF1C3B1084969F614DA4B8D17229262B9A437859A
417A25A7F6C1ACAAAADA83A2968153406142A393DB78053CB5D420C4BF05B916E6C3832E40FC91F1
7D90A30213B23715E3DA141B9BE07FADCFF1F034ECE18B5FDCF9FB55D55BB445142D21EC10A8A136
91292C1812F0D60B6B023DC6F0A457772538A2B4F9A728F44068EE54517FE17528B4F941870124E1
D0B1003176E083D0954ABF6705BD15BE2062E6814D1D97B7983B43F4B53E6545648AA2E9485C0FA3
2EF3701DDC19BFF5A8F73FA9B2F078794A9C8D906DF0A324CC0D9A8F82FB500424FFC37E3ACF7F27
9B8EE0D80A7F6F53098EBF45BF384888CD3C3C835350D49D792BFF78C27F8729F674E281418A457C
AAAAAABA110B1656DA50795F30C97BA75E2FFDE0793254C72F78B6E02C175758FA91B673844AC3B3
05ABA178B859B875146472BC0C3B3D3C2322428227B1F47234106A81ED449D1DDD4005AB3665EF92
51277EE1D1289258CC73D7C3F92E0E070CB524716B804A1CFE70F18112F1DCF1C3CD3AD502ECB859
0DEC70E0DFC1DD29BC0EF859464C649A8E4FB091E3E217F30BDE05AFBB844983C20A69D60ACC78F1
D7F18793151E7E9D3B772EFFF0530EB00293921869F0A369527138A7E169EF23E19537AE4664DEC1
58E59F2FFD05FFBE672380090E744FDC8743CB1CCCCFF0E2D372261E8E1F1658419A0F0FEE90F964
F18F278C25CCF8C0DE1359B5457CAAAAAABAE16A1B1B636F634625B00C8FE9E495B74A52C1840F4D
4D2FF82EFE6C98D24D6B5C78E6AB57AF5643B16715E00EF1F5DE9DE3B887FC67C4DB47C138CC98CF
787F00EE61605739DADE67CAD594CFEBEADA63EBD1C8C948C9B73E2844ACF4F5301860EB7EF482FD
D1D32957C407CFB6A9E99A7E5DFC12C1CB889FACE6EF0DA69D97FF8FFF3358C5AAAAAAAA6EDCDAD4
46ABF7213BFC4F61CD7DE1CFDBE61E557B5F1B19FCC7F0EFAFDAD5EBBCCE3BB29177F3371EFCD4C6
0FC047FC61DB1E995E3FBEED5F9D6F2576FBFA545555554DA48E6C23672E31CF5E7D7E1D37C5AA89
D73AE7939987AFF367D44665F44BB7C1FB7B0F4FCE7B6B762519E1DB8D555BFDC86CE9771DE53357
5555551DFF0AE6C0E47EE6F8CCA2FFEA77BE34A31FCF392168EF737CFEB6AAE3599A03736C418D9D
FF0155EBD7E8371166B3F3BF791E89FAB7DFFE21E133507594B52914A3AAAAAAEA46A92DC101A37F
5D937FBE66F14CBD532D9BD9AAE9D4D1AFEF6B7E52F43D6B8BB2C15A93F9BFABDF7BE2C4093EBF1D
C3D76A7B773EFAFF650F7E6F555555D511D76E49A19B7DDA9FB2B8B1D863B66A5FEBC896EF63BE7D
AABAB12A3342D0E46AE299E9DB7B7937FEF0FAD457555555EDAA36350357BA5055B2763E4E76FE07
EC716DDB0D6F27F5FAFC6BE7AFEDB1AADD2AB577F8B7555555555555555555555555555555555555
55555555555555555555555555555555555555555555555555555555555555555555555555555555
5555555555555555555555555555555555555555555555555555555555556DAAFE7F44EB0FC6
}
\caption{Performance of DNN-based speaker classification. (The total number of DNN training data  was roughly $1.5\times10^5$ for HAFM, and $2.5\times10^4$ for MFCC and C-NSGT.) }
\label{FIG:ACC1}
\end{figure}

Fig. \ref{FIG:ACC1} shows performance of the harmonic-aligned frame mask (HAFM) in the vowel-dependent speaker classification tasks.
For comparison, the mel-frequency cepstral coefficients (MFCC) \cite{Fang2001} and the NSGT coefficients (C-NSGT) were also evaluated in the same way.
We also tested the condition that $f_0$ was included as an extra feature dimension.
It can be seen from the results that C-NSGT mostly performed the worst.
On the other hand, HAFM which is established based on C-NSGT outperforms the others with noticeably higher accuracy.
This implies that with the comparison way of feature extraction, the HAFM feature is more effective to capture and represent the speaker variations.
The accuracy of HAFM is $83\%$ for the ``DNN/iy/+DNN/u/'' case (i.e., DNNs of both vowels were combined for decision).  
It can also be noticed that to include $f_0$ as extra feature seems beneficial for MFCC. However, such benefit is generally not observed for both C-NGST and HAFM,
as $f_0$ related information has already been well incorporated in these features.

In the evaluation, it was also observed that the frame mask based DNNs performed extremely well in distinguishing the reference speaker from the rest of the speakers.
As the frame mask features were obtained by exhaustive comparison to the reference speaker, the resulted DNN were inherently good verification models for the reference speaker.
One of our future directions is to combine the verification models of all enrolled speakers to construct a more comprehensive system.

\section{Conclusions}

The frame mask approach has been extended from instrumental sound analysis to voiced speech analysis.
We have addressed the related issue by developing non-stationary Gabor transform (NSGT) with pitch-dependent and time-varying frequency resolution.
The transform allows effective harmonic alignment in the transform domain.
On this basis, harmonic-aligned frame mask has been proposed for voiced speech signals.
We have applied the proposed frame mask as similarity measure to compare and distinguish speaker identities,
and have evaluated the proposal in a vowel-dependent and limited-data setting.
Results confirm that the proposed frame mask is feasible for speech applications.
It is effective in representing speaker characteristics in the content-dependent context and shows a potential for speaker identity related applications, specially for limited data scenarios.

~\\

\end{document}